\documentclass[structabstract]{aa}  

\usepackage{graphicx}
\usepackage[varg]{txfonts}
\usepackage{amsmath}
\usepackage{longtable}
\usepackage{longtable,lscape}
\usepackage[singlelinecheck=off]{caption}
\usepackage{lscape}
\usepackage{epsfig}

\newcommand\be{\begin{equation}}
\newcommand\en{\end{equation}}

\usepackage{txfonts}
\begin{document} 

\title{The low-mass stellar population in the young cluster Tr~37}
 
\subtitle{Disk evolution, accretion, and environment\thanks{Observations reported here were obtained at the MMT Observatory, a jointfacility of the Smithsonian Institution and the University of Arizona.}\thanks{Based on observations collected at the German-Spanish Astronomical Center, Calar Alto, jointly operated by the Max-Planck-Institut f\"{u}r Astronomie Heidelberg and the Instituto de Astrof\'{i}sica de Andaluc\'{i}a (CSIC).}}

\author{Aurora Sicilia-Aguilar\inst{1}, Jinyoung Serena Kim\inst{2}, Andrej Sobolev\inst{3}, Konstantin Getman\inst{4}, Thomas Henning\inst{5}, Min Fang\inst{1}}

\institute{	\inst{1}Departamento de F\'{\i}sica Te\'{o}rica, Facultad de Ciencias, Universidad Aut\'{o}noma de Madrid, 28049 Cantoblanco, Madrid, Spain \\
	\email{aurora.sicilia@uam.es}\\
	\inst{2}Steward Observatory, University of Arizona, 933 North Cherry Avenue, Tucson, AZ 85721-0065\\
	\inst{3}Astronomical Observatory, Ural Federal University, Lenin Avenue 51, 620000 Ekaterinburg, Russia\\
	\inst{4}Department of Astronomy \& Astrophysics, 525 Davey Laboratory, Pennsylvania State University, University Park PA 16802\\
	\inst{5}Max-Planck-Institut f\"{u}r Astronomie, K\"{o}nigstuhl 17, 69117 Heidelberg, Germany\\
}

   \date{Submitted 9 May 2013, accepted 30 July 2013}

\abstract
{}
{We present a study of accretion and protoplanetary disks around M-type stars in the 
4~Myr-old cluster Tr~37. With a well-studied solar-type population, Tr~37 is a benchmark
for disk evolution.}
{We used low-resolution spectroscopy to identify and classify 141
members (78 new ones) and 64 probable members, mostly M-type stars. H$\alpha$ emission provides
information about accretion. Optical, 2MASS, Spitzer, and WISE data are used to trace the 
SEDs and search for disks. We construct radiative
transfer models to explore the structures of
full-disks, pre-transition, transition, and dust-depleted disks.}
{Including the new members and the known solar-type stars, we confirm that
a substantial fraction ($\sim$2/5) of disks show signs of evolution, either as radial dust evolution
(transition/pre-transition disks) or as a more global evolution (with low small-dust masses,
dust settling, and weak/absent accretion signatures).
Accretion is strongly dependent on the SED type. About half of 
the transition objects are consistent with no accretion, and dust-depleted disks have
weak (or undetectable) accretion signatures, especially among M-type stars.}
{The analysis of accretion and disk structure suggests a parallel evolution of
dust and gas. We find several distinct classes of evolved disks, based on SED type and
accretion status, pointing to different disk dispersal mechanisms and probably different evolutionary paths. 
Dust depletion and opening of inner holes appear to be independent processes: most transition disks
are not dust-depleted, and most dust-depleted disks do not require inner holes.
The differences in disk structure between M-type and solar-type stars in Tr~37 (4~Myr old) are not
as remarkable as in the young, sparse, Coronet cluster (1-2~Myr old), suggesting 
that other factors, like the environment/interactions in each cluster, are likely to play 
an important role in the disk evolution and dispersal. Finally, we also find some evidence of clumpy star formation
or mini-clusters within Tr~37.}


\authorrunning{Sicilia-Aguilar et al.}

\titlerunning{Low-mass stars in Tr~37}

\maketitle


\section{Introduction \label{intro}}
The evolution of accretion disks around low-mass stars, and the way 
protoplanetary disks cease to accrete and disappear, presumably after 
forming planets, is since long matter of discussion. Protoplanetary 
disks around solar-type stars have typical lifetimes of few Myr (Haisch 
et al. 2001; Sicilia-Aguilar et al. 2006a; Hernandez et al. 2007;
Fedele et al. 2010), but the high variety of disk 
morphologies at a given age suggests that disk evolution is controlled 
by several mechanisms. Various physical processes
can be invoked for disk removal, including grain growth, photoevaporation, and planet formation.
In addition, other external parameters may also
contribute to define the path followed by a dispersing disk, such as
age, stellar mass, stellar and planetary
companions, initial conditions, cluster environment, crowdedness in the
star-forming region, and angular momentum of the collapsing 
core (Hartmann et al. 2006; Bouwman et al. 2006; Alexander \& Armitage 2009;
Fang et al. 2013a; Sicilia-Aguilar et al. 2013; Dullemond et al. 2006). Further processes 
(ejection in multiple systems, photo-erosion of cores by massive nearby 
stars; Bate et al. 2005, 2012; Whitworth \& Zinnecker 2004) may also play a role in 
the formation of low-mass systems. The unexpected and not yet understood
relation between 
stellar mass and accretion rate (dM/dt $\sim$ M$^{2-3}$; Natta et al. 2004; 
Hartmann et al. 2006; Fang et al. 2009; but also dM/dt $\sim$ M$^{1-1.2}$; 
Barentsen et al. 2011) is consistent with these multiple parameters affecting disk 
evolution (Clarke \& Pringle 2006; Gatti et al. 2008; 
Herczeg \& Hillenbrand 2008). 

Multiwavelength observations of disks around faint objects are a challenge for 
current instrumentation. There is a lack 
of combined optical photometry, spectroscopy, and complete IR data
for large samples of very low-mass objects. The first observations of disks 
around very low-mass stars and brown dwarfs (BD) suggested that they are
lower-mass analogs of the typical T Tauri disks.
Their disks are flared, show active accretion with strong H$\alpha$ 
emission, silicate emission from small grains in the disk atmosphere, 
processed and crystalline silicates, and dust continuum emission down to
the far-IR and millimeter wavelengths (Muench et al. 2001; Klein et al. 2003;
Mohanty et al. 2004; Jayawardhana et al. 2005; Apai et al. 2005; Scholz et al. 2007; Scholz 
\& Jayawardhana 2008; Harvey et al. 2010, 2012a,b;). Many of these observations were 
biased towards the most luminous disks and stronger accretors. Spitzer 
data suggested that there is an important fraction of harder-to-detect,
settled, low-mass, and transitional disks with inner gaps around the very low-mass
objects (masses $<$0.2 M$_\odot$ down to the BD regime; Morrow et al. 2008).
IR silicate spectroscopy of M-type stars also suggested differences in 
innermost disk evolution (Kessler-Silacci et al. 2007; 
Sicilia-Aguilar et al. 2007, 2008; Pascucci et al. 2009). Differences in the 
disk structure, dead zones, and accretion mechanisms for the
lower-mass objects could also change the evolution of the disk and the 
formation of planets around very low-mass stars (Hartmann et al. 2006). 

More recent observations suggest that the stellar mass is 
not the only parameter that controls the disk structure and subsequent 
evolution. Studies of stars in sparse clusters vs. more populous star-forming
regions point to differences in the disk fraction vs. age trend. Sparse clusters would have relatively
lower disk fractions at an early age, but these disks may survive for longer timescales
compared to more massive regions (Fang et al. 2013a). The Herschel Space Telescope has traced the 
typical sizes and structures where young stars are formed (Arzoumanian et al. 2011; Hacar et al. 2013),
revealing the details of star-forming filaments. Surprisingly, some
sparse associations, instead of being quiet low-mass star-forming regions,
appear to be very crowded, active, and interactive already at a very early stage (Sicilia-Aguilar et al. 2013).
The cluster dynamics, interactions, and angular momentum in the collapsing cloud 
could also affect the initial mass function (IMF) and the disk properties and subsequent 
evolution (Hsu et al. 2012, 2013; Becker et al. 2013; Dullemond et al. 2006).

With this work we want to address disk evolution and its
dependency on stellar mass/spectral type and environment by studying the
low-mass members in the Tr~37 cluster. Tr~37 is located at 870 pc distance (Contreras et al. 2002)
and part of the Cep OB2 complex (Platais et al. 1998; Patel et al. 1995, 1998).
The cluster is a key region for disk evolutionary
studies due to its intermediate age ($\sim$4 Myr; Sicilia-Aguilar et al. 2005) 
compared to the typical disk lifetimes (3-10 Myr; Sicilia-Aguilar et al. 2006a; Hern\'{a}ndez et al. 2007; Fedele et al. 2010).
Multiwavelength studies (Sicilia-Aguilar et al. 2004, 2005, 2006a,b; from now on
SA04, SA05, SA06a, and SA06b) have targeted the solar-type population in
the region, finding evidence of substantial disk evolution and dust
processing (SA06a; Sicilia-Aguilar et al. 2007, 2011b, from now on SA11).
Detailed H$\alpha$ photometry surveys (Barentsen et al. 2011) revealed an
extended population of accreting stars, some of which could be younger than
the main cluster (SA05, Getman et al. 2012). 

The disk structure for the solar-type population has been already extensively
discussed in SA06a and SA11. Using a combination of optical (photometry and
spectroscopy), Spitzer (IRAC, MIPS, and IRS), and millimeter-wave (IRAM) data, together
with simple radiative transfer models (RADMC; Dullemond \& Dominik 2004), we constrained
disk parameters and deviations from typical, uniform, flared disks.
Here we present a detailed multiwavelength study of the low-mass population
in Tr~37. By using optical spectroscopy, we were able to classify more than 200
objects among cluster members and probable members, including 78 newly identified 
members, most of them M-type stars. Combining this
information with our previous optical photometry and Spitzer IRAC/MIPS data, we analyze the disk
characteristics of the objects and put them in context comparing them to
our previous study of the solar-type stars. All observations are presented in 
Section \ref{obs}. In Section \ref{analysis} we examine the membership of the
candidates and derive their fundamental stellar properties. In Section \ref{discu}
we explore the implications of the newly discovered objects for disk
dispersal and evolution. Finally, Section \ref{conclu} summarizes our results.

\section{Observations and data reduction \label{obs}}

\subsection{Optical and IR data and candidate selection for spectroscopy}

This study aimed to complete the previous work of SA05/SA06a/SA11
on the disks around solar-type stars in the Tr~37 cluster, which was approximately
complete for stars with spectral types K4-M2, by addresing the M-type
population in the cloud.
Optical photometry and spectroscopy are required to confirm the cluster
membership, to obtain spectral types, to detect accretion, 
and to estimate extinction, age, and stellar mass.
The starting point of the target selection was the deep optical photometry
obtained at Calar Alto Observatory with the LAICA camera on the 3.5m telescope,
consisting of
deep observations with the standard UVRI Johnsons filters
(see Sicilia-Aguilar et al. 2010 for a detailed
description of the observations and data reduction). The data had a large
dynamical range, being complete in the range
U$\sim$15-21, V$\sim$13-21, R$_J \sim$12-21, and I$_J \sim$11-20
mag. The low-mass candidates relevant for our survey were in general too
faint to be detected with the U band filter. The requirement for the 
objects to be preferably candidate M-type members resulted in a selection of objects
with R$_J$=17-20.5 mag, considering that the extinction over Tr~37 is
moderate and relatively uniform (A$_V$=1.56$\pm$0.55 mag).
Examining the 2MASS  (Cutri et al. 2003) counterparts of our optical targets allowed 
to refine the selection of objects consistent with diskless late-type
stars and the locus of classical T Tauri stars (CTTS)  in the J-H vs H-K diagram (Bessell \& Brett 1988; 
Meyer et al. 1997), although for
M-type stars with evolved disks, the excesses in H and K bands are usually very 
small or negligible. The 2MASS coordinates were also used for the later
spectroscopy, given the strong requirements of multiobject spectrographs.

To complete the spectral energy distributions (SEDs) of the candidates, we
re-reduced the Spitzer IRAC and MIPS observations (from both our previous
datasets in SA06a, and Spitzer archival data), following the
method described in SA11. These new photometry is not
significantly different from that of SA06a, although the use of
a more recent pipeline with improved flat fielding, 
and smaller apertures result in more accurate
magnitudes for the in-cloud sources and a better detection of 
very faint objects in the MIPS 24$\mu$m maps. To ensure that contamination
by cloud emission or ghosts remains minimal, we visually inspected all the candidates.
In particular, detailed inspection of the 24$\mu$m data was necessary to remove 
objects suffering from nebular emission. 

For the spectroscopic followup, we selected the sources with ages under 
100 Myr  that were consistent with solar- and M-type
stars at 870 pc distance according to the V vs V-I and V-R diagrams
and the Siess et al. (2000) isochrones. This produced a list of about 400 targets, of
which approximately 100 had IR excesses consistent with circumstellar disks
and that lacked spectroscopic characterization.  
We assigned priority to
allocating the fibers and slits to objects with IR excess, followed by 
those with isochrone ages $<$10 Myr, and finally those between 10-100 Myr.
All the observed targets are listed in the Appendix \ref{tables-appendix}, Table \ref{spec-table},
together with the relevant information about H$\alpha$, Li I emission, presence of disks,
spectral type, extinction, and membership (see Section \ref{analysis} for details). 
A few remaining fibers were asigned to previously known objects, 
or to objects with strong IR excesses but lacking optical data.
This study is thus biased towards objects with disks, so
we cannot estimate the disk fraction. The main advantage
is that we covered more than 90\% of the 
candidate M-type stars with IR excesses, obtaining a superb dataset to explore
the various disk structures.
The photometry data (optical, 2MASS, and Spitzer) are listed
only for the members and probable members in the Appendix \ref{tables-appendix}, 
Tables \ref{opt2mass-table} and \ref{spitzer-table}.

For completeness, after the survey we also checked the WISE database for counterparts to
our objects. The Spitzer data are preferred because the larger WISE beam
often suffers from strong contamination due to cloud emission, resulting in
overestimated fluxes, especially in bands 3 and 4. Nevertheless,
band 3 (12$\mu$m), together with IRAC 8$\mu$m, offers valuable information
regarding the silicate emission.
Whenever there was a good agreement between IRAC and WISE bands 1 and 2,
MIPS 24$\mu$m and WISE band
4 (22$\mu$m), we 
assume there is no significant contamination and thus use the 12$\mu$m WISE data. 
The WISE fluxes were obtained 
applying the Wright et al. (2010) color corrections depending on the Spitzer
SED shape. The WISE data used in this project are also listed in  the Appendix \ref{tables-appendix},
Table \ref{wise-table}.

The spectroscopic properties of the observed objects and membership criteria are described in Section \ref{analysis}.
Thanks to the strong candidate selection criteria, the efficiency of the
survey was very high, with about 90\% of 
candidates with isochronal ages younger than 10 Myr and IR excesses
being confirmed as cluster members. The success rate fell down 
to $\sim$5-20\% among the objects without excess.
A number of objects with very low S/N could not be ruled out nor confirmed
as members (see Table \ref{spec-table}).
A few examples of SEDs of members with IR excesses
are shown in Figure \ref{exampleseds-fig}. The SEDs of all members and probable members with IR
excesses are displayed in the Appendix \ref{seds-appendix}, Figures \ref{cttsseds1-fig}-\ref{classIseds-fig}, and 
the SEDs classification criteria are discussed in Section \ref{accdisk}. 

\begin{figure*}
\centering
\begin{tabular}{ccccc}
\epsfig{file=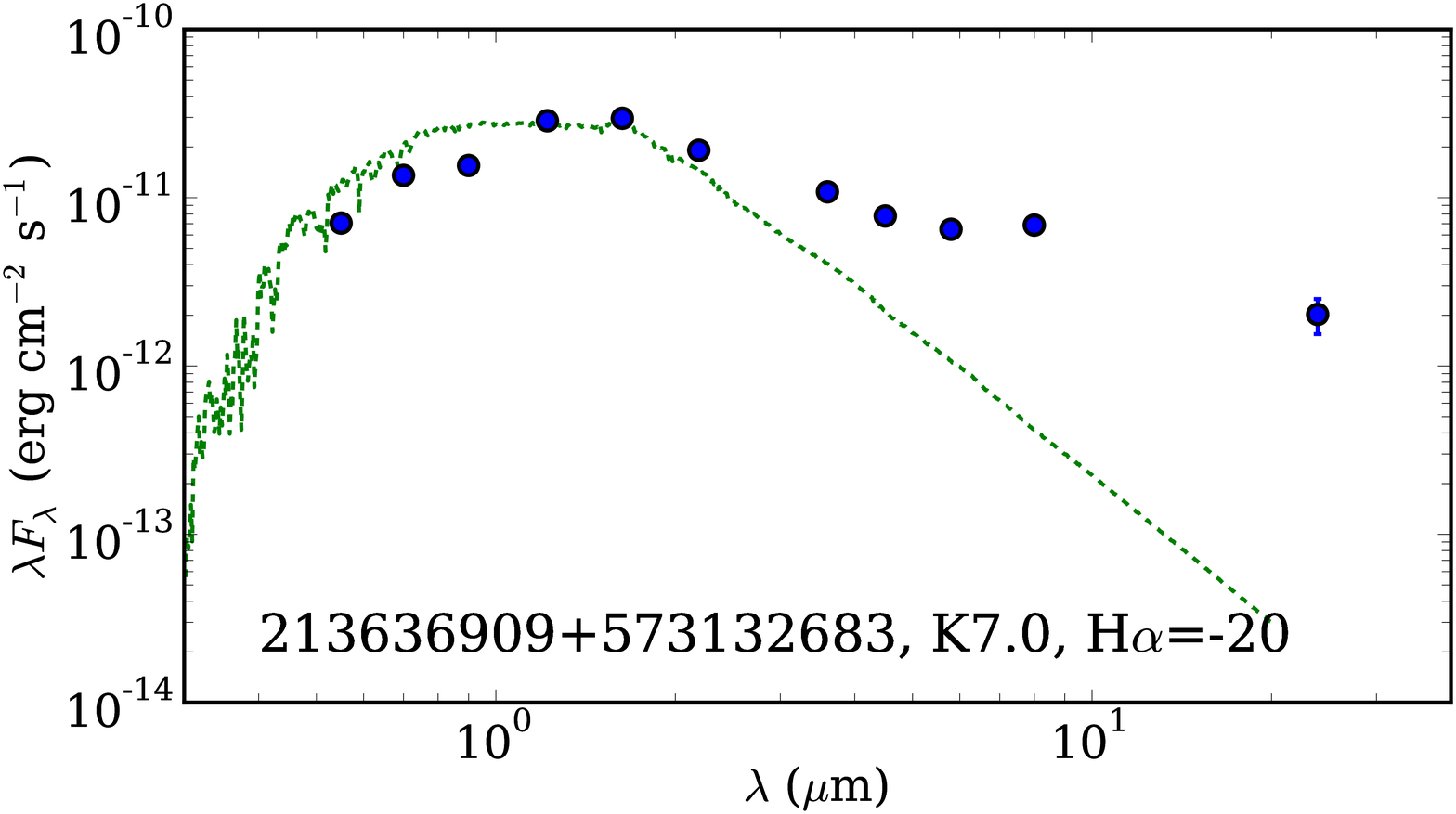,width=0.24\linewidth,clip=} &
\epsfig{file=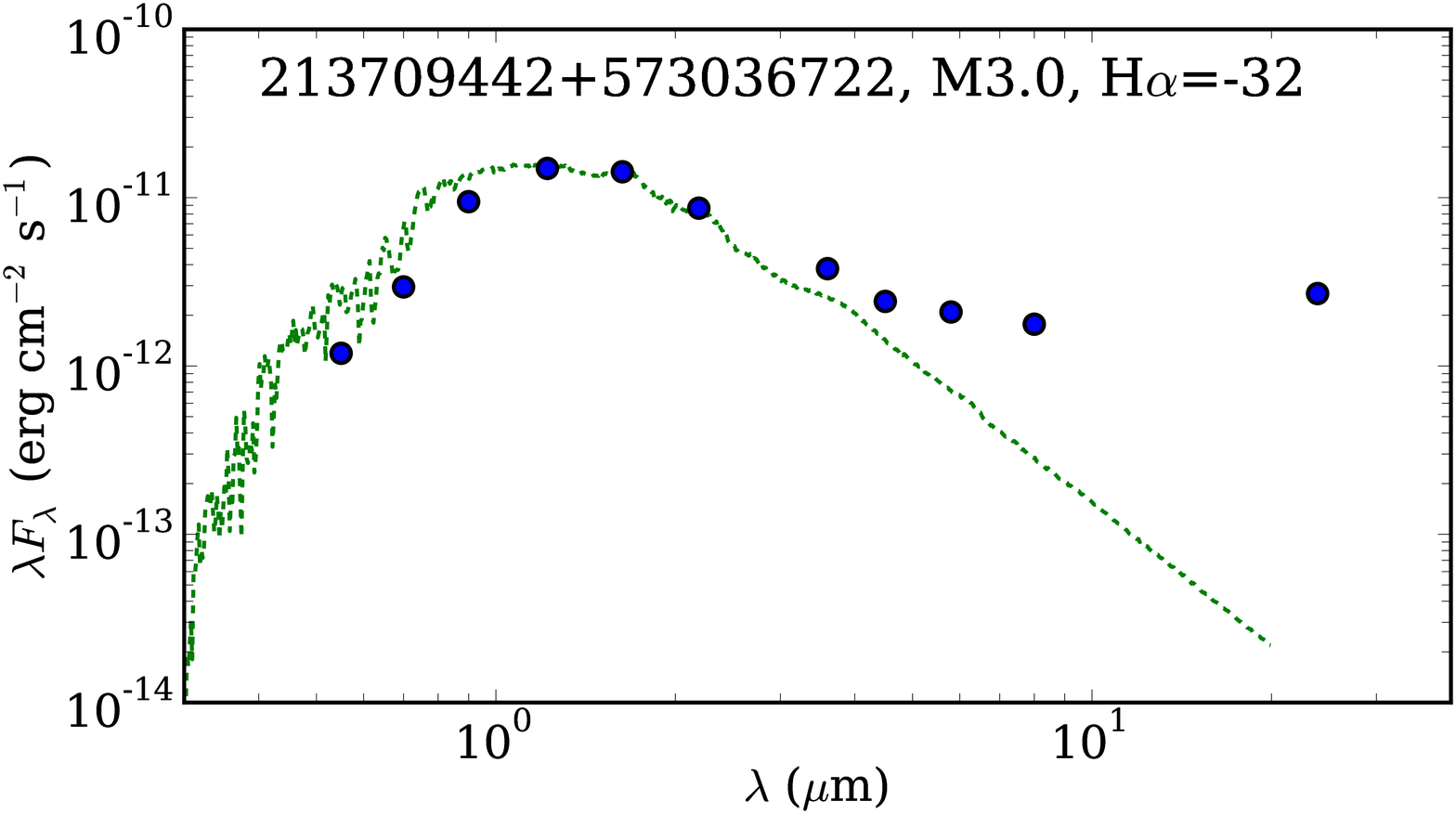,width=0.24\linewidth,clip=} &
\epsfig{file=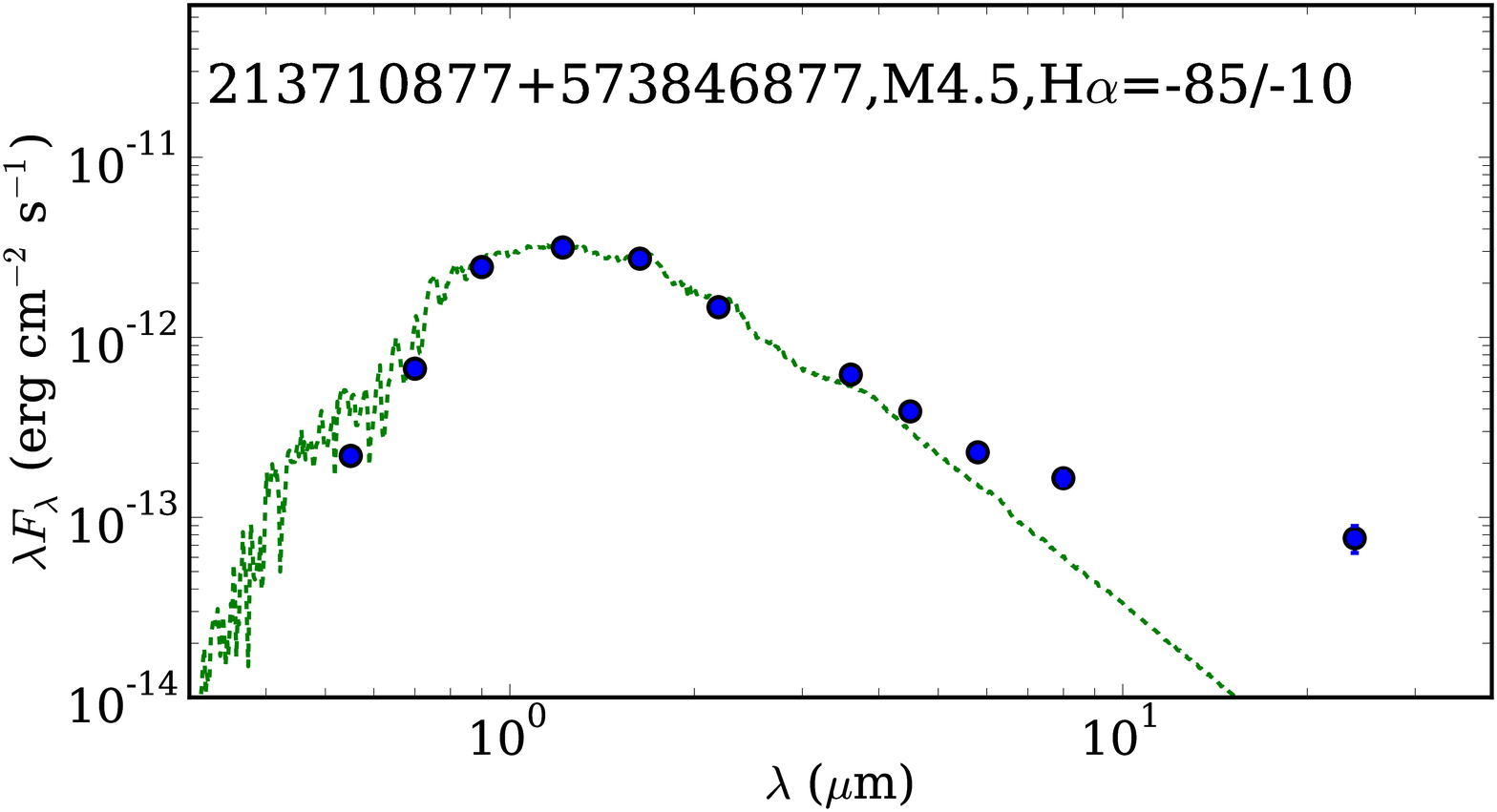,width=0.24\linewidth,clip=} &
\epsfig{file=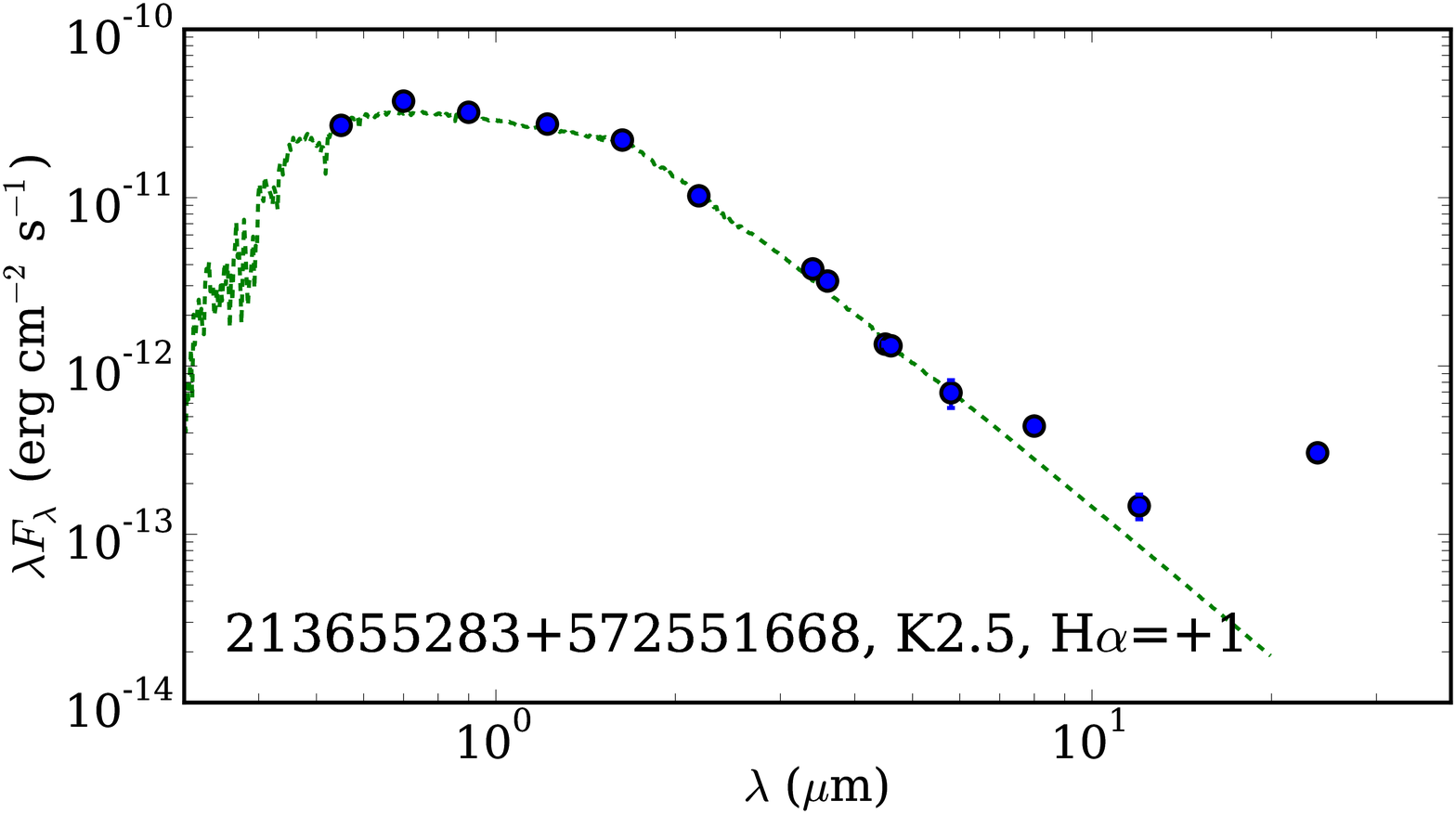,width=0.24\linewidth,clip=} \\
\epsfig{file=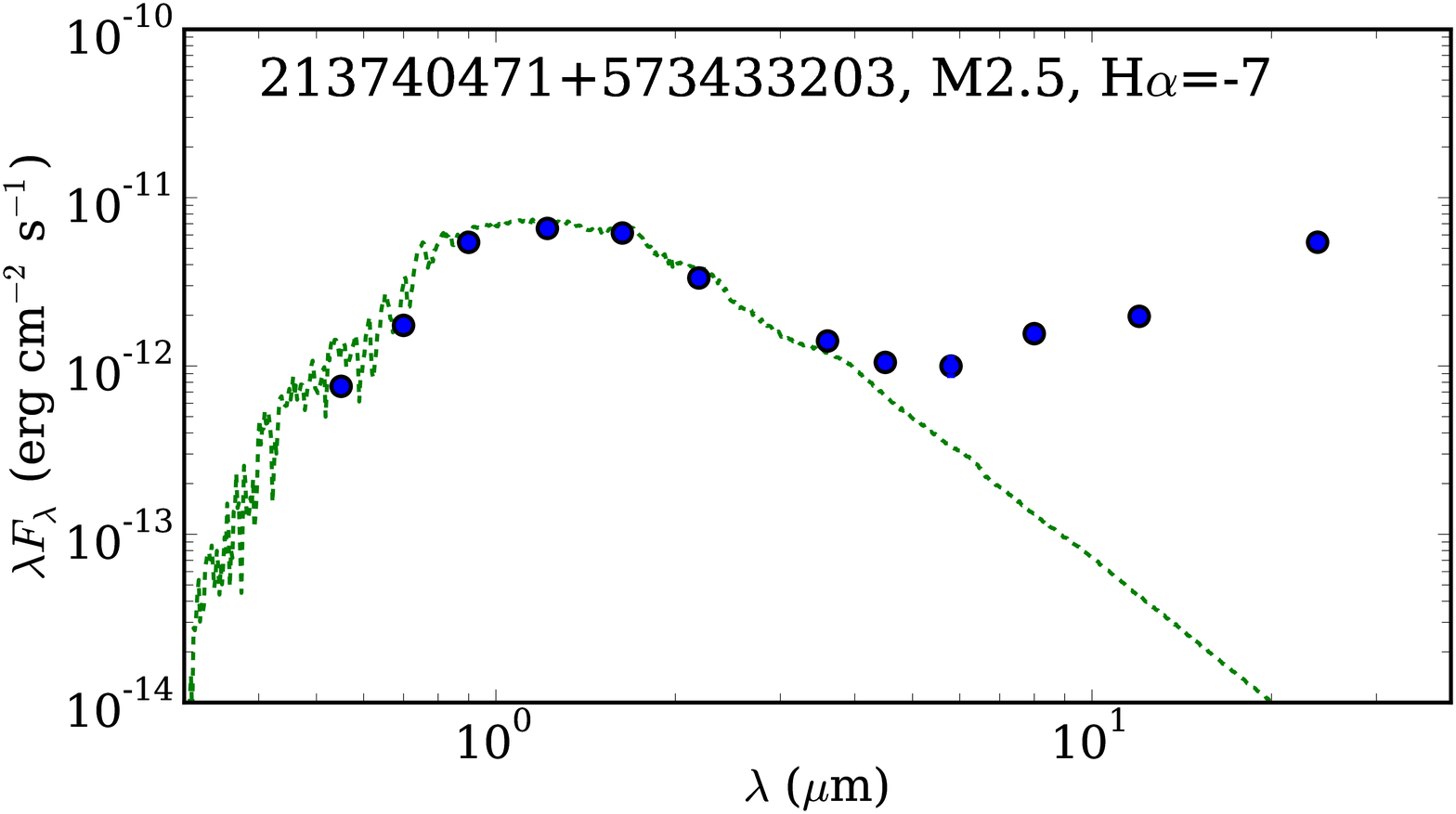,width=0.24\linewidth,clip=} &
\epsfig{file=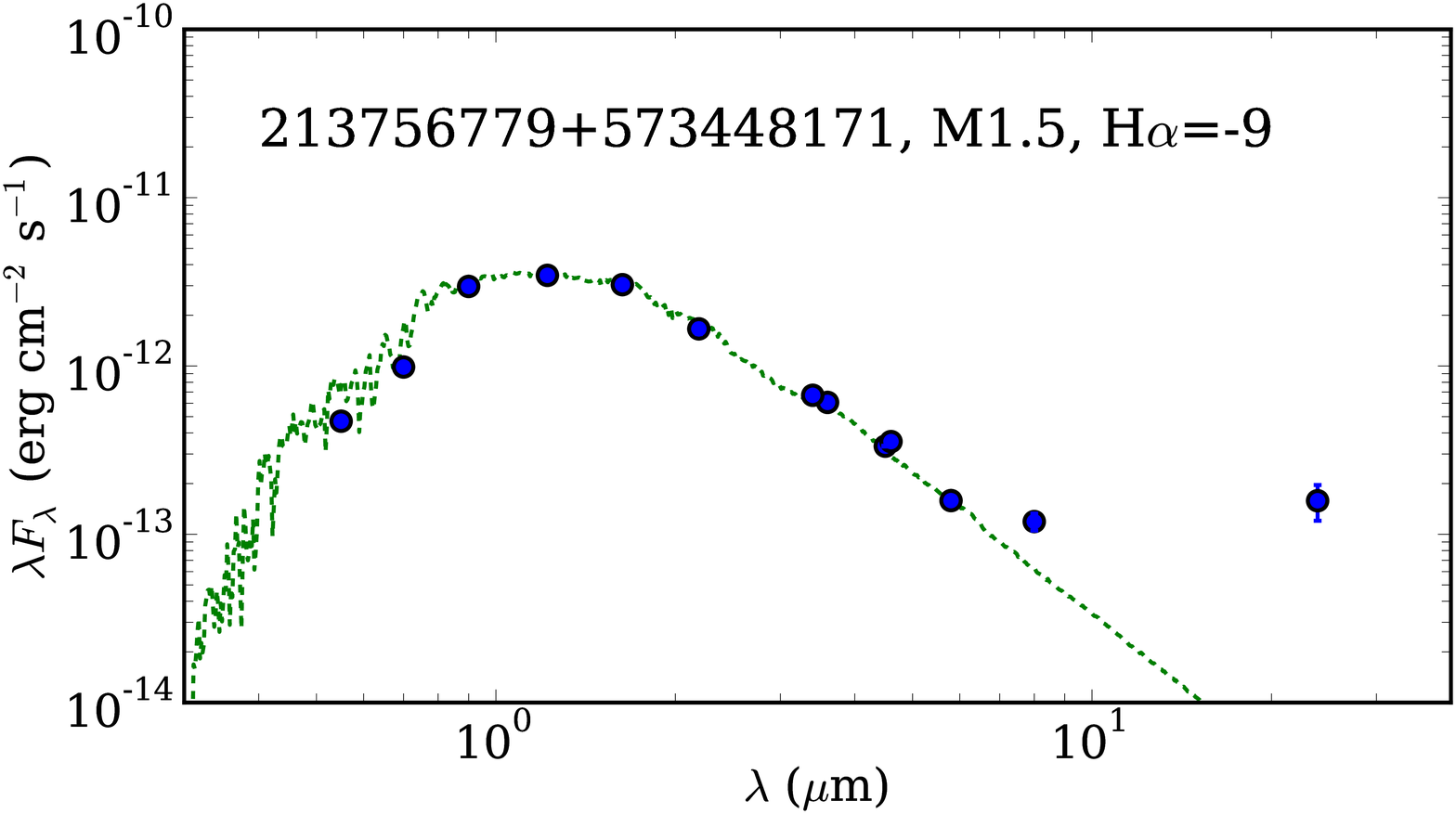,width=0.24\linewidth,clip=} &
\epsfig{file=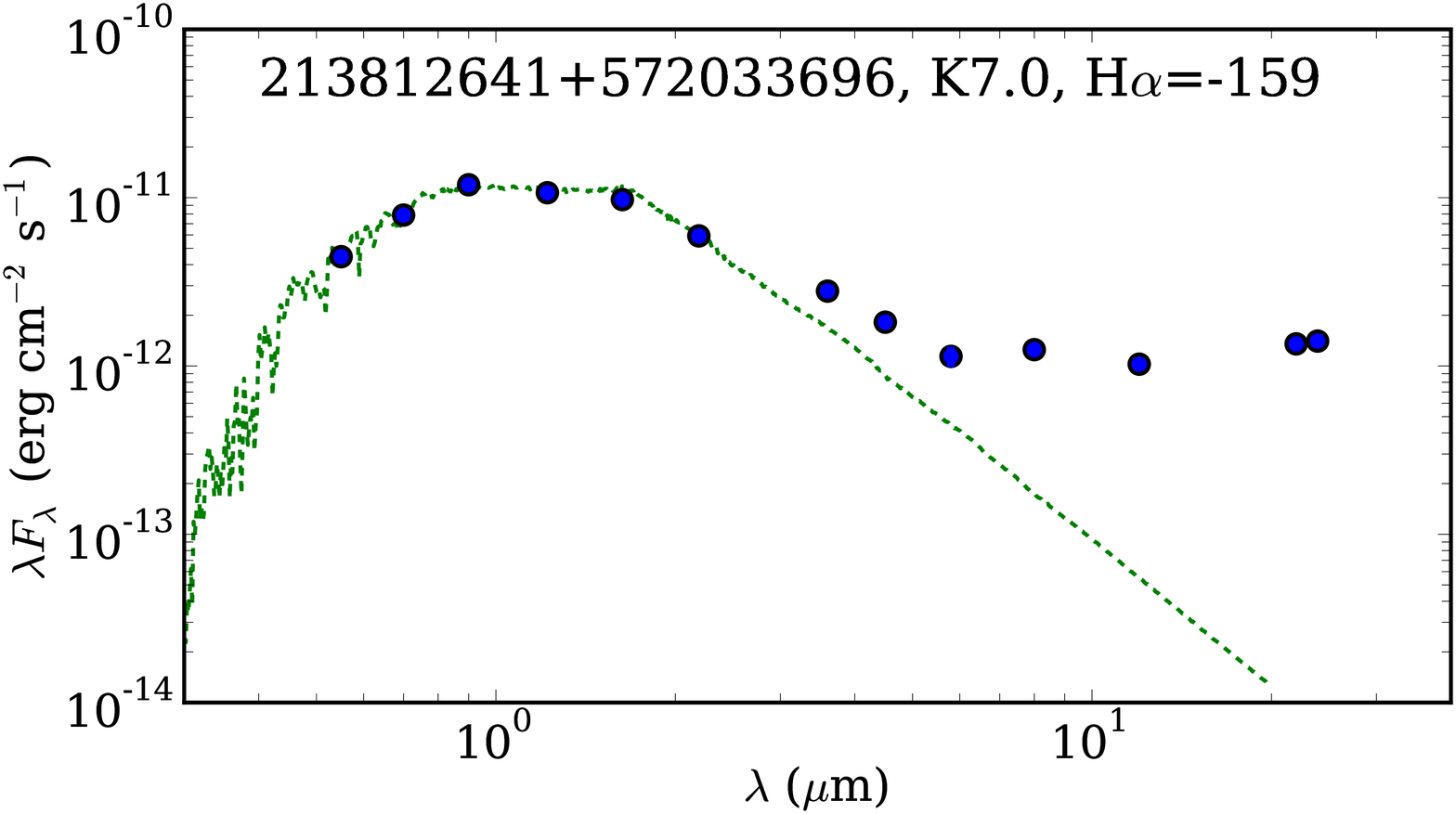,width=0.24\linewidth,clip=} &
\epsfig{file=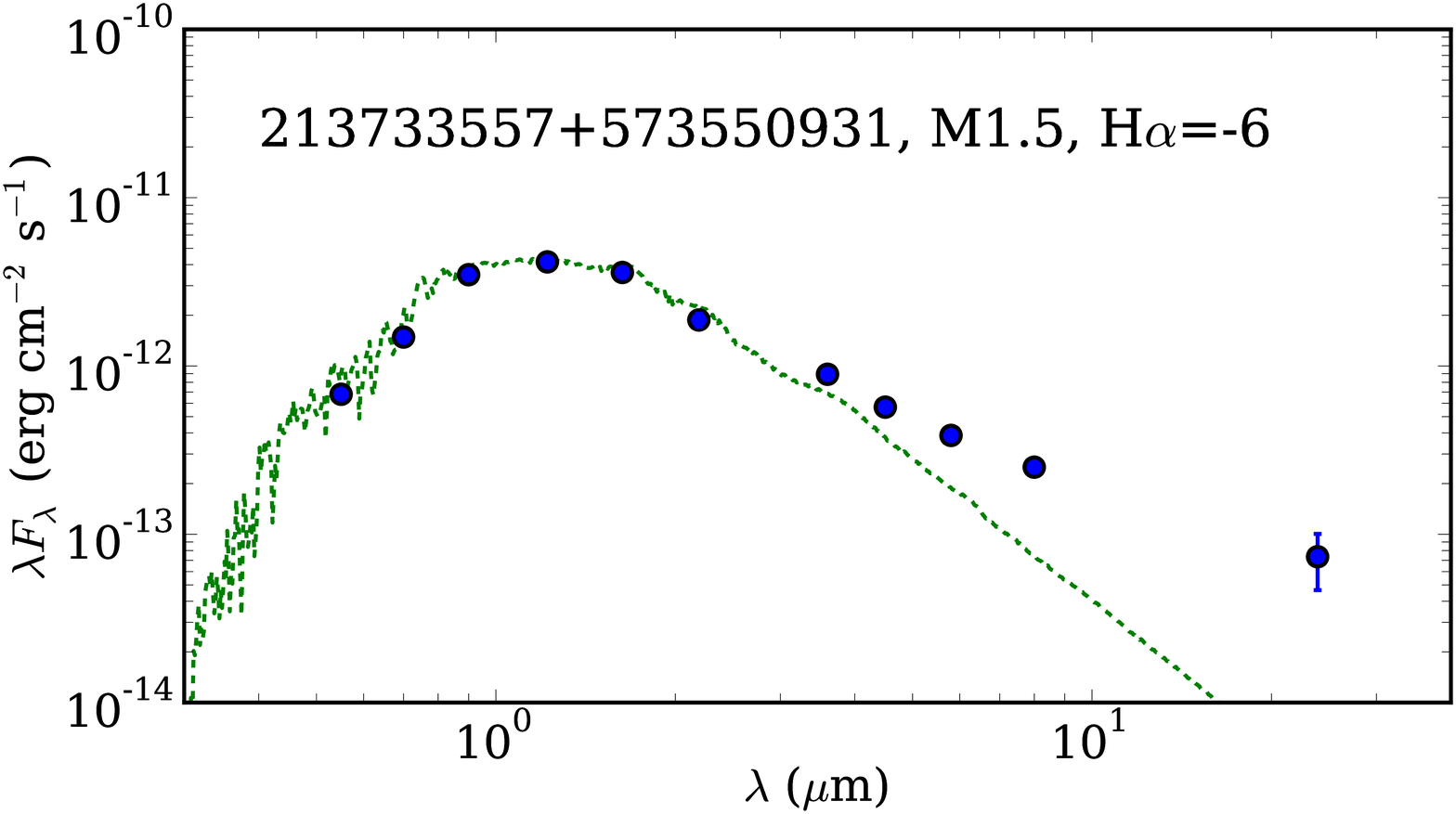,width=0.24\linewidth,clip=} \\
\end{tabular}
\caption{Examples of SEDs of some of the members with IR excess. 
For comparison, the photosphere of a star with the same spectral type
from the MARCS models (Gustafsson et al. 2008) is displayed. All
datapoints have been extinction-corrected according to their individual
values of A$_V$ and assuming a standard extinction law (see Table \ref{spec-table}.
Information about the H$\alpha$ EW (in \AA) is also
displayed. The rest of SEDs of objects with IR excesses
are displayed in the online material.\label{exampleseds-fig}}
\end{figure*}

\subsection{SCORPIO multislit/6m Russian SAO RAS Telescope}

One observational campaign took place with the SCORPIO  
(Spectral Camera with Optical Reducer for Photometric and Interferometric Observations) 
multislit spectrograph (Afanasiev \& Moiseev 2005) at the
6m telescope of the Special Astrophysical Observatory of the Russian Academy of
Sciences, in the Northern Caucasus. The observations were obtained during 2008 October 4, 5,
and 6. SCORPIO 
is a multislit spectrograph with 16 movable long slits  (each of 18'') that can be placed over a 2.9$\times$5.9
arcmin$^2$ field. We observed 3 configurations, with slits allocated to cover between 12 and 16 candidate 
objects. We used the grism VPHG1200r, with a spectral coverage 5700-7500 \AA\ and a 5 \AA/pix
resolution. Due to the poor weather conditions, we had to reduce the exposure time to 3$\times$15-10 min per field.

The observations were reduced following standard IRAF\footnote{IRAF is
distributed by the National Optical Astronomy Observatories,
which are operated by the Association of Universities for Research
in Astronomy, Inc., under cooperative agreement with the National
Science Foundation.} procedures within the $specred$ package, in particular, routines within
the $twodspec.apextract$ package. The spectra were bias-corrected, flat fielded, and extracted. The wavelength
calibration was performed with a Ne lamp, observed with the same slit configuration as the datasets. 
The sky subtraction was done by using the sky spectra adjacent to the star in the slit. 
A total of 18 new member candidates were identified
among the observed objects. The main limitation of these observations was the poor S/N of the data,
due to the poor weather conditions, that did not allow us to clearly detect the Li I absorption in most
of the candidates. We therefore re-observed many of these objects in our subsequent spectroscopic
program with the MMT. The advantage of using SCORPIO is that the
long slit data allow unambiguous subtraction of the nebular emission 
around the sources.

\subsection{Hectospec/MMT Spectroscopy}

The candidate objects were observed during two campaigns with the multifiber spectrograph
Hectospec on the 6.5m MMT telescope in Mount Hopkins, AZ. The first set of observations 
were taken on 2009 July 14 and 19, and the second one on 2010 May 17 and 19. The weather conditions
were fair during both campaigns. Hectospec is a multifiber spectrograph with 300 fibers
that can be allocated to individual objects or sky over a 1 degree field of view.
We took at total of 3 configurations, 
each one including between 180 and 220 candidate objects, plus numerous sky positions.
We used the 270 gpm grating, with a spectral coverage from $\sim$3650-9200\AA\
and a typical resolution of 5\AA. This is the same setup we used for the observations of
the solar-type stars (SA05). For each field, we took 3$\times$30 min exposures, followed by
sky observations obtained by shifting the whole configuration by 5" and exposing 3$\times$10 min.
These sky observations are particularly useful to attain a good sky subtraction in the cases
of faint targets or objects in regions with complicated background, since they provided the
sky spectrum in a region very close to the objects, and obtained with the same fiber.

\begin{figure*}
\centering
\epsfig{file=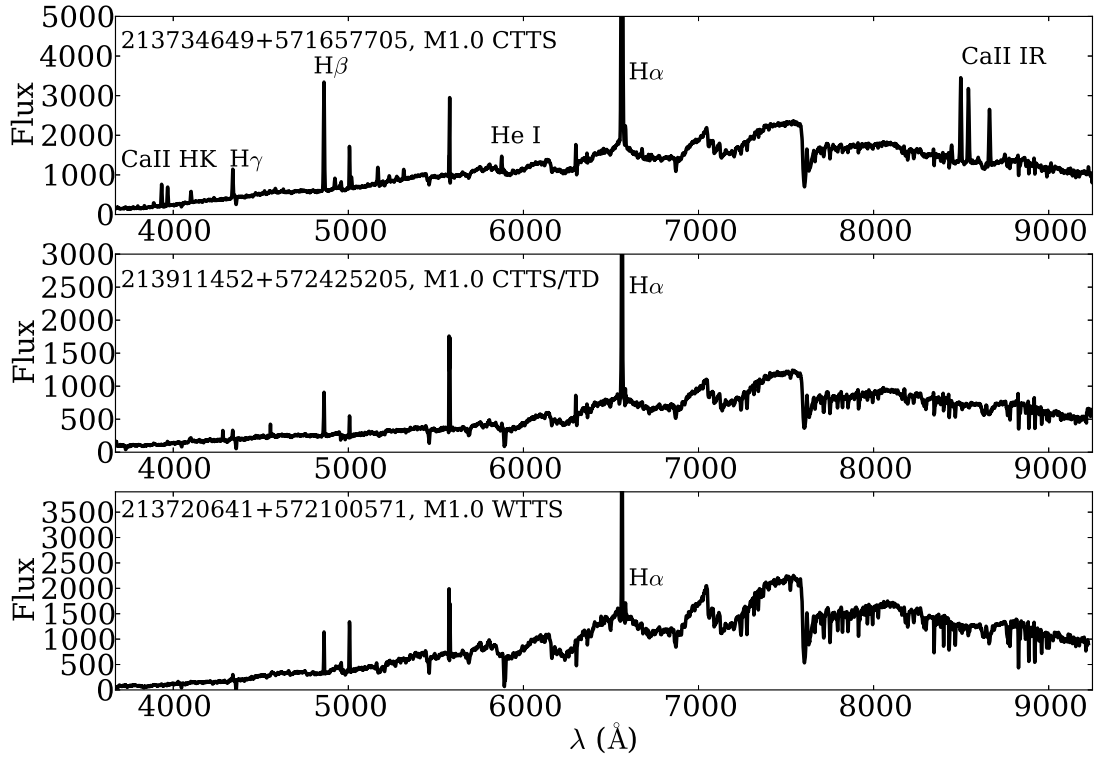,width=0.9\linewidth,clip=}
\caption{Some examples of spectra taken with the MMT/Hectospec. The H$\alpha$ features are cut to 
reveal more details of the whole spectra. Note that the spectra still show some telluric lines in emission (mostly 
at 5577 and 5004\AA).  The flux is in arbitrary
instrumental units.\label{example-fig}}
\end{figure*}

The bias, flat fielding, and wavelength calibration were done with standard IRAF
tasks within the $dofibers$ package. For the wavelength calibration,
we used a HeNeAr calibration lamp. Due to the offsets between the even- and odd-numbered fibers,
the wavelength calibration was done separately for the even and odd fibers.
The one-dimensional spectra were extracted and combined before the sky subtraction.
The sky subtraction can be a problem in regions like Tr~37, where substantial, spatially-variable
emission from the H II region can affect the H$\alpha$ line, one of our main
membership diagnostics. 
For each night, we classified all the sky spectra in "bright", "medium", and "faint", according
to the intensities of the nebular emission lines (especially, H$\alpha$, [N II], and [O I]),
combining the three sets to create bright, medium, and faint sky templates.
Each template resulted from a minimum of 20 spectra. The three templates were subtracted from
all the objects, and we then examined the success of each sky subtraction by checking
the night sky and nebular emission lines. We then selected 
the best-subtracted spectrum. For about a fourth of the
objects, none of the three sky subtractions provided a good result (either by excess or by defect).
In those cases, we used the wavelength-calibrated individual sky spectra taken in the
proximity of the star, which improved the subtraction of the nebular and night sky lines,
although the difference in exposure time resulted in a higher noise than in the cases where
the sky template spectra were applied. Some examples of spectra are shown in Figure \ref{example-fig}.

\subsection{CAFOS/2.2m Calar Alto narrow band imaging \label{cafos}}

Since our previous spectroscopic surveys had revealed substantial forbidden
line emission in the Tr~37 globule, we also performed a narrow
band imaging in the [S II] lines using the Fabry-P\'{e}rot interferometer
with the CAFOS wide-field imager mounted on the 2.2m telescope in Calar Alto.
The data were obtained as part of a Director Discretionary Time 
program on 28 August 2009.

\begin{figure*}
\centering
\epsfig{file=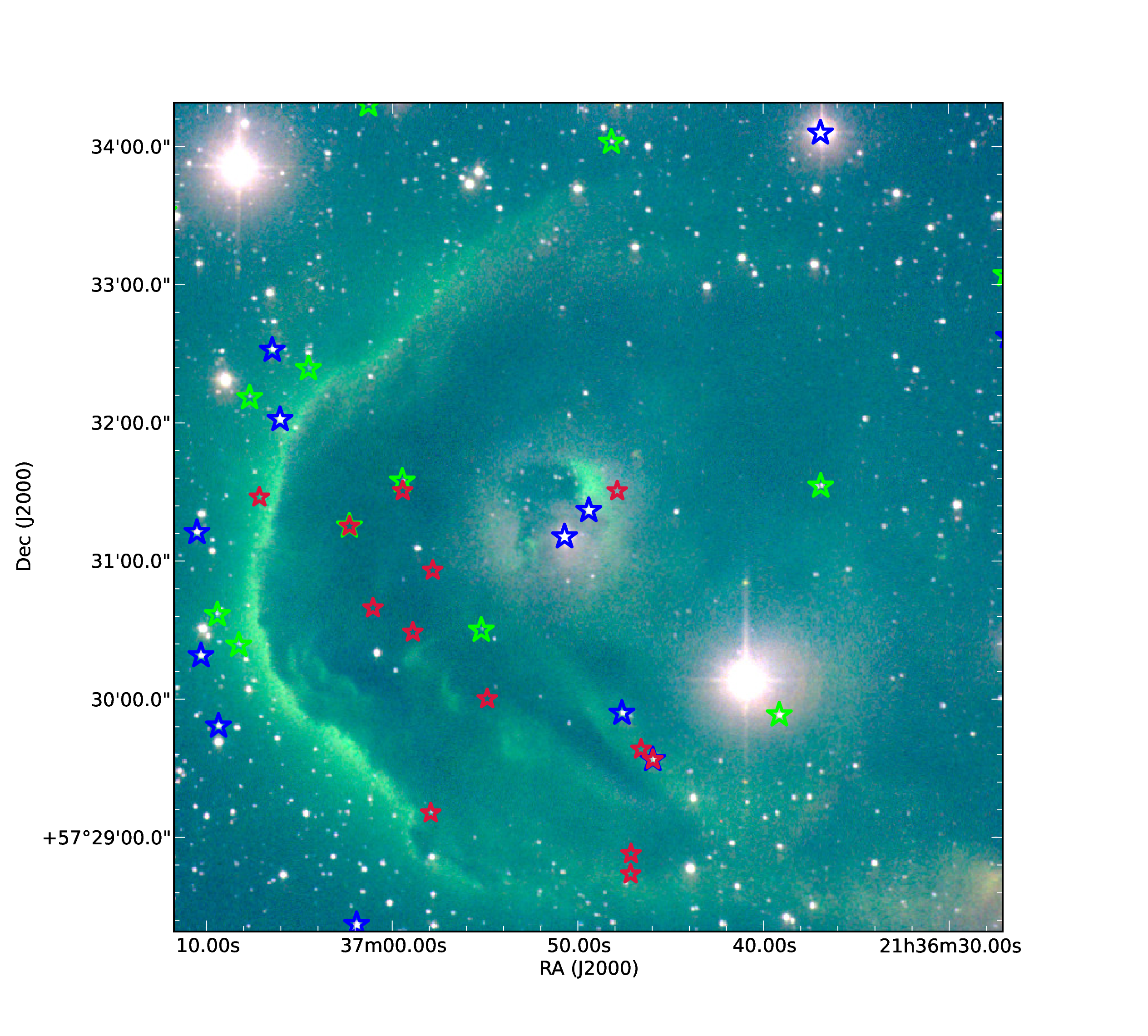,width=0.49\linewidth,clip=}
\epsfig{file=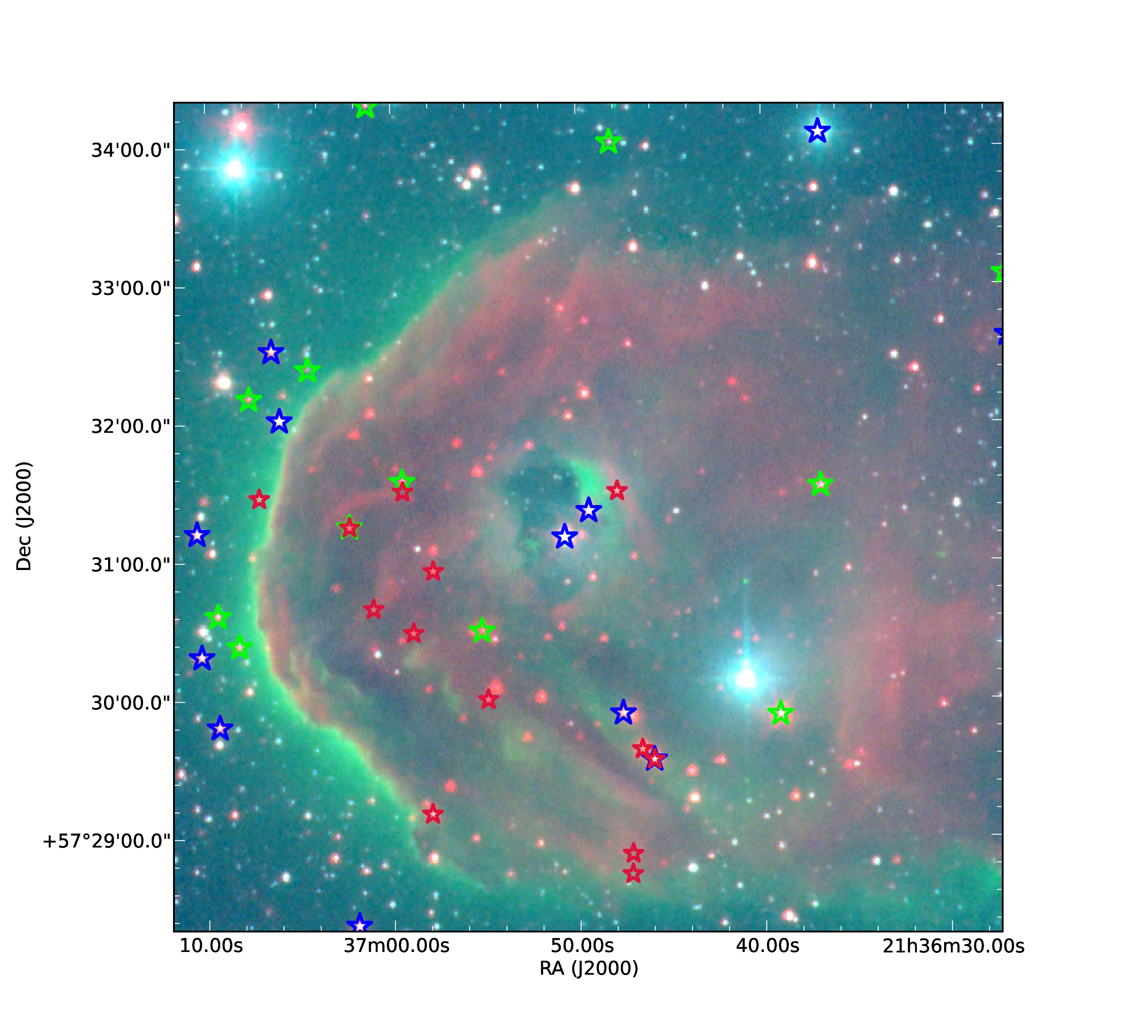,width=0.49\linewidth,clip=}
\caption{Two 3-color images of the IC 1396 A globule. Left: narrow band at 6750\AA, [S~II] combined at 6716 and 6730\AA, and R band as red, green,
and blue, respectively. Right: IRAC1, [S~II] combined at 6716 and 6730\AA, and R band as red, green,
and blue, respectively. Known members detected via optical spectroscopy are marked as
blue stars, new members are marked as green stars. Spitzer embedded candidates (SA06a) are marked by red stars.
\label{forbidden-fig}}
\end{figure*}

We obtained images centered in the Tr~37 globule, around the CTTS
14-141, RA(2000) DEC(2000) 21:36:49.41 +57:31:22.0, which showed remarkable variable forbidden 
line emission in our previous spectroscopic surveys.  
We obtained 3 dithered 600s exposures with the Fabry-P\'{e}rot interferometer at 
6716, 6730, and 6750 \AA, resulting in narrow-band images centered on the two [S II] lines
and a continuum observation with the same width in the line-free region around 6750\AA.
An additional 3$\times$10s Johnsons R band image was also obtained for comparison.
The data were reduced (bias, flat field) and combined according to the standard 
procedures using IRAF packages. Finally, a line-only image was obtained by subtracting the
continuum 6750 \AA\ image from the [S II] ones. To improve the S/N, we combined both [S II] lines for the final result.
We did not attempt any flux calibration, which is difficult due to the properties of the filter
and to the fact that there is a wavelength drift over the CAFOS field. The resulting data clearly reveals
several shocks in the field, not only associated to 14-141, but also
to several of the protostars and embedded objects in the cloud (see Figure \ref{forbidden-fig}).

\section{Analysis\label{analysis}}

\subsection{Spectral types and extinction \label{stype}}

As a preliminary step to determine the membership and properties, we 
obtained spectral types for all the observed objects, which are listed in
the Appendix \ref{tables-appendix}, Table \ref{spec-table}.
The spectral types of the objects with good S/N were derived using the combinations
of indices employed for the classification of K-M3 stars in Cep OB2 by Sicilia-Aguilar
et al. (2004), and a similar scheme for M3-M8 stars as it was used for classification of
the Coronet cluster members (Sicilia-Aguilar et al. 2008). The classification of
M3-M8 stars was 
based on the indices defined by Kirkpatrick et al. (1995), Mart\'{\i}n et al. (1996), and
Riddick et al. (2007). 
From the indices therein, we selected the best ones that
did not suffer from strong problems due to atmospheric features, extinction, or lack of S/N in the blue
(see Table \ref{index2-table}).
The calibration in Sicilia-Aguilar et al. (2004) is
done in effective temperature (see Table \ref{index1-table}).
The effective temperatures (T$_{eff}$, in units of 10$^3$ K) are derived as T$_{eff}$=T$_0$ + b$\times$(I-I$_0$), 
where I is the measured index, and T$_0$, I$_0$, and b are the zero-point values of effective
temperature and index, and the slope of the relation resulting from the fit of several standard stars
(Sicilia-Aguilar et al. 2004). The index I is obtained as a function of the flux in different wavelength ranges 
(F$_{\lambda1-\lambda2}$ is the flux between the wavelengths $\lambda_1$ and $\lambda_2$ in \AA).
The values of T$_{eff}$ are transformed to spectral types using the calibration by Kenyon \& Hartmann (1995).
The spectral types  of objects observed with SCORPIO were derived by direct comparison to standard
stars, given their reduced wavelength coverage. We also note that since the SCORPIO 
spectra do not include the main features for classification
of G/K stars, stars other than M-type observed by SCORPIO have very uncertain spectral types.

\begin{sidewaystable}
\caption{Indices used to derive the spectral type for stars later than M2.5.\label{index2-table}} 
\begin{tabular}{l c c c l l}
\hline\hline
Name & $\lambda$ Numerator & $\lambda$ Denominator& Range   & Calibration & Reference \\
\hline
PC1 & 7030-7050 & 7525-7550 &  M3-M9 & -0.06+2.95 X	 & 1,2,4 \\
PC2 & 7540-7580 & 7030-7050 &  M4-M8 & -0.63+3.89 X      & 1,2,4 \\
R1  & 8025-8130 & 8015-8025 &  M2.5-M8 & 2.8078+21.085(X-1.044)-53.025(X-1.044)$^2$+60.755(X-1.044) & 3 \\
R2  & 8415-8460 & 8460-8470 &  M3-M8 & 2.9091+10.503(X-1.035)-14.105(X-1.035)$^2$+8.5121(X-1.035) & 3 \\
R3  & (8125-8130)+(8415-8460) & (8015-8025)+(8460-8470) & M2.5-M8 & 2.8379+19.708(X-1.035)-47.679(X-1.035)$^2$+52.531(X-1.035) & 3 \\
TiO 8465 & 8405-8425 & 8455-8475 & M3-M8 & 3.2147+8.7311(X-1.085)-10.142(X-1.085)$^2$+5.6765(X-1.085) & 3 \\
VOa & (7350-7370)+(7550-7570) & 7430-7470 & M5-M9 & 5.0705 + 11.226(x-0.982)+6.7099(x-0.982)$^2$ & 3 \\
VOb & (7860-7880)+(8080-8100) & 7960-8000 & M4-M9 & 3.4875 + 29.469(x-1.017)-156.53(x-1.017)$^2$+394.28(x-1.017)-325.44(x-1.017)$^4$ & 3 \\
\hline
\end{tabular}
\tablefoot{Indices used for spectral typing of mid- and late-M type stars. The wavelengths are given in \AA. 
In the calibration, X represents the index obtained by dividing the numerator and
denominator, and the resulting number indicates the M subtype. 
References: 1= Kirkpatrick et al. (1995); 2= Mart\'{\i}n et al. (1996); 3= Riddick et al. (2007).
4= Sicilia-Aguilar et al. (2008)}
\end{sidewaystable}

\begin{table*}
\caption{Indices used to derive the spectral type for G,K,M-type stars.\label{index1-table}} 
\begin{tabular}{l c c c c l }
\hline\hline
Name &T$_0$ (10 K) & I$_0$  & b (10 K) & Range (T$_{eff}$/10$^3$ K, Spec. Type)  & Index (I) \\
\hline
TiO~6185 & 3.40	& 1.58$\pm$0.03 &	-0.94$\pm$0.11 & 3-3.9, M5-M0 & F$_{6165-6210}$/F$_{6100-6150}$  \\
TiO~7140 & 3.40	& 1.99$\pm$0.02	&       -0.40$\pm$0.03 & 3-3.7, M5-M2 & F$_{7125-7155}$/F$_{7020-7050}$  \\
MgI & 5.35	& 0.84$\pm$0.02 &	5.50$\pm$0.28  & 4.8-6, K2-G0 & (2$\times$F$_{5160-5180}$)/(F$_{5005-5025}$+F$_{5225-5245}$)  \\
CaI & 	3.22	& 1.43$\pm$0.01	 & 	-1.18$\pm$0.05 & 3-3.6, M5-M3 &  (F$_{6000-6200}$+F$_{6300-6320}$)/(2$\times$F$_{6155-6175}$)  \\
CaI &	4.10	& 1.16$\pm$0.03	 &	-8.74$\pm$0.80 & 3.6-5.1, M3-K1 & (F$_{6000-6200}$+F$_{6300-6320}$)/(2$\times$F$_{6155-6175}$)  \\
\hline
\end{tabular}
\tablefoot{Indices used for spectral typing of G, K, and early-M star (Sicilia-Aguilar et al. 2004).}
\end{table*}

The indices are not applicable out of a given spectral range. 
Therefore, we did a preliminary visual classification of the stars as "earlier than G", "GK-type",
and "M-type" before computing the spectral types. Due to our selection criteria, 
most of the stars found to be earlier than G are
unlikely members, and they are simply listed
in Table \ref{spec-table} as "early type" (E). For the objects classified as "GK-type", we 
used the MgI and CaI indices.
The number of indices applicable to G-type and early-K stars
is also very low, so the spectral types of objects earlier than K5 were refined by comparison
to standard stars. 
For the objects classified as "M-type", we used first the TiO~6185 index to determine
whether the object was earlier or later than M3. For objects earlier than M3,
we then used the CaI and TiO71 indices to refine the spectral type. For M3 or later
types, we used the PC1, R1, and R3 indices to derive a preliminary spectral type,
which was refined by using the TiO~8465, VOa, VOb, and PC2 indices, if within their applicable
range. 
For each star, the final spectral type is calculated as the average of the types obtained
with the applicable indices. In all cases, we performed a visual comparison
with standard stars to check that the estimated spectral type is in agreement with
the appearance of the spectrum. The few cases where there was a discrepancy between the
derived type and the visual inspection could be tracked to different problems (presence of 
stellar or sky emission lines, lack of S/N in part of the spectrum, contamination by scattered
light) that were addressed by removing the discrepant indices, 
and using the non-contaminated ones to derive the final spectral
type. 

In case of the objects with more than one spectra, we derived the spectral type for each
one and consider the result of the best S/N spectrum (if the quality of
both was very different). There is
a very good agreement between
all estimates (within 1-2 subtypes, depending on spectral type and data quality).
Spectral types had been previously derived for 14 among the 17 objects in common with 
our previous campaigns. Among these, we recover the previous spectral type ($\pm$1 subtype) in 9
cases, 2 cases are off by up to 2 subtypes, and 3 are off by up to 3 subtypes (corresponding mostly
to faint M-type stars with low S/N in this or in previous surveys and/or 
limited spectral coverage). This confirms our typical
estimated error of 1-2 subtypes, strongly depending on the S/N.
In some cases, real spectroscopic variations were observed between 2009 and 2010,  
probably due to variations in the accretion activity and in the stellar
spots. Two examples of this behavior are
213701319+573418289 (M3.0/M1.0, 2009 spectrum later than the 2010 one) and
213710877+573846877 (M4.5/M2.0, later in 2009).

We obtained the extinction of each object by
comparing the observed optical data at VRI with the theoretical colors for young
Taurus stars with the same spectral type (Kenyon \& Hartmann 1995), and 
applying the relations of Cardelli et al. (1989)
to derive A$_V$ from E(V-I) and E(V-R) for the Johnsons filters. Whenever VRI photometry was available, we
used both E(V-I)=0.521 A$_V$ and E(V-R)=0.249 A$_V$, estimating A$_V$ as the average, with
an error corresponding to the standard deviation. In the cases where one optical band was missing, 
we used the available data (V-I, V-R, or R-I) to compute the extinction, and estimated the error considering the
typical spectral typing error. In the cases where no optical data (or only one optical
band) were available, the extinction was derived from the 2MASS photometry, considering the
theoretical colors from Bessell \& Brett (1988) and the corresponding relations for
E(J-H)=0.092 A$_V$ and E(H-K)=0.076 A$_V$.  For the few objects that are confirmed to
be members but have no reliable spectral types, we take the average cluster extinction
(A$_V$=1.6 mag; SA05). 
The final extinctions are listed in the Appendix \ref{tables-appendix},
Table \ref{spec-table}.

\subsection{Membership \label{membership}}

\begin{figure}
\centering
\epsfig{file=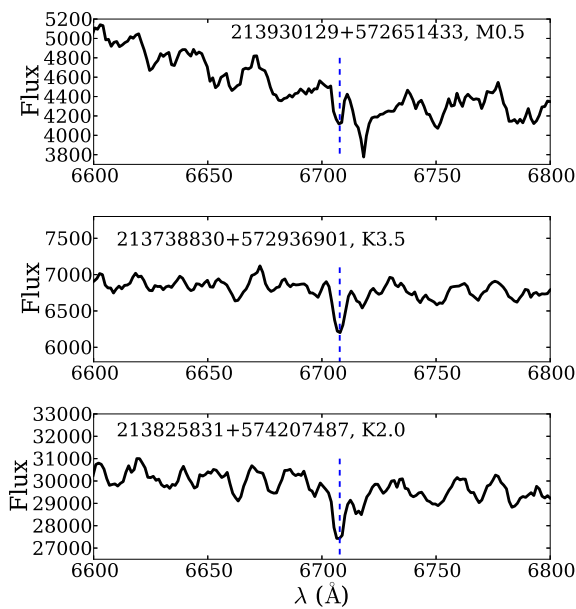,width=0.9\linewidth,clip=}
\caption{Some examples of Li I detections with the MMT/Hectospec. The flux is in arbitrary
instrumental units.
The broad photospheric lines in 213825831+574207487 are also typical of a 
relatively fast rotator. \label{Li-fig}}
\end{figure}

The membership of the stars was established by a combination of several indicators.
Typical youth indicators are the Li I absorption at 6708\AA\ and the H$\alpha$ equivalent
width (EW).
The EW of both H$\alpha$ and Li I are listed in Table \ref{spec-table}.
Table \ref{emission-table} lists other lines that have been observed in some
members, mostly strongly accreting CTTS where the whole Balmer series is visible,
together with the Ca II IR triplet, He I lines, Ca II H and K lines, and some
Fe I and O I emission (Hamann \& Persson 1992;
Appenzeller et al. 1986), plus a few objects with forbidden [S II] and [O I] lines related
to shocks (Hamann 1994).
The most reliable indicator of youth for late-type stars is the
Li I absorption at 6708\AA\ (Figure \ref{Li-fig}). Objects with clear Li I detection are labelled as sure members,
and objects with good S/N at 6708\AA\ and no evidence of Li I absorption as labeled as
non-members. Unfortunately, the Li I line is not
always detectable in these faint objects, so we need additional membership criteria.

The H$\alpha$ line in emission is the main characteristic of T Tauri stars.
In general, we followed the criteria of White \& Basri (2003) regarding accretion
and the distinction between classical and weak-line T Tauri stars (CTTS/WTTS). 
For the objects observed with the multifiber spectrograph Hectospec,
sky subtraction is complicated due to the strong nebular emission
in Tr~37, so low H$\alpha$ EW values are uncertain. Young, non-accreting M-type stars can be hard to
classify based on their H$\alpha$ emission since one of the main sources
of contamination are older dMe stars in the field.
Therefore, late-type stars with strong H$\alpha$ emission
($>$10\AA) were classified as members, especially if the Spitzer data show
an excess consistent with a disk. Stars with weak H$\alpha$ emission
($<$10\AA) and no disk were considered as potential members, requiring additional information 
to confirm membership. Stars with H$\alpha$ absorption,
especially those with late K or M spectral types\footnote{Some early K and late G WTTS may have
nearly zero emission, despite their youth, with the White \& Basri (2003) criterion for CTTS
being as low as H$\alpha$ EW$>$3\AA.}, were rejected as non-members.

Our previous study of the solar-type stars
revealed that the extinction towards Tr~37 is moderate, relatively constant over the
cluster (except in the globules), and mostly due to foreground material. The distance
of the cluster (870 pc, Contreras et al. 2002) imposes a minimum extinction around 
A$_V$=1 mag, and the study of the solar-type population revealed an average extinction
of A$_V$=1.56$\pm$0.55 mag, with nearly all the bona-fide members outside the
globule having A$_V<$3.5 mag (SA05).
This allows us to use the extinction as a further membership criterion, by
defining the extinction ranges expected for the cluster
members. Since this survey targeted fainter and thus potentially more extincted
members, we explored the extinction
distribution of our newly found objects with
Li I absorption (42 in total). The average extinction and 
standard deviation is A$_V$=1.94$\pm$0.69 mag. There are no Li I-detected members with A$_V>$4 mag, 
and only 7\% have extinctions $\geq$3.5 mag. There are also no Li I members with
extinctions lower than 0.8 mag, and only 5\% of the members have extinctions below 1 mag,
while 29\% have A$_V \geq$1.5 mag.
Therefore, we consider as candidate members all the objects with extinctions
in the range 1-4 mags. Objects in this range with additional membership indicators (strong
H$\alpha$ emission, IR emission from a disk) are considered as sure members, while objects with 
no additional indicators of youth (no Li I absorption, no disk, weak H$\alpha$, but
extinction consistent with the cluster values) are considered as probable members.
Objects with extinctions out of the cluster range but strong indicators of
membership (strong accretion-related lines and disk emission) are considered as members. 

Finally, since this survey was fully independent of the X-ray observations
of the IC~1396A globule by Getman et al. (2012), we also checked the objects in common with X-ray
observations. A total of 23 of the X-ray YSO candidates and 16 of the non-Xray YSO with disks
were found in common with our member
selection, all labeled as sure and probable members. Four probable members with X-ray
detections were thus upgraded to sure members. Three further objects (one probable member, two
probably non-members) were listed among uncertain X-ray detections, and we are keeping their original
membership classification.

\begin{figure}
\centering
\epsfig{file=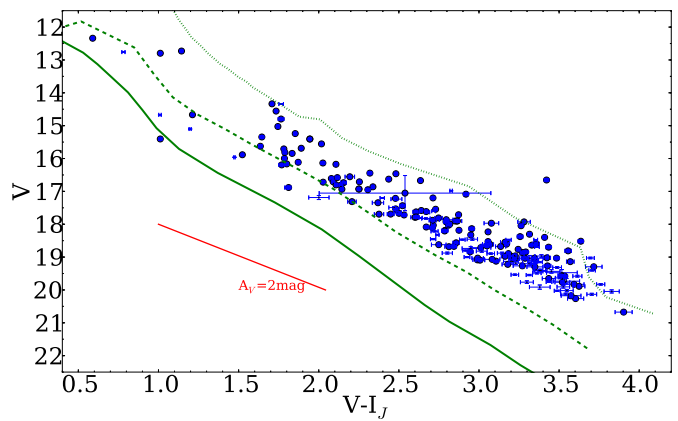,width=1.0\linewidth,clip=}
\caption{A V vs V-I (Johnsons system) diagram of the members (large filled circles) and probable members (small points) found
in this survey. For comparison, the Siess et al. (2000) isochrones for 1, 10, and 100 Myr are displayed,
transformed into the Johnsons system according to Fernie (1983) and Getman et al. (2012). An extinction vector (Cardelli et al.
1989) is also displayed. Errorbars (considering the photometric and A$_V$ errors) are also displayed,
although they are often smaller than the symbols.
\label{vvi-fig}}
\end{figure}

Placing the new members and probable members in the V vs. V-I diagram (Figure \ref{vvi-fig}), we find that their ages
are consistent with the mean age of 4 Myr derived previously (SA05). 
Since the Siess et al.
isochrones are given in the Cousins system, we transformed them into the Johnsons system 
following the prescriptions of Fernie (1983, for blue stars, V-I$_c <$1.5 mag), and Getman et al. (2012, for
red stars,  V-I$_c >$1.5 mag). A few of the objects appear
below the 10 Myr isochrone, but in most cases they are stars with flared disks and high 
(or even anomalous) extinction 
being thus probably UXor candidates (all labeled in Table \ref{spec-table}). 
In addition,  early K and G stars tend to
deviate from typical isochrone models, probably due birthline uncertainties (Hartmann 2003).
Some of the probable members that have anomalous positions in the V vs. V-I diagram may be
contaminants, giving us an idea of the uncertainty in the membership of the objects marked with 'P'.

Taking into account the five different criteria (Li I absorption, H$\alpha$ emission,
disk excess, extinction, and X-ray), from the initial collection of 565 objects observed,
we arrive to a collection of 141 sure members, 
plus 64 probable members and 33 probable non-members. The rest of
the 308 observed stars are rejected as non-members or marked as uncertain cases on the basis of
low S/N (19 in total). A total of 28 identified members 
are in common with the H$\alpha$ survey of Barentsen et al. (2011), and 17 objects corresponded
to re-observations of previously identified members in SA05 and SA06b\footnote{These include 11-1499, which was
labeled as probably non-member following the criteria in this study, 
but had Li I absorption in our better S/N spectra
from SA05 and X-ray emission according to Getman et al. (2012), so 
we consider it as member, and the previously identified probable member 21-851, 
now rejected as cluster member.}. One further object was suggested to be a YSO by
Morales-Calder\'{o}n et al. (2009). Excluding the objects that had been previously suggested to
be members by the different authors, we arrive to
78 newly identified members and 64 probable members
(to be confirmed with future followup observations), mostly M-type stars and GK members with
extinctions higher than the cluster average and with protoplanetary disks.

\section{Discussion \label{discu}}

\subsection{Disk structure and the accretion/disk connection \label{accdisk}}

\begin{figure*}
\centering
\epsfig{file=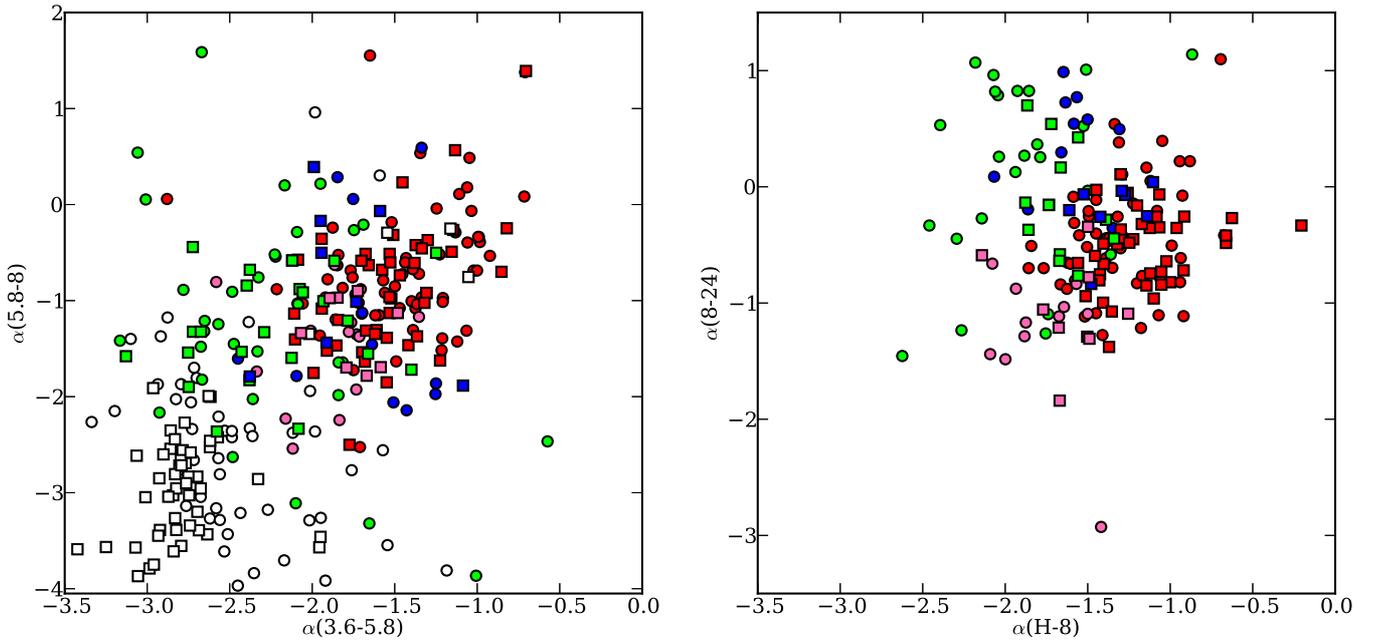,width=1.0\linewidth,clip=}
\caption{Different disk slopes for the objects with different spectral types. Circles mark the
newly identified members, squares denote the previously known ones. Diskless objects are open symbols
(none of the diskless objects is detected at 24$\mu$m, and in general diskless objects with apparent
excesses are due to contamination and/or uncertain photometry).
Red symbols are the typical full-disks. Green symbols mark the TD. Blue symbols mark the PTD. Pink symbols mark the
dust-depleted disks. 
\label{alphadisk-fig}}
\end{figure*}

As we had found in our previous papers (SA06a; Sicilia-Aguilar et al. 2007), there is a large variety of disk morphologies in
Tr~37. We have classified the new disks following the scheme developed in our Spitzer IRS-based
study (SA11). The main criteria for defining the disk evolutionary state
are the presence of inside-out evolution, as in transitional and pre-transitional disks, 
and the evidence of generalized small-dust depletion, as in
homologously depleted disks (Currie et al. 2009). 
A comparison of the disk slope at short and long wavelengths is fundamental to
distinguish disks with evidence of inside-out evolution, although the silicate feature at 10$\mu$m also plays
an important role. Objects with inner gaps or partially cleared disks, such as pre-transitional
disks (Espaillat et al. 2010), have weak near-IR fluxes, strong silicate features, and strong mid-IR excesses, 
are especially hard to identify without spectroscopic information. Low fluxes at 24$\mu$m and longer wavelengths
are a key to identify small-dust-depleted disks.

To quantify inner disk evolution, we consider $\alpha$($\lambda_1-\lambda_2$), defined as 
the SED slope between two wavelengths:
\begin{equation}
	\alpha(\lambda_1-\lambda_2)=\frac{\log(\lambda_1 F_{\lambda_1})-\log(\lambda_2 F_{\lambda_2})}{\log\lambda_1 -\log\lambda_2}.
\end{equation}

Objects with nearly photospheric colors ([3.6]-[4.5]$<$0.2 mag, $\alpha$(3.6-4.5)$\sim$-3.0$-$-2.13) but significant
excess at longer wavelengths are classified as transition disks (TD).
Disks with moderate
to low near-IR excess ([3.6]-[4.5]$\sim$0.2-0.4 mag, $\alpha$(3.6-4.5)$\sim$-2.13$-$-1.30) 
and a change in the sign of the slope between 8-12 and 24$\mu$m are
labelled as ``kink'' disks (according to SA06a terminology) and are good candidates for pre-transitional disks (PTD).

Regarding the small-dust mass, and based on our previous results (SA11), we consider objects 
with low IR fluxes at all wavelengths 
and reduced 24$\mu$m fluxes as globally dust-depleted disks. 
Dust-depleted disks have near-IR in the PTD range or lower, but with similarly steep slopes at
all other wavelengths and thus very low 24$\mu$m fluxes (for K- and M-type stars, 
$\lambda$F$_\lambda \leq$ 2-3$\times$10$^{-13}$ erg cm$^{-2}$ s$^{-1}$).  We labelled the objects in Table \ref{spitzer-table}
according to the disk classification. Figure \ref{alphadisk-fig} 
is color-coded to show the different disk types vs. the SED slopes
at different wavelengths. The final classification also requires visual inspection of all the SEDs to detect
the few cases that satisfy the mentioned criteria but display SEDs that evidently do not correspond
to the presumed disk class.
Objects with uncertain photometry/excesses are also excluded from the following analysis and discussion.

\begin{figure*}
\centering
\epsfig{file=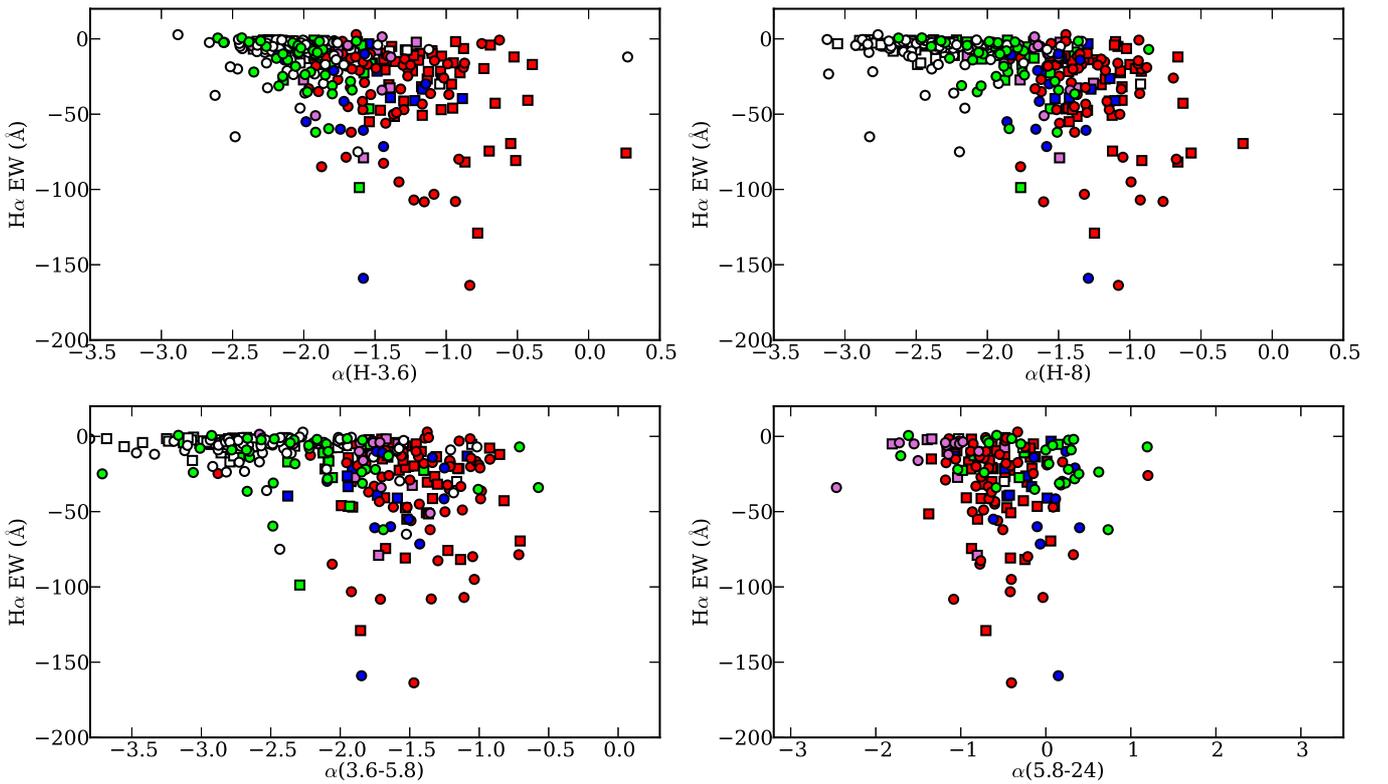,width=1.0\linewidth,clip=}
\caption{H$\alpha$ EW vs. disk slope for the members of Tr~37. Known members from previous work are
marked as squares (SA05,SA06a,b). New sure members from this work are displayed as circles. Diskless
stars are open symbols, typical full-disks are colored in red, PTD are marked in blue, TD are colored in green, and
dust-depleted objects appear in pink.
\label{alphaha-fig}}
\end{figure*}

\begin{figure*}
\centering
\epsfig{file=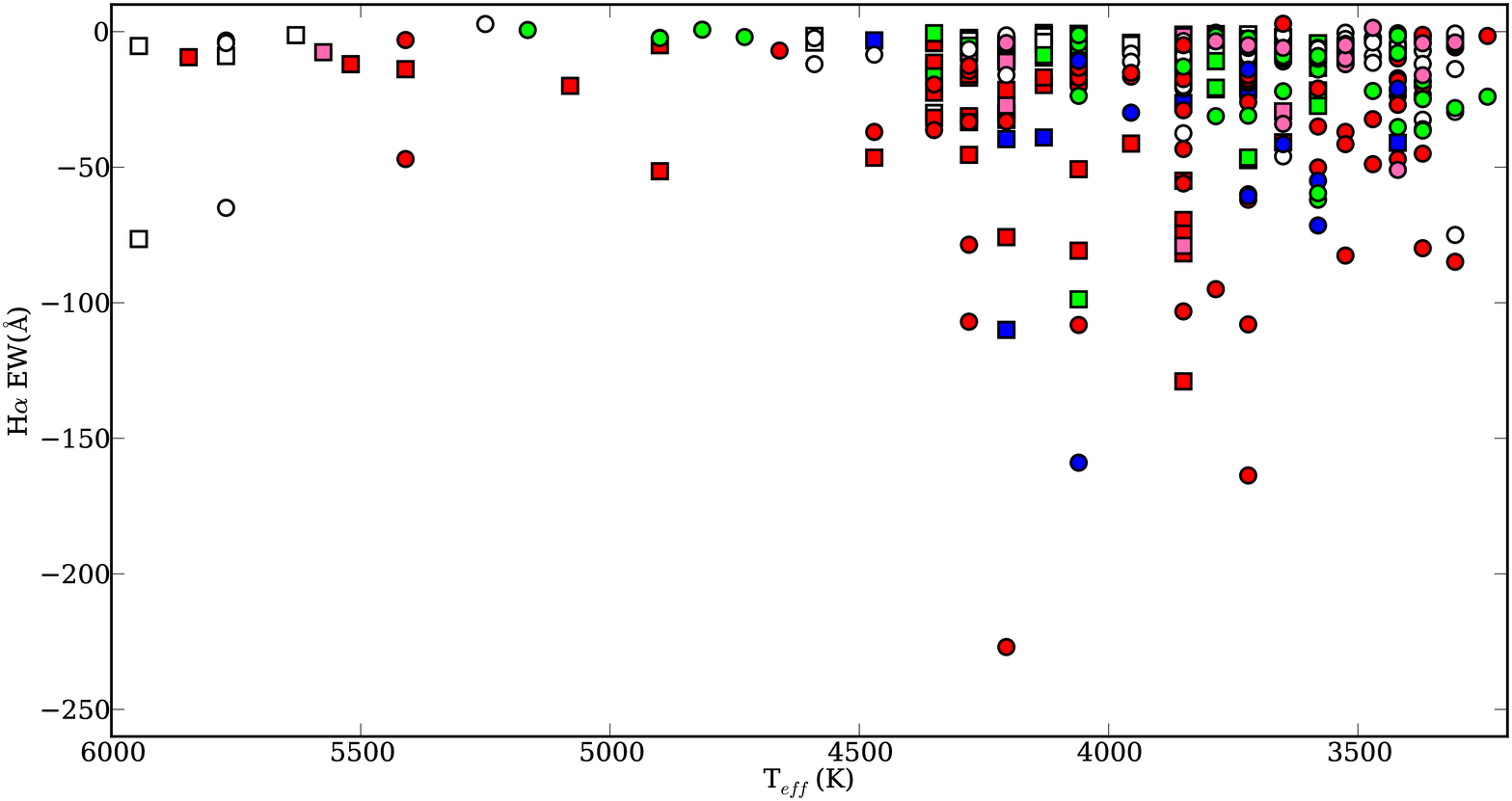,width=1.0\linewidth,clip=}
\caption{H$\alpha$ EW vs. effective temperature for the objects with different types of disks. Squares denote
the stars observed in our previous programs, while new members from this work are marked by circles. Diskless
stars are open symbols, full-disks are colored in red, PTD are marked in blue, TD are colored in green, and
dust-depleted objects appear in pink.\label{stydiskha-fig}}
\end{figure*}

Several studies have suggested a strong connection
between the IR excesses typical of protoplanetary disks and active accretion (e.g. Sicilia-Aguilar et al. 2006b, 2010;
Fedele et al. 2010) while some others suggested that the differences between non-accreting, 
WTTS and CTTS  in terms of disks are minimal (e.g. Cieza et al. 2007; Oliveira et al. 2013).
Differences in accretion between transitional and non-transitional disks have also been
matter of debate (Najita et al. 2007; Muzerolle et al. 2010; Sicilia-Aguilar et al. 2010; Fang et al. 2009, 2013b;
Mer\'{i}n et al. 2010).
Tr~37 is an interesting region, having a large number of disks with signs of evolution and a
disk fraction slightly below 50\% (SA06a).
A first approach including all known low-mass stars with spectral types G, K, M,
suggests a high correlation between the H$\alpha$ EW, the SED slopes, and our disk classification 
(Figure \ref{alphaha-fig}; the disk classification is shown in the color-code).
Figure \ref{stydiskha-fig} displays the H$\alpha$ EW vs. the effective temperatures (T$_{eff}$)
for the whole sample of low-mass members, also color-coded according to their disk types.
Instead of separating the objects in CTTS/WTTS according to their H$\alpha$ EW, we 
directly considered the measured H$\alpha$ EW within each disk class, 
and performed a double-side Kolmogorov-Smirnov (KS) test to find out the
probability that the H$\alpha$ EW of different types of objects are drawn from the same sample.

\begin{table*}
\caption{Results of the double-sided KS probability test for the H$\alpha$ EW for objects with different types of disks.\label{ksha-table}} 
\begin{tabular}{l c c c c c l }
\hline\hline
Disk types 		& 2006 Sample 	& 2013 Sample 	&  All Objects  & M-type	& G/K-type     	& Comments \\
\hline
Full-disks/Diskless   	& 1E-17		& 6E-8		& 1E-22		& 7E-12		& 2E-11		& Strong difference\\
Full-disks/PTD		& 0.47		& 0.63		& 0.28		& 0.62		& 0.57		& No significant difference\\
Full-disks/TD		& 0.0001	& 0.02		& 5E-6		& 0.009		& 5E-5		& Significant difference\\
Full-disks/Depleted	& 0.07		& 0.004		& 3E-4		& 0.001		& 0.24$^a$	& Significant difference (only M-type stars)\\
Diskless/PTD		& 1E-5		& 4E-4		& 1E-8		& 1E-5		& 5E-4		& Strong difference\\
Diskless/TD		& 0.007		& 0.07		& 0.001		& 0.0001	& 0.28		& Significant difference (only M-type stars)\\
Diskless/Depleted 	& 0.006		& 0.61		& 0.16		& 0.82		& 0.005$^a$	& Potential difference (only GK stars)\\
Disked/Diskless		& 1E-14		& 4E-6		& 4E-18		& 2E-10		& 1E-8		& Strong difference\\
\hline
\end{tabular}
\tablefoot{The 2006 sample corresponds to the members identified in our previous work, the
2013 sample includes the newly detected objects in this work. $^a$ Small number statistics,
only 6 depleted disks among G/K stars.}
\end{table*}

Table \ref{ksha-table} summarizes the results of the KS test. We find unmistakable evidence of a different
distribution of H$\alpha$ EW between objects with disks (including all types of disks) and without disks
(no evidence of excess emission at any wavelength), with probabilities nominally 0 that both distributions
are drawn from the same sample. We also find significant differences between the H$\alpha$ EW of TD 
and full-disks, mostly related to the fact that between 40 and 60\%  of the TD have H$\alpha$ 
values consistent with non-accreting
stars\footnote{Note that although H$\alpha$ EW is a good indicator of accretion, further data like high-resolution
H$\alpha$ spectroscopy or U band photometry is needed to rule out cases with very low accretion rates}, 
while this is extremely rare among full-disks. 
Fang et al. (2009, 2013b) found that the fraction of strong accretors (H$\alpha$ EW$>$2$\times$H$\alpha_{threshold}$
for a given spectral type) in the Orion Lynds 1641 and 1630N clouds is around 60\% among full disks, 
while only 21-26\% of the TD have comparable high accretion.
Fang et al. (2013b) inventoried  a sample of $\sim$1390~YSOs in Lynds~1641. Their sample includes more 
than 60 PTD/TD objects which have observed with spectroscopy measuring H$\alpha$ EW. Among their PTD/TD 
sample, 36 objects can be classified as TD, and 25 sources as PTDs according to our criteria. The fractions of 
accretors in both samples are 42$\pm$11\% and 52$\pm$14\%, respectively. Therefore, the results in the Orion Lynds clouds 
are consistent with Tr~37. 

Dust-depleted disks have also significantly lower
H$\alpha$ EW, and thus significantly lower accretion rates (sometimes consistent with no accretion) 
than normal full-disks.
This supports the idea of a strong connection between gas and dust evolution, where both signs
of inside-out evolution (in TD) and global dust depletion appear to be related to lower accretion rates.
Observations of evolved disks and accretion signatures in older clusters, like $\gamma$ Velorum 
(Hern\'{a}ndez et al. 2008; Jeffries et al. 2009) and $\eta$ Cha (Lawson et al. 2004; Bouwman et al. 2006; 
Sicilia-Aguilar et al. 2009) show that, despite the differences in environment and disk classes, there is 
a clear trend for evolved disks to display reduced or even zero accretion rates, consistent with parallel 
dust and gas evolution.
On the other hand, the H$\alpha$ EW and thus the accretion rates of PTD 
and full-disks are not significantly different. This is remarkable, since
PTD are good candidates to host clean gaps in their disks, maybe related to planet formation
(Espaillat et al. 2010). It could be due to difficulties to identify
disks with gaps from unresolved observations and lacking spectra of the 10$\mu$m region,
or be a direct consequence of most of the full-disks in Tr~37 having moderate accretion rates, compared
to younger regions like Taurus (SA06b; Sicilia-Aguilar et al. 2010). 
Although half of the TD have H$\alpha$ EW consistent with no accretion, 
and most of the dust-depleted disks also show relatively low H$\alpha$ EW,
the difference between disked and
diskless objects is still remarkable even if TD and dust-depleted disks are included.

We also explored whether the differences in accretion vs. disk type depend 
on the spectral type. Table \ref{ksha-table} reveals that the differences are subtle.
Non-accreting TD seem to be more frequent around
G/K-type stars. One explanation could be that for a given excess, the holes in a G/K-type star will be much
larger than in M-type objects, although it could also indicate that accretion can survive for longer time in
the disks around M-type objects, once the inner disk starts dispersing. 
The data also suggests that M-type dust depleted disks are consistent with no accretion, which
is in agreement with our observations of M-type stars in the 
Coronet cluster (Sicilia-Aguilar et al. 2008, 2011a). Nevertheless, a remarkable difference is that
while most of the disks around M-type stars in the Coronet cluster are dust-depleted or
have very low disk mass compared to their stellar mass (M$_d$/M$_*<$10$^{-4}$; Sicilia-Aguilar et al. 2013),
strongly dust-depleted disks are a minority among M-type stars in Tr~37, at least considering our
detection limits at 24$\mu$m. In general, the differences in accretion and disk
structure between G-K type and M-type stars are smaller in Tr~37 than observed in the Coronet cluster 
 (Currie \& Sicilia-Aguilar 2011).

\subsection{Radiative transfer models of the disks\label{radmc}}

To extract quantitative information from our previous disk structure
classification, we followed the scheme in SA11, using the RADMC radiative 
transfer code (Dullemond \& Dominik 2004) to model a subset of the disks 
within each class (normal full-disks, PTD, TD,
and dust depleted disks). 
The basic parameters in the RADMC code are the inner disk radius,
the outer disk radius (R$_{disk}$), the disk vertical scale height, and the dust
distribution. The inner disk radius is set to the
place where the temperature is 1500 K, approximately the dust sublimation radius,
and the outer disk radius is taken to be 100 AU, although our data offers a poor constrain
on this parameter and there is substantial degeneration between disk radius and
total disk mass. The disk vertical scale height is given as H/R$\propto$R$^{1/7}$,
although we also computed the scale given by hydrostatic equilibrium. Finally, 
the dust distribution is taken to be a collisional power law with exponent -3.5,
between a typical minimum size of 0.1$\mu$m and a typical maximum size of 100-10000$\mu$m.
The gas to dust ratio is 100, and gas and dust are always assumed to be well-mixed,
with all dust grains having the same temperature distribution. The stellar parameters
(mass, radius) are derived from the optical data and spectral types, using
the transformations in Siess et al. (2000) and Kenyon \& Hartmann (1995).
The dust is considered as amorphous silicate dust with equal amount
of magnesium and iron (J\"{a}ger et al. 1994; 
Dorschner et al. 1995\footnote{See http://www.astro.uni-jena.de/Laboratory/OCDB/newsilicates.html}). 
We also include 25\% of carbon grains with the
same size distribution than the silicate dust. The SEDs are calculated for an intermediate
disk inclination (45 degrees), although for these relatively flat disks, there is little
difference unless the disk is viewed edge-on.
 
As in our previous papers, our models are not aimed at a perfect reconstruction of the SED, 
but at exploring the parameter space. Starting 
with the most simplified model possible, we check whether any structural differences
are required to obtain SEDs similar to the ones we observe by varying only
the disk mass and the vertical scale height (with values not far from hydrostatic equilibrium). 
Only if no reasonable
fit is attained in this way, we proceed to change the inner disk radius,
the dust grain distribution (reducing the content of small
dust grains or increasing the maximum grain size), and also to introduce
radial variations in the vertical structure and/or the dust composition.
There is a very strong degeneration between the flaring of the outer disk, the
total disk mass, the grain distributions in the inner and outer disk, and
the radius (or disk temperature) at which the disk properties (flaring, density, grain size distribution) 
change. Constraining the precise properties of
individual disks with unresolved observatios is highly degenerated, so our purpose is to 
prove that a simple disk structure cannot reproduce
the observed SEDs for TD and PTD, unless some radial changes are included, even if
the precise values of the varying parameters cannot be determined.
Table \ref{model-table} summarizes the model disk parameters 
within each disk class. A brief discussion on each class and the most appropriate
models follows.

\subsubsection{The typical full-disks\label{ctts-models}}

\begin{figure*}
\centering
\begin{tabular}{ccc}
\epsfig{file=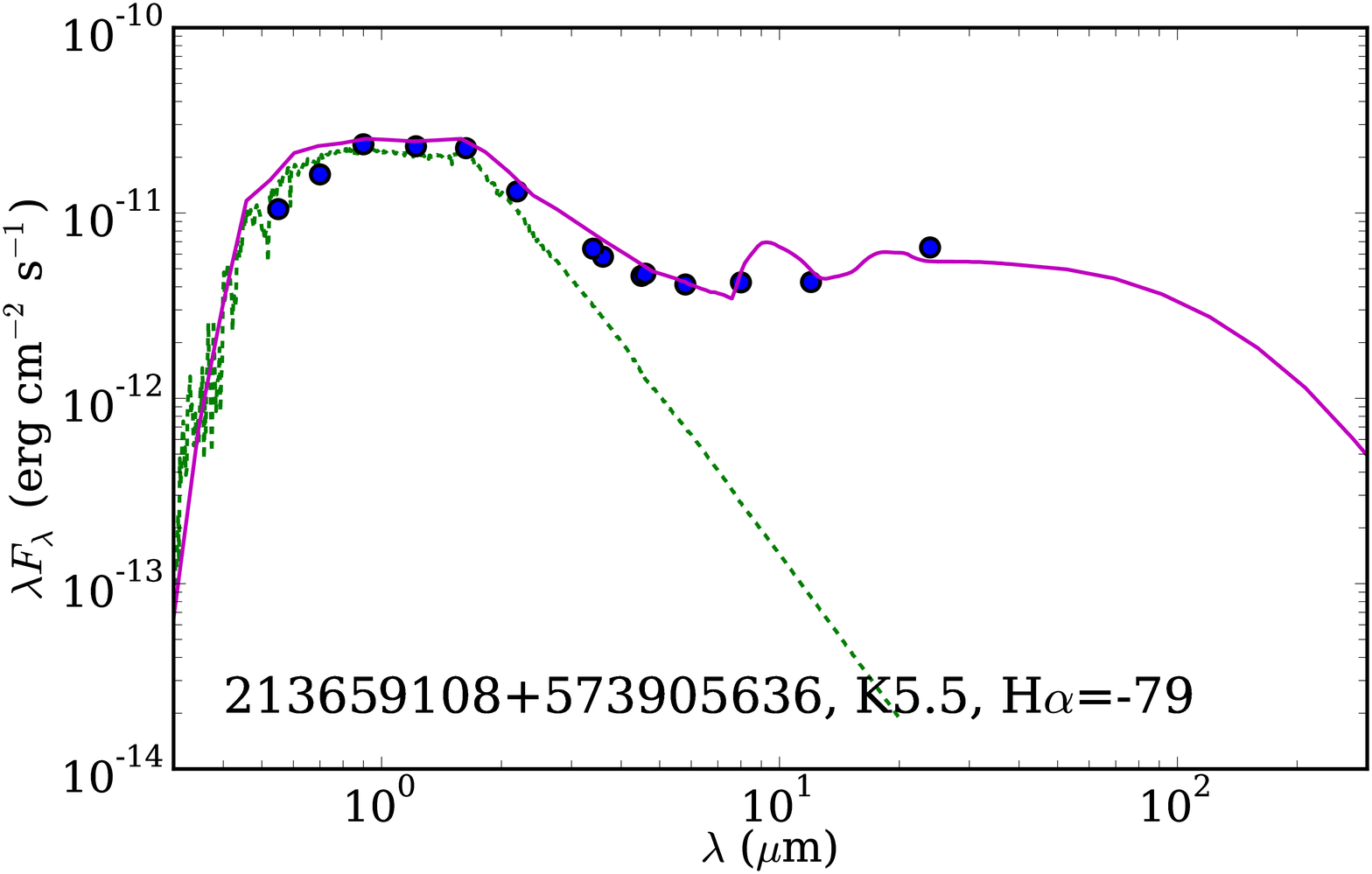,width=0.33\linewidth,clip=} &
\epsfig{file=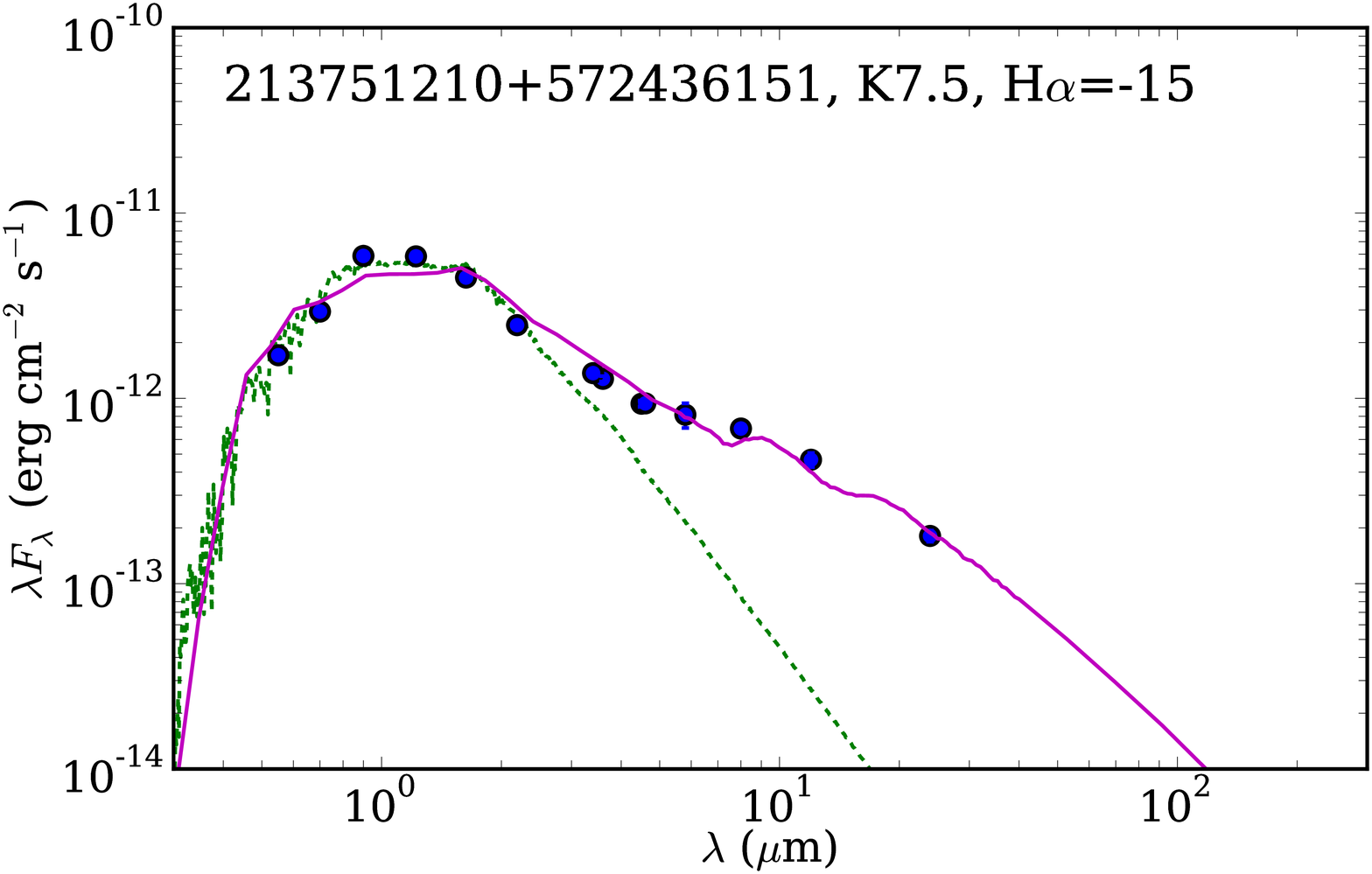,width=0.33\linewidth,clip=} &
\epsfig{file=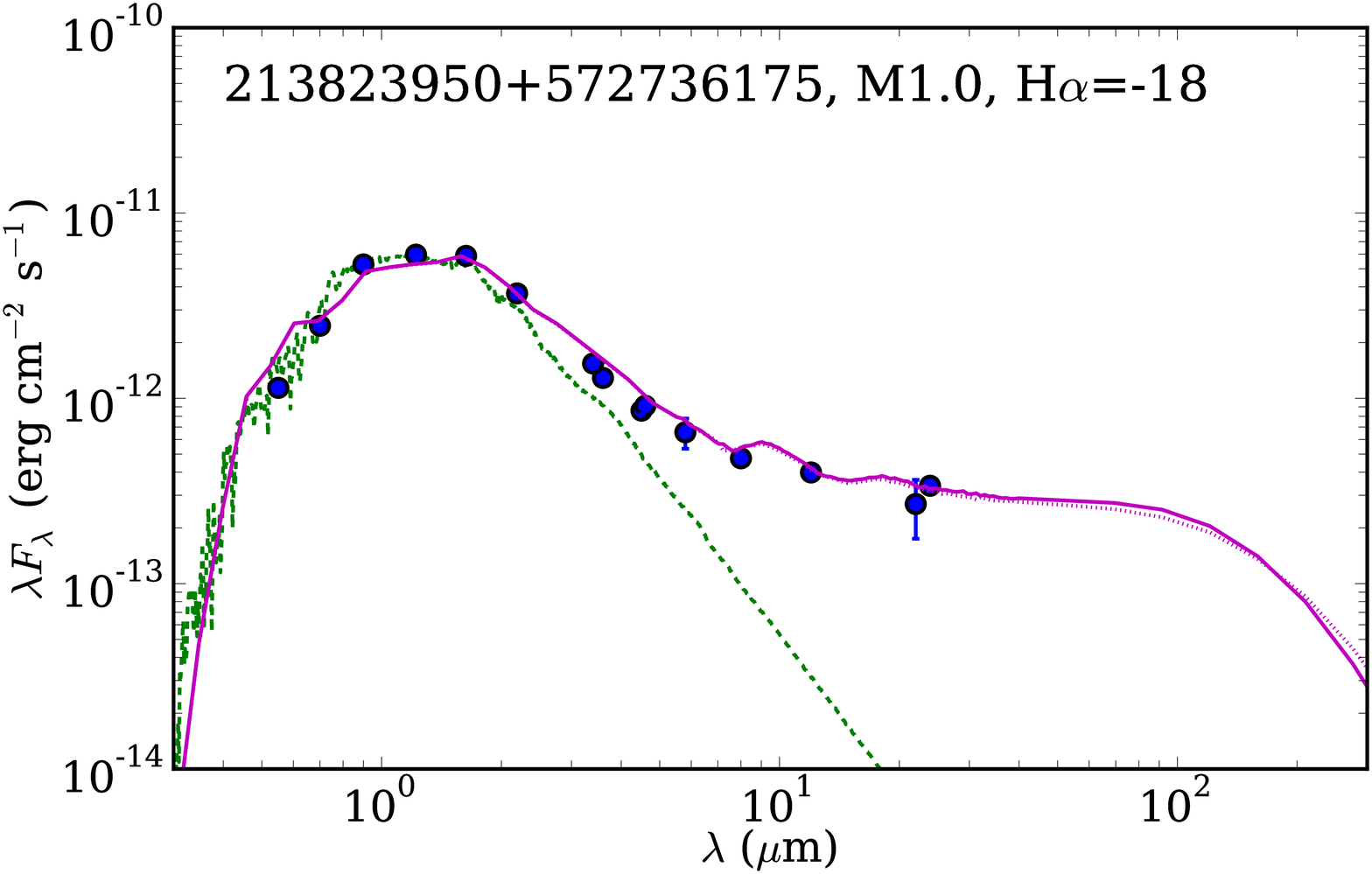,width=0.33\linewidth,clip=} \\
\end{tabular}
\caption{Disk models for selected SEDs of typical full-disks (magenta lines). See
Table \ref{model-table} for information on the individual models. 
For comparison, the photosphere of a star with the same spectral type
from the MARCS models (Gustafsson et al. 2008) is displayed. All
datapoints have been extinction-corrected according to their individual
values of A$_V$ and assuming a standard extinction law (see Table \ref{spec-table}.
Information about the H$\alpha$ EW (in \AA) is also
displayed.\label{modelCTTS-fig}}
\end{figure*}

The full-disks (Figure \ref{modelCTTS-fig}) are easy to reproduce by regulating the disk mass
to fit the 24$\mu$m point, and changing the vertical scale height, either assuming a global power law for the
disk flaring at all radii or hydrostatic equilibrium. The required
vertical scale heights do not depart much from hydrostatic equilibrium, although we find
disks that appear higher than predicted by hydrostatic equilibrium (especially, among
the higher mass disks) while other disks tend to be flatter, as expected
from dust settling. We modeled three objects, with different spectral types and
different disk masses, 213659108+573905636, 213751210+572436151, and 213823950+572736175, 
in representation of this broad and numerous class.

\begin{figure*}
\centering
\begin{tabular}{ccc}
\epsfig{file=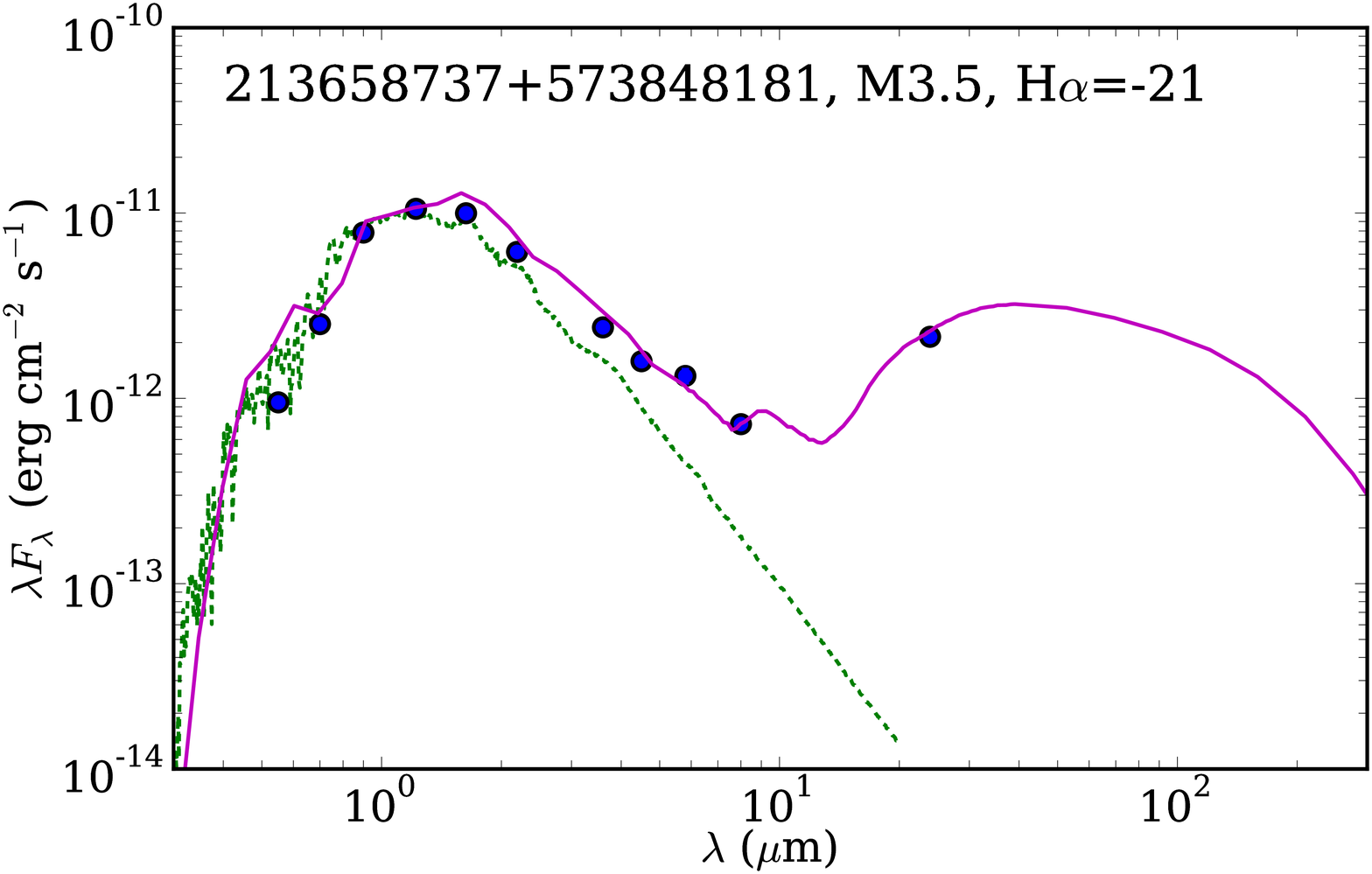,width=0.43\linewidth,clip=} &
\epsfig{file=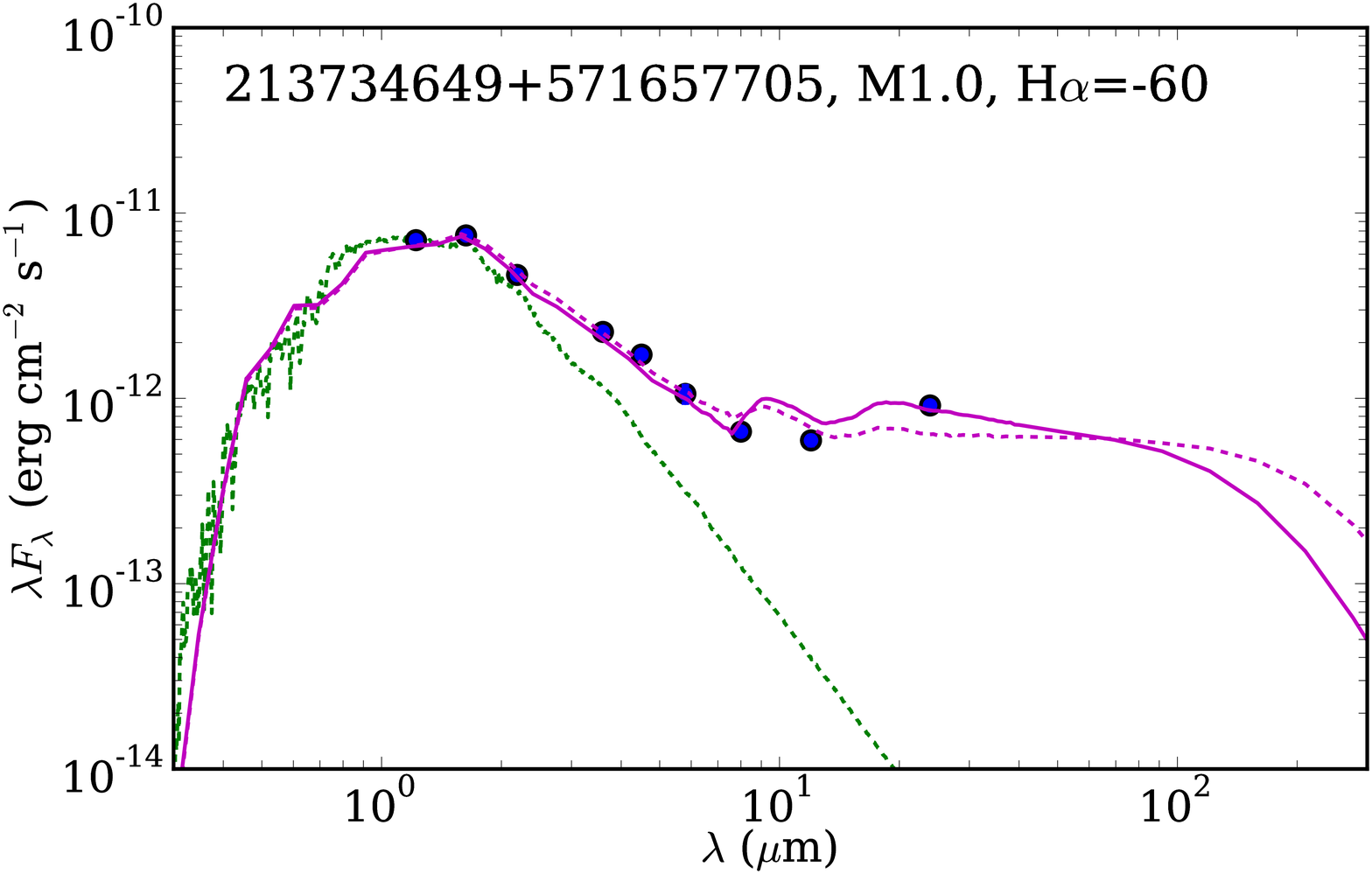,width=0.43\linewidth,clip=} \\
\epsfig{file=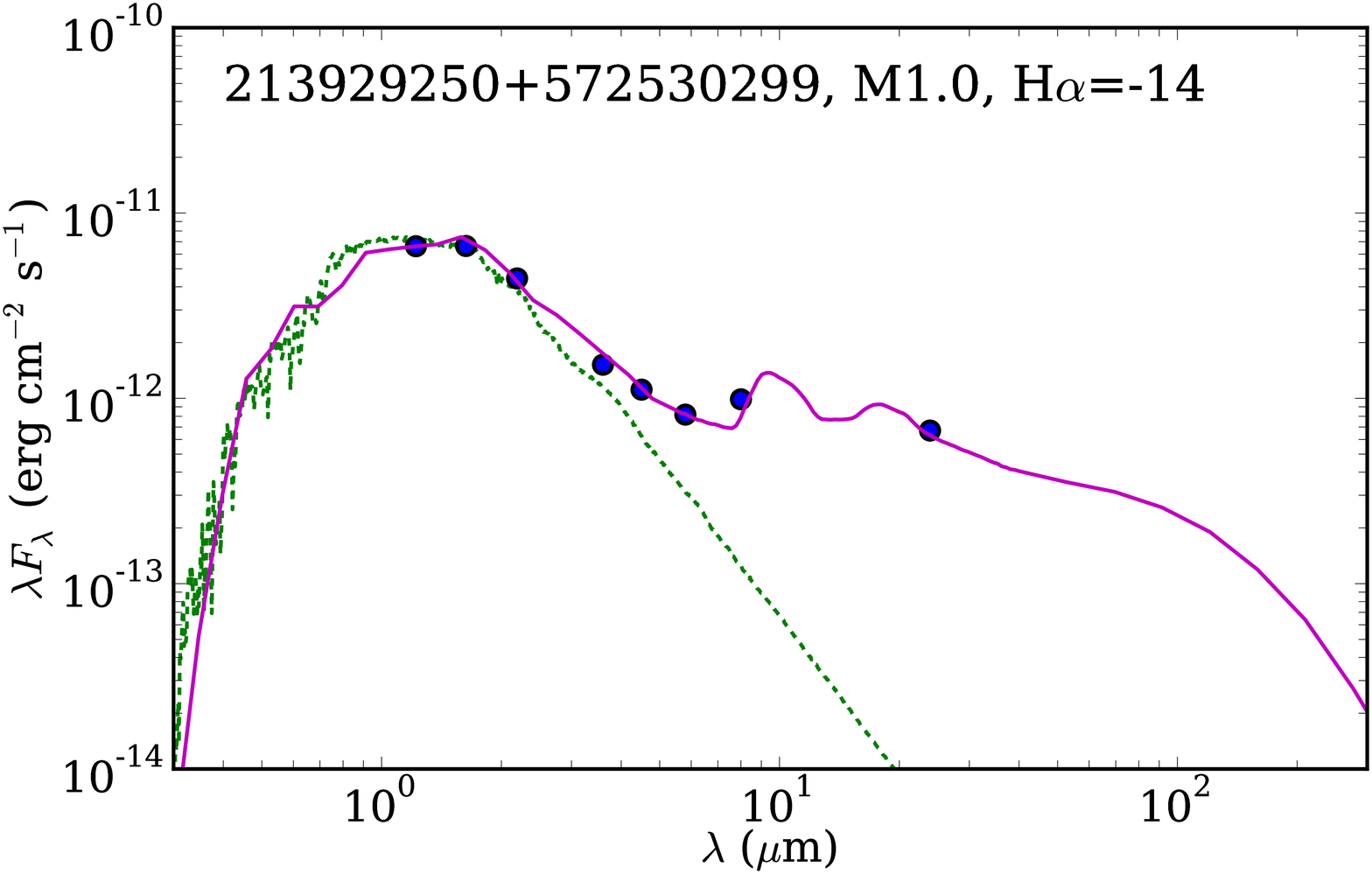,width=0.43\linewidth,clip=} &
\epsfig{file=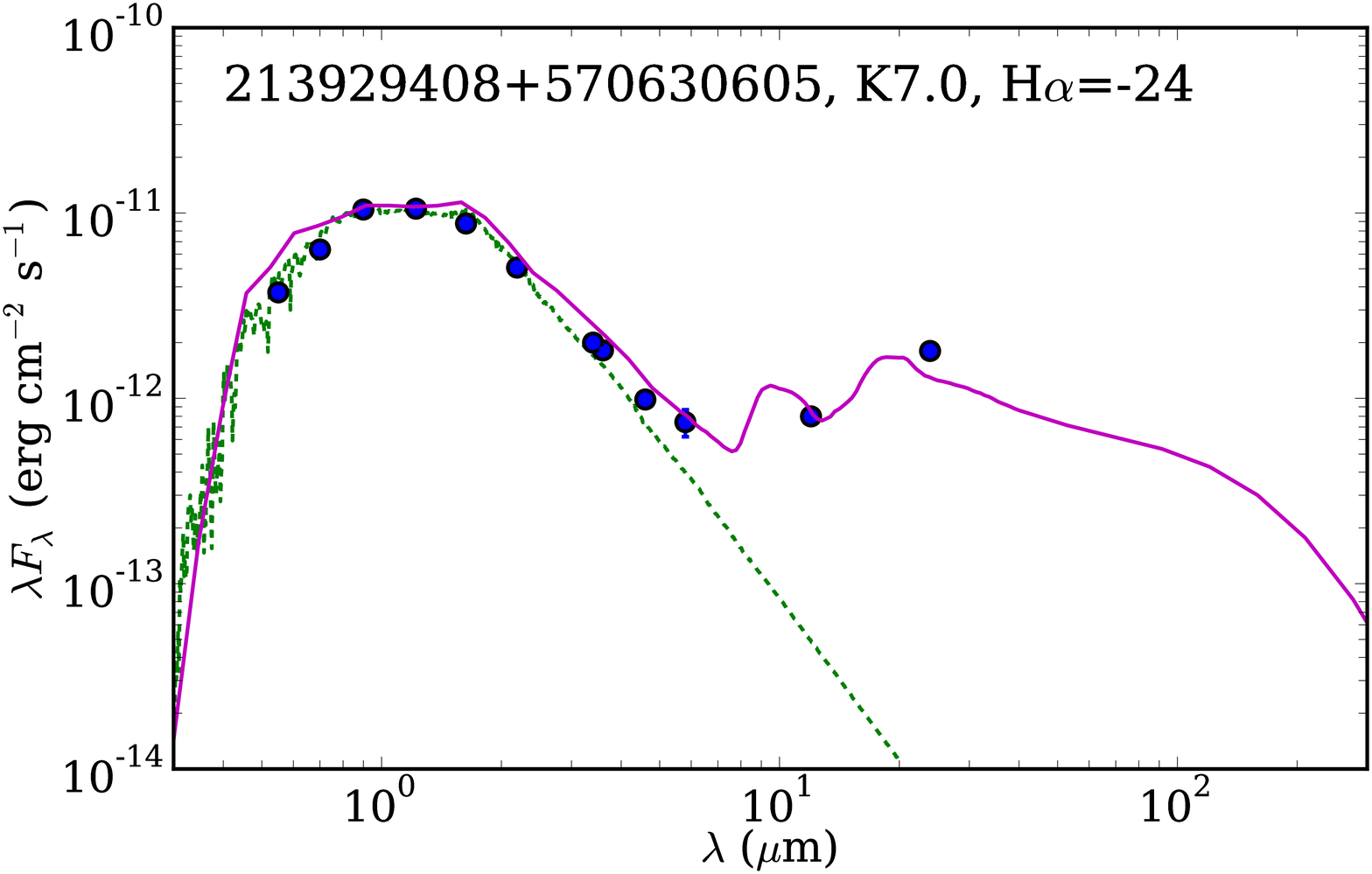,width=0.43\linewidth,clip=} \\
\epsfig{file=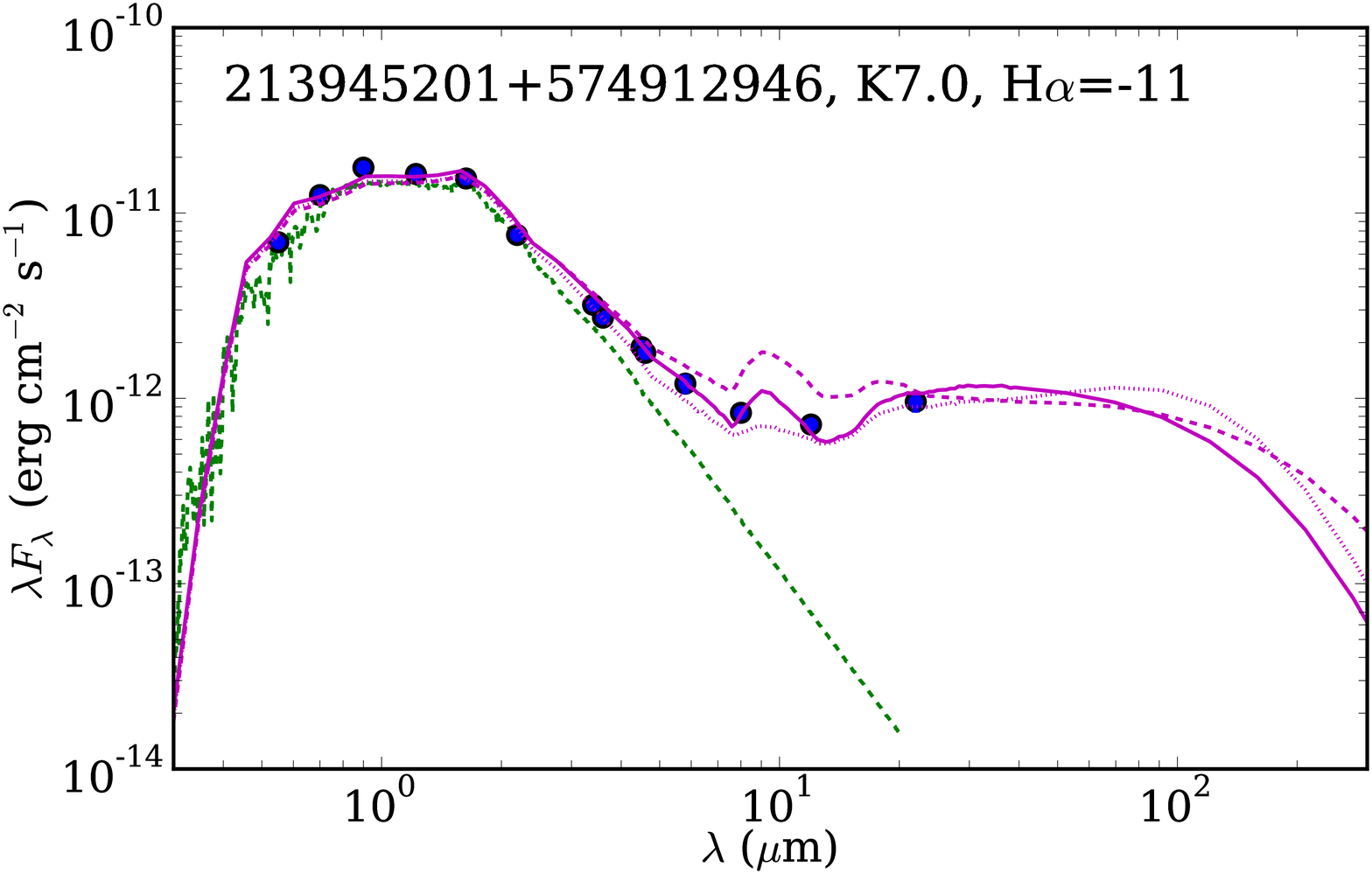,width=0.43\linewidth,clip=} &
\epsfig{file=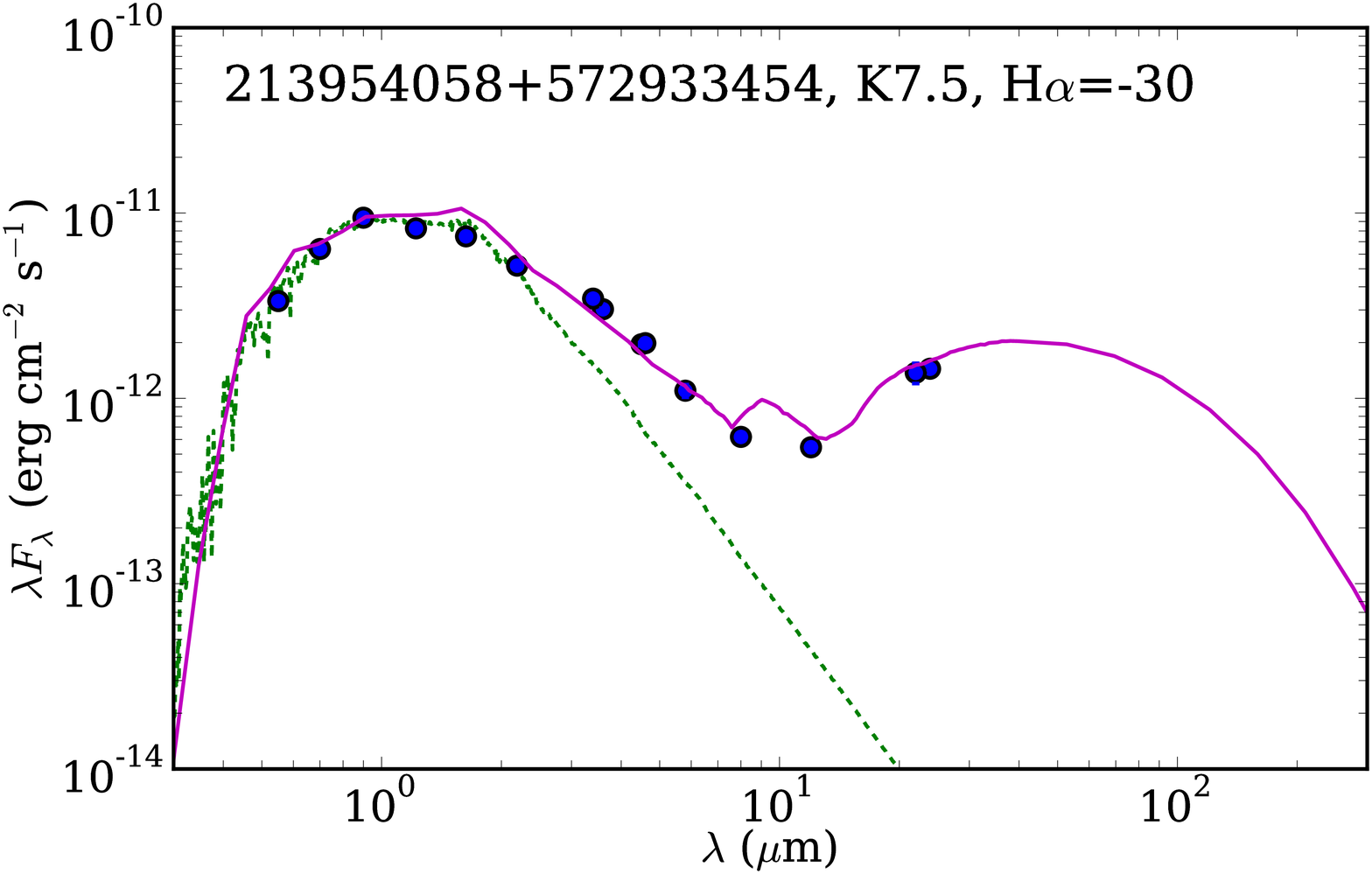,width=0.43\linewidth,clip=} \\
\end{tabular}
\caption{Disk models for selected SEDs of pre-transitional disks (magenta lines). See
Table \ref{model-table} for information on the individual models. 
For comparison, the photosphere of a star with the same spectral type
from the MARCS models (Gustafsson et al. 2008) is displayed. All
datapoints have been extinction-corrected according to their individual
values of A$_V$ and assuming a standard extinction law (see Table \ref{spec-table}.
Information about the H$\alpha$ EW (in \AA) is also
displayed. \label{modelPreTD-fig}}
\end{figure*}

Disks with large excesses, like 213659108+573905636, require a very large amount of mass in small
dust grains. Although the maximum grain size is unconstrained, if we assume a collisional distribution with
maximum grain sizes over 100$\mu$m and a gas to dust fraction of 100, we would require disk masses 
about 10\% of the estimated stellar mass.
This would be on the upper limit according to submillimeter and millimeter observations (Andrews \& Williams
2005, 2007). On the other hand, disks like that around 213823950+572736175, with a low-to-moderate mass,
probably require large grains to account for mass enough to sustain the 
accretion rates we expect from their strong H$\alpha$ emission.

Other disks have very strong near-IR excesses but a lower 24$\mu$m than expected, suggestive of
strong flaring but a very low dust mass, even if we consider substantial grain growth (e.g. 213751210+572436151,
213809997+572352782). This is
similar to other disks in the Coronet cluster (CrA-159 and HBC 677; Sicilia-Aguilar et al. 2013).
Further observations at longer wavelengths are needed
to clarify the total dust content and other special
structures that could affect their morphology, like truncation or gaps at larger radii. 
As a first approach, we consider these disks within the full-disk class.

\subsubsection{The pre-transitional disks \label{pretd-models}}

The pre-transitional disks (Figure \ref{modelPreTD-fig}) are the hardest class to identify with our available
data. The lack of a mid-IR spectrum and information on the silicate feature
are a strong constraint to identify disks with partially optically thin regions in their
inner disks (Espaillat et al. 2010; SA11). Nevertheless, the 
high 8-12$\mu$m fluxes and sharp kinks in the mid-IR
suggest disks where the properties in the innermost and outer disk change.
Our models do not include 
gaps, although a gap created by a companion or by photoevaporation can offer a natural
barrier to explain the radial changes in properties.
Since this class is the most problematic, we selected 6 objects as examples (213658737+573848181,
213734649+571657705, 213929250+572530299, 213929408+570630605, 213945201+574912946, 213954058+572933454)
and ran different models to show the degeneration of the
various parameters. Our list of models is far from complete, but it shows that,
despite the degeneracy, the SEDs reveal that some radial variations in disk properties are needed.
The radial changes, reflected in Table \ref{model-table}, are a signature of a distinct
inner and outer disk, with different flaring, inner radius, grain sizes, and mass.

The case of 213734649+571657705 shows that a simple full-disk model fails to both reproduce
the inner and outer disk properties (dashed line in Figure \ref{modelPreTD-fig}), while a change in grain size distribution
and flaring offers a good compromise to reproduce the SED (bold line).
213945201+574912946 offers a practical example of the impossibility to fit a pre-transitional
SED with a simple full-disk model (dashed line), that would require extreme flattening far
beyond hydrostatic equilibrium and a very large mass to account for the turn up at 24$\mu$m,
but would still overestimate the fluxes at 8-12$\mu$m. A model with a differentiated
inner disk and hydrostatic equilibrium comes close to the result (dotted line), but we note that the
best fit is attained by assuming the inner disk has a larger vertical scale height than predicted
by hydrostatic equilibrium (bold line).

\subsubsection{The transitional disks \label{td-models}}

\begin{figure*}
\centering
\begin{tabular}{cc}
\epsfig{file=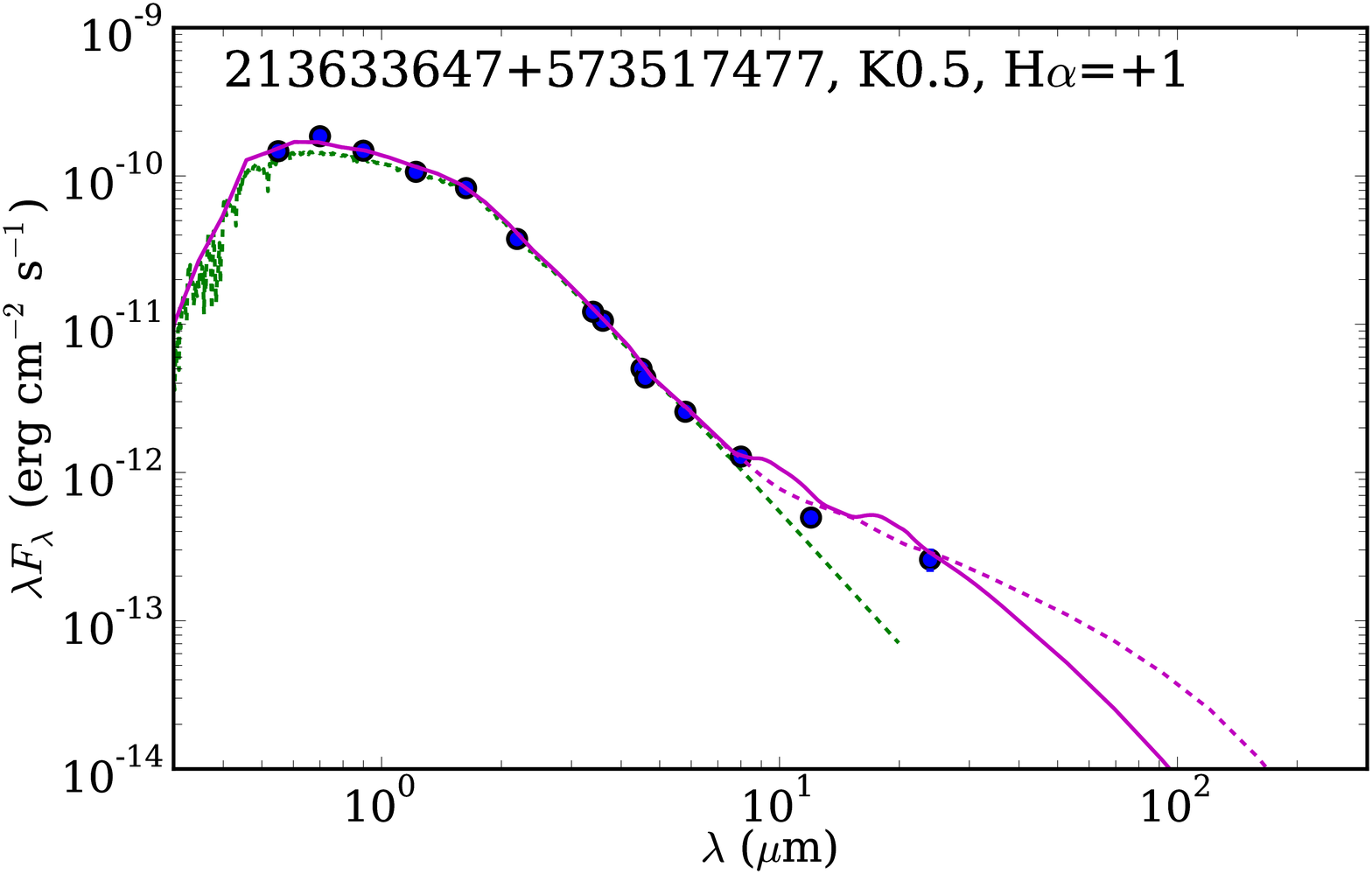,width=0.43\linewidth,clip=} &
\epsfig{file=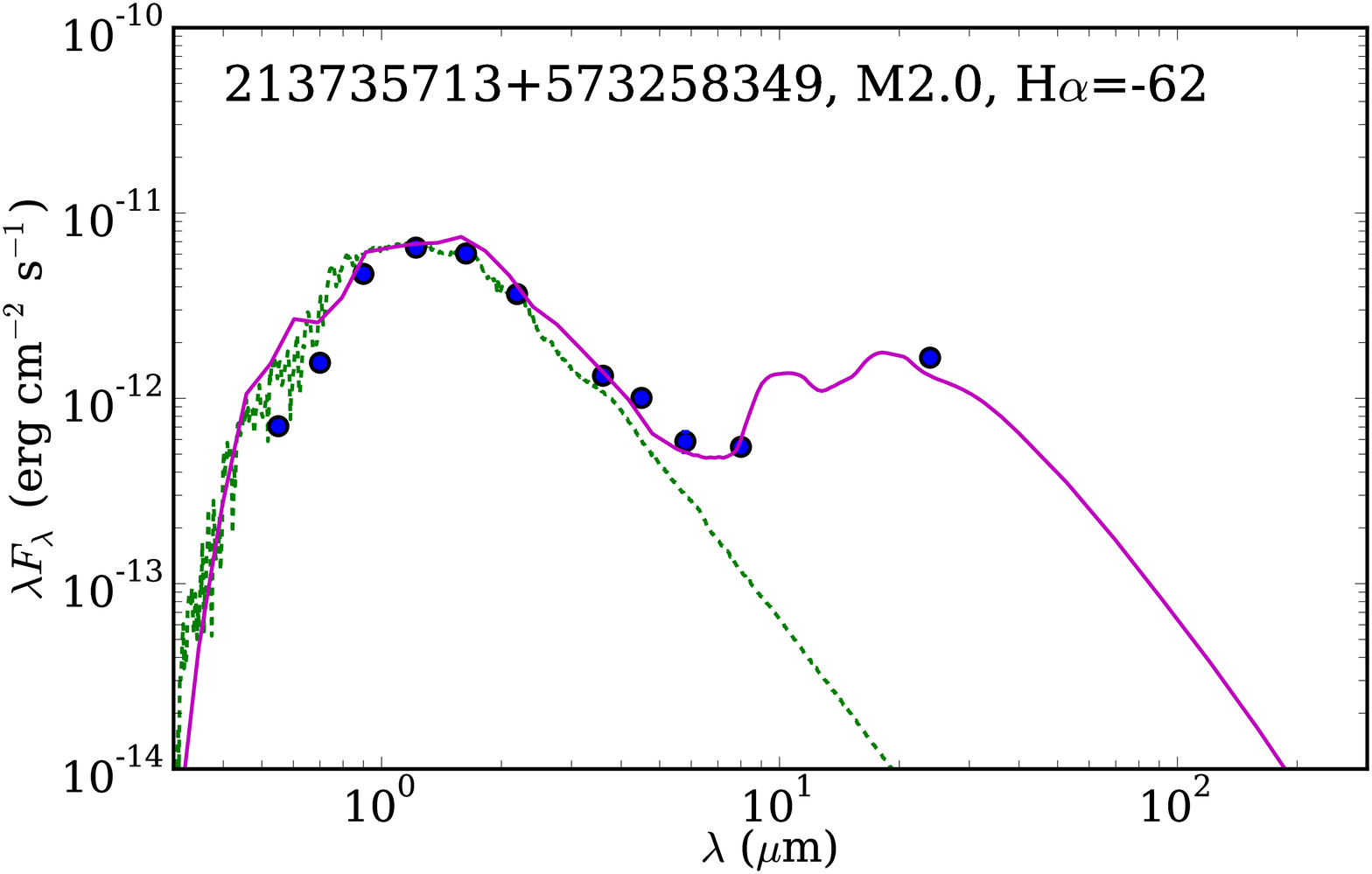,width=0.43\linewidth,clip=} \\
\epsfig{file=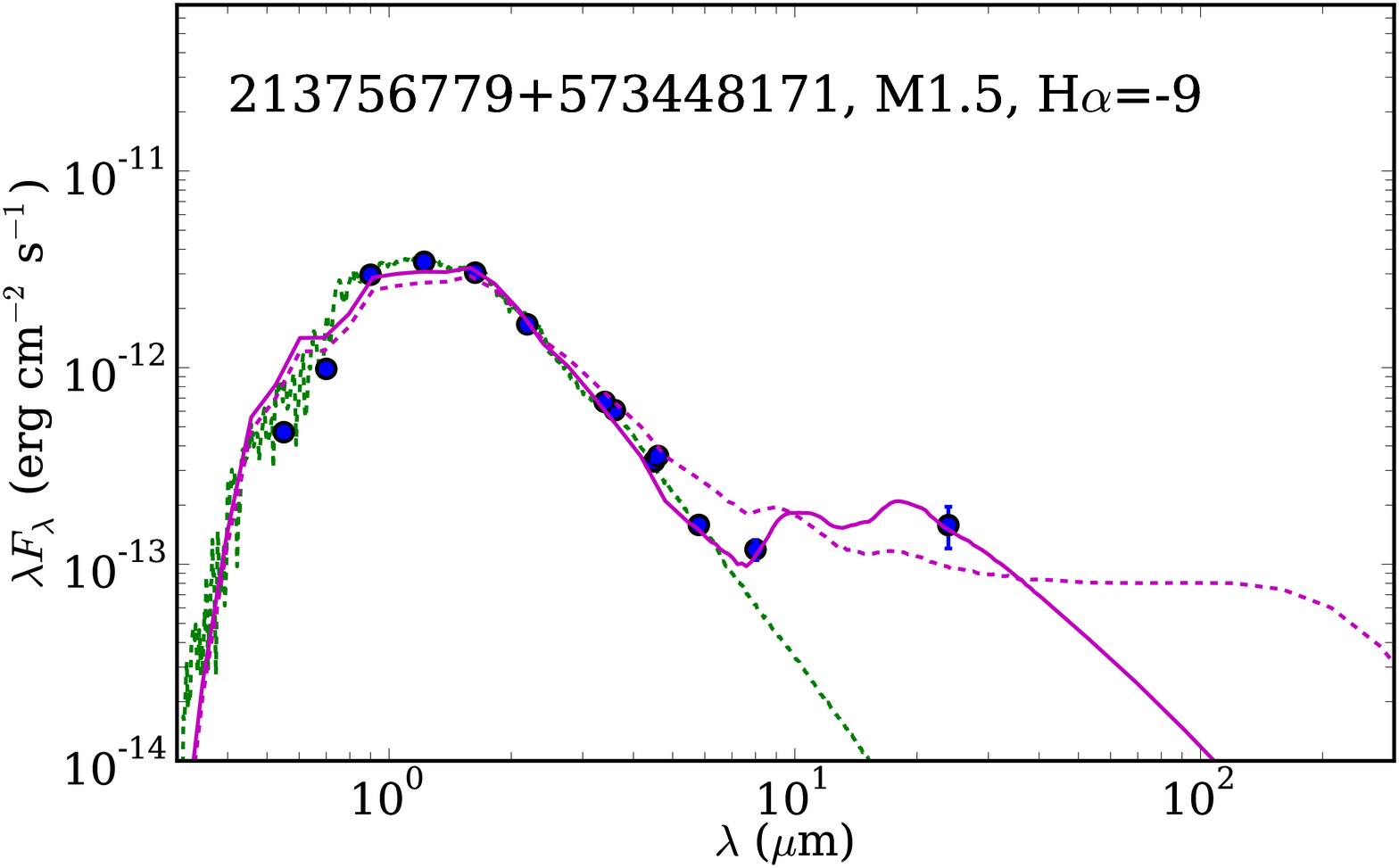,width=0.43\linewidth,clip=} &
\epsfig{file=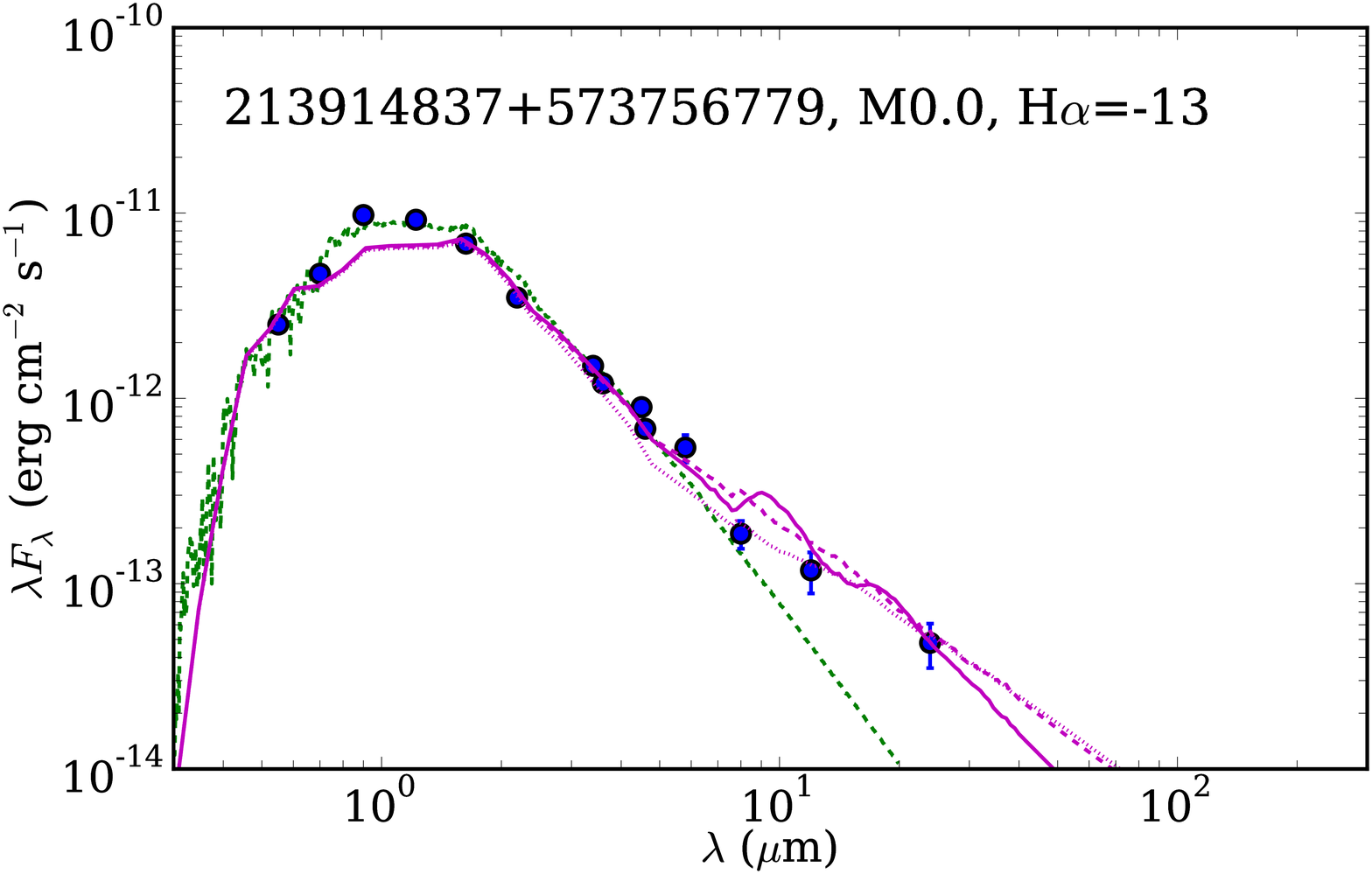,width=0.43\linewidth,clip=} \\
\end{tabular}
\caption{Disk models for selected SEDs of transitional disks (magenta lines). See
Table \ref{model-table} for information on the individual models. 
For comparison, the photosphere of a star with the same spectral type
from the MARCS models (Gustafsson et al. 2008) is displayed. All
datapoints have been extinction-corrected according to their individual
values of A$_V$ and assuming a standard extinction law (see Table \ref{spec-table}.
Information about the H$\alpha$ EW (in \AA) is also
displayed. \label{modelTD-fig}}
\end{figure*}

For transitional disks  (Figure \ref{modelTD-fig}), we have evidence for an inner hole that is most likely devoid
of small dust. We constructed models for objects with different spectral types and different size holes,
parameterized by the temperature of the inner disk rim. As for pre-transitional disks, there is
a great degeneracy between disk scale height, dust composition, and size of the hole, although the
low near-IR fluxes cannot be explained without cutting the dust distribution at a temperature 
significantly lower than the dust destruction temperature (see for instance the large grain model
for 213756779+573448171, which fails to reproduce the low near-IR fluxes).

Without information on silicate emission it is hard to constrain the minimum grain size.
Dust distributions with larger minimum size grains produce lower
near-IR emission. The total disk mass is also highly dependent on the outer disk
radius and on the vertical scale height. Some of the disks could be close to hydrostatic
equilibrium, with only minimal differences in some cases (e.g. 213633647+573517477). In others, the strong excesses 
require a vertical scale height over the hydrostatic equilibrium (e.g. 213735713+573258349), or
the weak emission asks for a reduced vertical scale height (e.g. 213756779+573448171).
The mass of certain TD, even assuming a significant grain growth, is well below 10$^{-4}$M$_*$,
which means that they also appear to be substantially dust-depleted (e.g. 213914837+573756779). 
Nevertheless, the main
difference with this class, as we explain in \ref{dustdepleted-models}, is that not all dust-depleted disks
need to have an inner hole, and not all TD need to have very low small-dust-grain masses.

\subsubsection{The dust depleted disks \label{dustdepleted-models}}

\begin{figure*}
\centering
\begin{tabular}{cc}
\epsfig{file=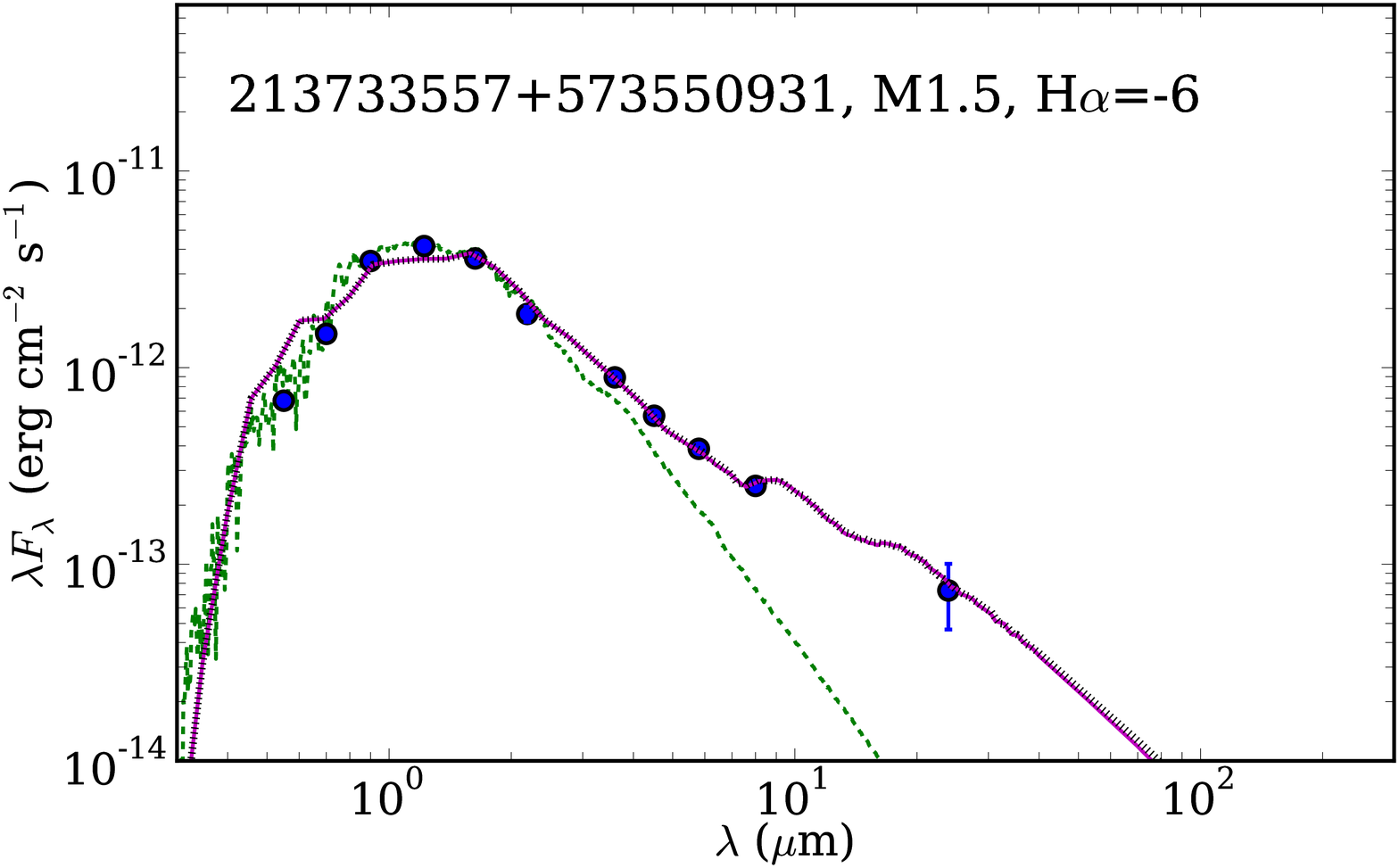,width=0.43\linewidth,clip=} &
\epsfig{file=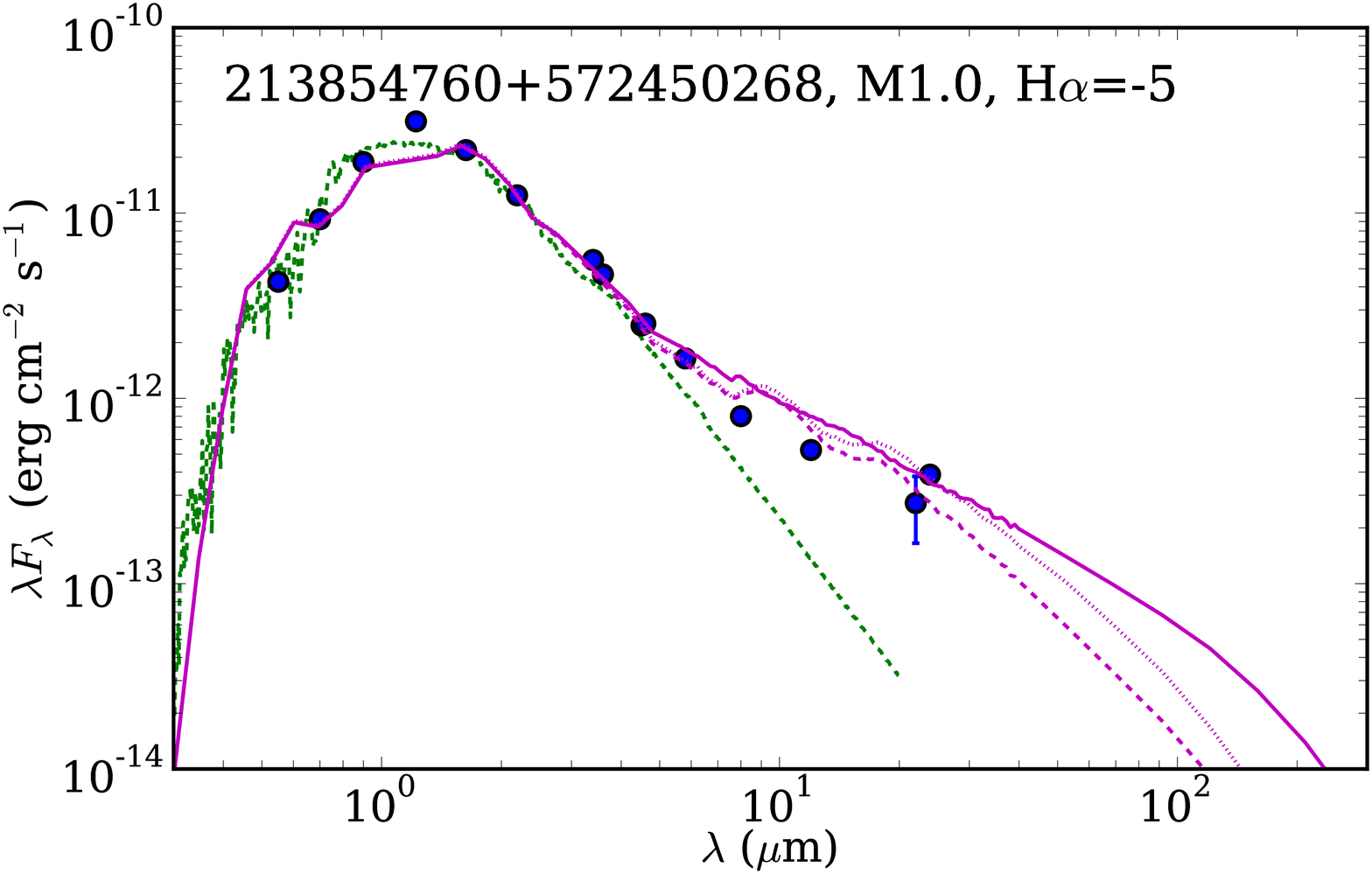,width=0.43\linewidth,clip=} \\
\epsfig{file=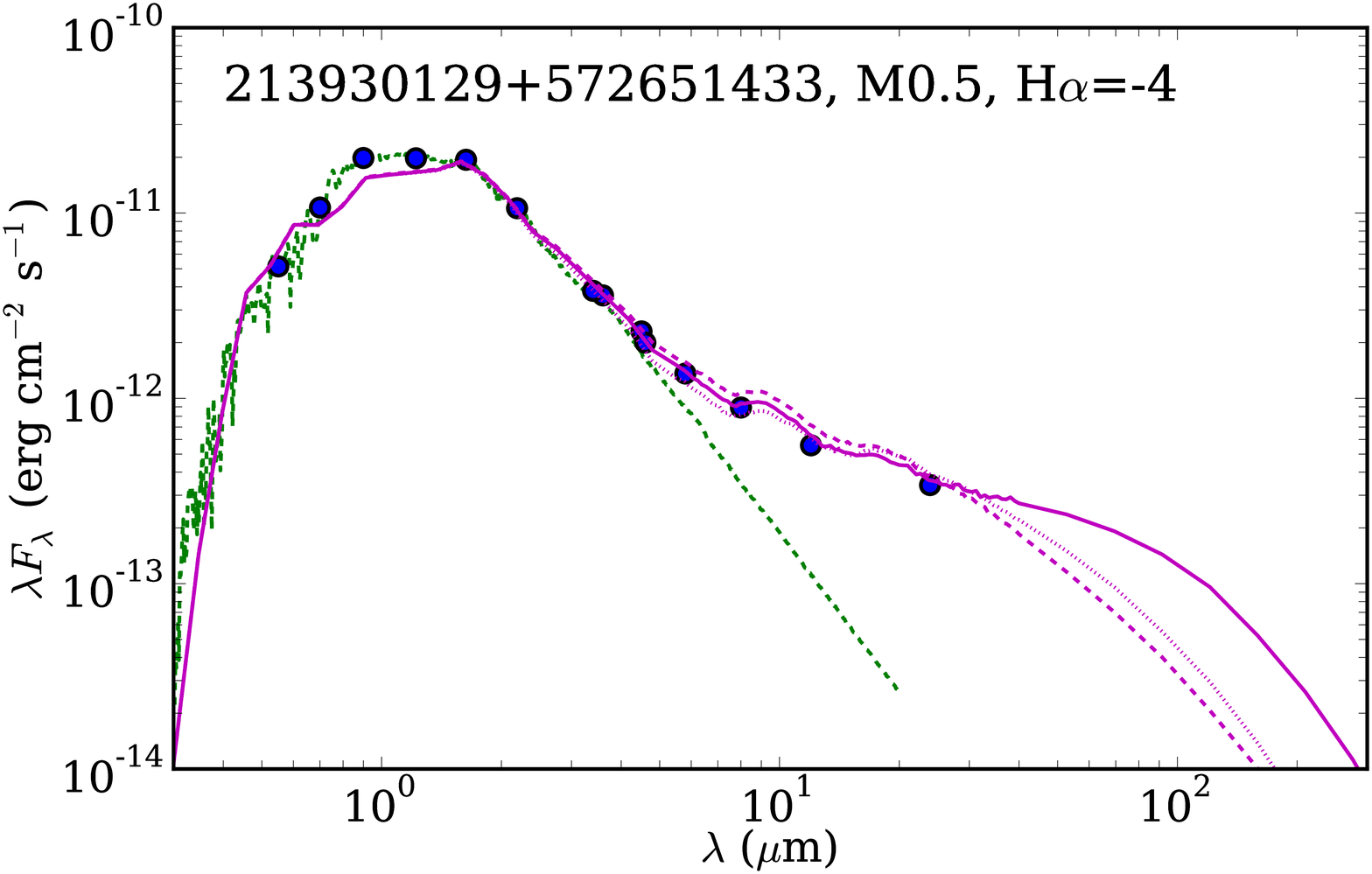,width=0.43\linewidth,clip=} &
\epsfig{file=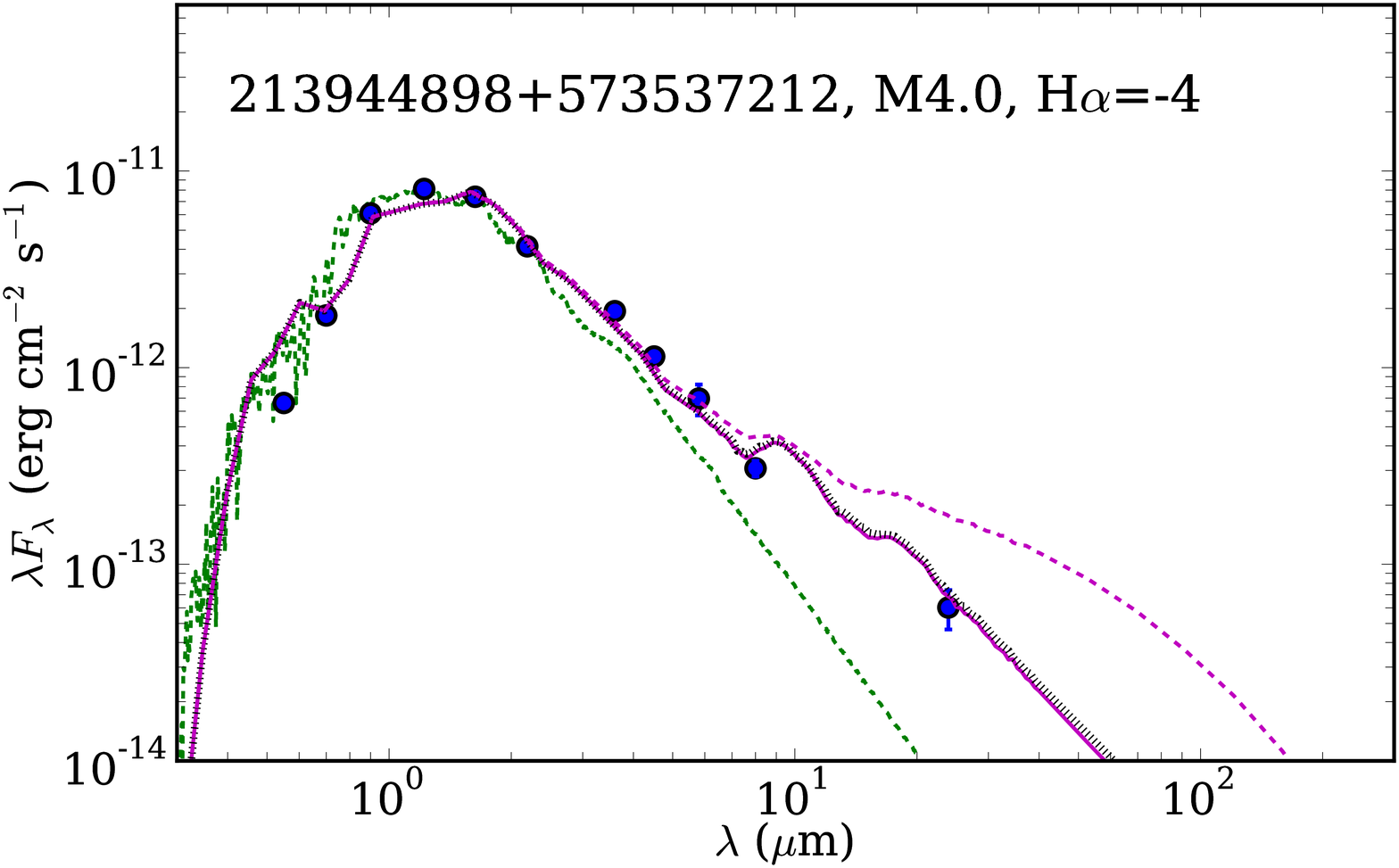,width=0.43\linewidth,clip=} \\
\end{tabular}
\caption{Disk models for selected SEDs of dust-depleted disks (magenta lines). See
Table \ref{model-table} for information on the individual models. 
For comparison, the photosphere of a star with the same spectral type
from the MARCS models (Gustafsson et al. 2008) is displayed. All
datapoints have been extinction-corrected according to their individual
values of A$_V$ and assuming a standard extinction law (see Table \ref{spec-table}.
Information about the H$\alpha$ EW (in \AA) is also
displayed. \label{modeldepleted-fig}}
\end{figure*}

Dust-depleted disks  (Figure \ref{modeldepleted-fig}) are also relatively simple. They can be reproduced with
model without radial variations, but require a very low small dust mass (at least, with
very low mass in the submicron to 20-30$\mu$m range) and/or very strong settling. There is a strong 
degeneracy between mass and vertical scale height. In general, if we assume that the dust-depleted
disks are in hydrostatic equilibrium, we require a total mass several orders of magnitude
below the typical values of normal full-disks. If we impose strong settling and disk
flattening, the small dust grain depletion does not need to be so dramatic, although in general
an increase of mass over one order of magnitude (for the same grain size distribution) 
cannot be compensated by flattening the disk, since the 24$\mu$m fluxes would increase
far beyond what is observed (e.g. 213030129+572651433).
Some of the depleted disks may also have inner holes; 
213854760+572450268 is one example where the failure of various models to reproduce simultaneously the 24 and 8 $\mu$m flux
may indicate the presence of a small hole. However, this is not always necessary and most of them
do not satisfy the criteria for TD.

The mass of the disk also depends strongly on the disk radius and on the maximum grain
size, but even assuming millimeter-sized maximum grain sizes requires a very low mass for the disk, compared to
typical full-disk values (see for instance 213733557+573550931). This implies that even if our data does not allow us to 
fully rule out a total higher disk mass in the disks classified as dust-depleted, the mass content in grains 
smaller than 20-30$\mu$m needs to be very low, which is in any case a strong sign of evolution.
From the analysis in Section \ref{accdisk}, the dust depleted disks clearly show accretion differences
with respect to full disks. Therefore, despite the degeneration between disk mass and disk flaring/settling,
which has also been discussed in the literature (e.g. Espaillat et al. 2012), our study suggests that
these disks are in a different evolutionary stage from typical full disks. Even if we cannot constrain 
the mass of the disks, and even if the case is particularly hard for M-type stars that have naturally lower
near-IR excesses, the low probabilities that M-type dust depleted disks and M-type full disks are drawn from the
same collection of objects, and the fact that M-type dust depleted disks and M-type diskless stars are consistent
with the same H$\alpha$ values (see Table \ref{ksha-table}) strongly point to a different evolutionary status compared
to full disks.
Accretion termination/gas depletion may be a sign of parallel dust and gas evolution, and also favor stronger  
dust settling within the disk.
The study of silicate features (as in the case of 01-580; Sicilia-Aguilar et al. 2011a), and far-IR/submillimeter
observations are a key to better constrain the disk mass of these objects , their structure, 
and the presence of settling (see for instance the models for 213930129+572651433 and 213944898+573537212). 

\begin{landscape}
\begin{table}
\caption{Disk models for selected members, according to their disk structure. The comments also include a reference to the
models plotted in the figures. }
\label{model-table}
\centering
\begin{tabular}{l c c c c c c c l}
\hline\hline
Object & T$_{eff}$ & R$_*$        & M$_*$        & M$_{disk}$  & a$_{min}$-a$_{max}$ & H$_{Rdisk}$/R$_{disk}$ & T(R$_{in}$) & Comments/Model reference in plots \\
       &  (K)      &  (R$_\odot$) &  (M$_\odot$) & (M$_\odot$) &  (~$\mu$m)          &                        & (K)         &  \\
\hline
{\bf Full-disks}\\
213659108 & 4330 &  1.49 & 1.0  & 0.08   & 0.1-100 	& 0.12 & 1500 & Very massive, rich in small dust (bold line) \\
213751210 & 3955 &  0.80 & 0.7  & 2.4E-5 & 0.1-10000	& 0.25 & 1500 & Very low-mass, maybe smaller radius? (bold line)\\
213823950 & 3729 &  0.92 & 0.6  & 2.9E-4 & 0.1-100	& 0.15 & 1500 & Normal-to-low disk mass, mass depends on grain size (bold line)\\
" 	  & 3729 &  0.92 & 0.6  & 2.4E-3 & 0.1-10000	& 0.15 & 1500 & Normal-to-low disk mass, mass depends on grain size (dashed line)\\
{\bf PTD}\\
213658737 & 3500 & 1.40   & 0.3  & 3.0E-3 & 0.1-100       & 0.13/0.22 & 1500/120 & Less flared, less dense inner disk (bold line)\\
213734649 & 3720 & 1.00   & 0.45 & 4.5E-4 & 0.1-2/0.1-100 & 0.15/0.20 & 1500/300 & Change in grain distribution inner/outer disk (bold line) \\
"	  & 3720 & 1.00   & 0.45 & 4.5E-4 & 0.1-100	  & 0.16      & 1500 	 & Cannot fit mid-IR and near-IR properly (dashed line)\\
213929250 & 3720 & 1.00   & 0.40 & 2.0E-4 & 0.1-2/0.1-100 & 0.16/0.18 & 1500/400 & Radial change in grain size and vertical scale height (bold line)\\
213929408 & 4060 & 1.15   & 0.80 & 8.0E-4 & 0.1-2/0.1-100 & 0.13/0.12 & 1500/200 & Optically thin, small-grain inner disk (bold line)\\
213945201 & 4060 & 1.40   & 0.80 & 0.04	  & 0.1-100	  & 0.07      & 1500	 & An extreme attempt to fit with a uniform disk model (dashed line)\\
"	  & 4060 & 1.40   & 0.80 & 6.4E-4 & 0.1-100 	  & hydro     & 1500/200 & Needs more flaring in inner part (dotted line)\\
"	  & 4060 & 1.40   & 0.80 & 4.8E-4 & 0.1-100	  & 0.13/0.11 & 1500/200 & Best with puffed inner disk (bold line)\\
213954058 & 3960 & 1.10   & 0.80 & 4.0E-4 & 0.1-100	  & 0.14/0.23 & 1500/150 & Needs a thin inner disk and a thick, flared outer disk (bold line)\\
{\bf TD}\\
213633647 & 5165 & 2.90   & 2.00 & 4.0E-6 & 20-1000       & 0.32/hydro & 500     & No small dust, inner hole (dashed line)\\
"         & 5165 & 2.90   & 2.00 & 9.6E-8 & 0.1-100       & 0.32       & 500     & Small grains, lower mass, silicate feature (bold line)\\	
213735713 & 3580 & 1.10   & 0.35 & 7.0E-5 & 0.1-10000     & 0.20       & 250     & Low mass, low T(R$_{in}$), high vertical scale height (bold line)\\
213756779 & 3650 & 0.75   & 0.40 & 1.6E-5 & 0.1-10000     & 0.15       & 350     & Low mass, inner hole (bold line) \\
"         & 3650 & 0.75   & 0.40 & 1.2E-4 & 20-10000      & 0.08       & 1500    & Low scale height, large grains, no hole, too much near-IR excess (dashed line)\\
213914837 & 3850 & 1.05   & 0.50 & 2.5E-7 & 0.1-100       & 0.06       & 1500    & Small grains, very low mass, overpredicts 8-12$\mu$m excess (bold line)\\
"         & 3850 & 1.05   & 0.50 & 4.0E-6 & 20-1000       & 0.06       & 1500    & No small grains, better fit to 8-12$\mu$m but still high (dashed line)\\
"         & 3850 & 1.05   & 0.50 & 3.0E-6 & 20-1000       & 0.06       & 800     & Best fit includes small hole (dotted line)\\
{\bf Depleted}\\
213733557 & 3700 & 0.80   & 0.50 & 1.5E-6 & 0.1-100       & 0.12       & 1500    & Without strong grain growth, very low mass (bold line)\\
"         & 3700 & 0.80   & 0.50 & 1.5E-5 & 0.1-10000     & 0.12       & 1500    & Larger grain size, higher mass, but still depleted (dotted line)\\
213854760 & 3720 & 1.90   & 0.40 & 8.0E-5 & 20-1000       & 0.06       & 1500    & Too much 8-12$\mu$m excess (bold line)\\
" 	  & 3720 & 1.90   & 0.40 & 2.0E-6 & 0.1-100       & 0.05       & 1500    & Extremely low vertical scale height underpredits 24$\mu$m (dashed line)\\
" 	  & 3720 & 1.90   & 0.40 & 8.0E-5 & 20-1000       & hydro      & 1500    & Hydrostatic equilibrium overpredicts 8-12$\mu$m excess (dotted line)\\
213930129 & 3785 & 1.69   & 0.70 & 7.0E-4 & 0.1-100       & 0.05       & 1500    & Very settled disk with small grains (bold line)\\
"         & 3785 & 1.69   & 0.70 & 5.6E-6 & 0.1-100       & 0.08       & 1500    & Lower mass disk, close to hydrostatic equilibrium (dashed line)\\
"         & 3785 & 1.69   & 0.70 & 4.9E-6 & 0.1-100       & hydro      & 1500    & Hydrostatic equilibrium, lower disk mass (dotted line)\\
213944898 & 3500 & 1.21   & 0.26 & 1.3E-5 & 0.1-100       & 0.08       & 1500    & Very settled disk with low mass still has too much mid-IR excess (dashed line)\\
"         & 3500 & 1.21   & 0.26 & 3.9E-6 & 0.1-100       & 0.10       & 1500    & Lower mass, higher vertical scale height (bold line)\\
"         & 3500 & 1.21   & 0.26 & 3.9E-5 & 0.1-10000     & 0.10       & 1500    & If strong grain growth is assumed, the mass may be higher (black dotted line)\\
\hline
\end{tabular}
\tablefoot{RADMC disk models for selected members
with different disk structures (normal full-disks, pre-transitional disks [PTD], 
transitional disks [TD], and dust-depleted disks [Depleted]). The names of the objects have been shortened to
fit the table. The stellar parameters (T$_{eff}$, R$_*$)
are determined from the optical observations and spectroscopy. The disk parameters are
modified to reproduce the observed SED in the whole wavelength range, starting with the simplest 
possible model (see Section \ref{radmc}). The disk masses are estimated assuming a gas-to-dust ratio of 100
and a collisional distribution for the dust with grain sizes a$_{min}$-a$_{max}$.
The outer disk radius is taken to be 100 in all cases. A larger radius would also result in a
larger dust mass.
For pre-transitional disks with physically different inner and outer disks, we give the grain distributions,
vertical scale height, and inner rim temperature for both the inner and the outer disk.
In case of different models, the comments include a reference to the appropriate figure.}
\end{table}
\end{landscape}

\subsection{What drives disk evolution and dispersal in Tr~37? Clues from accretion, dust, and environment \label{evolution}}

Putting together the members found in our previous papers and the newly found ones, we sum up to 
361 low-mass (spectral types G, K, M) members in Tr~37. The current results are
consistent with our previous estimates of the disk frequency (48$\pm$5\%, including all objects with
excesses down to 24$\mu$m, SA06a), although since this survey was strongly biased towards
objects with disks, we cannot re-estimate the disk fraction.

Among the 218 stars with disks (excluding Class I and uncertain objects), 
the full-disk objects are the most numerous, with a total of 111 
stars (about 51\% of the total). The least numerous are the pre-transitional disks (21 disks
or nearly 10\% of the total) and depleted disks (25 objects, about 11\% of the total).
The detection rate of PTD is low probably because of the lack of silicate feature observations. 
Our study of IRS spectra (SA11) suggests that the group of PTD is likely more
numerous than estimated here from photometry alone. 
The group of dust-depleted objects has to be also regarded as a lower limit, since detection of low 24$\mu$m
excesses among the latest-type stars are also compromised. 
Finally, transitional disks sum up to 61 objects,
or about 28\% of the sample.
In total, objects with evidence for inside-out evolution (considering as such both transitional and pre-transitional
disks) conform nearly 38\% of the total sample, being a substantial fraction of the total population of 
disked objects. An important degree of dust evolution has thus occurred in Tr~37,  
also in agreement with the significantly low accretion rates observed among cluster members 
(SA06b; Sicilia-Aguilar et al. 2010), suggestive of parallel dust and gas evolution in the disks.

Our current spectra do not allow us to estimate the accretion rates of the observed objects due to 
the uncertainty in the sky subtraction. However, considering as a typical value that of
3$\times$10$^{-9}$ M$_\odot$/yr (Sicilia-Aguilar et al. 2010), and an age of 4 Myr, 
one would expect that the disks would have had a minimum
disk mass of 0.012 M$_\odot$. To reconcile this value with the typical disk masses
observed, in general lower by at least a factor of few to one order of magnitude, strong grain
growth needs to be taken into account. This is particularly important in case of the disks with very low dust masses
and relatively strong H$\alpha$ emission (like 213751210+572436151), that would be about to lose their
disks within less than 10$^4$ yr.
If we do not include strong grain evolution or an anomalous gas to dust fraction, we
would need to assume that are witnessing a very special moment of disk evolution
in Tr~37, where a large fraction of disks is expected to disperse within the next $\sim$10$^5$ years, which
is highly unlikely
from the statistical point of view and considering the age spread among Tr~37 members.
Dust evolution (and thus a higher disk mass than expected) would be the only way to ensure that formation of 
giant planets is still possible among Tr~37 disks, if it has not occurred yet.

We also find evidence that not all disks evolve along the same path. Some TD
have very large 24$\mu$m excesses, while others appear to be also dust-depleted, a sign that inner holes can
appear both in massive, small-grain-rich, disks and in disks that are substantially depleted
of small dust grains. Other disks are depleted of small dust grains but still have near-IR excesses, suggesting
that small-dust depletion alone does not trigger the immediate opening of a hole in the disk. Moreover,
although dust-depleted disks will also probably suffer inside-out evolution later on and enter the class of
dust-depleted TD, they will not go through the "classical" TD phase (with no near-IR excess followed by a
sharp increase at mid-IR wavelengths and a strong 24$\mu$m excess), given that they do not have enough mass nor
vertical scale height. This would need to be considered when estimating the lifetime of "classical" TD.
The diversity of objects is a sign of the different disk dispersal processes at work in a cluster as old
as Tr~37 ($\sim$4 Myr for the central population). The differences observed between the
accretion behavior of full-disks and TD/dust-depleted objects discussed in Section \ref{accdisk}
suggests that gas accretion and dust evolution evolve in a parallel way. However, the fact that
about 50\% of TD are consistent with no accretion, while this is exceptionally rare among normal full-disks,
is also a sign that not all disks with similar SEDs correspond to physically similar objects,
confirming our previous results (Sicilia-Aguilar et al. 2010, SA11).

\begin{figure*}
\centering
\epsfig{file=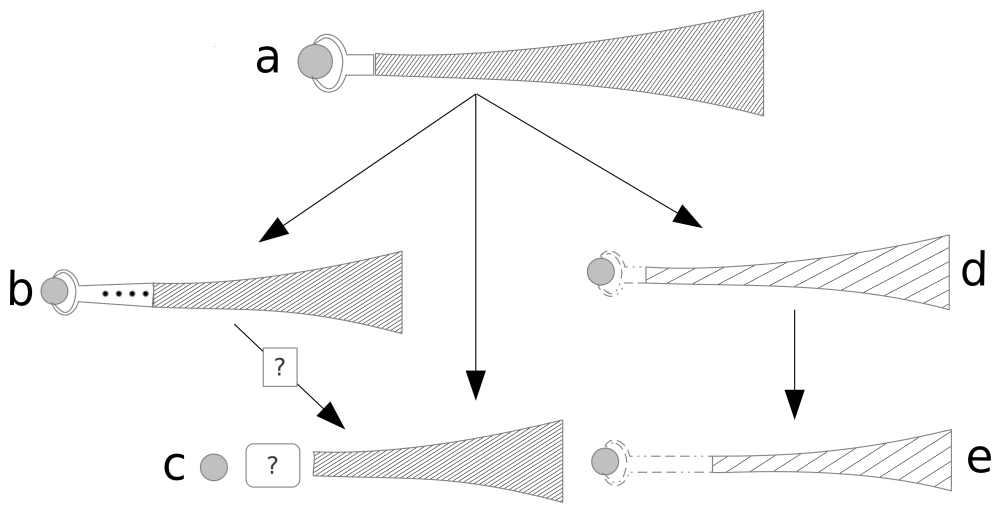,width=0.7\linewidth}
\caption{A sketch of the different evolutionary paths discussed in the text.   The full, accreting disk is
represented by the sketch 'a'. Such a disk can evolve in (at least) 3 different ways: suffering strong grain growth
and ending up as an accreting TD (b), suffering other inner-disk removal processes (e.g. photoevaporation, removal by
massive companions of stellar or planetary origin) to result in a non-accreting TD (c), or experiencing small-dust removal
at all radii (due for instance to generalized grain growth) and resulting in a small-dust depleted disk, typically with
reduced accretion (d). While accreting TD may later on transform into non-accreting TD or even dust-depleted TD (e), dust-depleted 
disks would not go through a typical TD stage (b,c) even if they later on also disperse from the inside-out (e).  
The list of evolved structures is not supposed to be complete, and the final outcome of each evolutionary path (e.g. star with
planetary system) is not constrained with our present data.
  \label{sketch-fig}}
\end{figure*}

The stars in Tr~37 reveal several paths of disk evolution (see a sketch of disk types
in Figure \ref{sketch-fig}). In general, evolution 
occurs from the inside-out, but the time scale when an inner hole develops in a disk appears to be different: in some
cases, substantial dust depletion, settling, or both, occur without the disk developing an inner hole, while in some other
cases, disks develop an inner hole while still being massive and very flared.
To open a hole by photoevaporation (e.g. Clarke et al. 2001; Alexander et al. 2006; Gorti et al. 2009),
the accretion rate or viscous transport through the disk needs to drop below a certain level first. Therefore, in
general objects with no significant accretion, low dust mass, and inner holes are good candidates to photoevaporation-related
holes or to very massive companions that can block accretion.
Nevertheless, very massive disks would also be less likely to disperse via photoevaporation until their mass and
accretion rate have decreased sufficiently to prevent refilling of the inner hole by viscous evolution. 
This would make it more likely that their holes are related to
other processes, like stellar companions or planet formation. Objects with transitional and dust-depleted
disks that still have significant accretion are good candidates to suffer strong grain growth (maybe combined
with settling), which would affect the IR excess but not the accretion rate, or very low-mass companions, that
would not block the flow of gas through the gap.
In the case of PTD, the differences in dust properties (especially, grain size) observed between the
inner and the outer disk could be due to dust filtering in a planet-formation scenario (Rice et al. 2006), as
we had also previously suggested in SA11. Recent spatially resolved observations show that the possibility of
dust filtering/dust trapping in certain parts of the disk is a real one, and that gas flow accross gaps appears
to occur (Van der Marel et al. 2013; Casassus et al. 2013). 
More data, in particular, silicate feature observations, and 
ideally also spatially resolved observations, would be needed to better define the class of PTD.

\begin{figure}
\centering
\epsfig{file=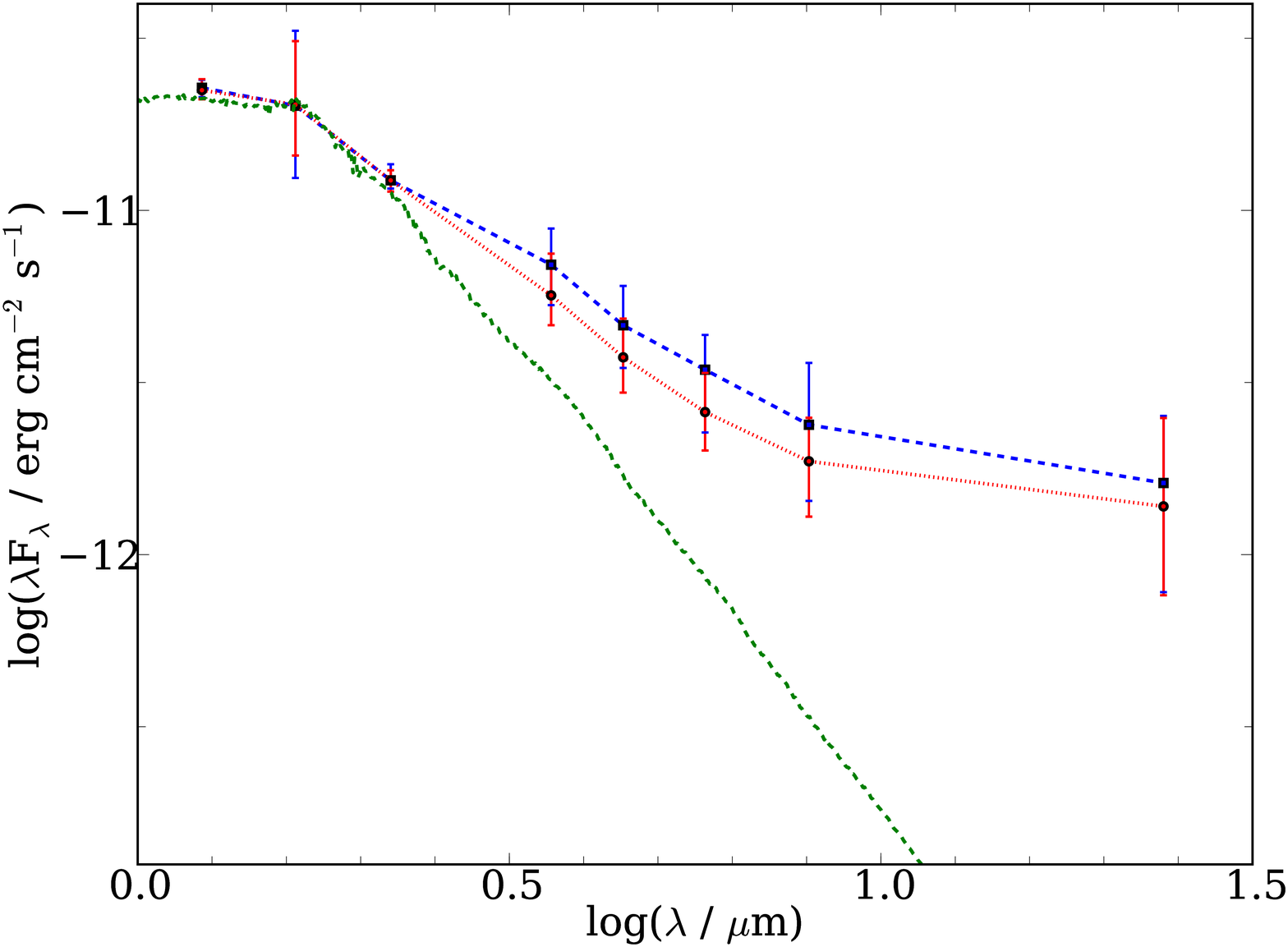,width=0.9\linewidth,clip=}
\caption{Median SED for the objects with disks for spectral type K (blue dashed line) and M (red dotted line),
with quartiles at each wavelength.  \label{mediandisk-fig}}
\end{figure}

We also explored the dependency of SED shape and disk structure with the spectral type/stellar mass.
Figure \ref{mediandisk-fig} shows a plot of the median SED for the K-type and
M-type objects with disks, scaled in both cases to the flux
in band H and corrected by the individual extinction values of the objects. The plot
includes the results of all the known members with well-defined IR excesses, excluding
those with anomalous SEDs and/or photometric problems. This sums up 
to 76 K-type stars and 109 M-type stars. Although M-type stars have systematically
lower IR excesses over the photosphere and the upper quartile for M-type stars hardly reaches
the mean values for the K-type objects, the difference is not as dramatic as observed in the Coronet
cluster (Currie \& Sicilia-Aguilar 2011). The 24$\mu$m data in Tr~37 are not complete, 
so we expect to miss a substantial part of the objects with low 24$\mu$m excesses, especially among the fainter M-type
stars.  Nevertheless, the most striking result in the young (1-2 Myr) Coronet cluster is not
the presence of disks with very low masses, but the absence of massive disks among
the M-type stars, which is not the case in Tr~37. The analysis of the H$\alpha$ EW in the different spectral type 
classes in Tr~37 also suggest that K- and M-type stars in Tr~37 are more similar in terms of accretion and disk 
properties than their counterparts
in the Coronet cluster. Disks around M-type stars in Tr~37 are apparently \textit{less evolved} than
in the Coronet cluster, although we would expect the opposite from their ages, 
a signature that time evolution alone cannot explain the observed disk 
properties in Tr~37 and the Coronet cluster.

Besides age, the main differences between the Coronet cluster and Tr~37 are the number of members and
the stellar density. In the Coronet cluster, we find about 50 stars (including resolved multiple
systems) within a radius of about 0.15pc (Sicilia-Aguilar et al. 2013). The membership count is probably close to
complete for spectral types between B9 and M6, considering that the cluster has been observed in the optical, X ray,
IR, and submillimeter (e.g. L\'{o}pez-Mart\'{i} et al. 2005, 2010; Groppi et al. 2004 2007; Peterson et al. 2011;
Sicilia-Aguilar et al. 2008, 2011a). Taking into account only the number of stars and not their mass, we 
arrive to a stellar density of about 3000 stars/pc$^3$.
In Tr~37, if we take a cluster radius of 5 pc, we count about 360 stars with spectral types O6 to M6, including
both disked and diskless objects. If we account
for a similar number of stars without disks that we may have missed in our surveys, 
and allow for a binarity rate of 50\%, we arrive to about 1000
stars in total, resulting in a density of 2 stars/pc$^3$. Even if our member number estimates are wrong by
a factor of few, the density difference between both regions would still be close to 3 orders of magnitude. 

The outskirts of 
the cluster may be less dense than the cluster center (e.g. Barentsen et al. 2011), but the difference in stellar
density between the Coronet and Tr~37 would still be very large. Even if we consider that Tr~37 has suffered substantial
expansion to arrive to its current stage (at a typical rate of 1pc/Myr), an initial cluster size of 2~pc would
only produce a density of 30 stars/pc$^3$. Moreover, Spitzer observations (SA06a) and isochrone
considerations (SA05; Getman et al. 2012) suggest that the population in the
outskirts is most likely younger and formed in-situ, not by expansion of the original cluster. 
According to Section \S 7 of Getman et al. (2012) there are 
235 (>60) stars down to 0.1M$_\odot$ near (inside) the IC~1396A globule, within 
the volume of roughly 75 pc$^3$ (7 pc$^3$). Assuming a binary rate of 50\%, 
this gives 5 stars/pc$^3$ (10 stars/pc$^3$) near (inside) the globule. This 
source density is still much lower than that of the Coronet cluster.
The immediate result is that, while the early formation phases of stars in a compact, dense cluster like the Coronet would
be heavily compromised by star-star interactions that could affect envelopes and disks, the environment in Tr~37
would have been much less interactive, allowing for larger envelopes and also larger disks masses. 
The initial disk structure may also pre-determine the evolutionary path that the disk will follow. Later on,
other phenomena like the effect of the central O6 star may also contribute to disk dispersal in Tr~37
(e.g. Mercer et al. 2009).

\subsection{The star-cloud interactions in Tr~37: Mini-clusters and sequential star formation}

\begin{figure}
\centering
\epsfig{file=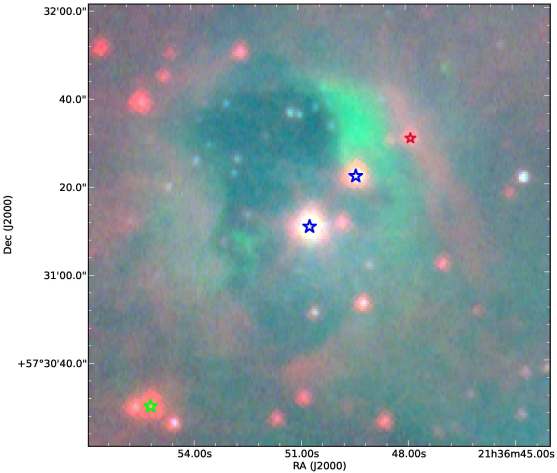,width=0.9\linewidth,clip=}
\caption{Detail of the 3-color (IRAC1, [S~II] combined at 6716 and 6730\AA, and R band as red, green,
and blue, respectively) image of the IC 1396 A globule, zooming around V390 Cep (in the center of the clearing) 
and 14-141 (marked in blue). Embedded
Spitzer-detected objects are marked as red stars.\label{forbidden3-fig}}
\end{figure}

Besides the fact that the O6 star HD~206267 shapes the large
IC~1396A globule and other structures in the outskirts of Tr~37, our
forbidden line images also reveal a smaller-scale interaction between some of the
low-mass members and the remaining cloud.
Figure \ref{forbidden3-fig} reveals a complex shocked structure in the
surroundings of V390 Cep (21:36:50.72 +57:31:10.7, also known as MVA-60; Marschall \& Van Altena 1987)
and the K6 star 14-141 (Sicilia-Aguilar et al. 2004). 
The star 14-141 is known to show variable accretion and to be associated to
variable forbidden line emission suggestive of time-variable shocks in the
proximities of the star (Sicilia-Aguilar et al. 2010).
In addition, the [S II] images reveal a large area emitting in forbidden lines,
associated to the rim of the cloud in the proximity of 14-141. It is not
possible to distinguish whether the shock is due to 14-141 or associated to
more embedded objects within the cloud. There is a very embedded, reddened object
that can be seen in the Spitzer images, although due to the variable background it is
not possible to extract accurate photometric information from it.

There is also evidence of a small jet or shock on the other side of the bubble surrounding
14-141, with no object related to it. From its location, it could be related to V390 Cep, which 
is classified as a variable Ae star (despite its lack of IR excess at any of the Spitzer wavelengths).
The shock emission would be located
at about 13000 AU projected distance from V390 Cep.
In addition, our Hectospec spectra reveal forbidden emission towards a few other
stars (213744131+573331130, 213810759+574013683, 213942378+573348653, and 
214059633+572210994; see Table \ref{emission-table}). All this suggest that even the low-mass stars
in the cluster contribute to shaping the surrounding cloud, which could have an effect on
local star formation.

\begin{figure*}
\centering
\epsfig{file=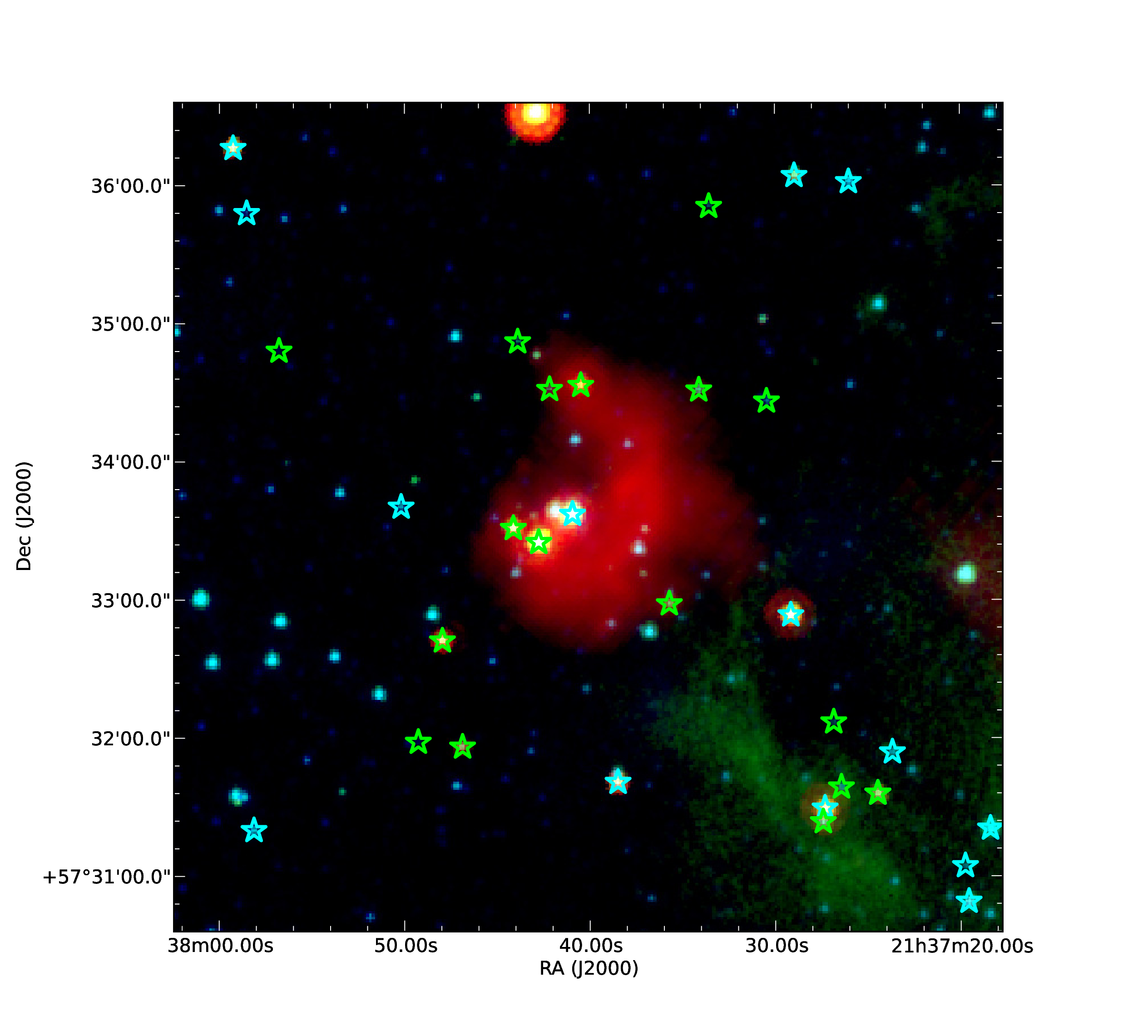,width=0.48\linewidth,clip=}
\epsfig{file=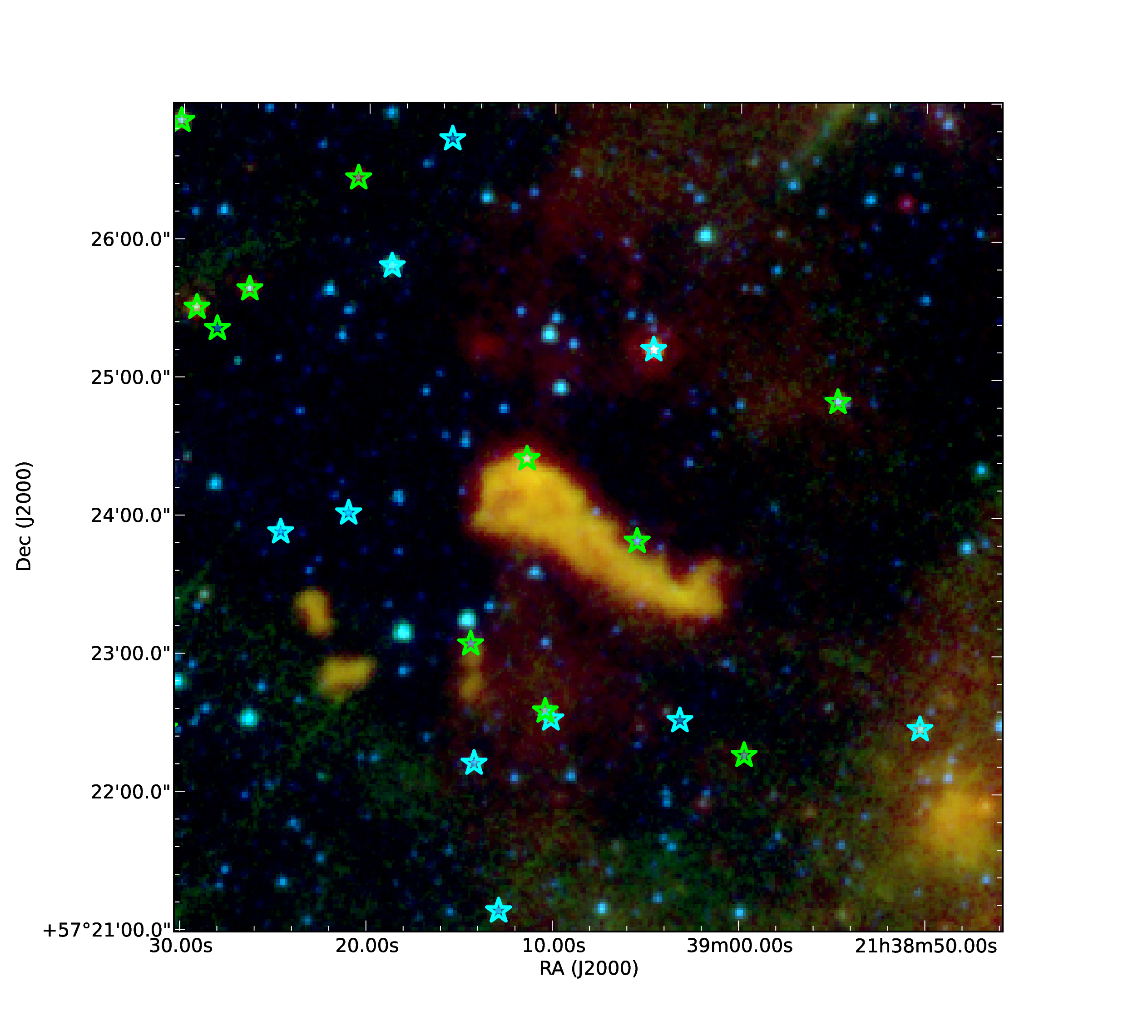,width=0.48\linewidth,clip=}
\caption{Two 3-color (3.6, 8, 24 $\mu$m) image of the small globules or mini-clusters in Tr~37.
Previously known members are marked as cyan stars, new members (from this work) are marked as green stars. \label{miniglobule-fig}}
\end{figure*}

We had already mentioned the age difference between the 
solar-type stars in the cluster center compared to the most embedded objects to the west
of Tr~37, located within the IC~1396A globule (SA05; SA06a; Barentsen et al. 2011; Getman et al. 2012). 
In this spectroscopic survey, we took advantage of our previous knowledge of the
cloud structure to obtain spectra of the objects associated with some smaller globules
(of the order of 20"-40", or approximately 0.1-0.2 pc at a distance of 870 pc;
Contreras et al. 2002). The most remarkable structures
contain 6 and 2 stars clearly associated with the extended
IR emission, respectively, plus other surrounding objects that could be either associated or projected
onto the nebula, forming what we call here "mini-clusters" (Figure \ref{miniglobule-fig}). 

The larger globule or mini-cluster (Figure \ref{miniglobule-fig}, left) is remarkable because of the very
strong H$\alpha$ emission of all its associated stars, which have 
very strong IR excesses as well. This small stellar group had been already noticed in the H$\alpha$ survey by
Barentsen et al. (2011) and in the X-ray/IR study of Getman et al. (2012), Section 8.3.2.
The region also host the binary B3+B5 star CCDM+5734,
which was found to have a transitional disk with a large hole and evidence of
small silicate dust (Sicilia-Aguilar et al. 2007).
The second structure (Figure \ref{miniglobule-fig}, right) contains the star 213911452+572425205. This object
appears as emerging from the cloud tip and shows
a large IR excess and a large number of accretion-related emission lines in
its spectrum. The globule also contains 213905519+572349596, a star without evidence of disk nor accretion.
Although obtaining ages of these embedded objects is problematic, the extreme H$\alpha$ values 
and IR excesses of all of them (except 213905519+572349596) could indicate that the stars are
younger than the general Tr~37 population, either related to triggered o sequential star formation
within small surviving globules in the cluster, or maybe to structures in the younger outskirts of the
cluster, seen in projection, or clumpy star formation.

\section{Summary and conclusions \label{conclu}}

We performed a spectroscopic survey targeting the low-mass 
members of the 4 Myr-old cluster Tr~37. The spectra allowed us
to identify a large number of new members and, together with our already
available data (optical photometry, 2MASS, and IR Spitzer IRAC/MIPS data), to
study the connection between accretion indicators and IR excesses from
protoplanetary disks. From our study, Tr~37 emerges as
a relatively evolved young cluster where its surviving disks display a variety of evolutionary
signs. The results of this work are summarized below:

\begin{itemize}
\item The survey provided data on spectral types, accretion, and extinction for
205 cluster members and probable members, leading to the discovery of 78 new cluster members, 
and 64 probable members. Most of these members are M-type (M0-M5) stars with IR excesses
typical of protoplanetary disks and accretion indicators. 

\item Gas accretion, detected via the H$\alpha$ EW, is strongly correlated with the
presence of disks. A KS
test including the new and the previously known members reveals
a nominally 0 probability that both disked (with all types of disks) and diskless
objects are drawn from the same distribution.
We also find significant differences between the accretion behavior of 
full-disks and disks with strong evidence of evolution (TD and dust-depleted disks).
About half of the TD are consistent with no accretion (being similar to diskless 
stars), while the other half have H$\alpha$ values similar to full-disks. 
Dust-depleted disks, in particular those around M-type stars, also show significantly
lower H$\alpha$ EW. This suggests that
dust and gas evolve in a parallel way: Objects with no evidence of dust
evolution are very unlikely to have suffered accretion termination, while accretion termination
(or very strong suppression of accretion) is common among transitional and dust-depleted disks.
The difference in accretion behavior between dust depleted
and full disks also suggests that objects with very low 24$\mu$m excesses are in a different
evolutionary stage, although it is not possible to tell if this is caused by parallel dust and
gas evolution, by strong dust settling after partial gas depletion, or by both.

\item We find a tentative trend 
(limited by low number statistics) that
accreting TD are more frequent among M-type stars than among GK-type stars, 
which could be either related to the larger holes in the disks around GK-type stars
for the same IR excess, or indicate that accretion last longer in the TD around M-type
stars. GK-type dust-depleted disks are generally accreting, in contrast with M-type
dust-depleted disks, pointing to a faster termination of accretion in M-type 
stars after global dust evolution.
Further observations of similar objects are required to confirm or reject these points.

\item We used the RADMC radiative transfer code to model 
a sample of disks with different types of SEDs. Full-disks can
be reproduced with simple models with different degrees of
grain growth, vertical scale height, and disk masses. Transitional disks require 
an inner hole to account for the lack of near-IR excess, although the
hole size (or the temperature of the inner disk rim) is highly
dependent on the dust properties and vertical scale height.
Pre-transitional disks models require a change in disk properties between the innermost
and outer disk, although there is a high degree of
degeneracy between location of the radius at which the properties change, dust distributions
on both sides, and vertical scale heights. Dust-depleted disks with very low 8-24$\mu$m fluxes
and reduced near-IR excesses require substantial mass depletion, if we assume the
disk is in hydrostatic equilibrium. Strong settling may contribute to low near-IR
excess, but a dust mass in small ($<$20$\mu$m) grains significantly lower than in
ful disks is still needed
to account for the low mid-IR fluxes and the steep shape of the SED. 

\item Including all the Tr~37 population identified in this and in the previous study
(over 200 objects with spectral types G,K, and M), we studied the prevalence of signs
of inside-out evolution and dust depletion in the disks. More than 1/3 of the disks
show signs of inside-out evolution (inner holes, or changes in disk properties 
between the inner and outermost disk). About 10\% of the disks have significant dust depletion,
resulting in very low mid-IR fluxes and typically low accretion rates. Nearly 20\% of the
disks with inner holes (TD) have also very low mid-IR excesses and are most
likely dust depleted as well. Among full-disks, substantial evolution in the
form of grain growth needs also to be assumed to obtain disk masses in
the appropriate range for the accretion rates observed. 

\item The diversity of disks in Tr~37 suggests that not all disks follow the same 
evolutionary path: while inside-out evolution seems to be the rule, we find that
some disks develop inner holes while keeping a relatively massive outer disk, while others
lose a significant fraction of their small dust grains before their inner parts are
cleared. This points to several effects contributing to disk dispersal.

\item Our study of Tr~37 does not reveal the striking differences
observed the young Coronet cluster, among K-type and M-type stars
(Currie \& Sicilia-Aguilar 2011).
Non-accreting and depleted disks among M-type stars
are not as frequent as in the Coronet, even though Tr~37 is substantially older. 
A possible explanation is the very different
environment and, in particular, the stellar density in both regions. While recent
studies suggest a density of the order of 3000 stars/pc$^3$ in the sparse Coronet
cluster, the density in Tr~37 is about 2-3 orders of magnitude lower.
This suggests that the differences in disk properties observed between both regions could
be related to very different environment and initial conditions, like early
interactions. Frequent close interactions at a protostellar level could result in lower
initial disk masses and a lack of massive disks among low-mass stars, which is what we observe
in the Coronet cluster.

\item Finally, we also studied the stellar content of
several small globules in Tr~37, finding that some of them harbor mini-clusters with
a few members. The stars within these mini-clusters show preferentially
full-disks with strong IR excesses and strong emission lines. This suggests that the objects
could be younger than the extended Tr~37 population, which could be indicative of
very small-scale sequential (or clumpy) star formation in the region. 

\end{itemize}

Acknowledgments: The authors are grateful to V.L. Afanasiev for conducting the SCORPIO observations, 
and V.V. Krushinsky and P.A. Boley for their help in the preparation of the observations with the 6m Telescope of SAO RAS,
and also to P. Berlind, N. Caldwell, and M. Calkins at the MMT observatory. We also
thank C. Dullemond for his help with the RADMC code, and the anonymous referee for his/her careful
comments that helped to clarify and organize this paper.

Based on observations collected at the German-Spanish Astronomical Center, Calar Alto, jointly 
operated by the Max-Planck-Institut f\"{u}r Astronomie Heidelberg and the Instituto de Astrof\'{\i}sica 
de Andaluc\'{\i}a (CSIC).
We also thank Calar Alto Observatory for allocation of director's discretionary time to this program.

This work is based on observations made with the Spitzer Space Telescope, which is operated by
the Jet Propulsion Laboratory, California Institute of Technology under a contract with NASA.
This publication makes use of data products from the Wide-field Infrared Survey Explorer, which 
is a joint project of the University of California, Los Angeles, and the Jet Propulsion Laboratory/California 
Institute of Technology, funded by the National Aeronautics and Space Administration. It
also makes use of data products from the Two Micron All Sky Survey, which is a joint project of the
University of Massachusetts and the Infrared Processing and Analysis Center/California Institute
of Technology, funded by the National Aeronautics and Space Administration and the National
Science Foundation.

ASA acknowledges support of the Spanish MICINN/MINECO "Ram\'{o}n y Cajal" program, grant number RYC-2010-06164,
and the action ``Proyectos de Investigaci\'{o}n fundamental no orientada", grant number
AYA2012-35008. MF is also supported by AYA2012-35008.

\onecolumn

\Online

\begin{appendix} 

\section{Complete data tables of observed objects, their classification, and the photometry of
members and probable members \label{tables-appendix}}

The following tables contain the full information (object type and membership) of the objects observed with
Hectospec/MMT (Table \ref{spec-table}), followed by the photometry data on members and probable
members (Tables \ref{opt2mass-table} and  \ref{spitzer-table}).

\begin{footnotesize}

\begin{longtable}{lcccccccl}
\caption{\label{spec-table} Observations and spectroscopic information. 
This table lists all observed objects, their Li and H$\alpha$ EW,
and membership criteria. The objects IDs are based on the spectrograph (optical) coordinates.
If H$\alpha$ appears in absorption, we label it as "Abs."
in the corresponding column.
If Lithium absorption can be excluded for EW larger than 0.1, we label it as `N' in the
corresponding column.
Membership and presence of disks are marked as `Y' (yes),
`N' (no), `P' (probable), and `pN' (probably not). In case the disk excess was observed
only at one wavelength, we label it (e.g. P(24) indicates that the object has probably an
excess, but only at 24$\mu$m). Undetermined values of the 
spectral type, uncertain membership, and uncertain IR fluxes
are marked by `X', objects with spectral types earlier than G are marked with `E'. 
The errors in the spectral type are of 0.5 subtypes for
objects K5 or later, 1 subtype for early K stars, and 2-3 subtypes for G-type
objects. The best spectral types (in case of multiple values due to different observations
with different S/N) are marked with $^b$. 
Objects previously identified by Sicilia-Aguilar et al. (2005, 2006a,b) are marked
with $^1$ (and their previous names are given in the comments), those identified by  Barentsen et al. (2011) are marked with "$^2$",
those in common with Morales-Calder\'{o}n et al.(2009) are marked with "$^3$", 
objects with X-ray counterparts in Getman et al.(2012) are marked with "$^4$", and
further IR-excess sources in the same paper are marked with "$^5$".
In the comments field, we mention if there is any problem due to IR nebular
emission (Neb.), nearby objects (Nobj.), if the SED suggest that the object is
variable (Var.), if the object appears older than 10 Myr in the V vs V-I diagram (oVVI),
or any other observation regarding the star/disk.} \\
\hline\hline
Spect. ID & H$\alpha$ EW &Li EW  & Sp. Type & A$_V$ & Disk? & Campaign & Member? & Comments\\
 &  (\AA) & (\AA) &  &  (mag) &  &  &  & \\
\hline
\endfirsthead
\caption{Continued.}\\
\hline\hline
Spect. ID & H$\alpha$ EW &Li EW & Sp. Type & A$_V$  & Disk? & Campaign & Member? & Comments\\
 &  (\AA) & (\AA) &  &  (mag) &  &  &  & \\
\hline
\endhead
\hline
213539203+573021561 & Abs. & --- & X    & --- & X & 2009-MMT & N   &      \\
213541592+573127753 & Abs. & --- & X    & --- & X & 2010-MMT & N   &      \\
213542993+573337049 & -16  & --- & M4.0 & 1.1$\pm$0.6 & Y & 2009-MMT & Y$^4$   &      \\
213554323+573443434 & -8   & --- & Late & --- & N & 2009-MMT & N   &      \\
213555607+573003406 & Abs. & N   & G9.0   & 6.4$\pm$1.9 & N & 2010-MMT & N   &      \\
213601910+573430689 & -1   & --- & K3.0 & 3.1$\pm$0.0 & Y & 2009-MMT & X   &      \\
213602343+573326502 & Abs. & --- & G9.0 & 2.8$\pm$0.2 & Y & 2009-MMT & N   &      \\
213602762+572834046 & Abs. & --- & X    & --- & N & 2009-MMT & N   &      \\
\hline
\end{longtable}

\end{footnotesize}

\begin{landscape}
\begin{longtable}{lccccccc}
\caption{\label{opt2mass-table} Optical and 2MASS data for confirmed members and probable members.
$^a$ Anomalous optical colors.$^b$Near HD206267. $^c$Near 21-840, likely contaminated. $^d$ Near
bright star, probably contaminated.} \\
\hline\hline
Spect. ID & V (mag) &R$_J$ (mag) &I$_J$ (mag)& 2MASS ID &J (mag) &H (mag)& K (mag)\\
\hline
\endfirsthead
\caption{Continued.}\\
\hline\hline
Spect. ID & V (mag) &R$_J$ (mag) &I$_J$ (mag)& 2MASS ID &J (mag) &H (mag)& K (mag)\\
\hline
\endhead
\hline
213542993+573337049 & 19.476$\pm$0.011 & 17.613$\pm$0.030 & 15.506$\pm$0.002 & 21354299+5733370 & 14.302$\pm$0.037 & 13.439$\pm$0.033 & 13.075$\pm$0.036 \\            
213607980+572637096 & --- & --- & --- & 21360798+5726371 & 18.366: & 17.484: & 14.507$\pm$0.088 \\            
213618699+573626870 & 19.516$\pm$0.011 & 17.693$\pm$0.030 & 15.870$\pm$0.003 & 21361870+5736268 & 14.707$\pm$0.046 & 13.911$\pm$0.056 & 13.600: \\            
213619969+573527805 & 21.154$\pm$0.036 & 19.230$\pm$0.032 & 16.871$\pm$0.004 & 21361996+5735278 & 15.721$\pm$0.065 & 15.084$\pm$0.089 & 14.724$\pm$0.089 \\            
213622159+572327898 & 19.269$\pm$0.013 & 17.468$\pm$0.023 & 15.416$\pm$0.002  & 21362216+5723278 & 14.263$\pm$0.044  & 13.496$\pm$0.050  & 13.173$\pm$0.039   \\
213622379+573141225 & 21.277$\pm$0.039 & 19.006$\pm$0.031 & 16.251$\pm$0.003 & 21362238+5731412 & 14.567$\pm$0.036 & 13.792$\pm$0.033 & 13.404$\pm$0.037 \\            
213625078+572750265 & 20.435$\pm$0.020 & 18.558$\pm$0.024 & 16.399$\pm$0.004 & 21362507+5727502 & 14.952$\pm$0.044 & 14.031$\pm$0.048 & 13.518$\pm$0.040 \\            
213626897+573304351 & 20.963$\pm$0.035 & 18.868$\pm$0.031 & 16.709$\pm$0.004 & 21362690+5733043 & 15.217$\pm$0.055 & 14.158: & 13.869: \\
\hline
\end{longtable}
\end{landscape}

\begin{footnotesize}

\begin{landscape}
\begin{longtable}{lccccccl}
\caption{ \label{spitzer-table} Spitzer IRAC and MIPS data for confirmed members and probable members.
Under 'SED' we specify the SED type (ND = no disk; ClassI= Class I object; Full= typical full disk; TD = transition disk; PTD = pre-transition
disk; Dep.= dust-depleted disk; Unc. = Uncertain SED type). Objects listed as "marginal detection" at 24$\mu$m have probably
excesses but their photometry is compromised by low S/N or different sources of contamination.} \\
\hline\hline
Spect. ID &IRAC-1 ($\mu$Jy) &IRAC-2 ($\mu$Jy) &IRAC-3 ($\mu$Jy) &IRAC-4 ($\mu$Jy)  &MIPS-1 ($\mu$Jy) & SED & Comments\\
\hline
\endfirsthead
\caption{Continued.}\\
\hline\hline
Spect. ID &IRAC-1 ($\mu$Jy) &IRAC-2 ($\mu$Jy) &IRAC-3 ($\mu$Jy) &IRAC-4 ($\mu$Jy)  &MIPS-1 ($\mu$Jy) & SED & Comments\\
\hline
\endhead
\hline
213542993+573337049 & 2683$\pm$72 & 1982$\pm$57 & 1808$\pm$222 & 1211$\pm$86 & 881$\pm$157  &  Dep.    &    \\
213607980+572637096 & 19571$\pm$94 & 39189$\pm$78 & 103489$\pm$512 & --- & 1159200$\pm$6406  &  ClassI    &    \\
213618699+573626870 & 1251$\pm$53 & 835$\pm$53 & 708$\pm$197 & 146$\pm$43 & ---  &   ND   &    \\
213619969+573527805 & 522$\pm$3 & 343$\pm$3 & 257$\pm$15 & 163$\pm$48 & ---  &  ND    &    \\
213622159+572327898 & 2220$\pm$53 & 1434$\pm$58 & 1244$\pm$273 & --- & --- &  Neb. contamination\\
213622379+573141225 & 1669$\pm$43 & 1066$\pm$55 & 1305$\pm$196 & 1984$\pm$230 & ---  &   ND   &    \\
213625078+572750265 & 3060$\pm$58 & 3013$\pm$61 & 3021$\pm$224 & 4872$\pm$26 & 9196$\pm$103  &  Full    &    \\
213626897+573304351 & 1392$\pm$54 & 750$\pm$54 & --- & --- & ---  &   ND    & Neb. \\
\hline
\end{longtable}
\end{landscape}

\end{footnotesize}

\begin{longtable}{lcccc}
\caption{ \label{wise-table} WISE data for confirmed members and probable members. Only good quality data in
agreement with the Spitzer fluxes are taken into account (see text). Uncertain data are marked with ':'. Data that were not
used in the SEDs (due to uncertainty or evident contamination) are marked with $^a$.} \\
\hline\hline
Spect. ID & WISE-1 (mJy) & WISE-2 (mJy) & WISE-3 (mJy) & WISE-4 (mJy)\\
\hline
\endfirsthead
\caption{Continued.}\\
\hline\hline
Spect. ID & WISE-1 (mJy) & WISE-2 (mJy) & WISE-3 (mJy) & WISE-4 (mJy)\\
\hline
\endhead
\hline
213633647+573517477 & 11.54$\pm$0.26 & 5.85$\pm$0.11 &   17.35$\pm$0.15 &  3.56$\pm$1.13$^a$    \\
213639147+572953326 & 148.01$\pm$3.54 & 164.49$\pm$3.03 &  1699.31$\pm$4.54 &  343.75$\pm$12.03$^a$    \\
213642470+572523186 & 1.63$\pm$0.04 &  1.27$\pm$0.03 &   8.44$\pm$0.09 &  2.38$\pm$0.99$^a$    \\
213655283+572551668 & 3.68$\pm$0.08 &  1.80$\pm$0.04 &  5.26$\pm$0.10 &   2.56:$^a$    \\
213659108+573905636 & 6.55$\pm$0.15 &  6.66$\pm$0.12 &  156.60$\pm$0.30 &   54.89$\pm$1.77$^a$    \\
213701319+573418289 & 1.66$\pm$0.06 &   1.25$\pm$0.04 &  27.49$\pm$0.13 &   10.13$\pm$1.65 $^a$  \\
213708137+573616213 & 2.36$\pm$0.06 & 2.14$\pm$0.05 &  28.03$\pm$0.19 &  9.78$\pm$1.04$^a$    \\
213708879+572107012 & 1.75$\pm$0.04 &   1.41$\pm$0.04 &  5.01$\pm$0.10 &  2.06:    \\
\hline
\end{longtable}

\begin{longtable}{lccc}
\caption{ \label{emission-table} Other emission lines observed.\\
Stars with multiple emission lines and their identifications.
The lines have been identified according to Sicilia-Aguilar et al.(2012), the NIST database,
Appenzeller et al. (1986), and Hamman \& Persson (1992). 
$^a$ Probably nebular. } \\
\hline\hline
Spect. ID &Line $\lambda$ (\AA)  &Obs. $\lambda$ (\AA)  &EW (\AA)  \\
\hline
\endfirsthead
\caption{Continued.}\\
\hline\hline
Spect. ID &Line $\lambda$ (\AA)  &Obs. $\lambda$ (\AA)  &EW (\AA)  \\
\hline
\endhead
\hline
213659108+573905636 & H$\alpha$ 6562.6  & 6562.5 & -76.5 \\ 
 " & H$\beta$ 4861.3 & 4860.9 & -22.0 \\ 
 " & H$\delta$ 4101.7 & 4101.6 & -7.2 \\ 
 " & H$\gamma$ 4340.5 & 4340.1 & -10.8 \\ 
 " & Ca~II 8498.0 & 8498.2 & -4.0 \\ 
 " & Ca~II 8542.1  & 8542.1 & -3.7 \\ 
 " & Ca~II 8662.1 & 8662.2 & -3.7 \\ 
213659472+573134908 & H$\alpha$ 6562.6 & 6564.5 & -108.3 \\
\hline
\end{longtable}

\section{SEDs of all the members with IR excesses \label{seds-appendix}}

\begin{figure*}
\centering
\begin{tabular}{ccccc}
\epsfig{file=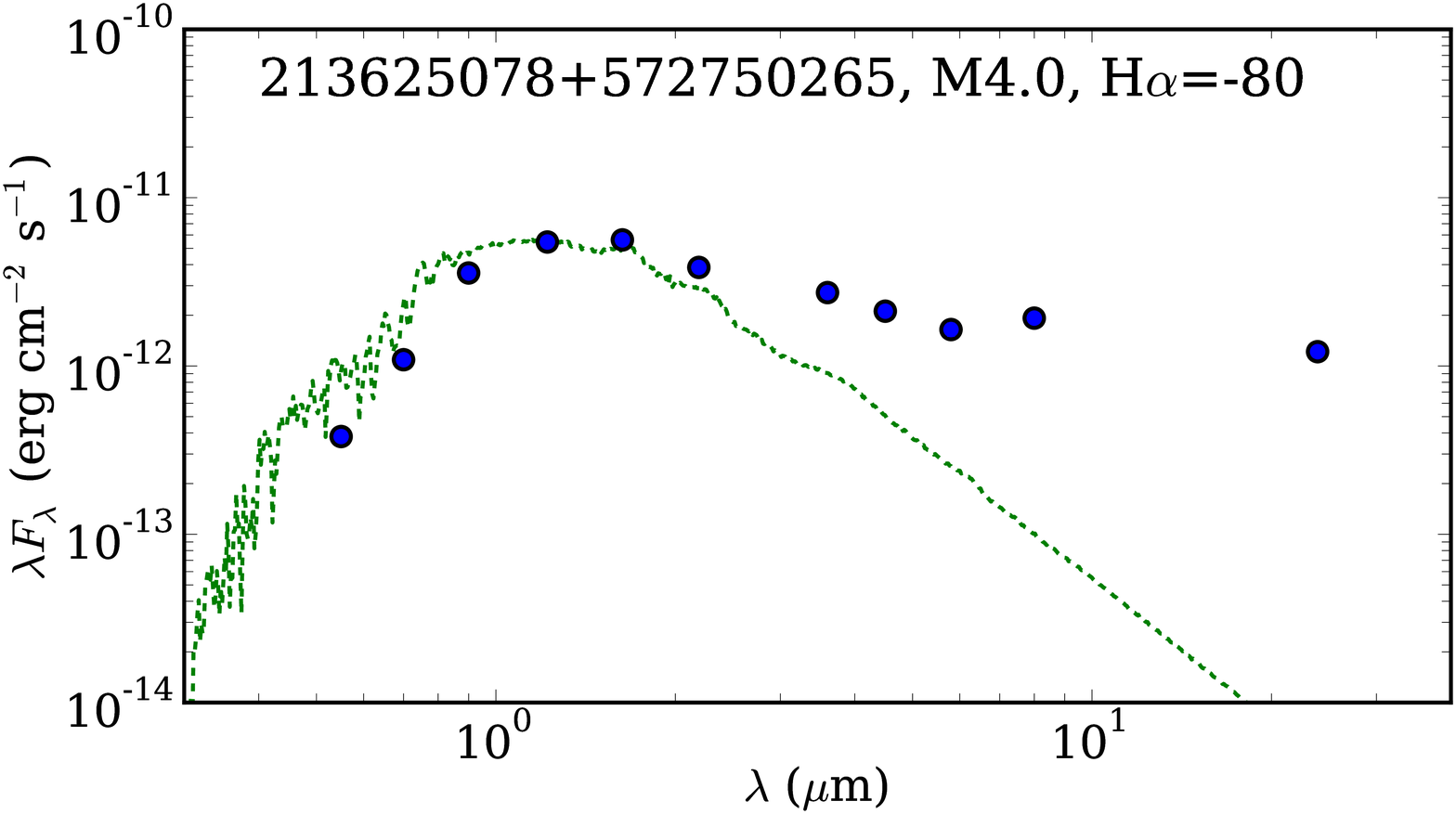,width=0.24\linewidth,clip=} &
\epsfig{file=213636909+573132683.eps,width=0.24\linewidth,clip=} &
\epsfig{file=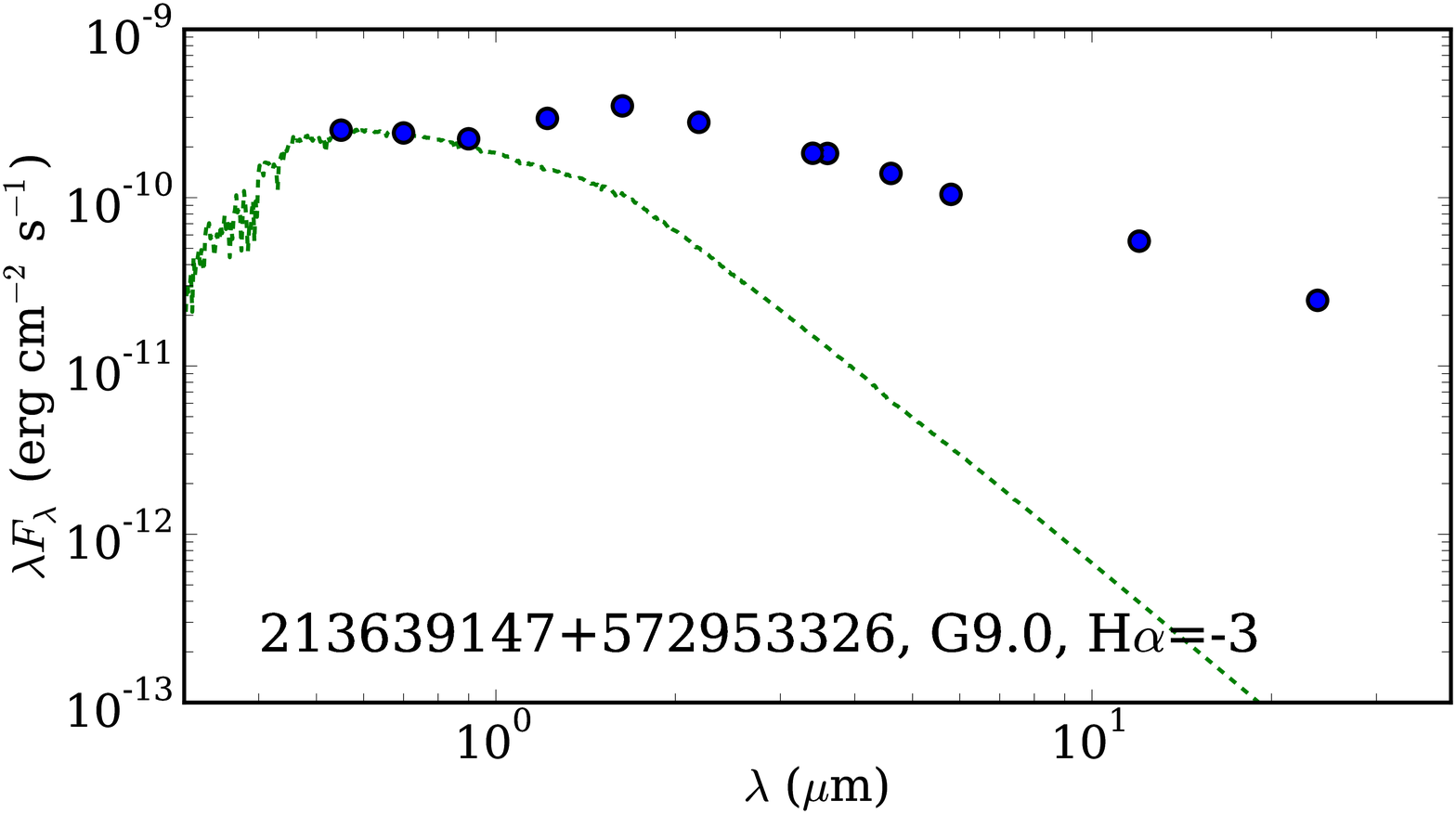,width=0.24\linewidth,clip=} &
\epsfig{file=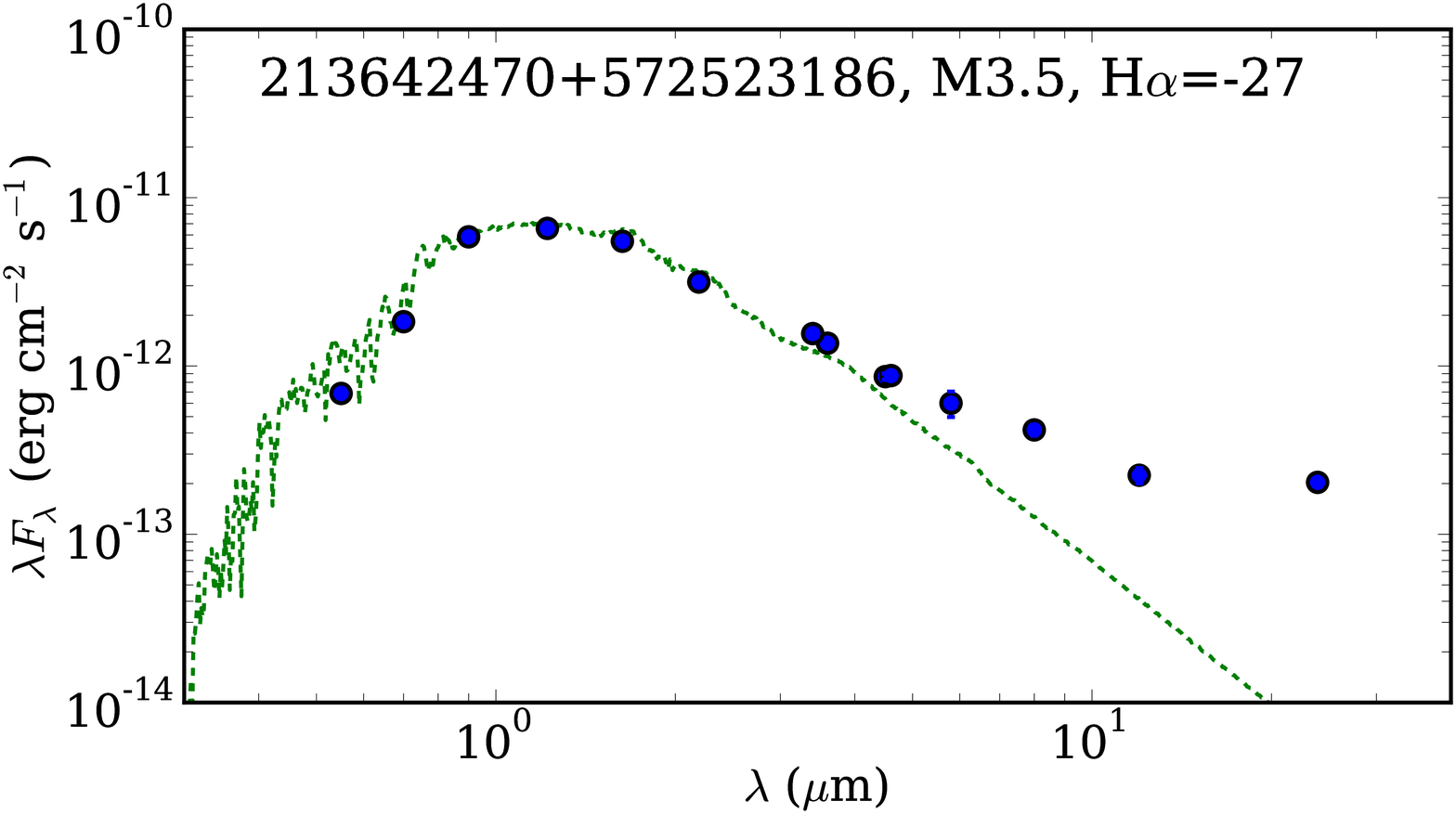,width=0.24\linewidth,clip=} \\
\epsfig{file=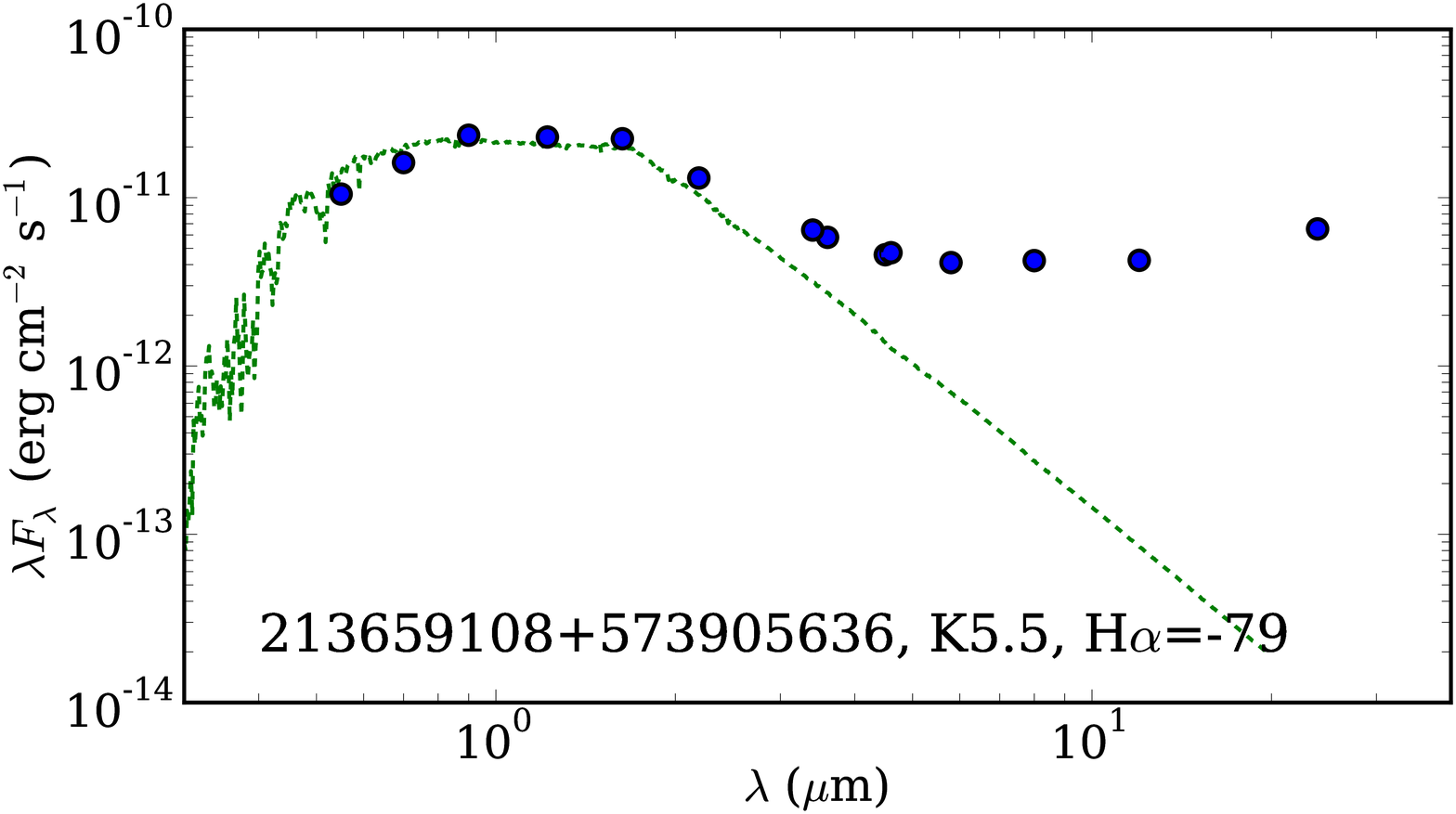,width=0.24\linewidth,clip=} &
\epsfig{file=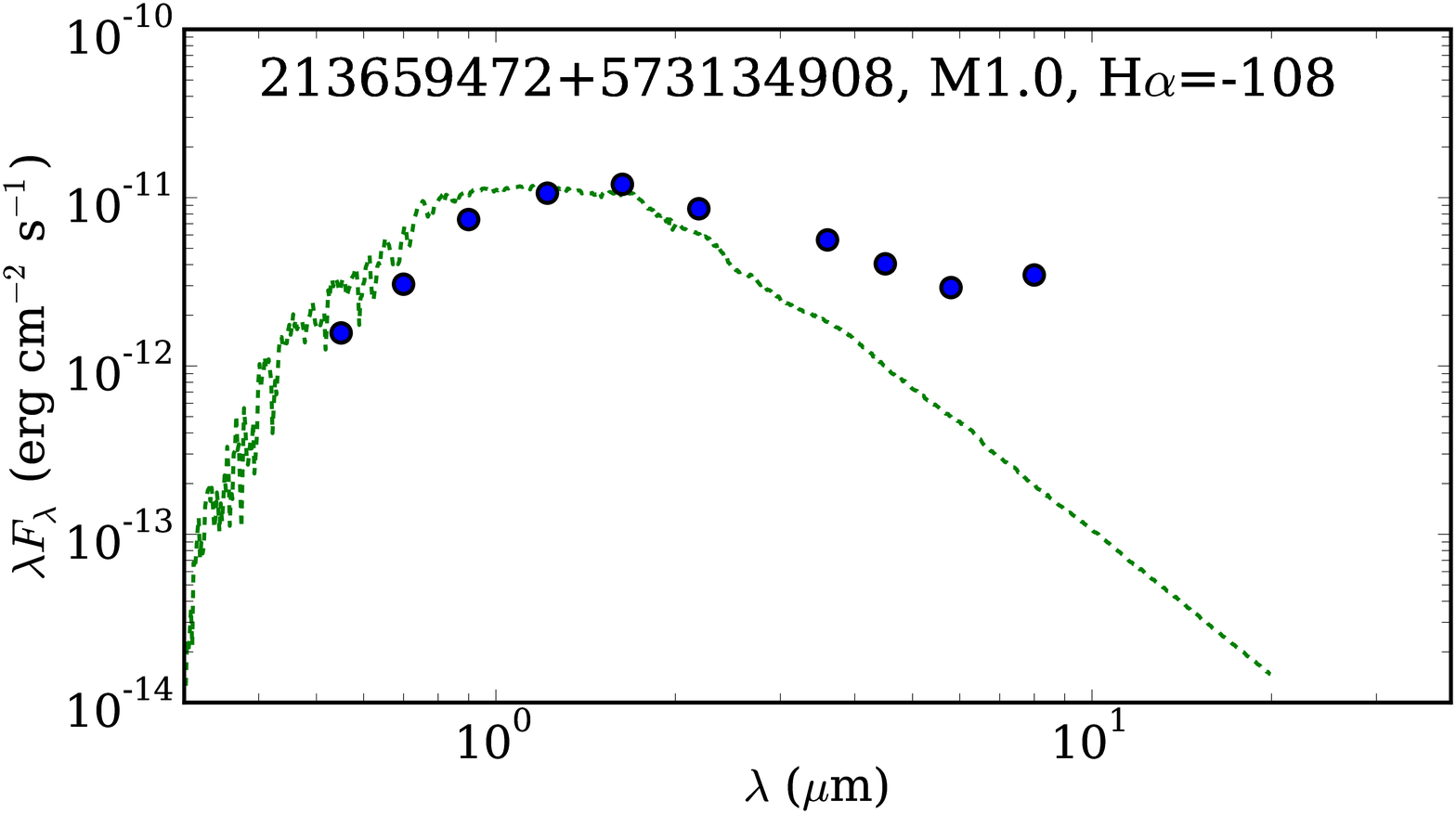,width=0.24\linewidth,clip=} &
\epsfig{file=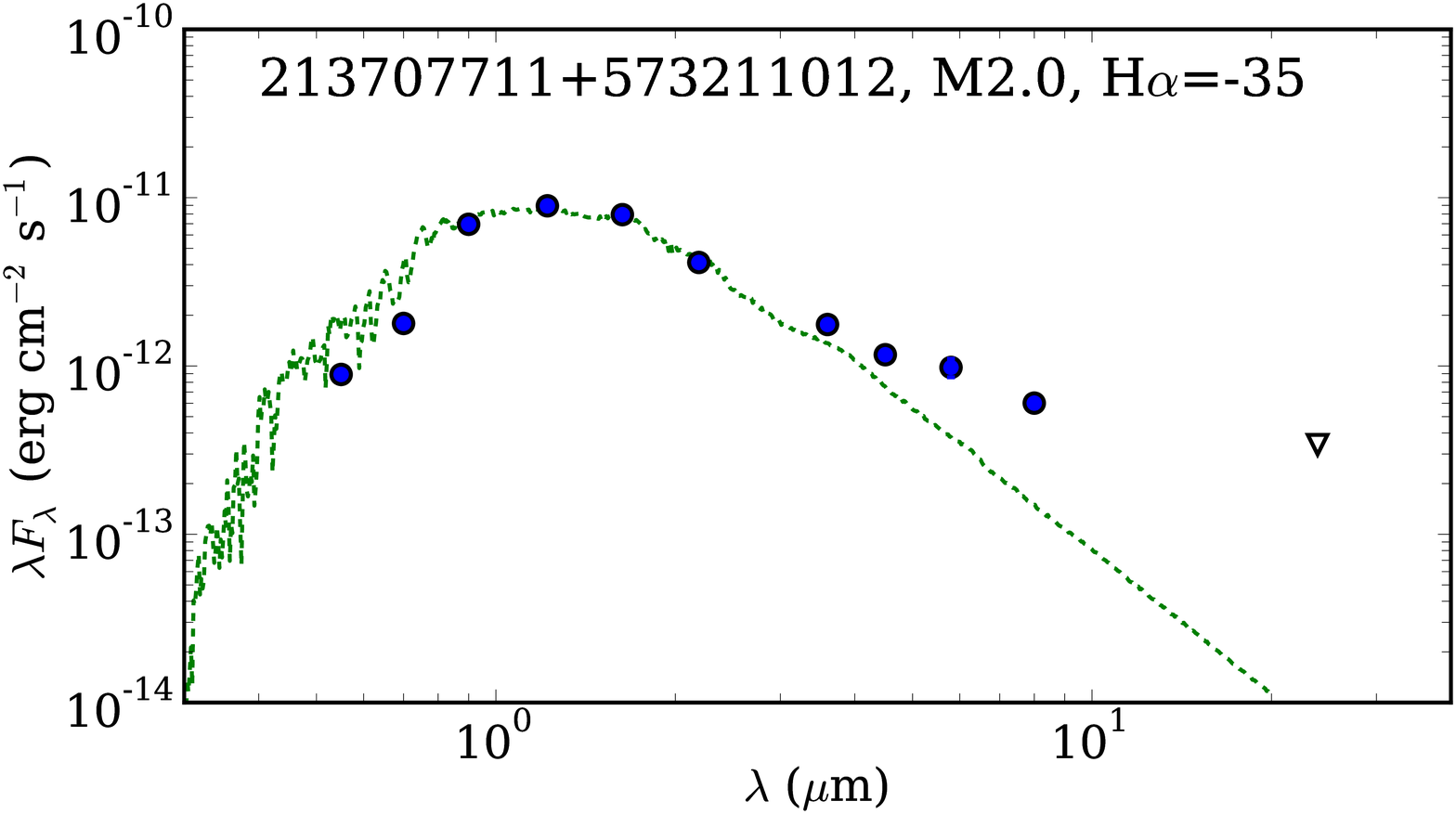,width=0.24\linewidth,clip=} &
\epsfig{file=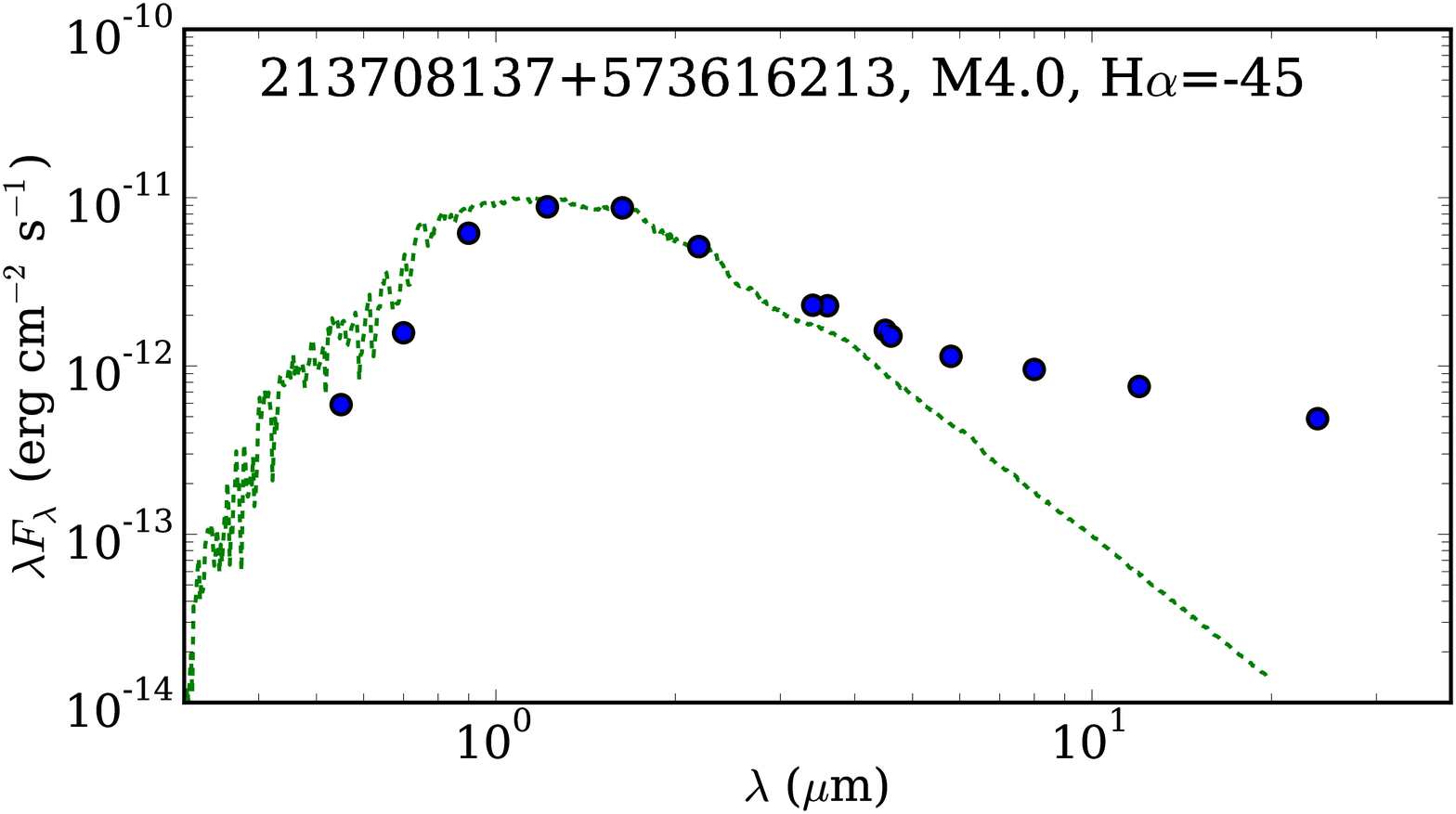,width=0.24\linewidth,clip=} \\
\epsfig{file=213709442+573036722.eps,width=0.24\linewidth,clip=} &
\epsfig{file=213710877+573846877.eps,width=0.24\linewidth,clip=} &
\epsfig{file=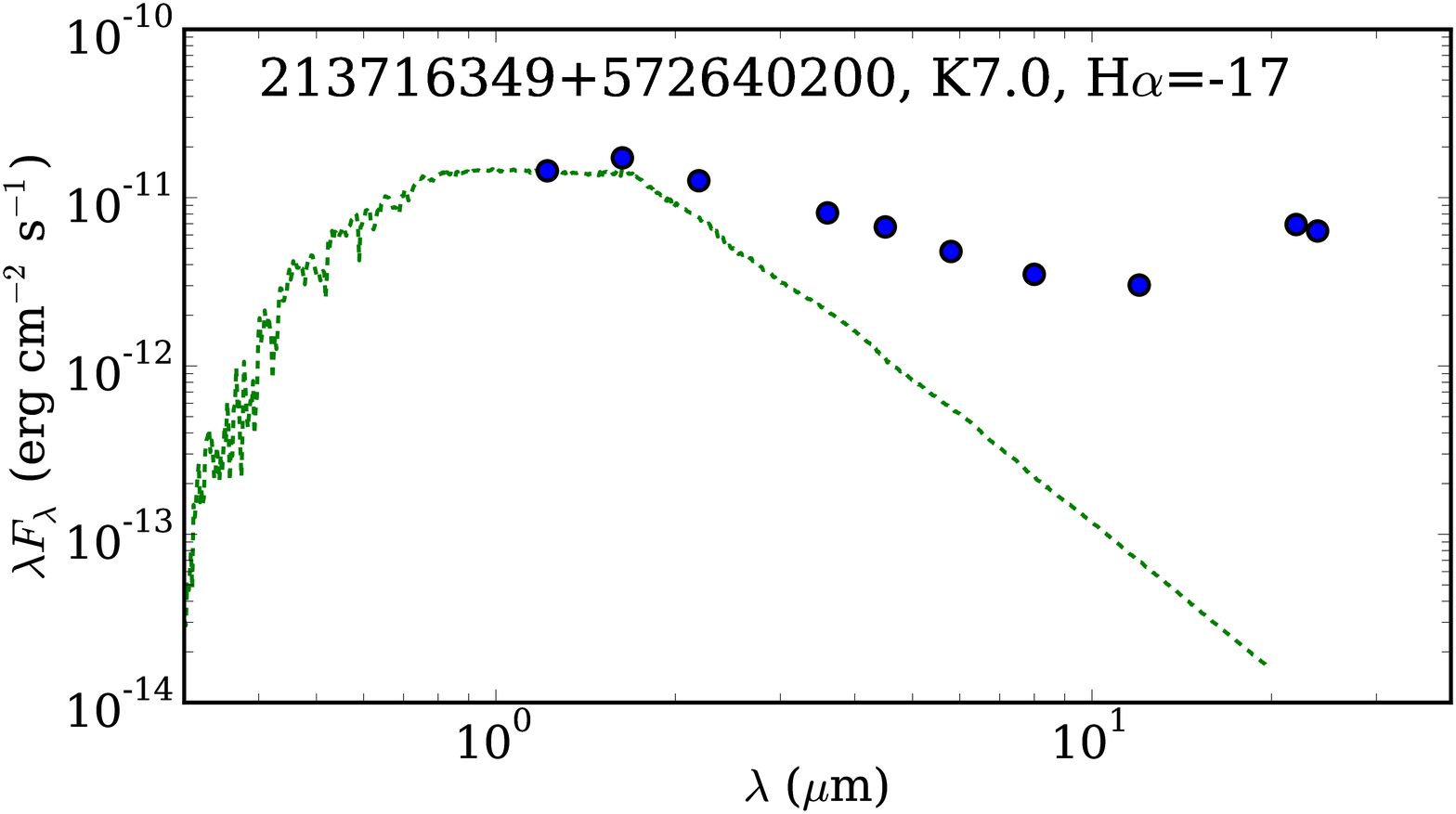,width=0.24\linewidth,clip=} &
\epsfig{file=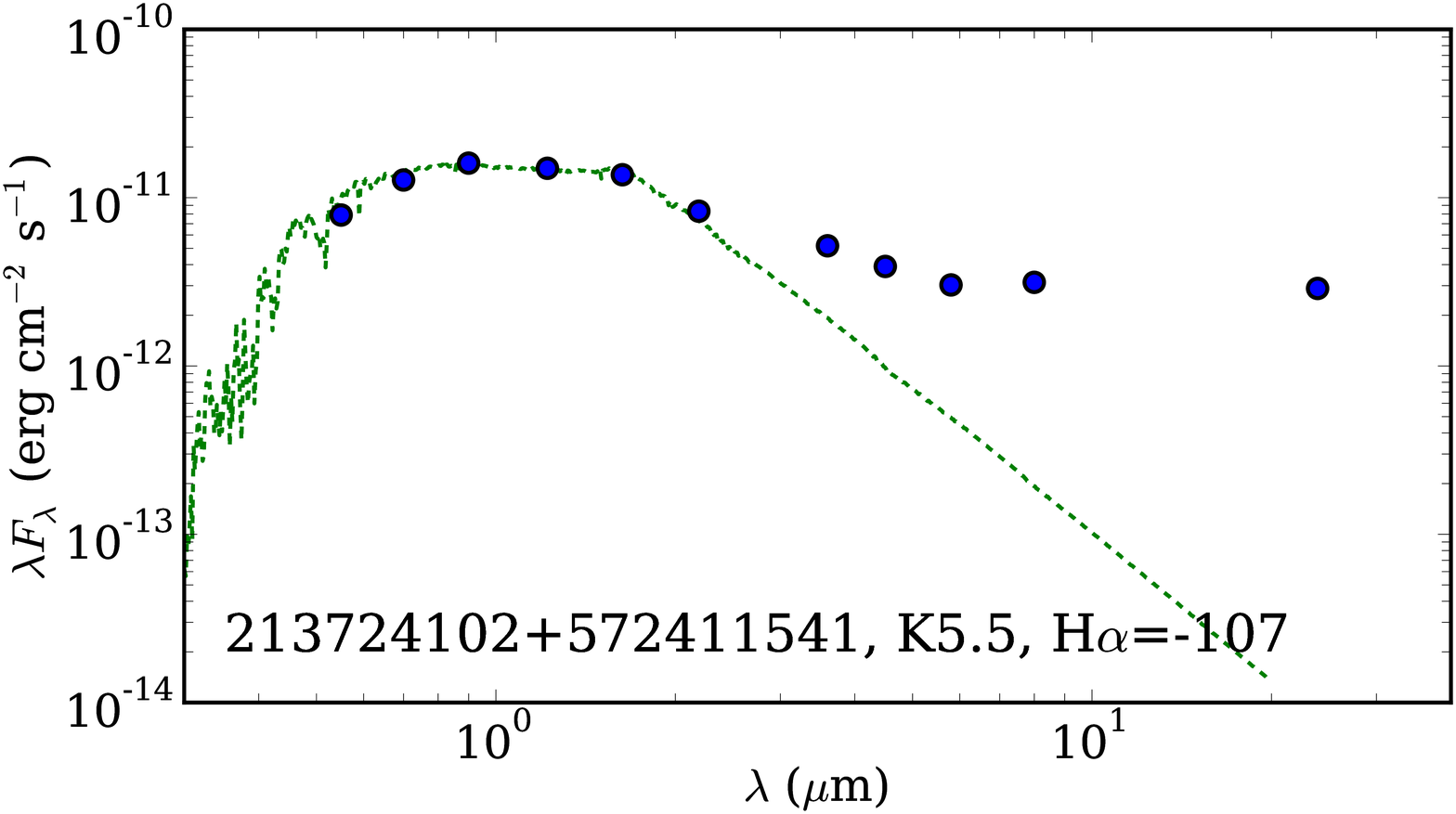,width=0.24\linewidth,clip=} \\
\epsfig{file=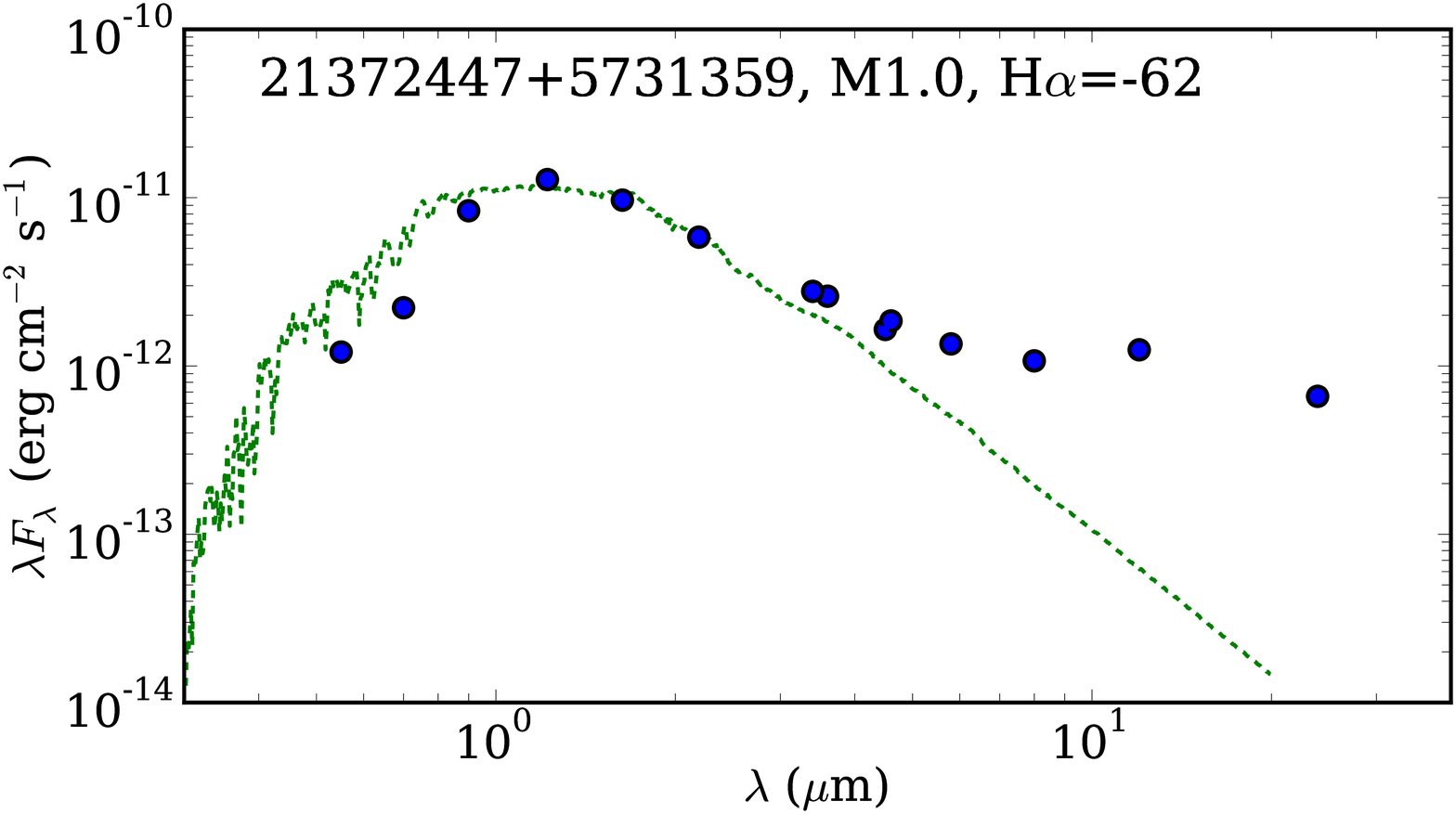,width=0.24\linewidth,clip=} &
\epsfig{file=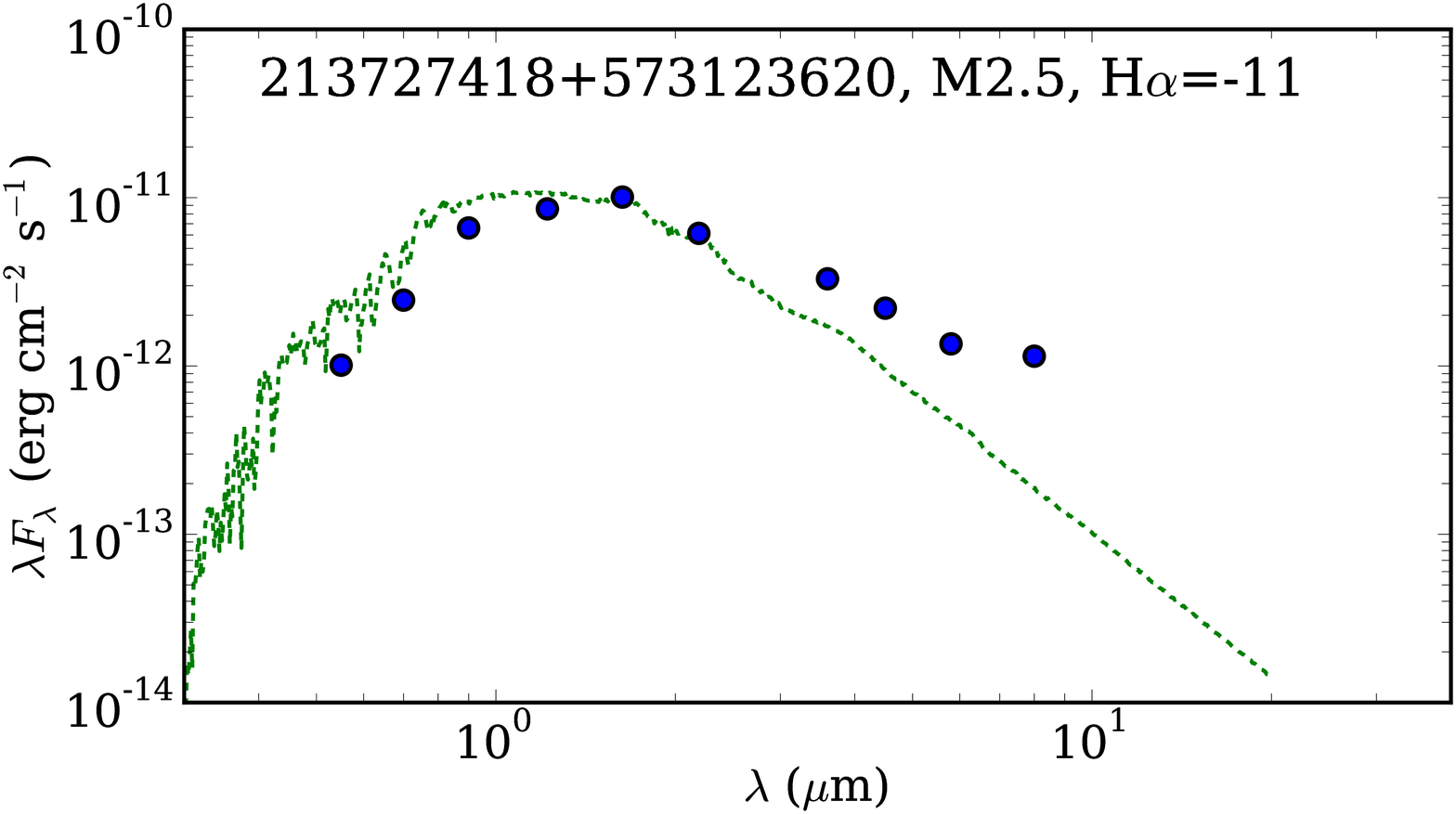,width=0.24\linewidth,clip=} &
\epsfig{file=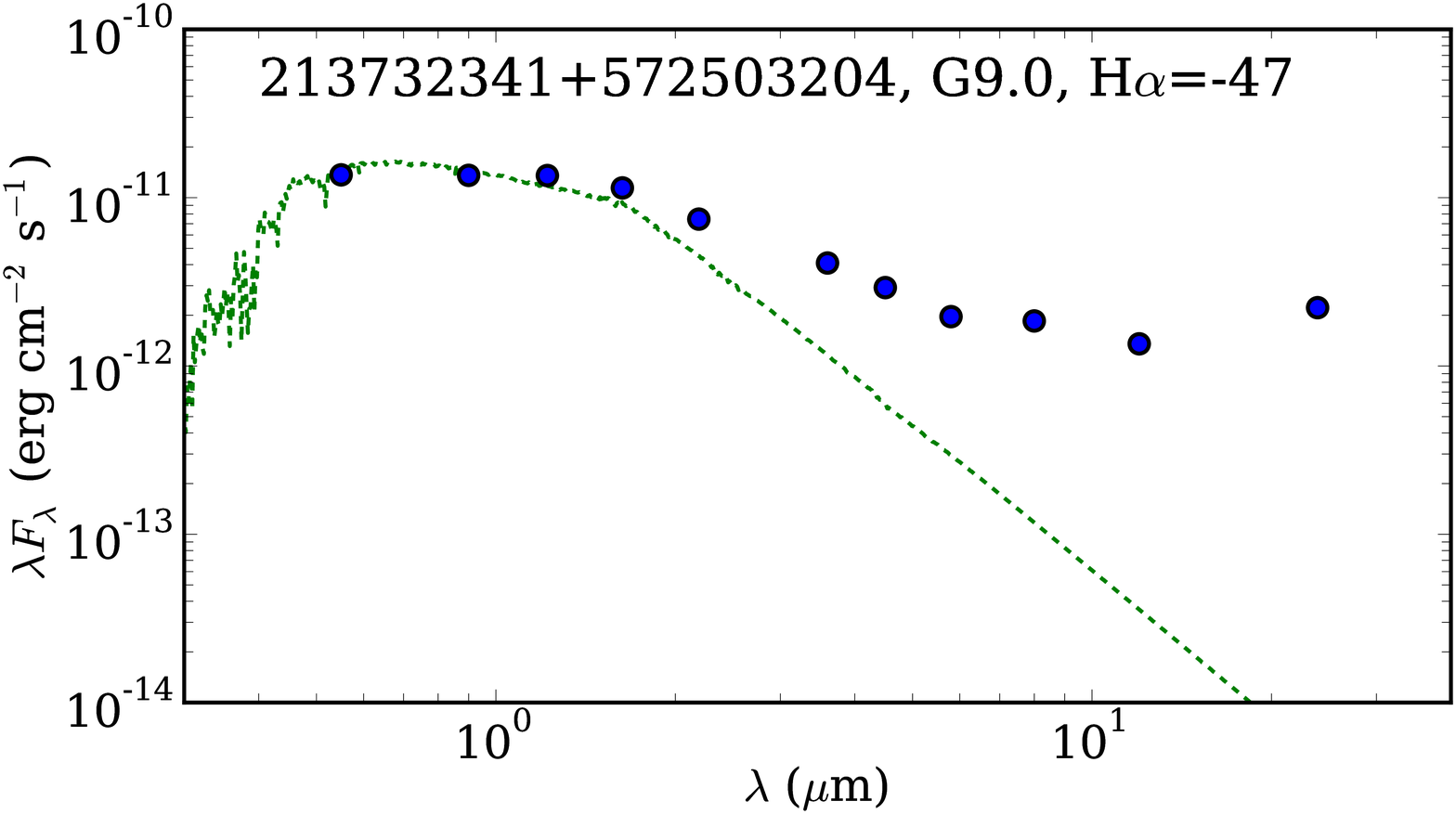,width=0.24\linewidth,clip=} &
\epsfig{file=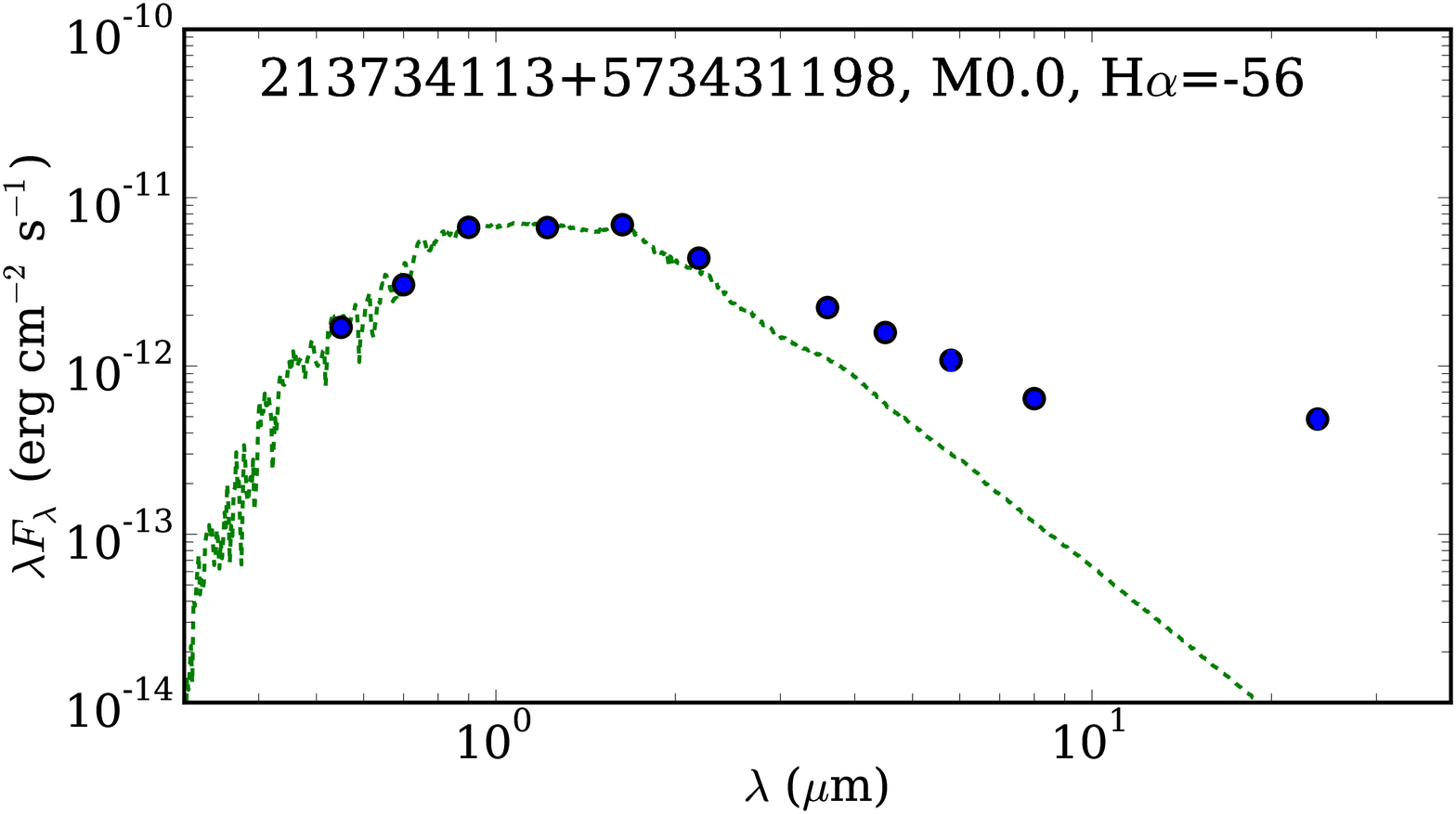,width=0.24\linewidth,clip=} \\
\epsfig{file=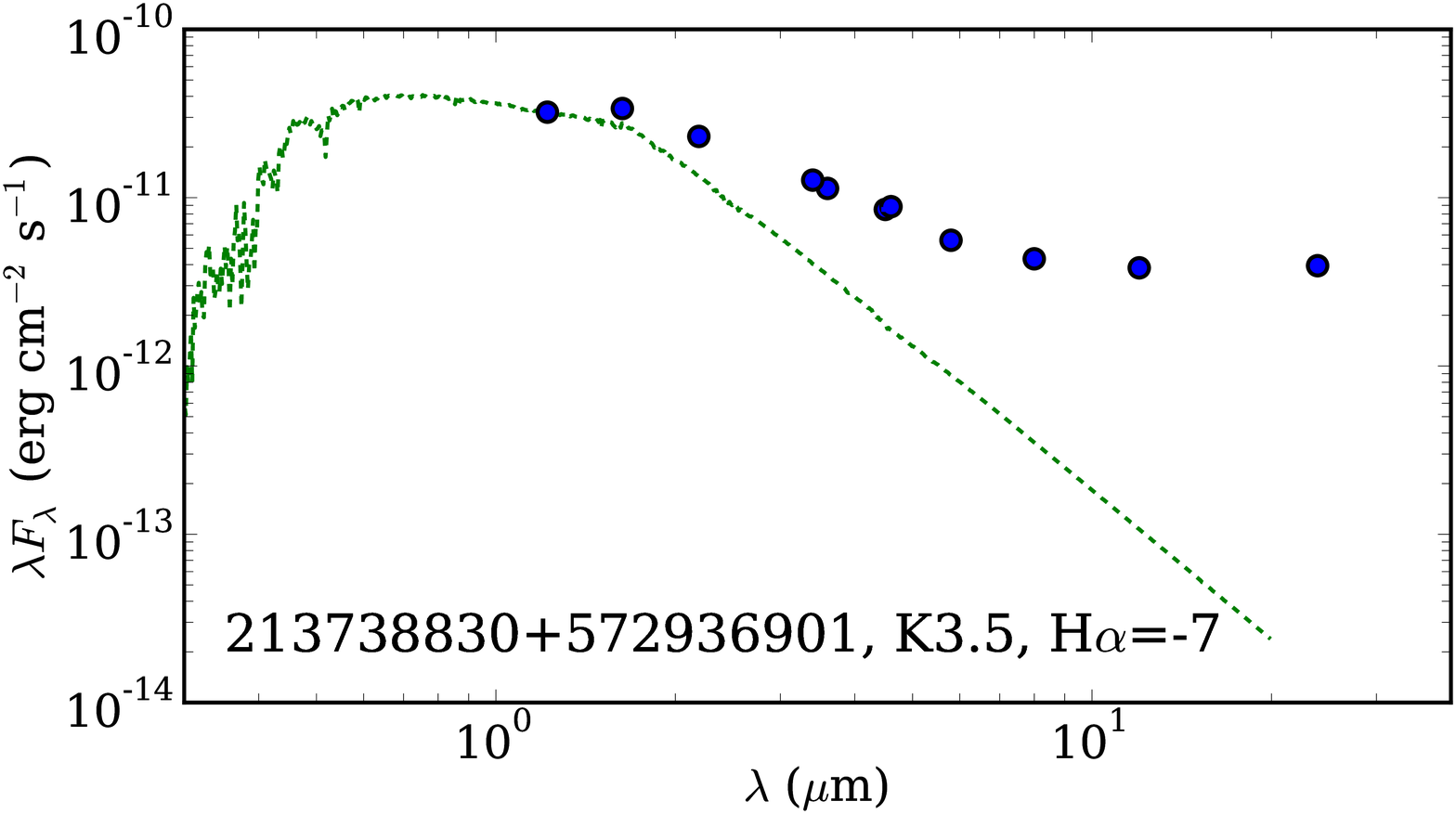,width=0.24\linewidth,clip=} &
\epsfig{file=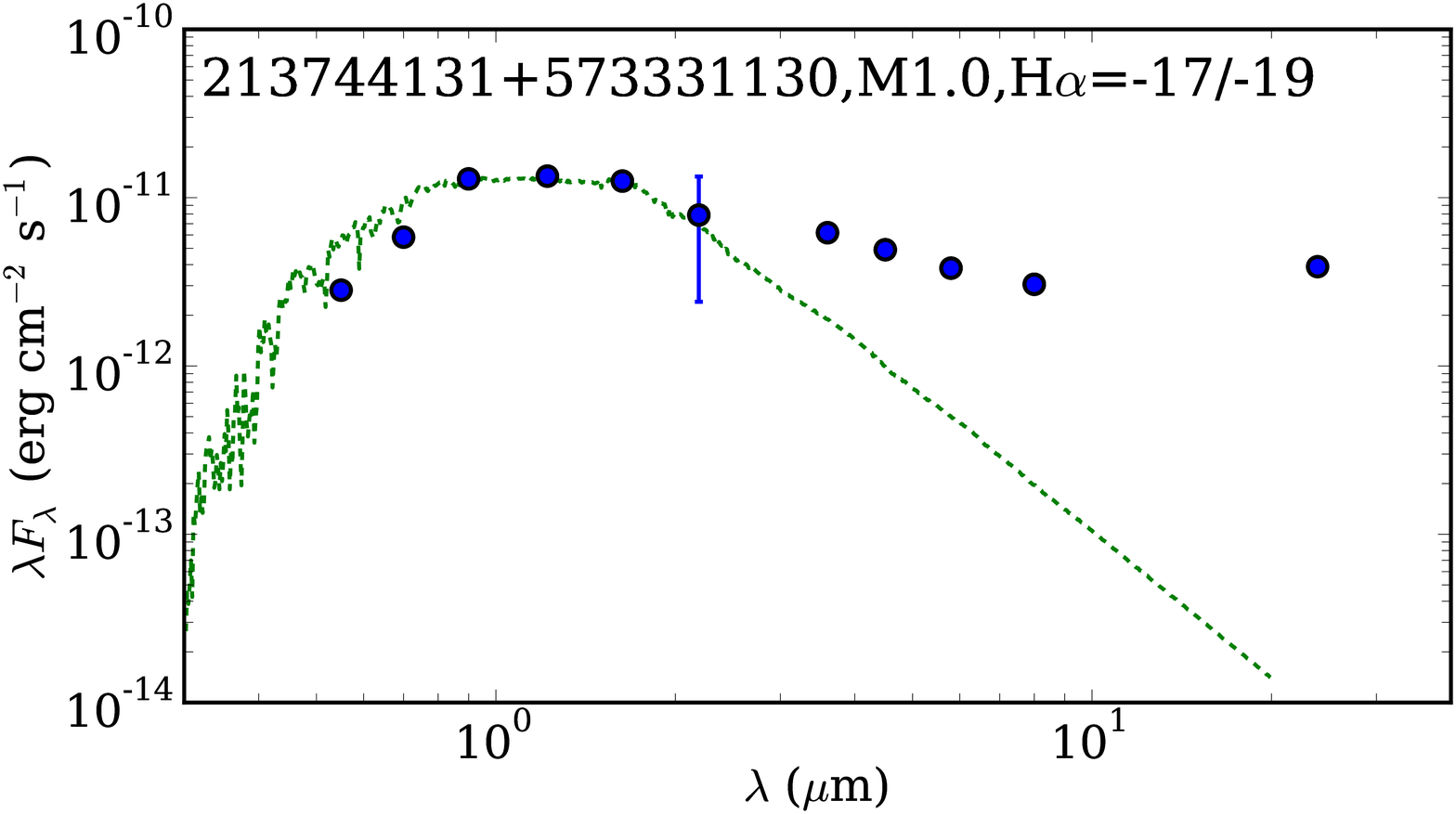,width=0.24\linewidth,clip=} &
\epsfig{file=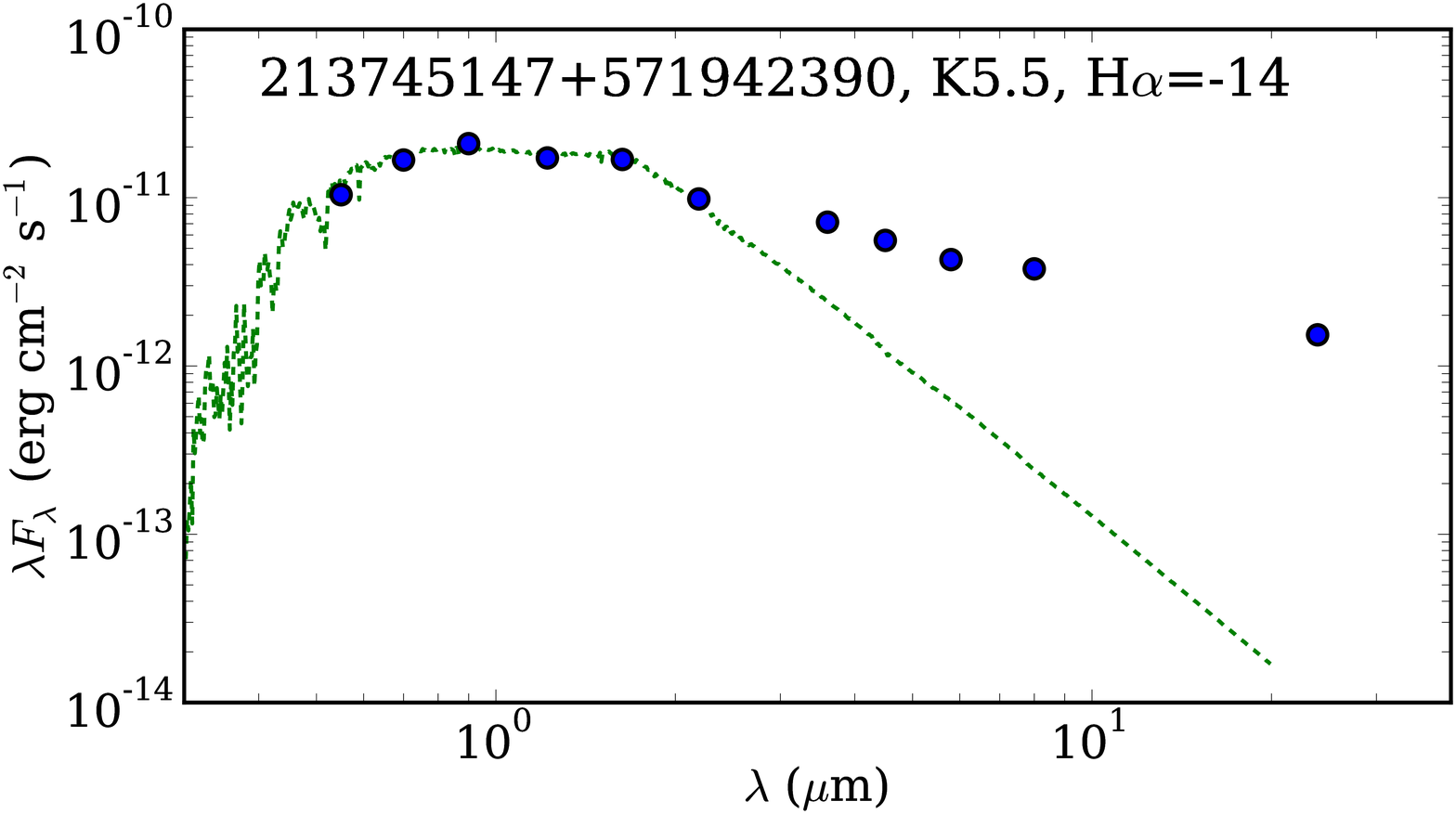,width=0.24\linewidth,clip=} &
\epsfig{file=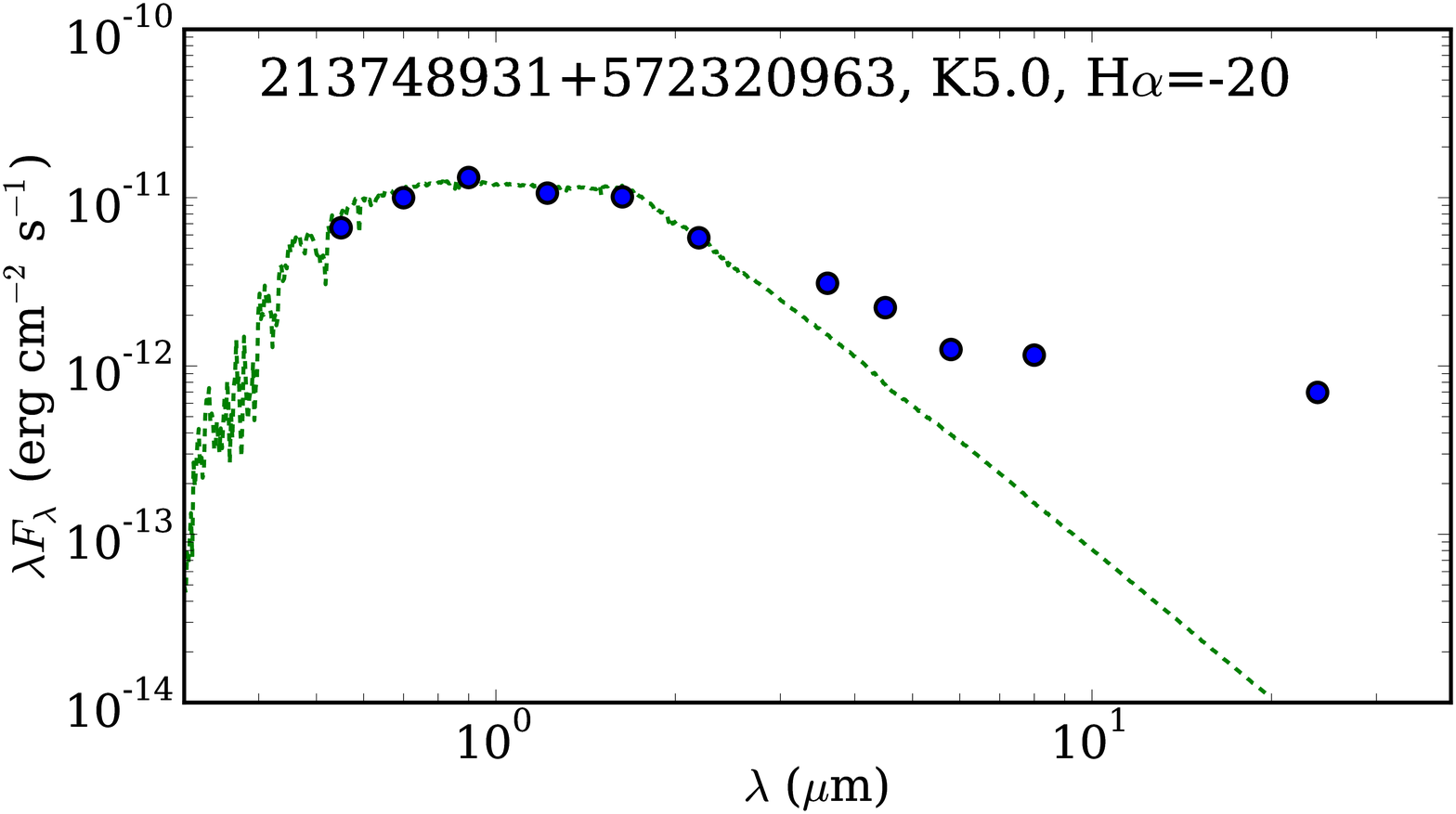,width=0.24\linewidth,clip=} \\
\epsfig{file=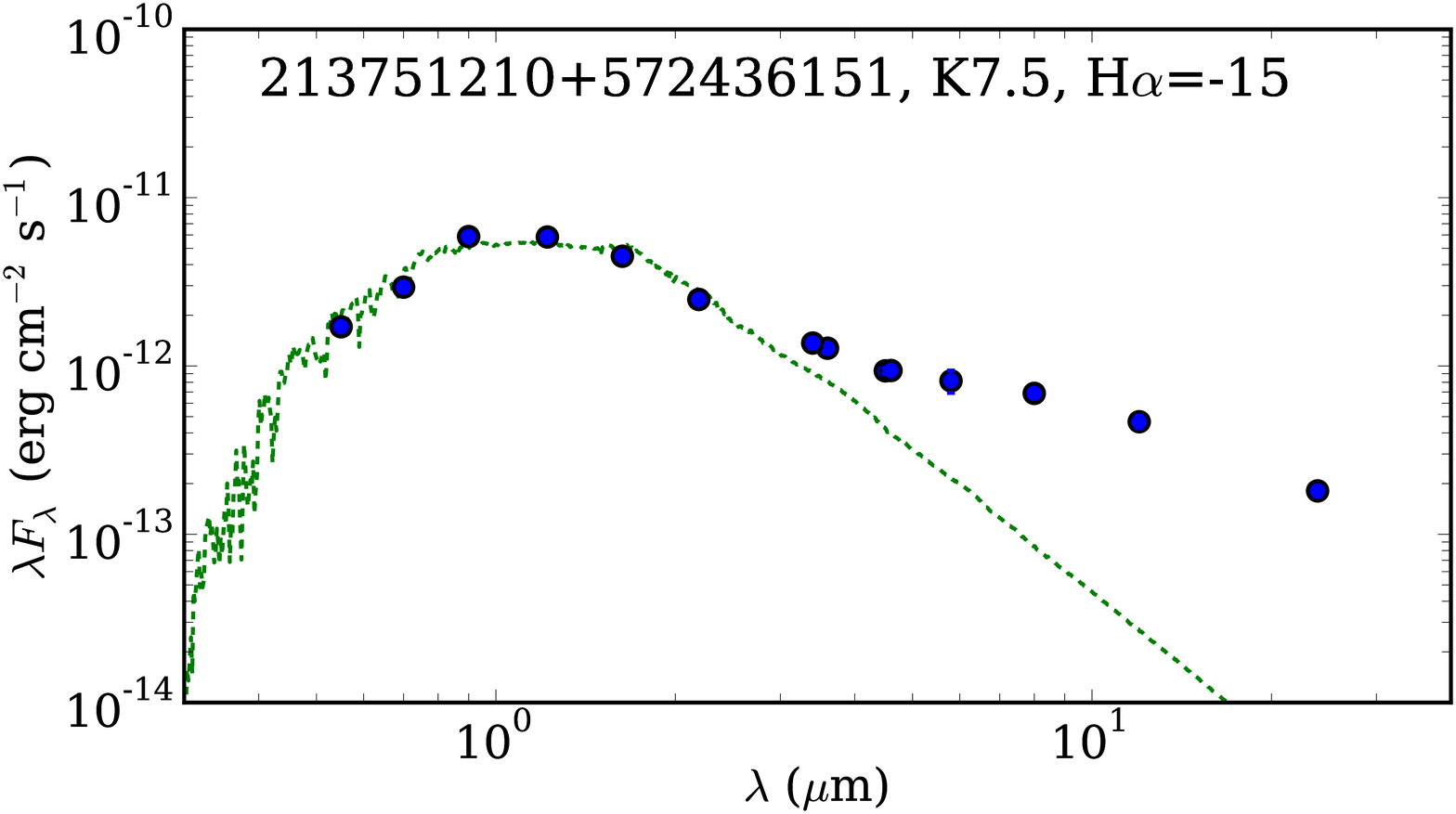,width=0.24\linewidth,clip=} &
\epsfig{file=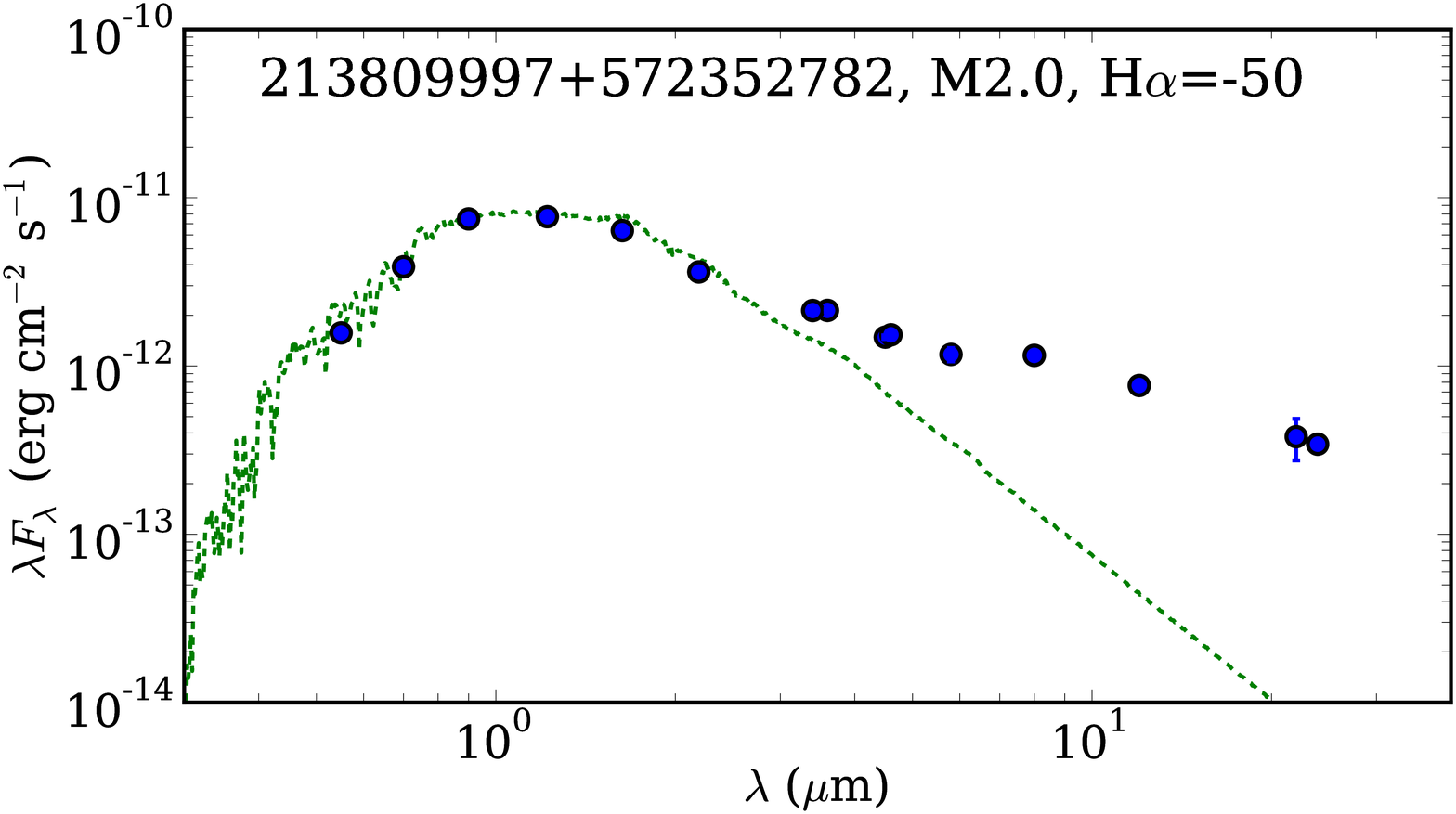,width=0.24\linewidth,clip=} &
\epsfig{file=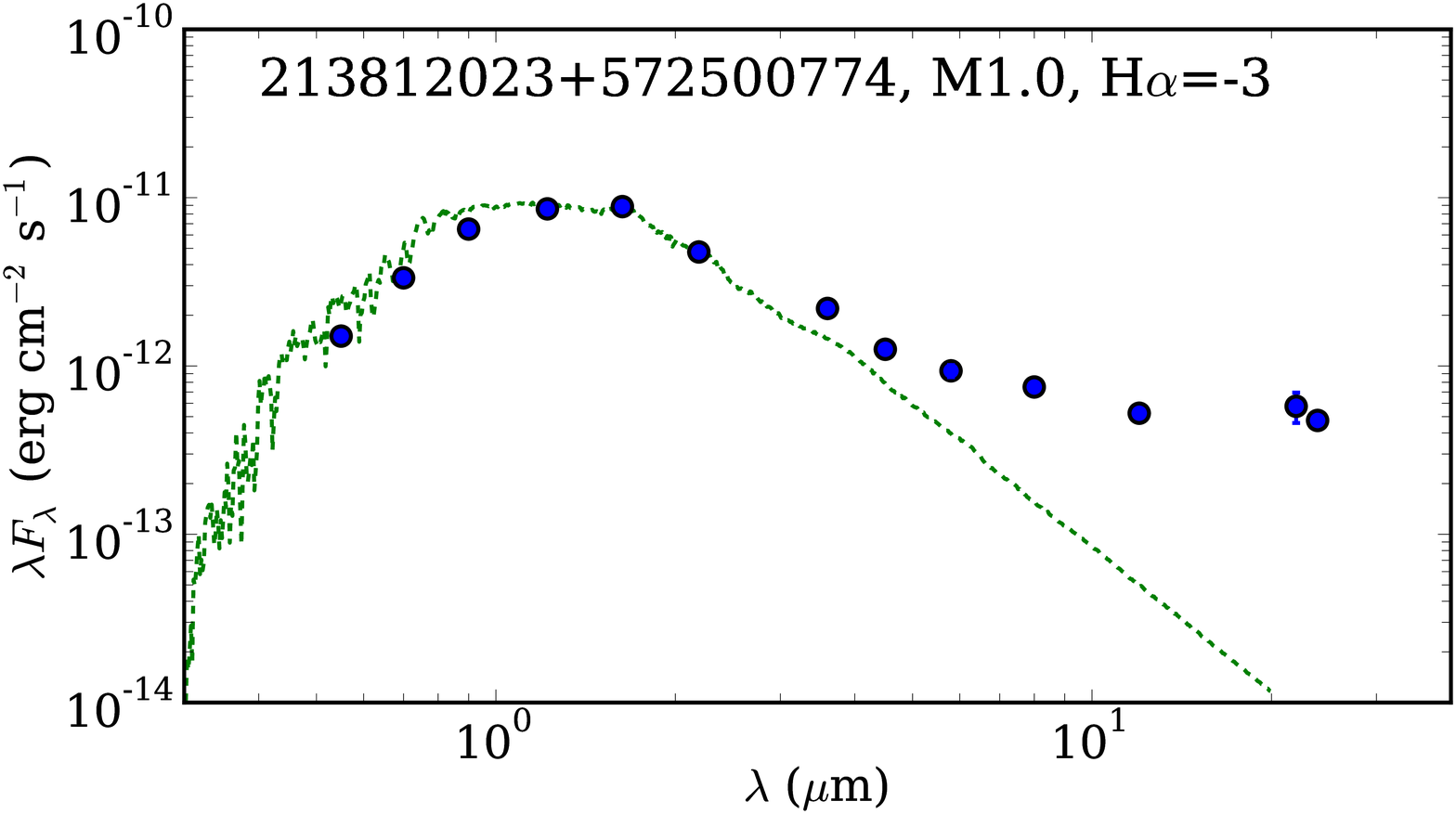,width=0.24\linewidth,clip=} &
\epsfig{file=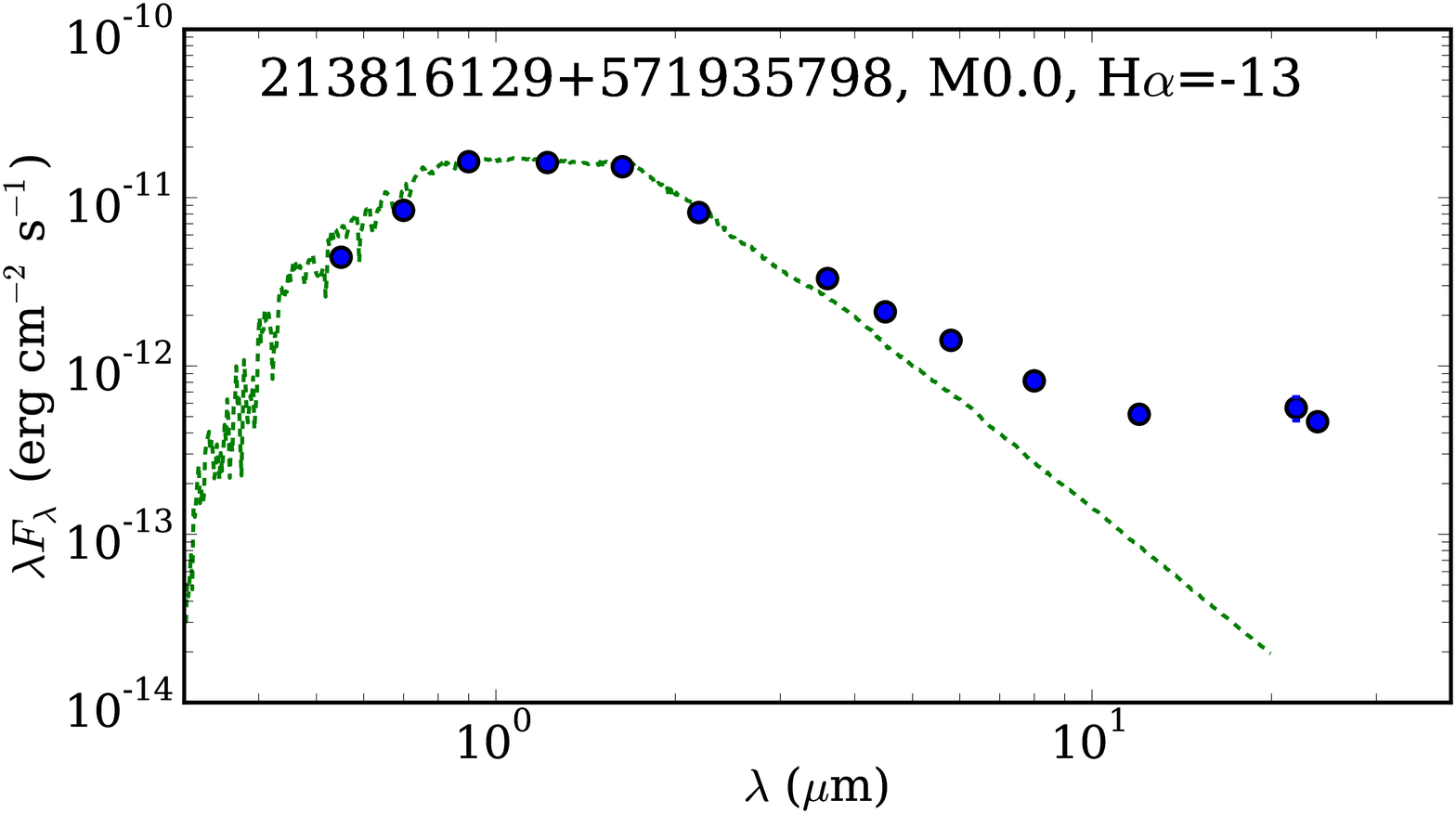,width=0.24\linewidth,clip=} \\
\epsfig{file=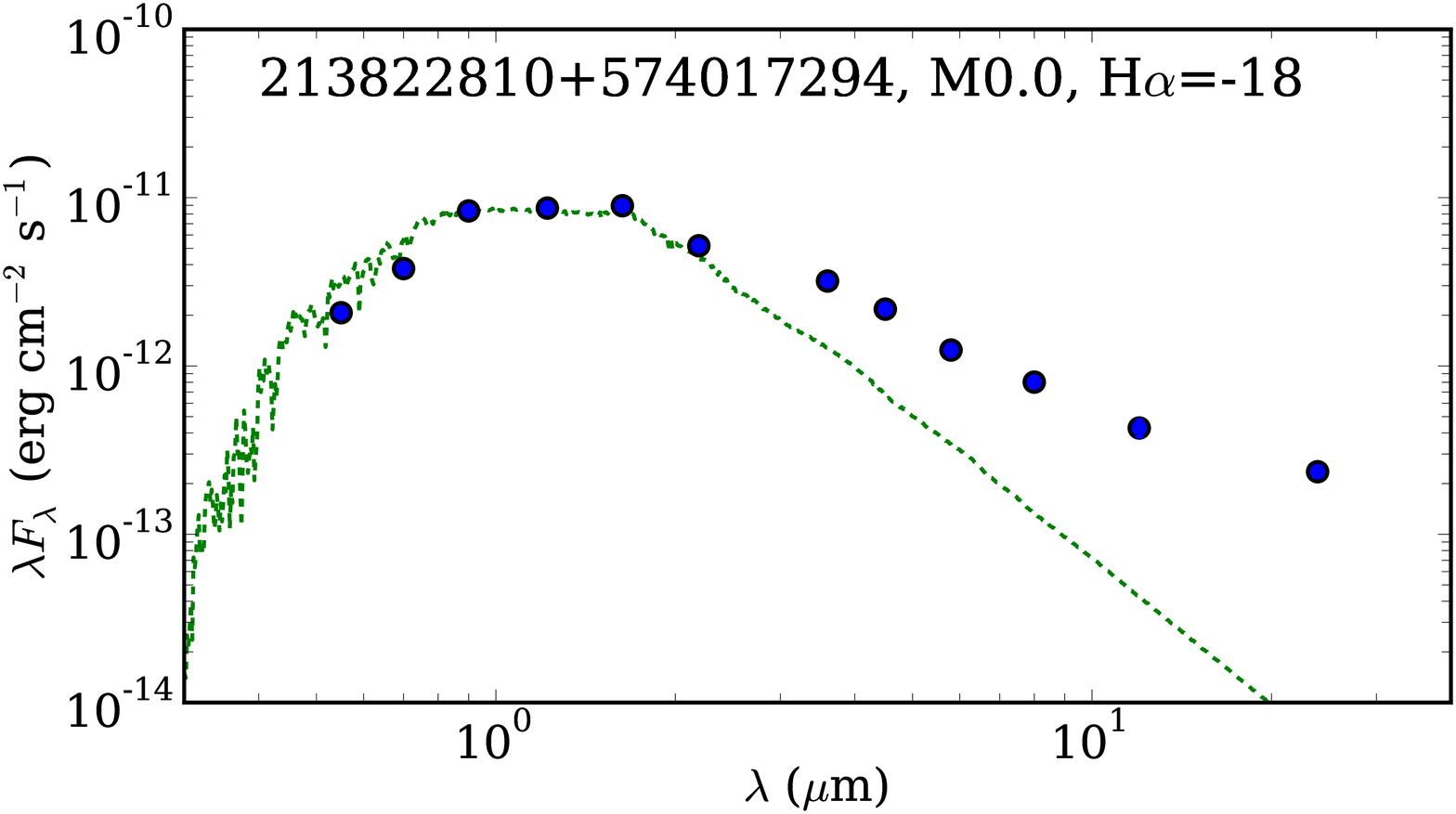,width=0.24\linewidth,clip=} &
\epsfig{file=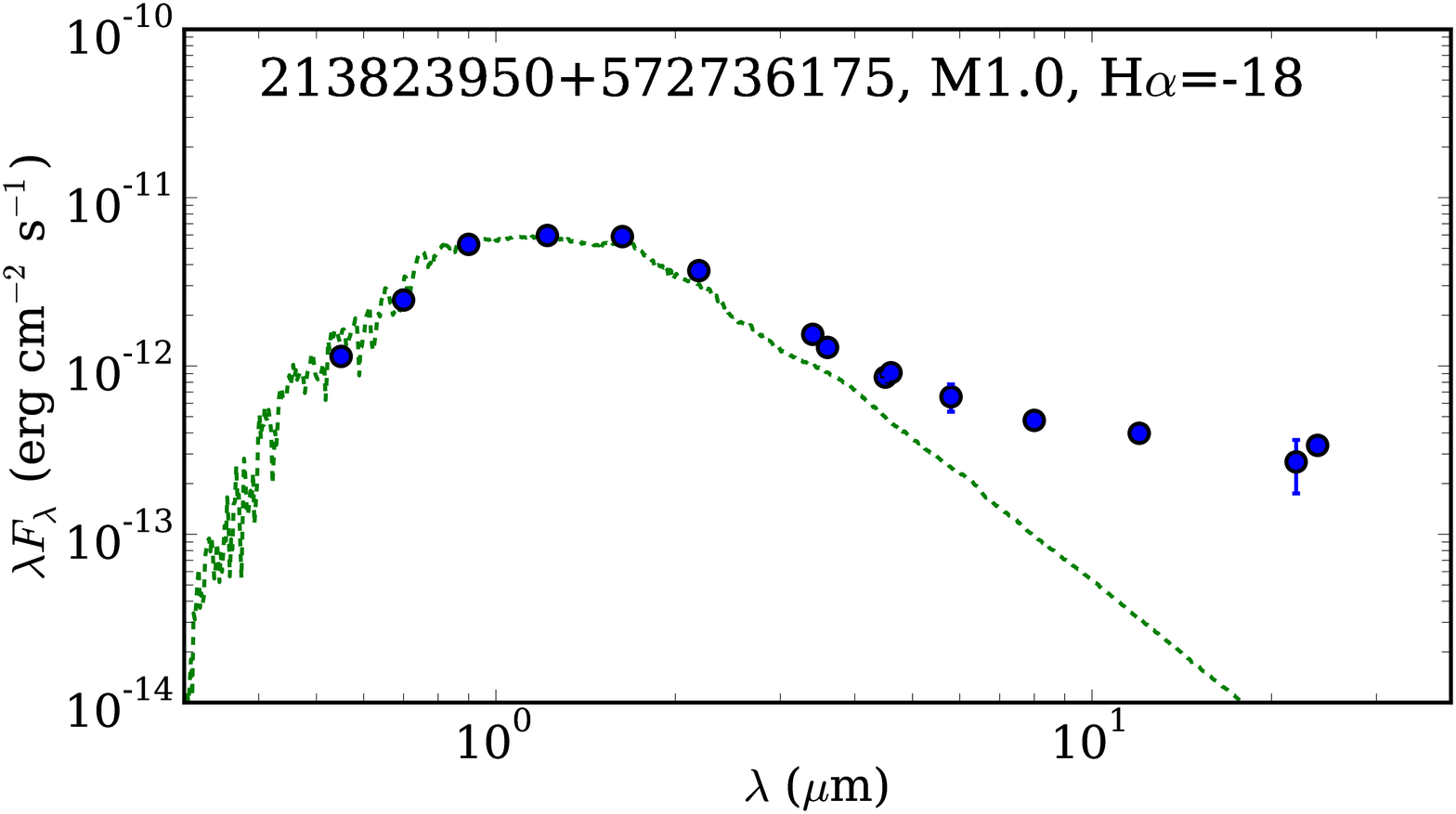,width=0.24\linewidth,clip=} &
\epsfig{file=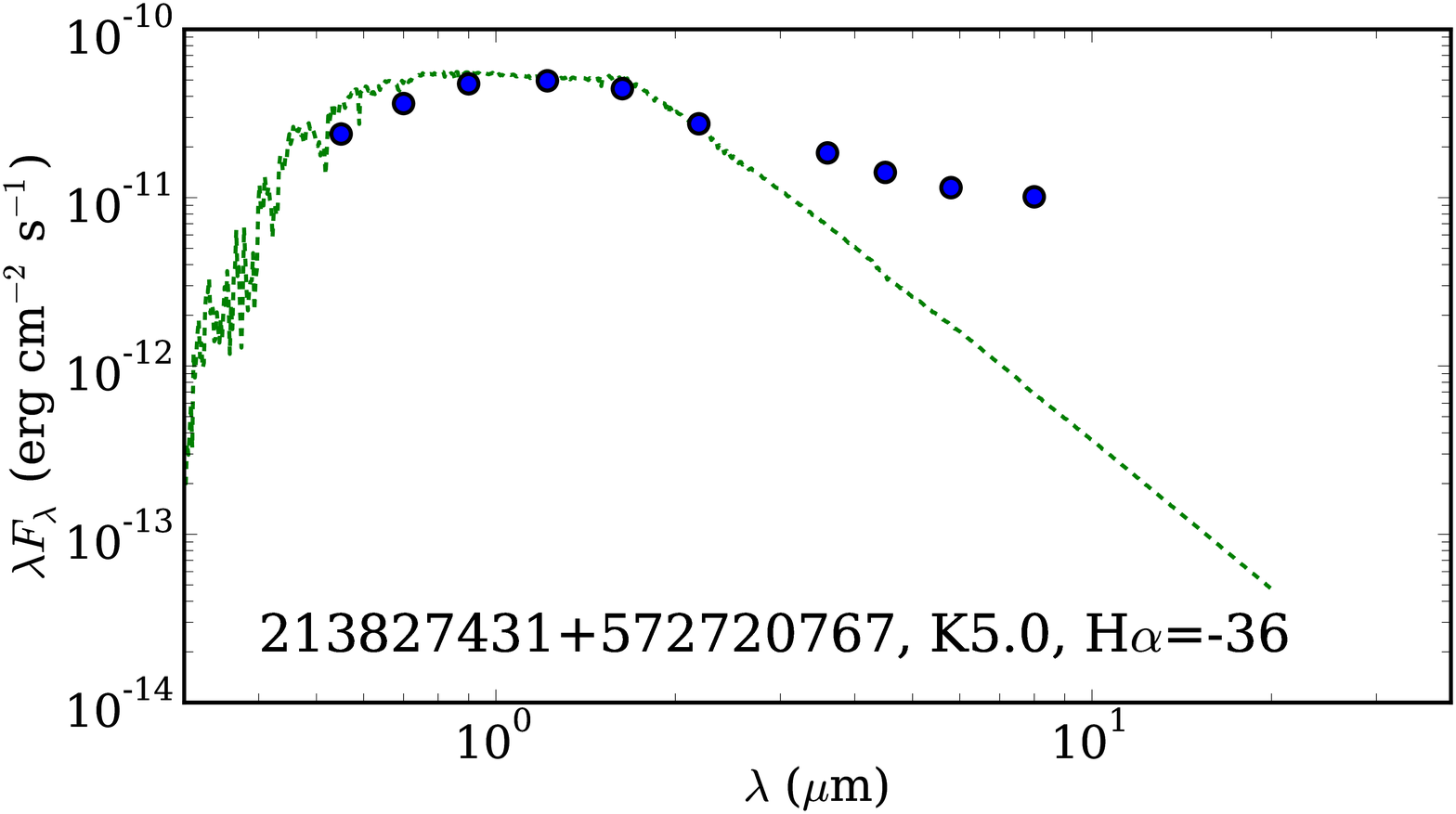,width=0.24\linewidth,clip=} &
\epsfig{file=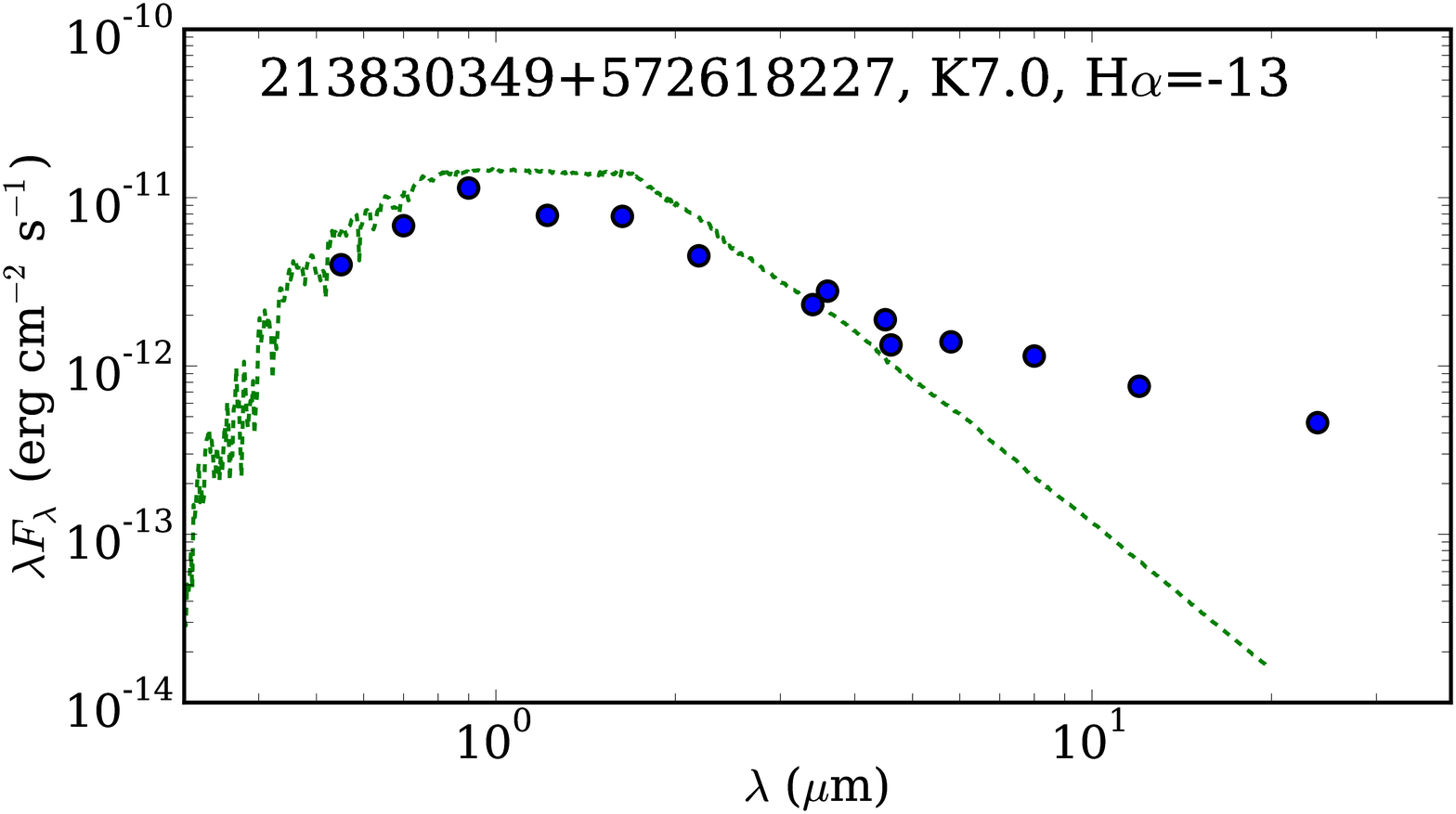,width=0.24\linewidth,clip=} \\
\epsfig{file=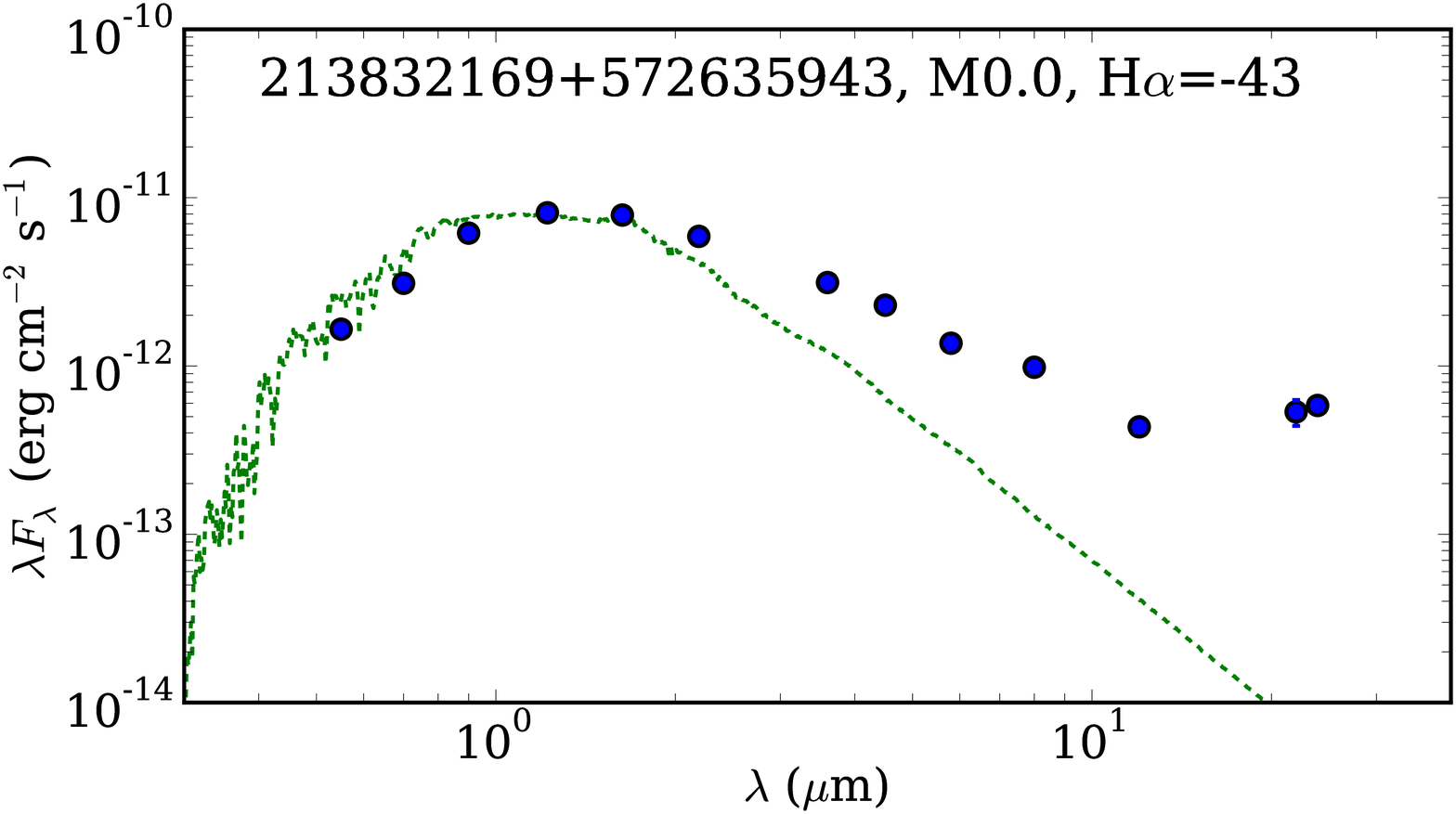,width=0.24\linewidth,clip=} &
\epsfig{file=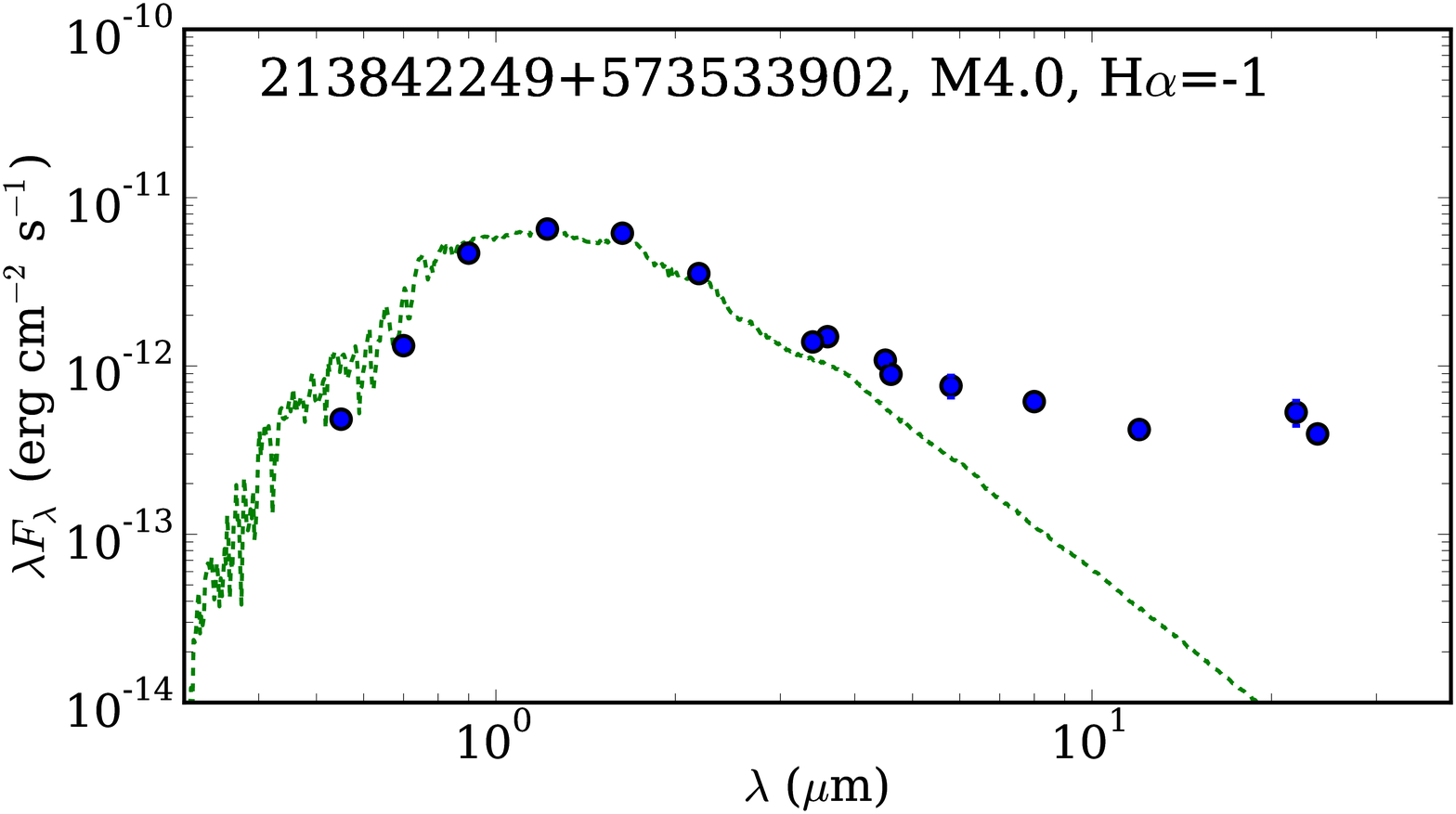,width=0.24\linewidth,clip=} &
\epsfig{file=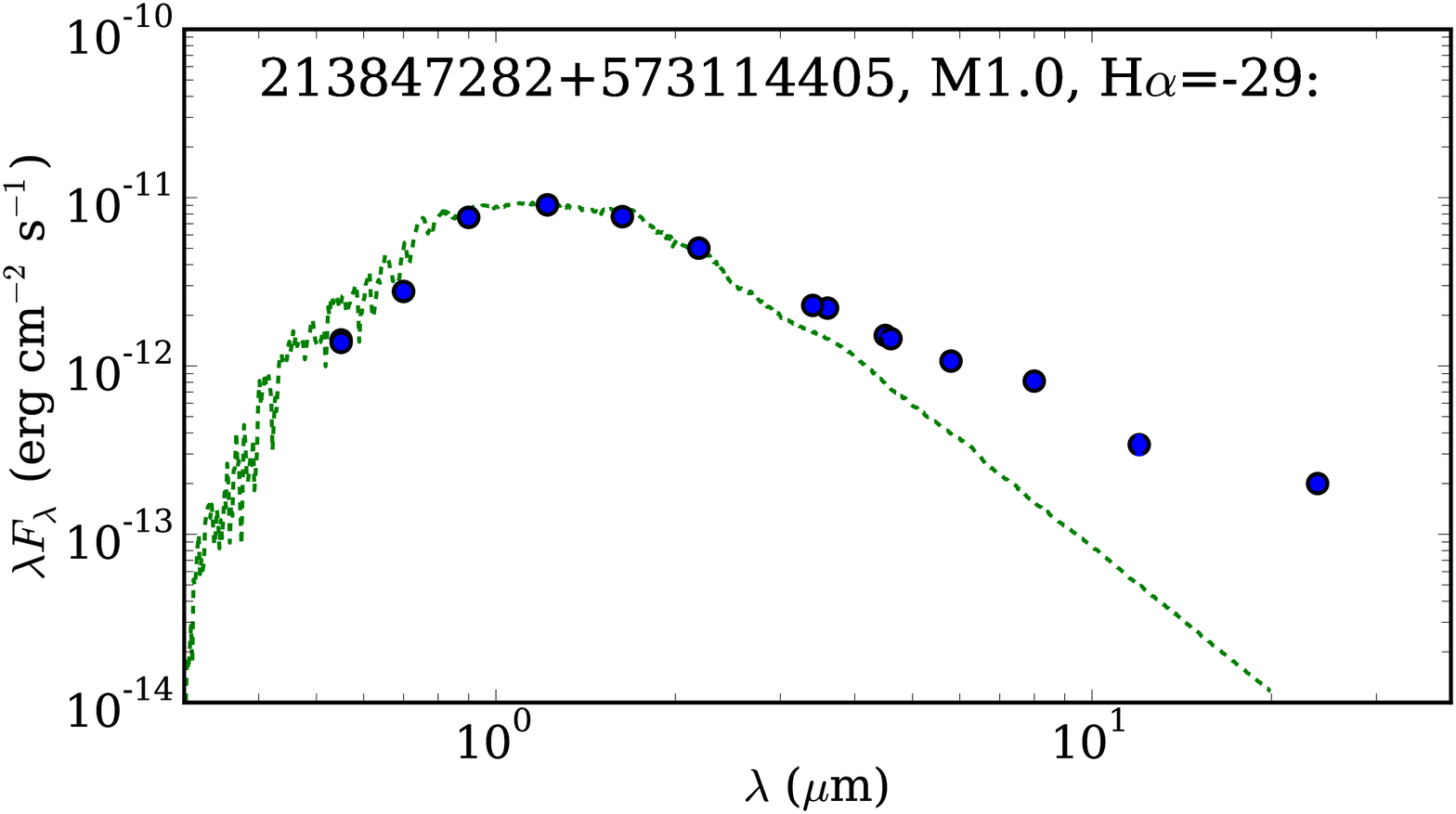,width=0.24\linewidth,clip=} &
\epsfig{file=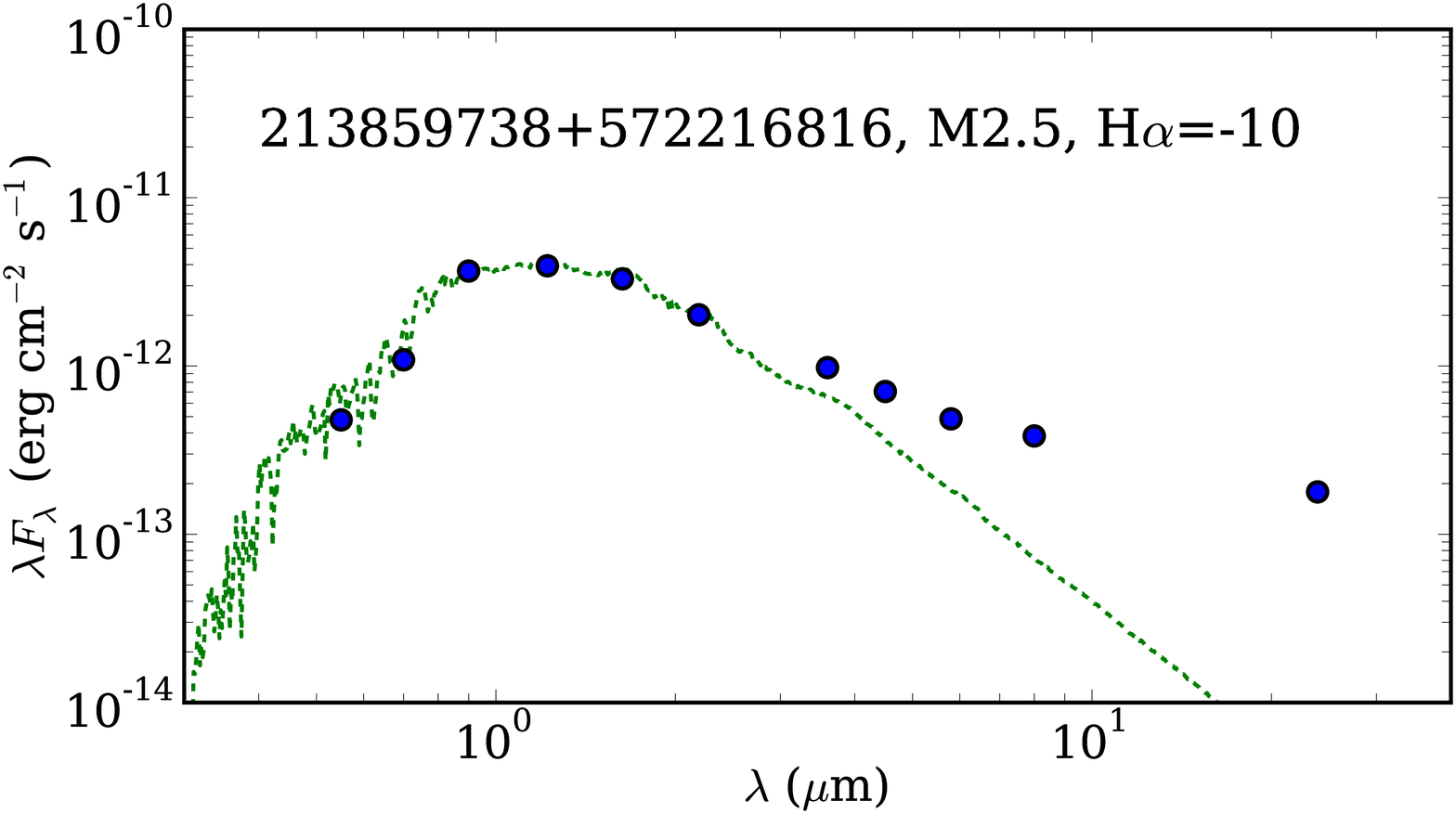,width=0.24\linewidth,clip=} \\
\end{tabular}
\caption{SEDs of the members and probably members with IR excess typical of full-disks. 
Inverted triangles represent upper limits. 
\label{cttsseds1-fig}}
\end{figure*}

\begin{figure*}
\centering
\begin{tabular}{ccccc}
\epsfig{file=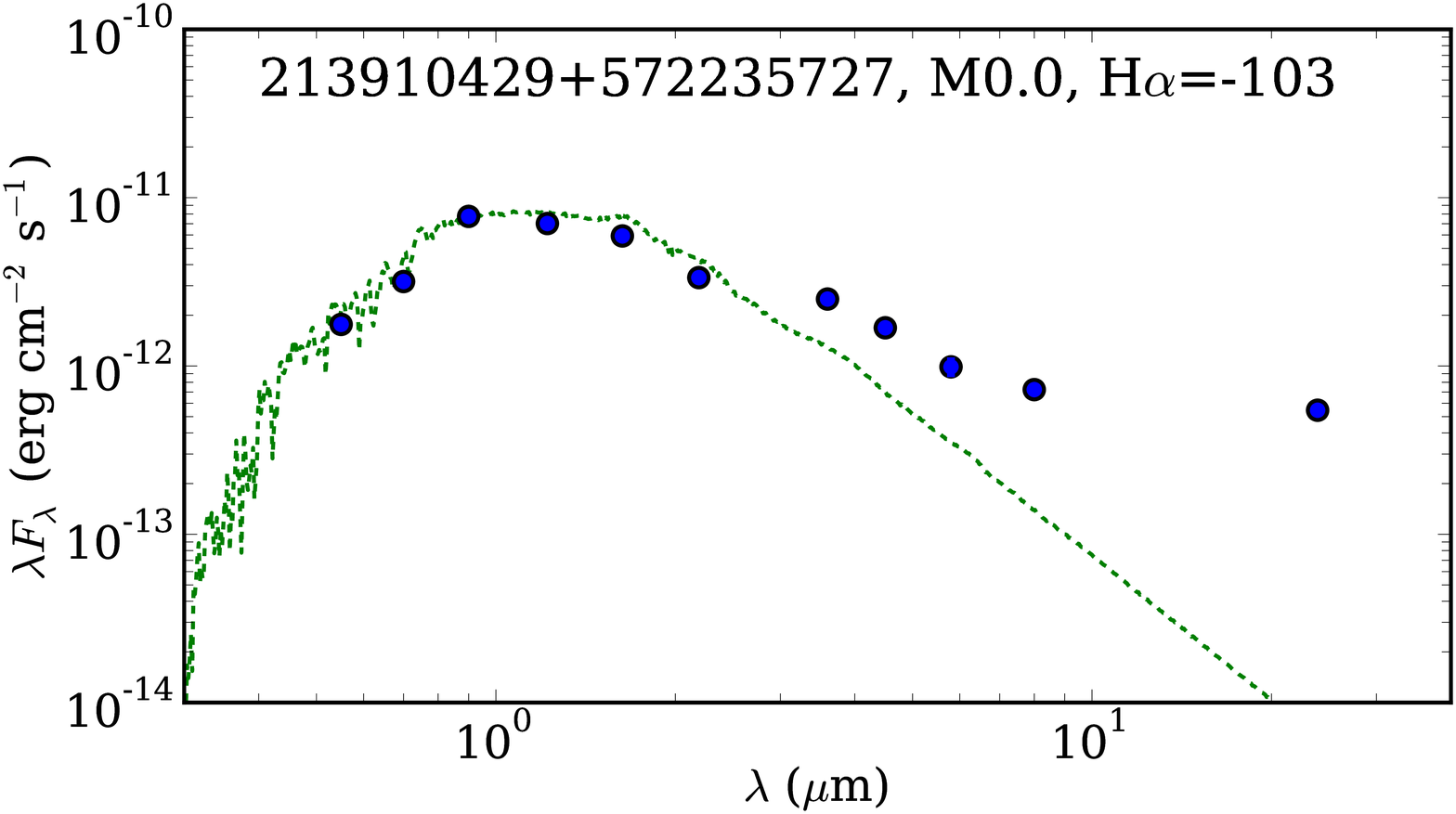,width=0.24\linewidth,clip=} &
\epsfig{file=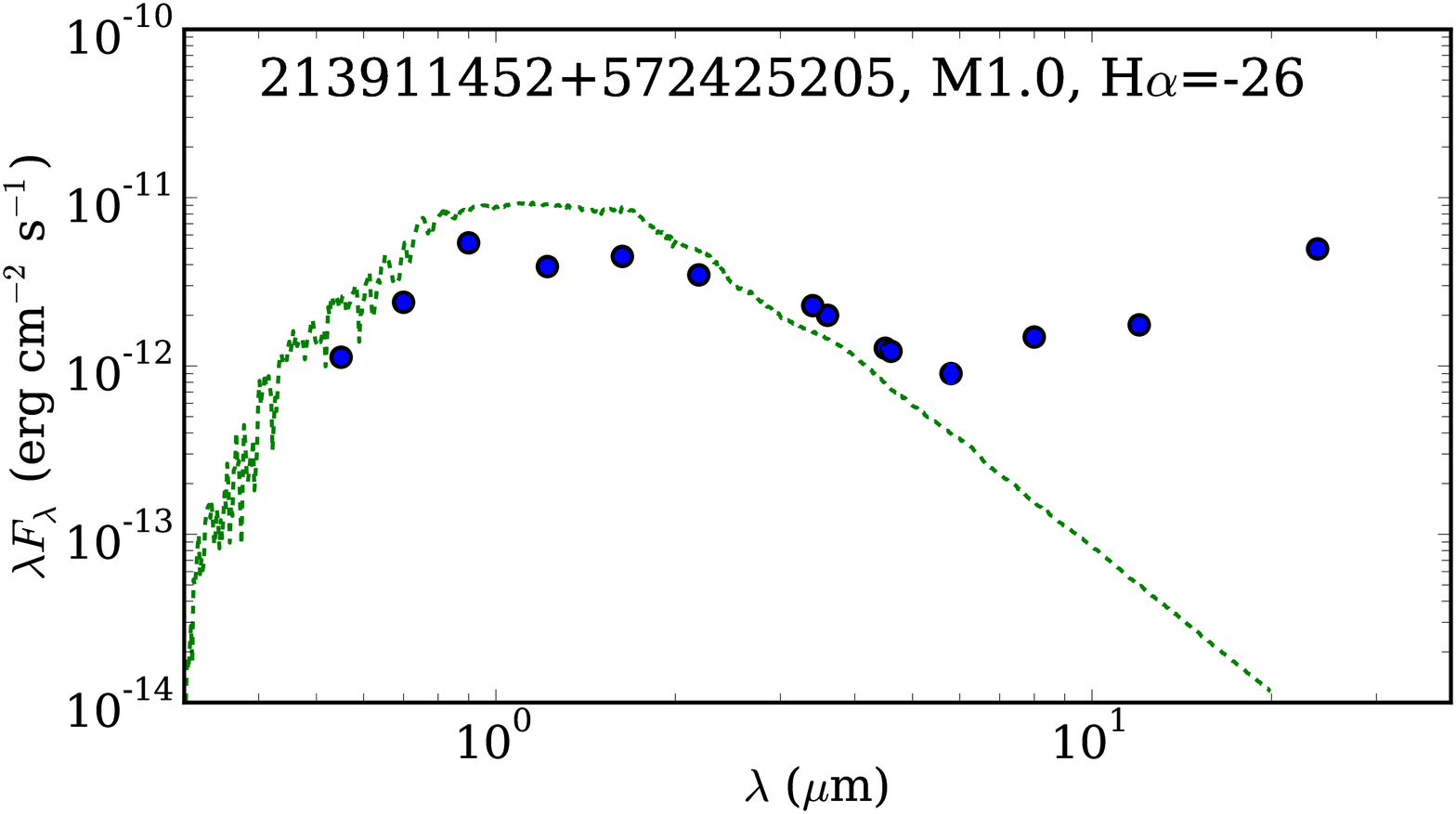,width=0.24\linewidth,clip=} &
\epsfig{file=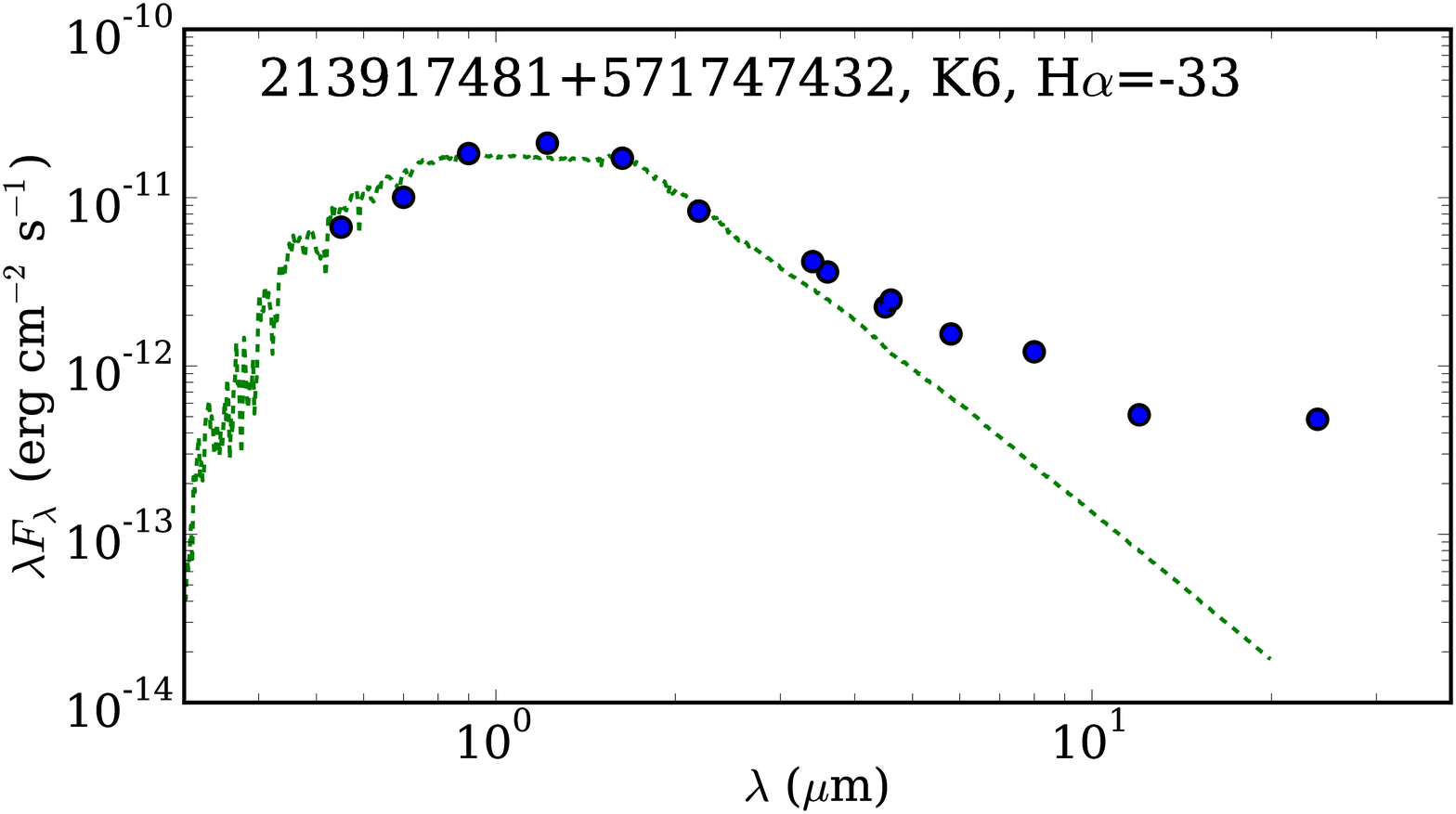,width=0.24\linewidth,clip=} &
\epsfig{file=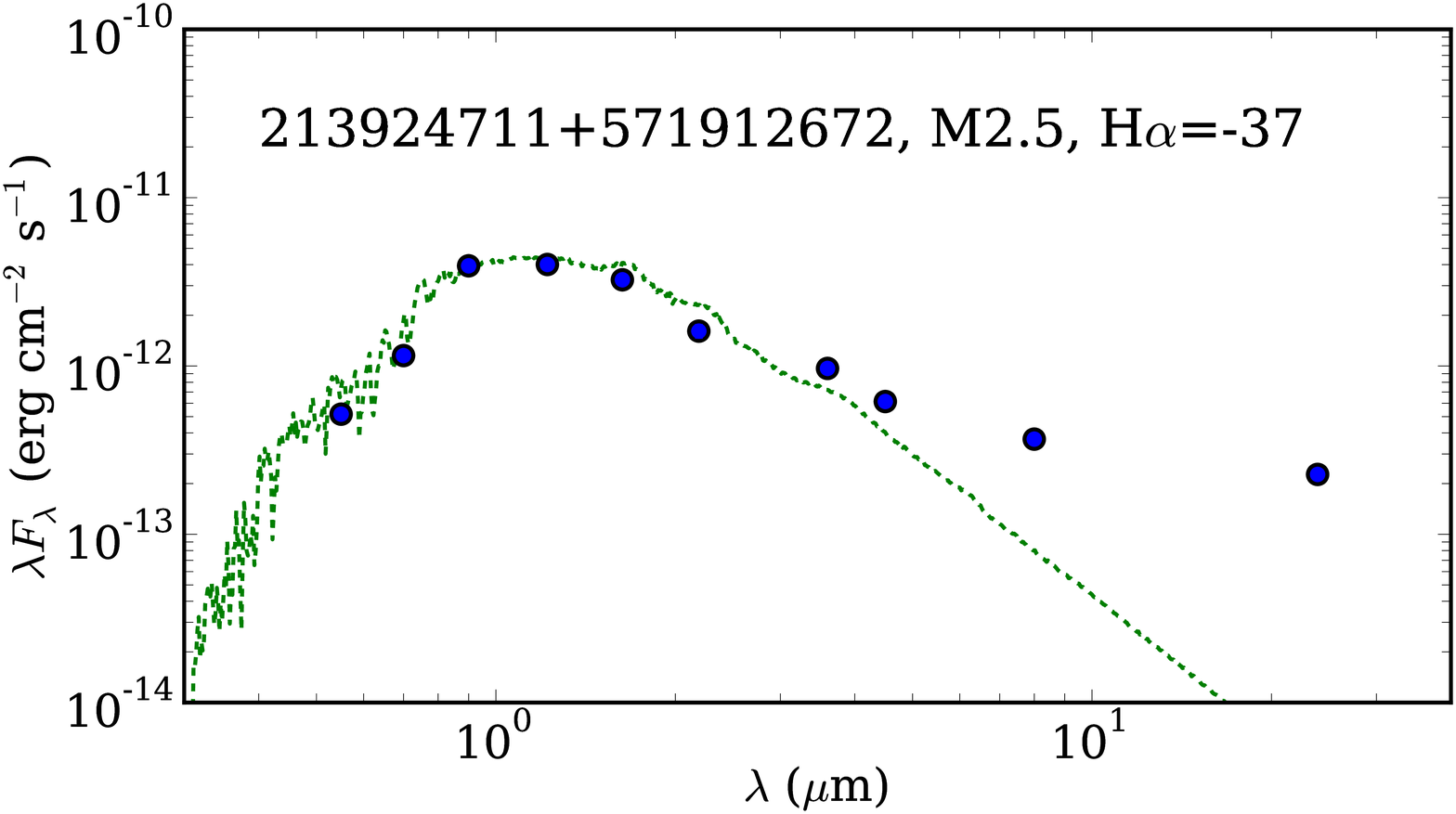,width=0.24\linewidth,clip=} \\
\epsfig{file=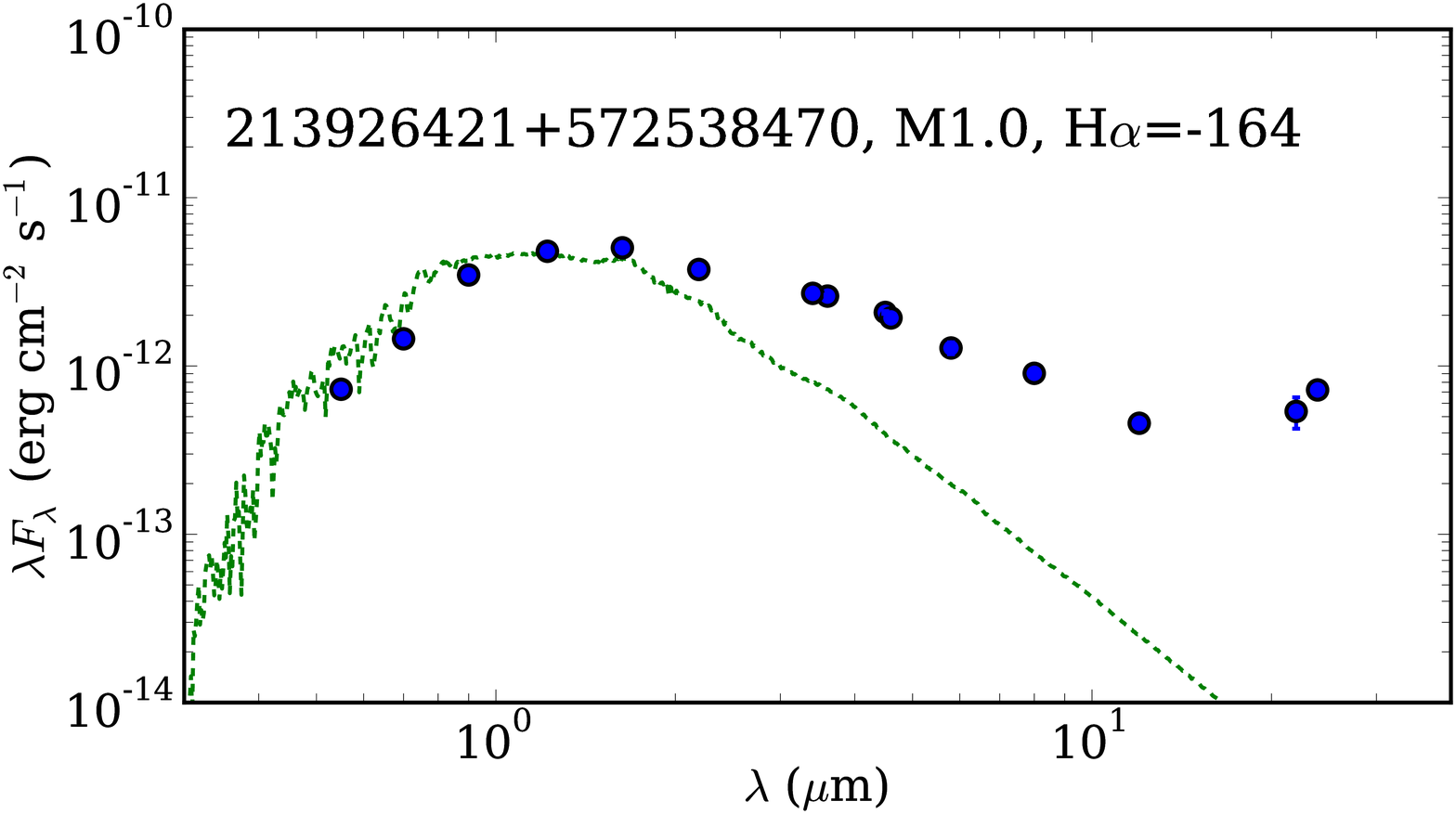,width=0.24\linewidth,clip=} &
\epsfig{file=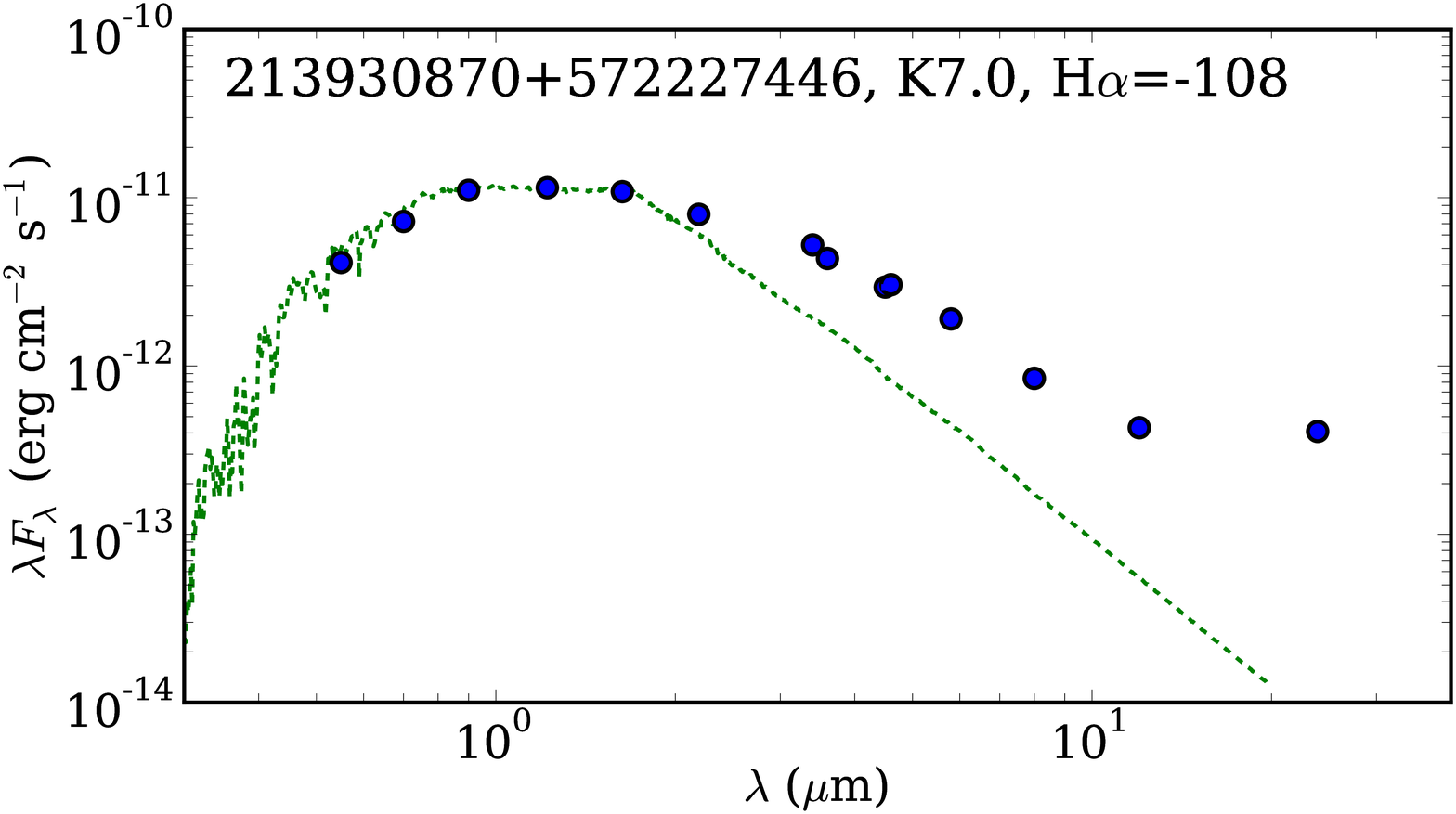,width=0.24\linewidth,clip=} &
\epsfig{file=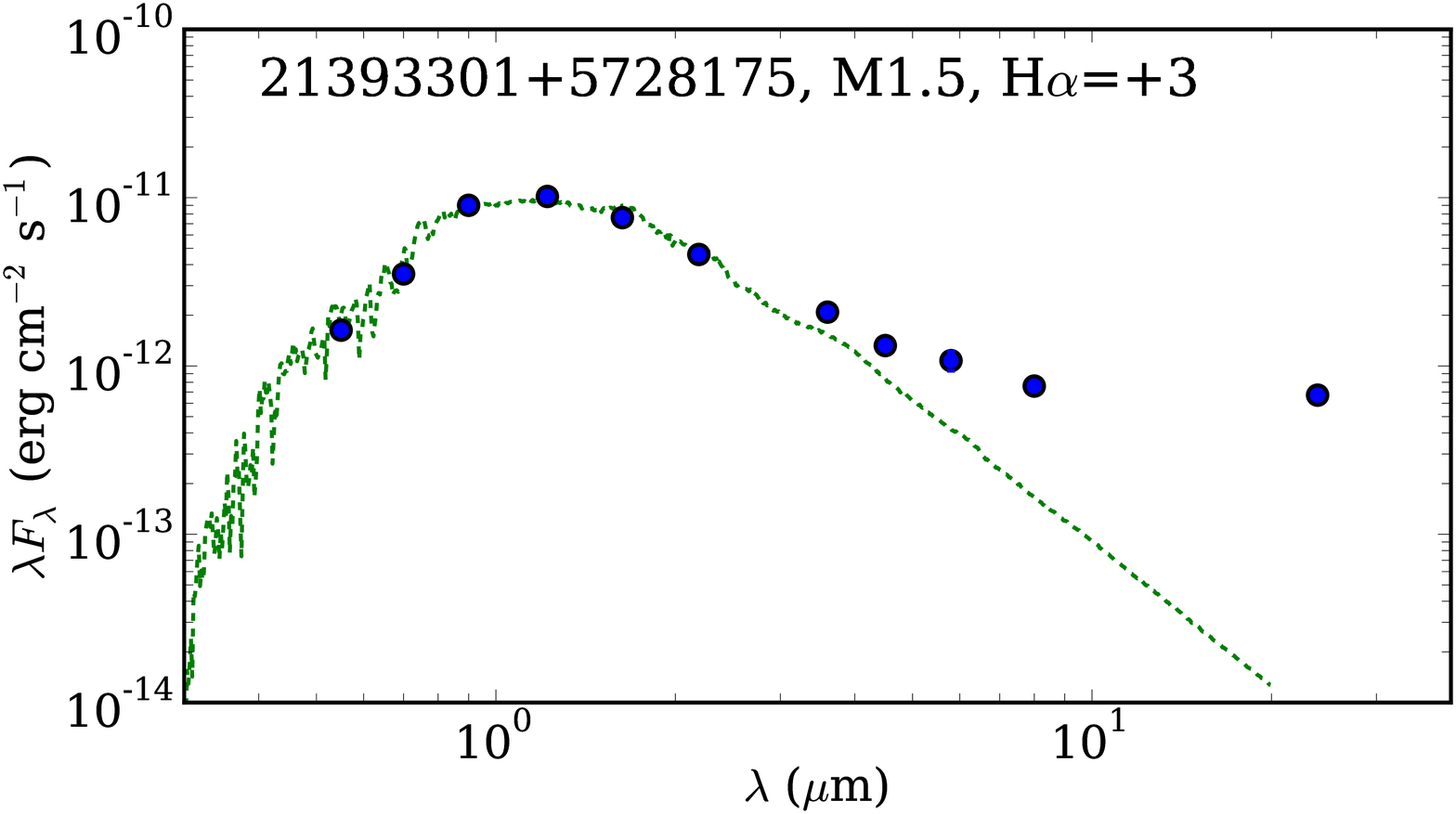,width=0.24\linewidth,clip=} &
\epsfig{file=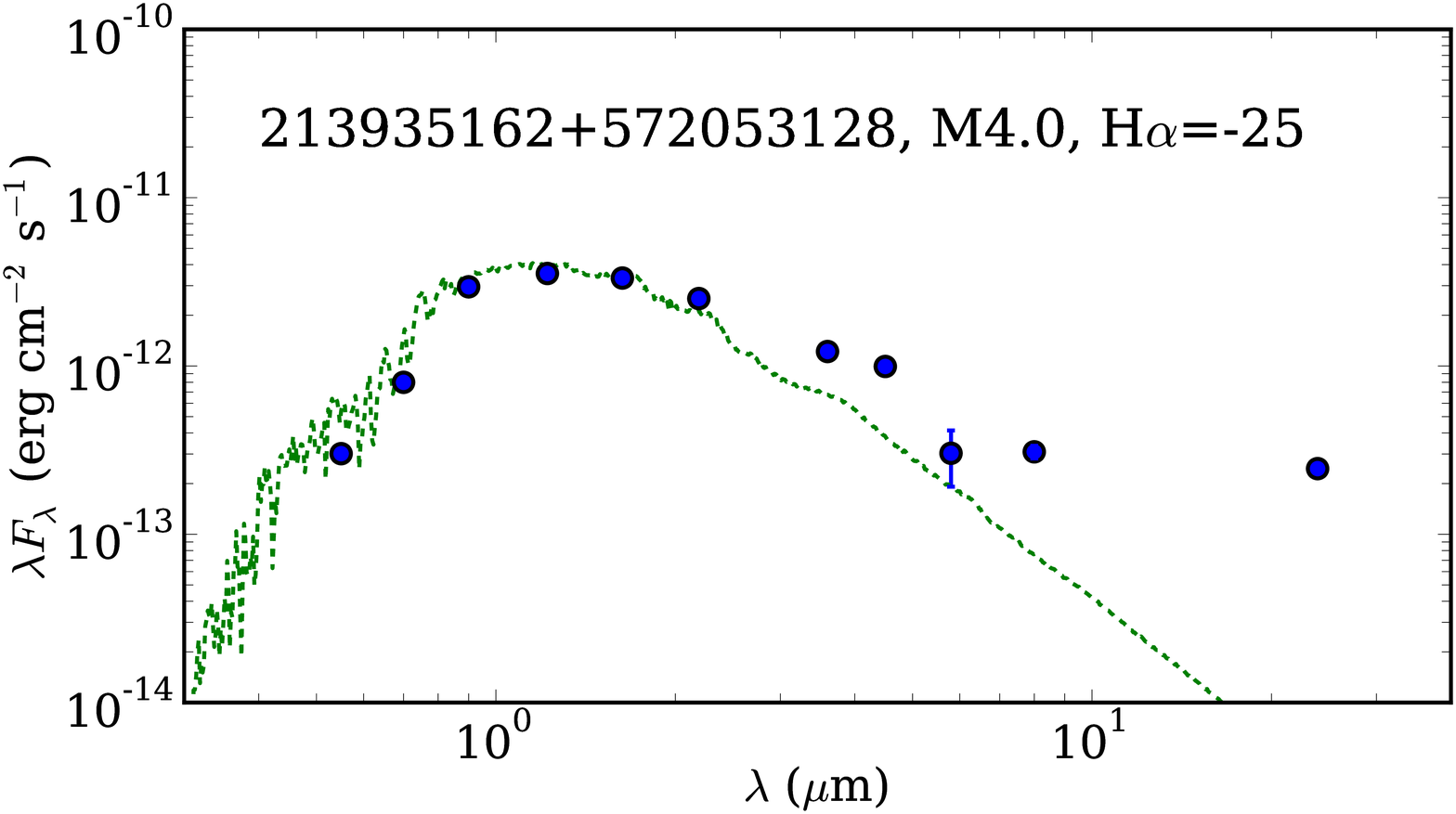,width=0.24\linewidth,clip=} \\
\epsfig{file=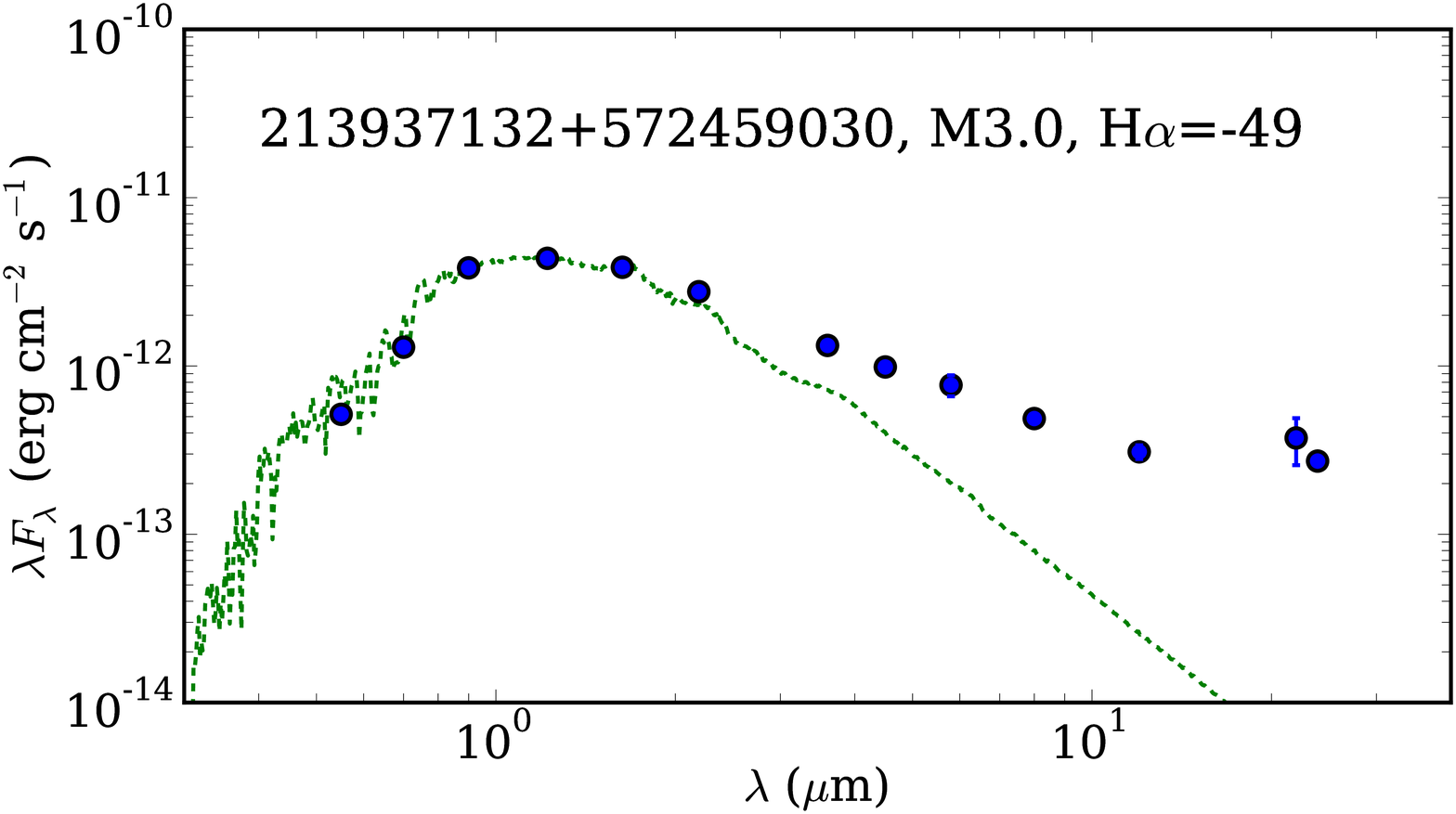,width=0.24\linewidth,clip=} &
\epsfig{file=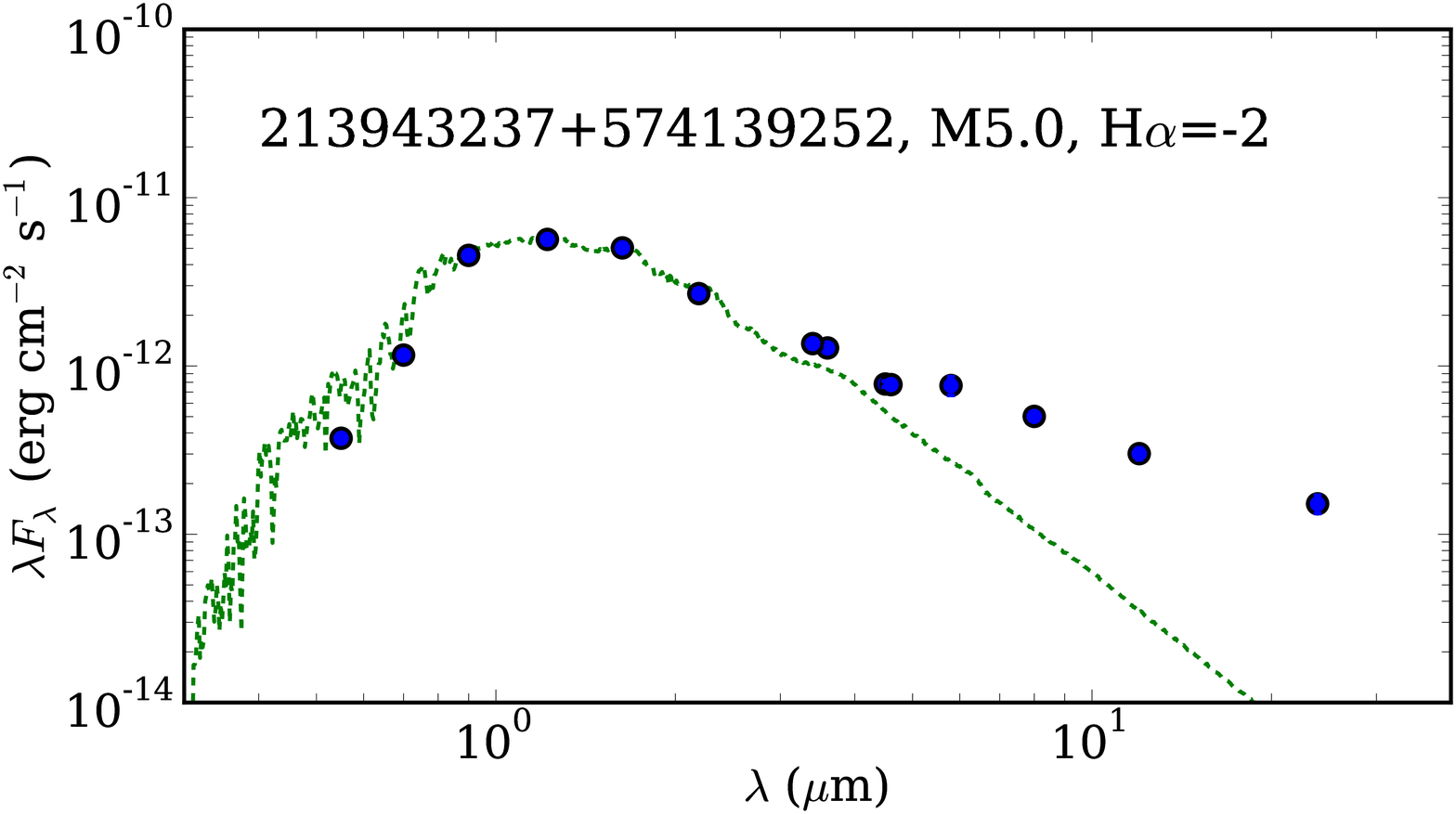,width=0.24\linewidth,clip=} &
\epsfig{file=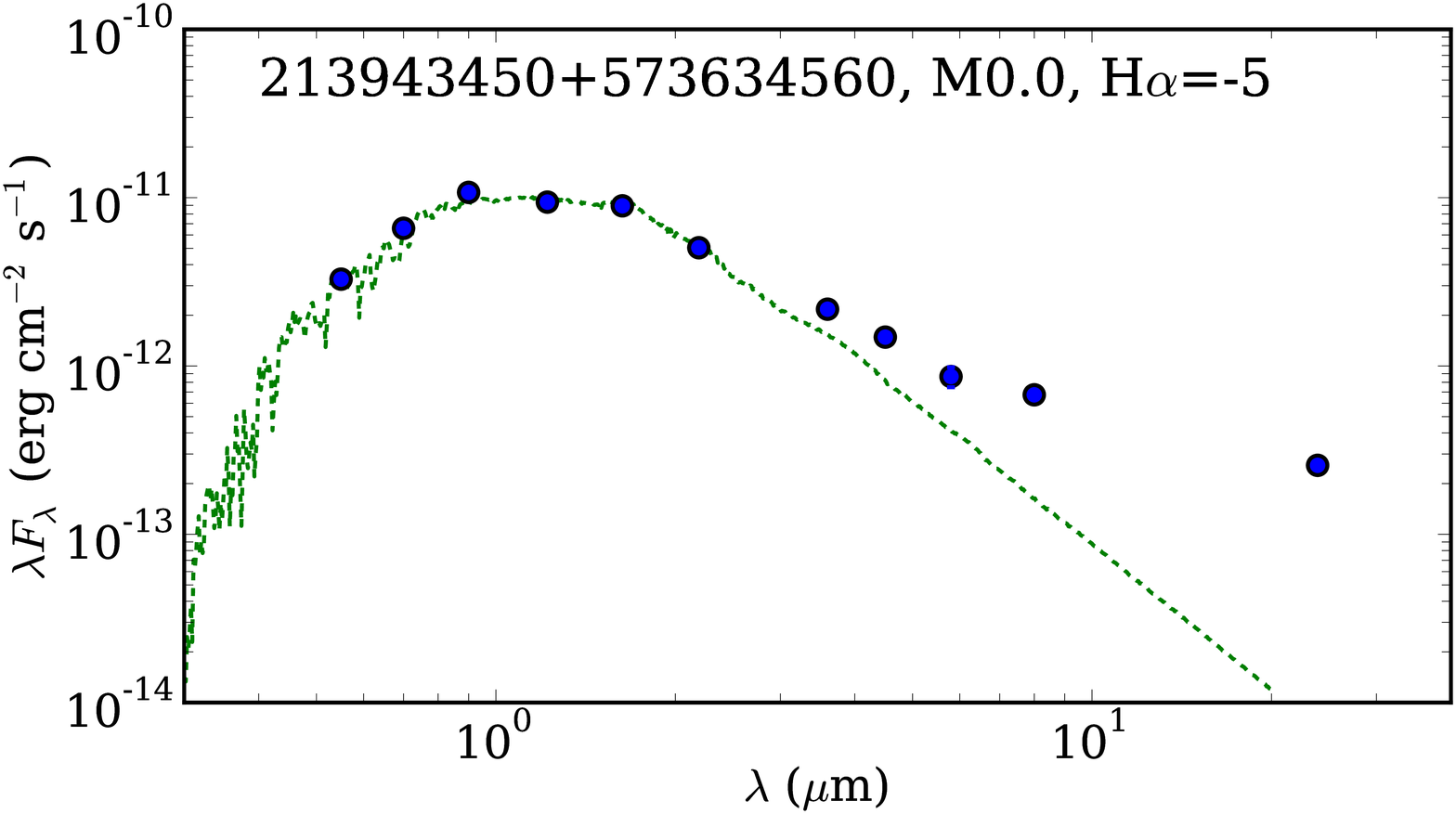,width=0.24\linewidth,clip=} &
\epsfig{file=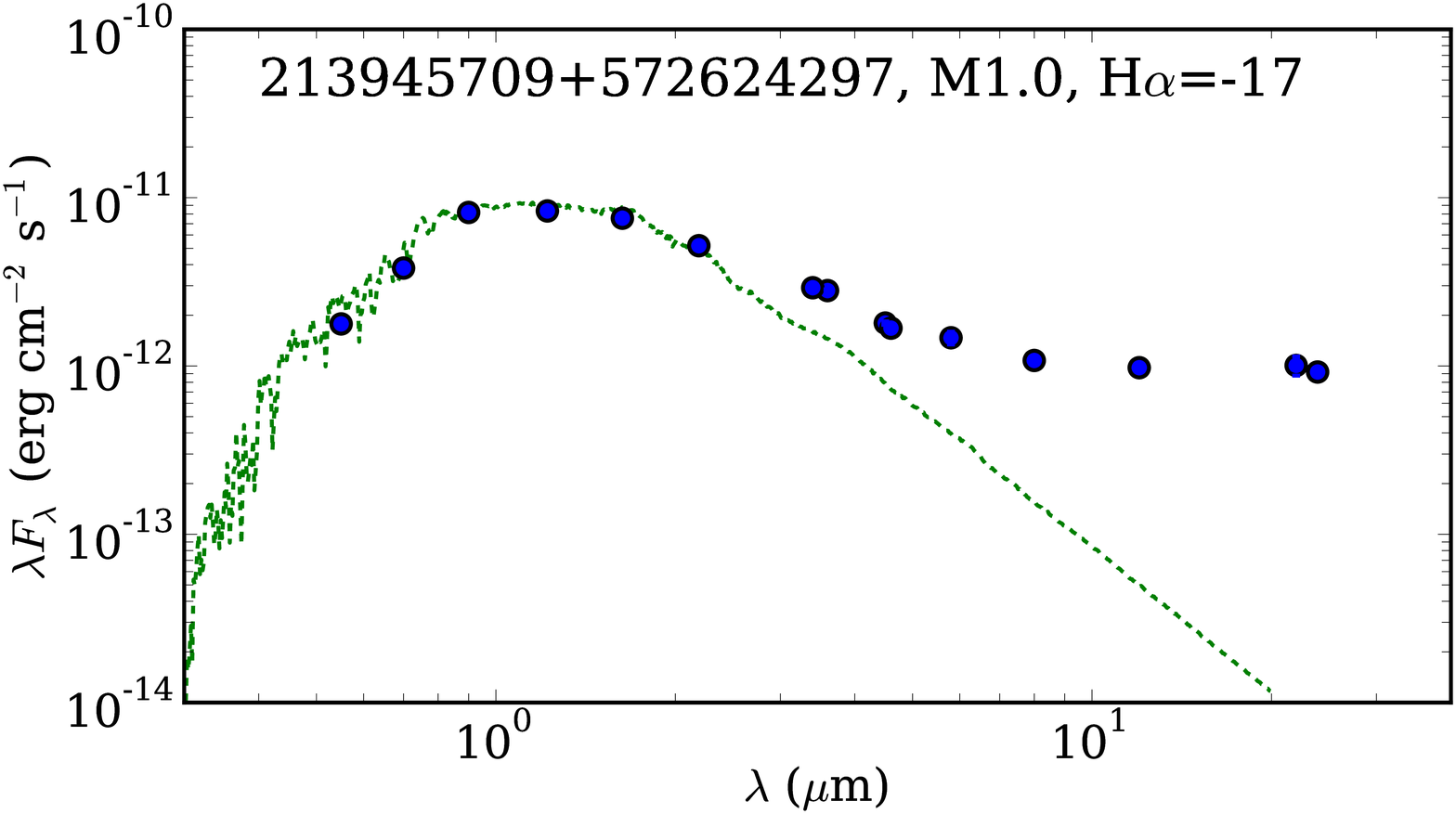,width=0.24\linewidth,clip=} \\
\epsfig{file=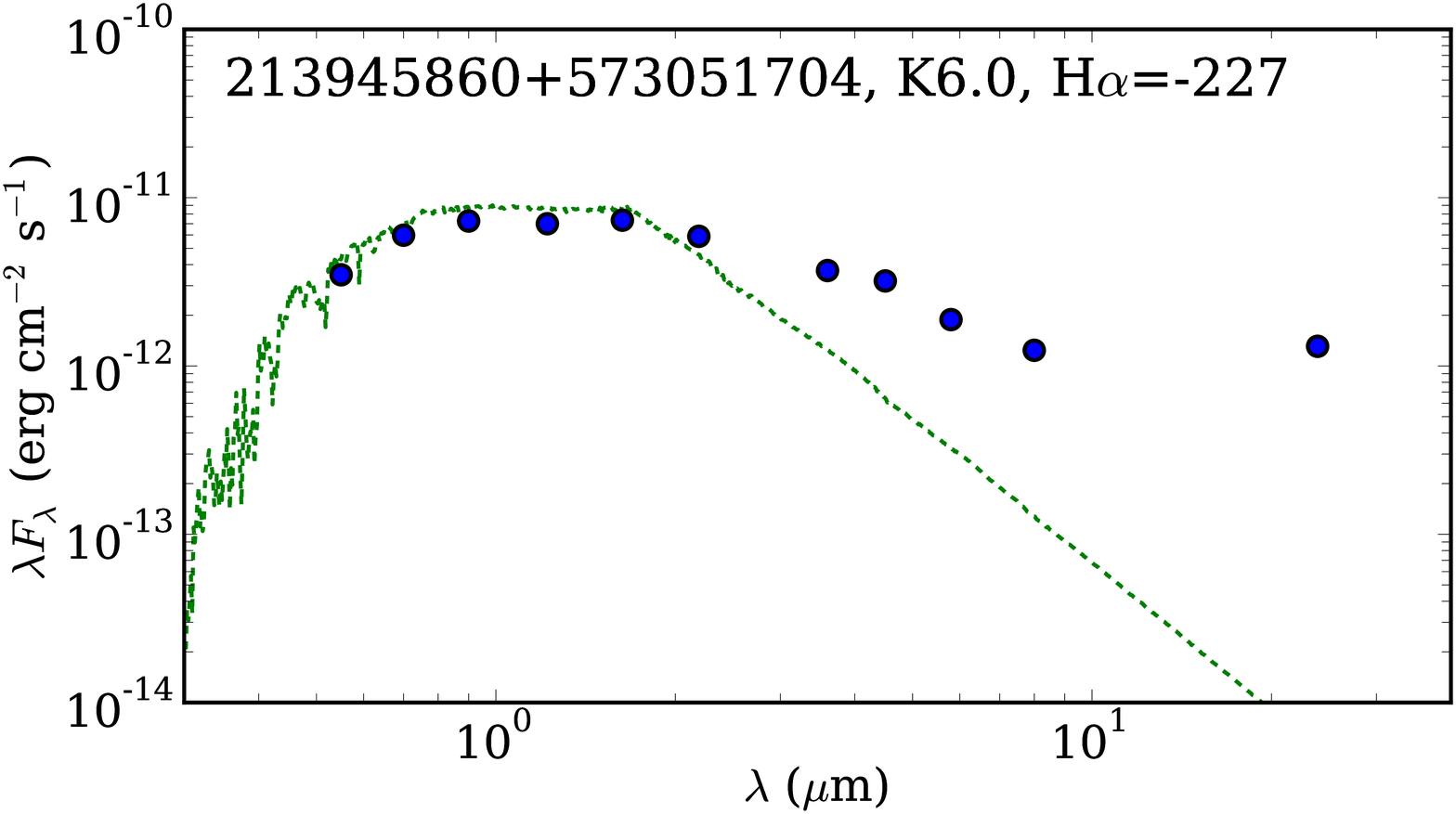,width=0.24\linewidth,clip=} &
\epsfig{file=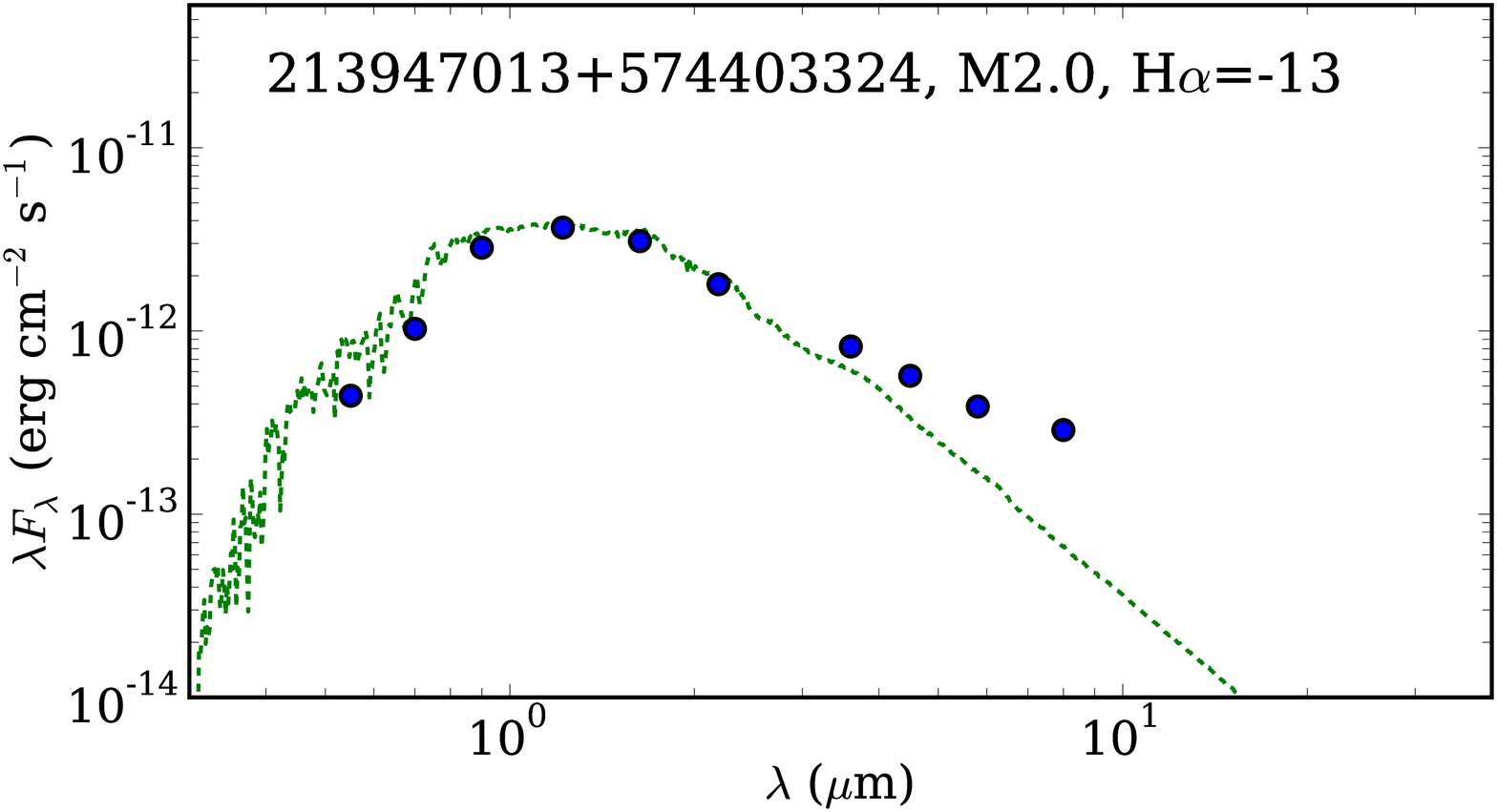,width=0.24\linewidth,clip=} &
\epsfig{file=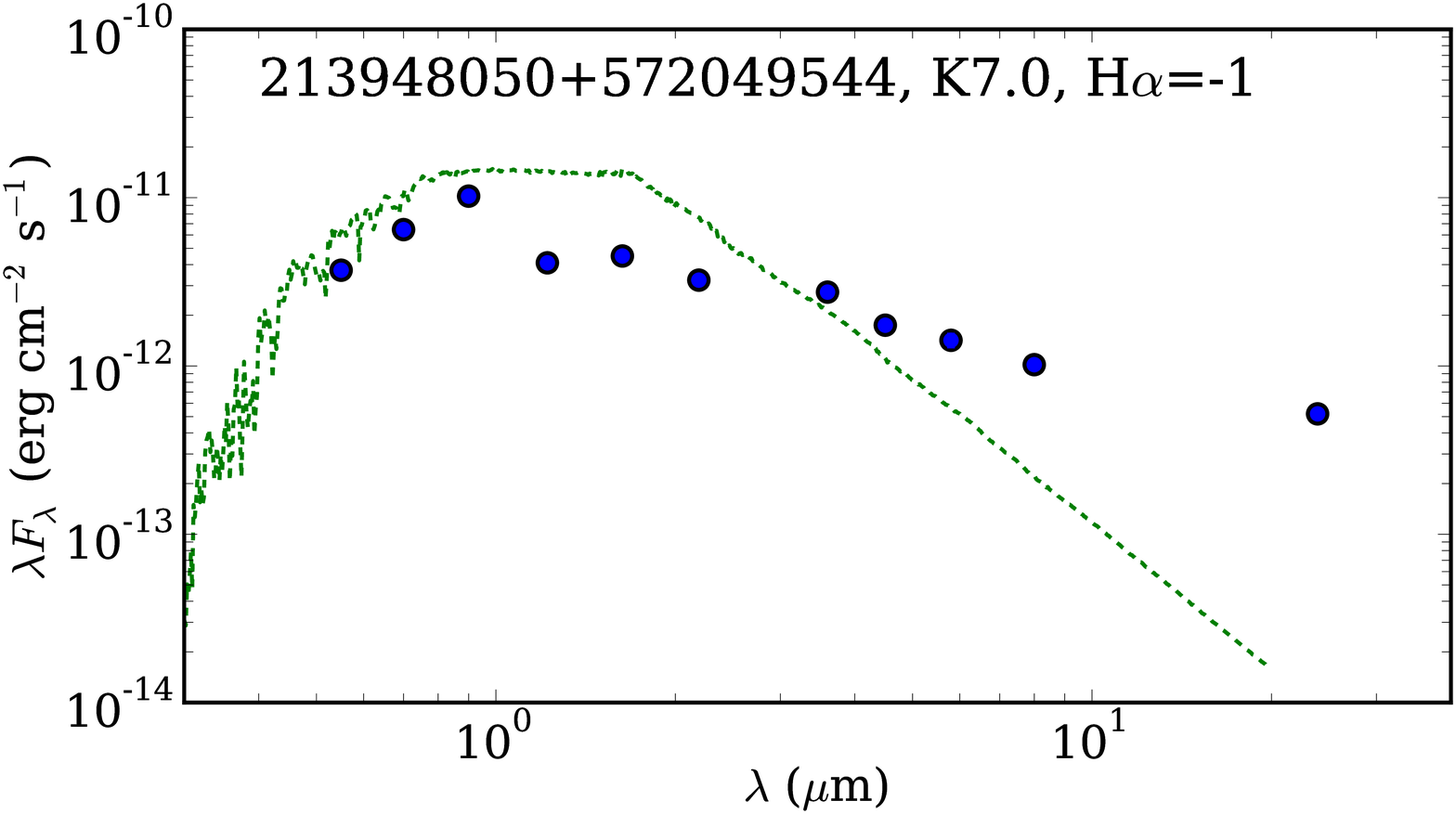,width=0.24\linewidth,clip=} &
\epsfig{file=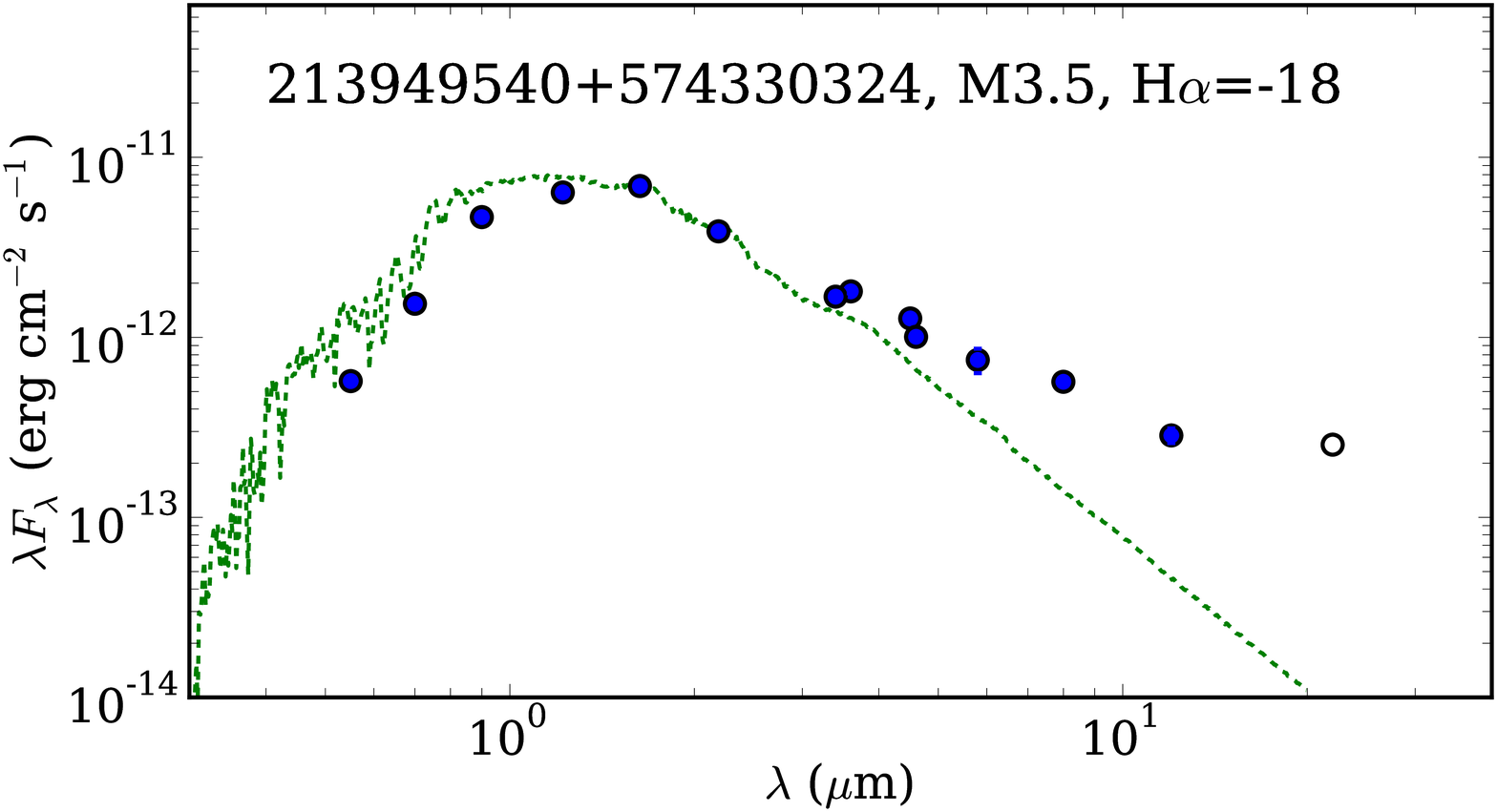,width=0.24\linewidth,clip=} \\
\epsfig{file=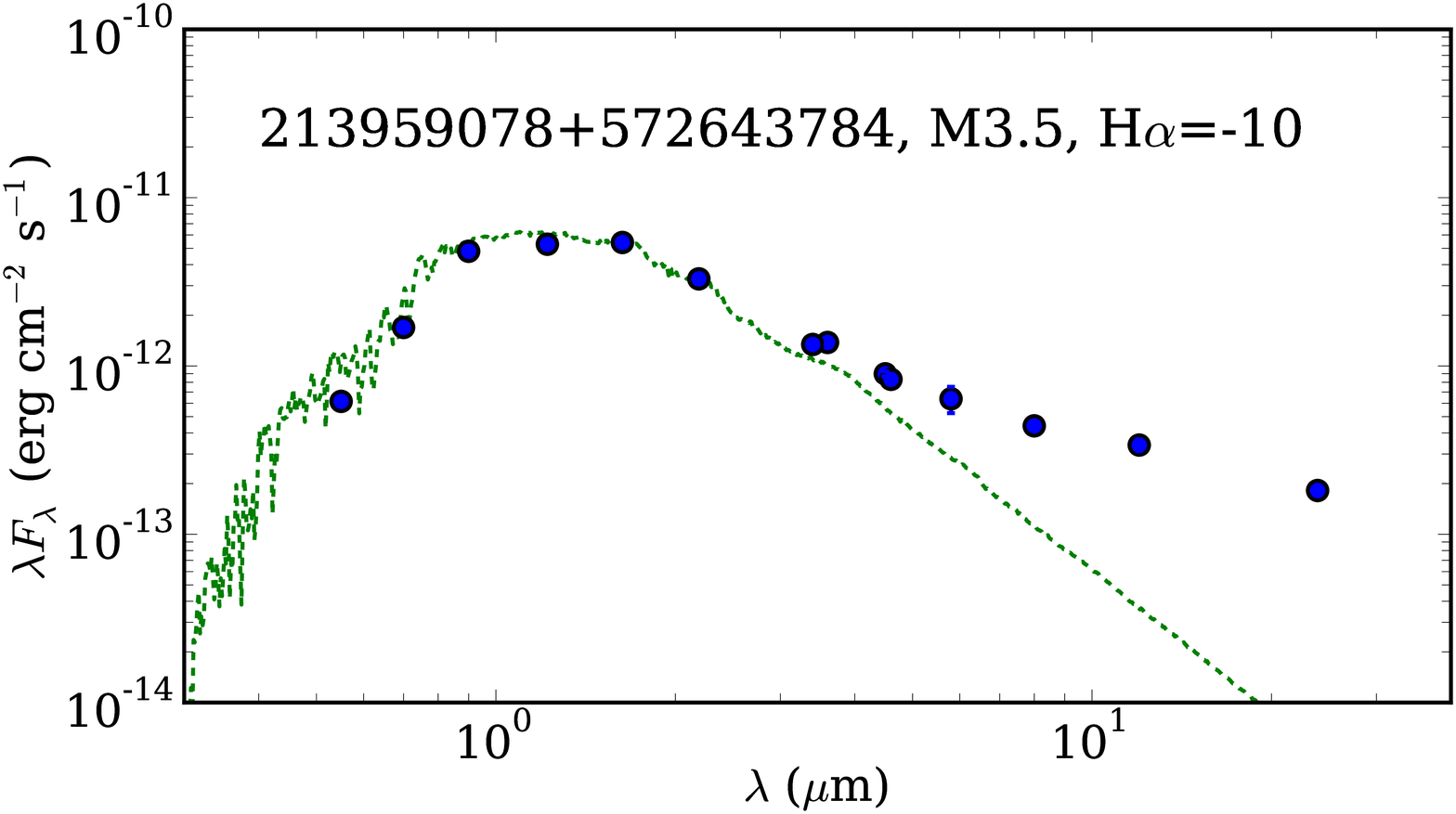,width=0.24\linewidth,clip=} &
\epsfig{file=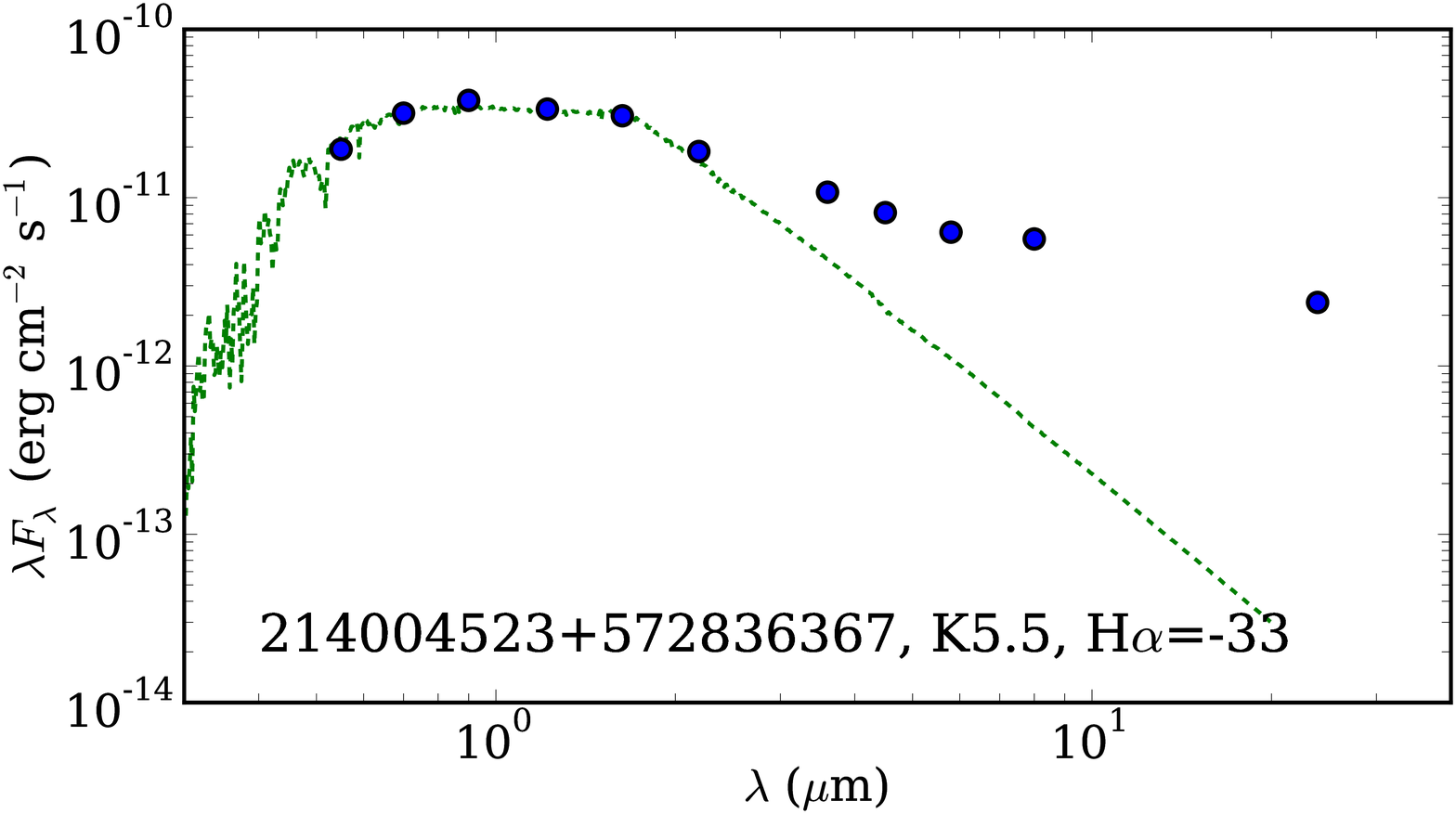,width=0.24\linewidth,clip=} &
\epsfig{file=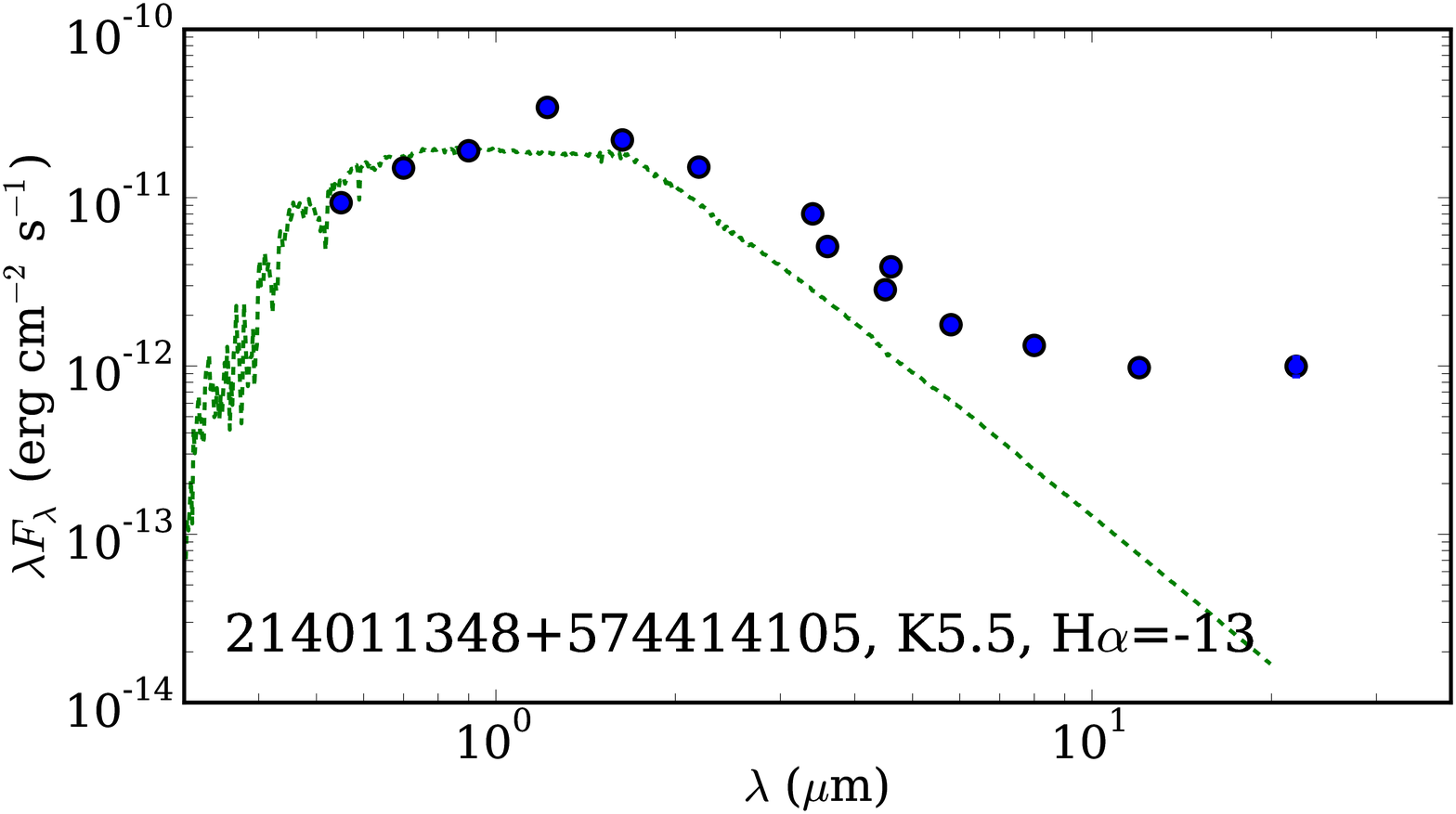,width=0.24\linewidth,clip=} &
\epsfig{file=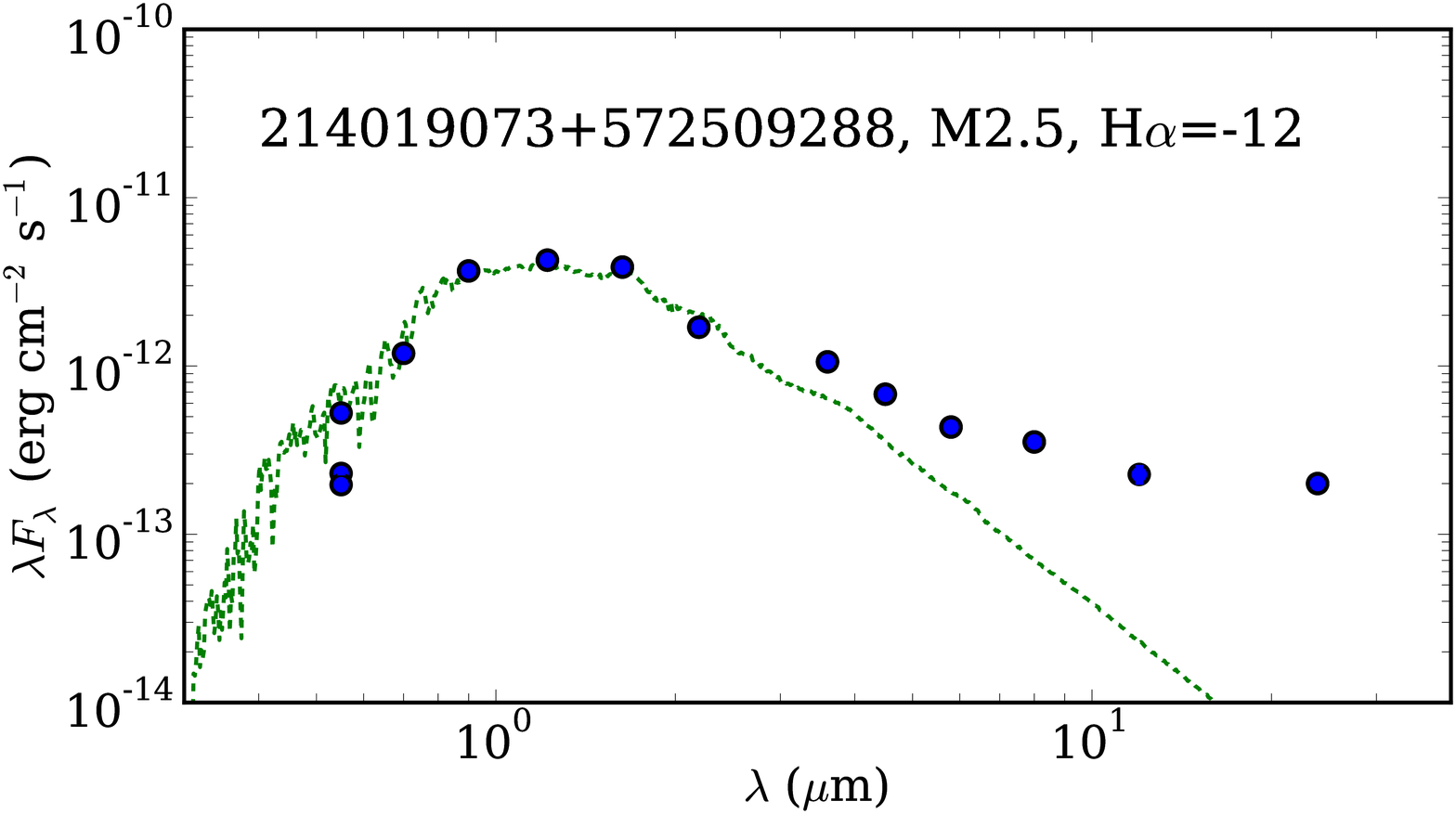,width=0.24\linewidth,clip=} \\
\epsfig{file=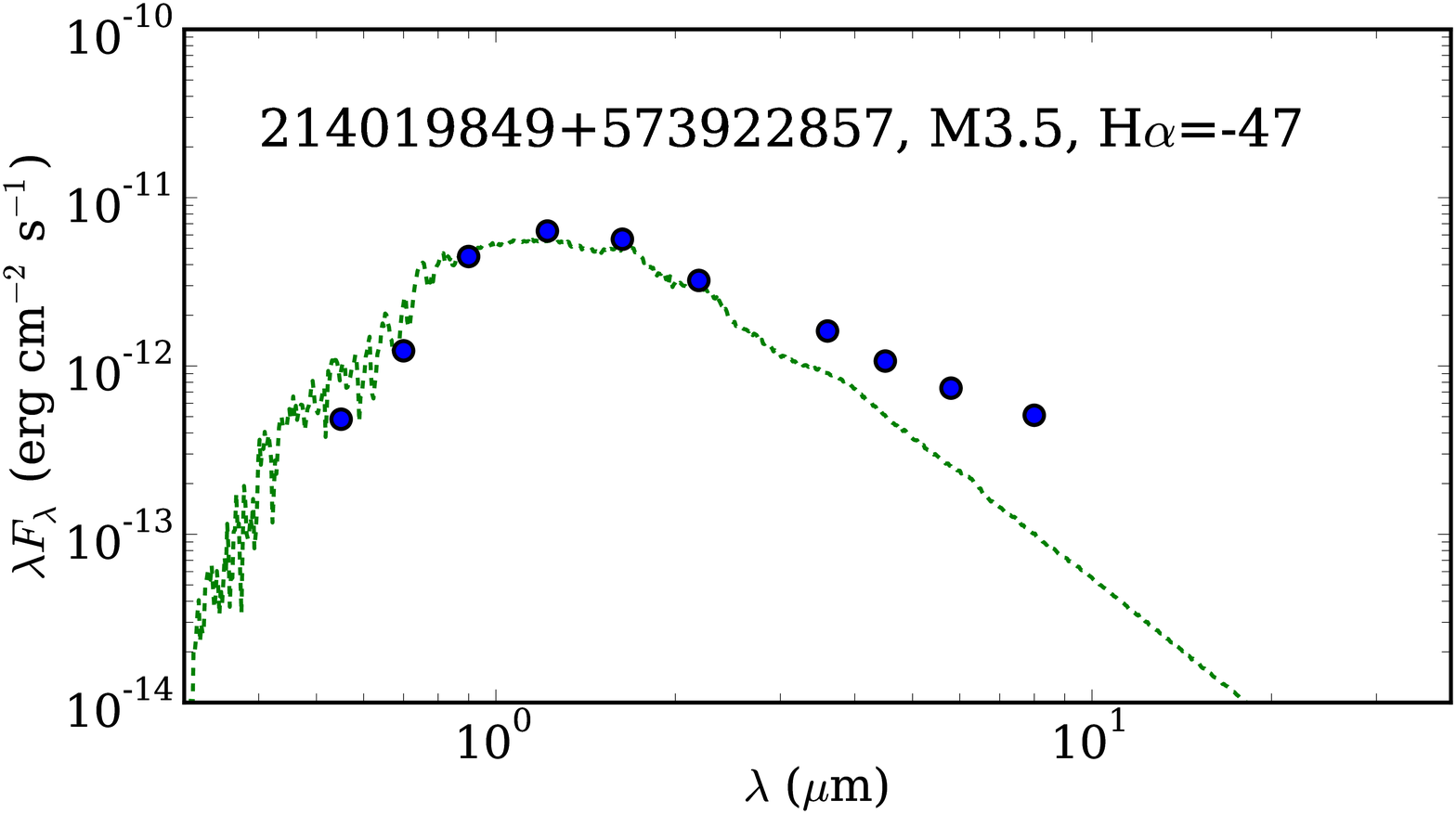,width=0.24\linewidth,clip=} &
\epsfig{file=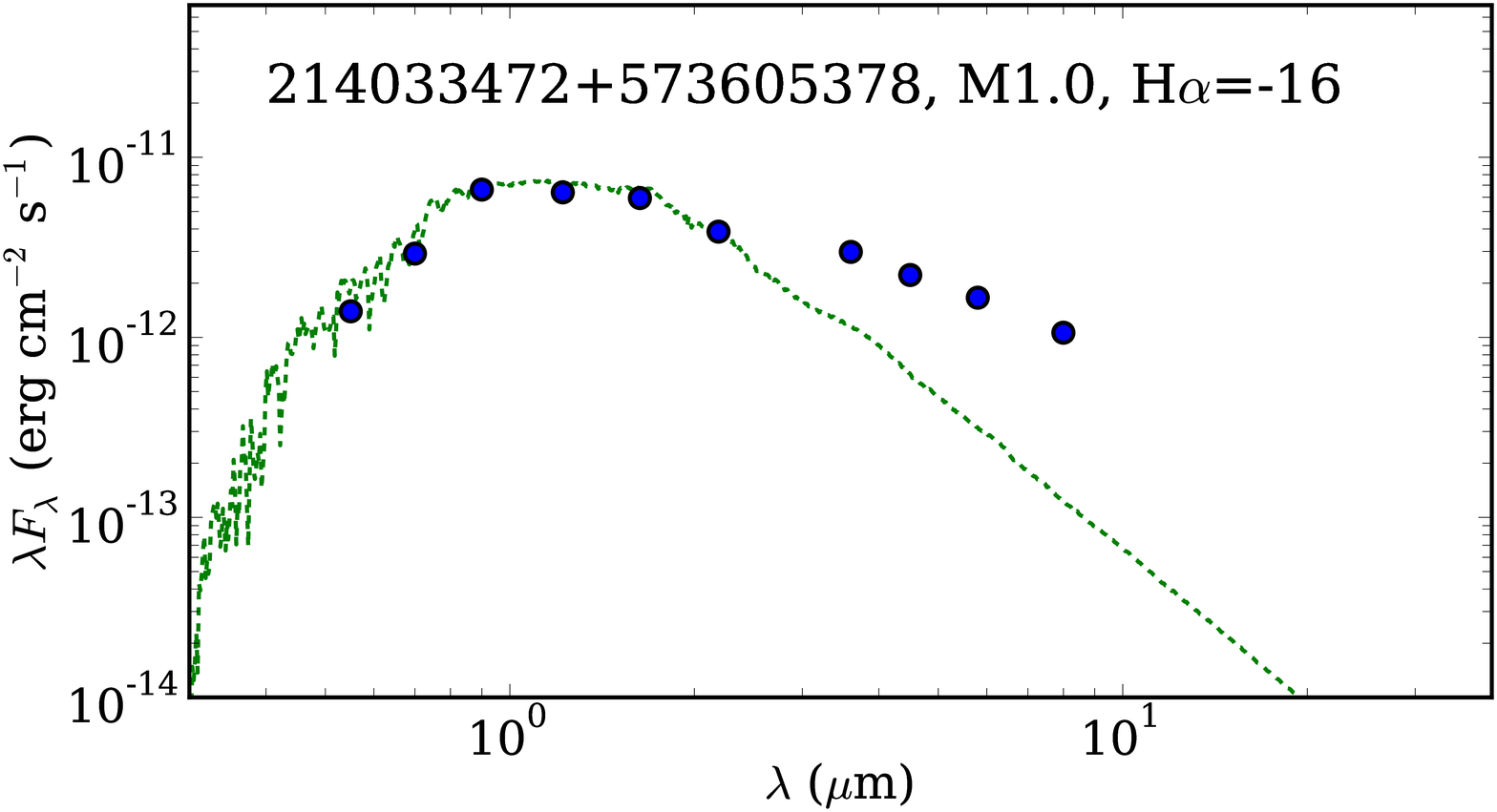,width=0.24\linewidth,clip=} &
\epsfig{file=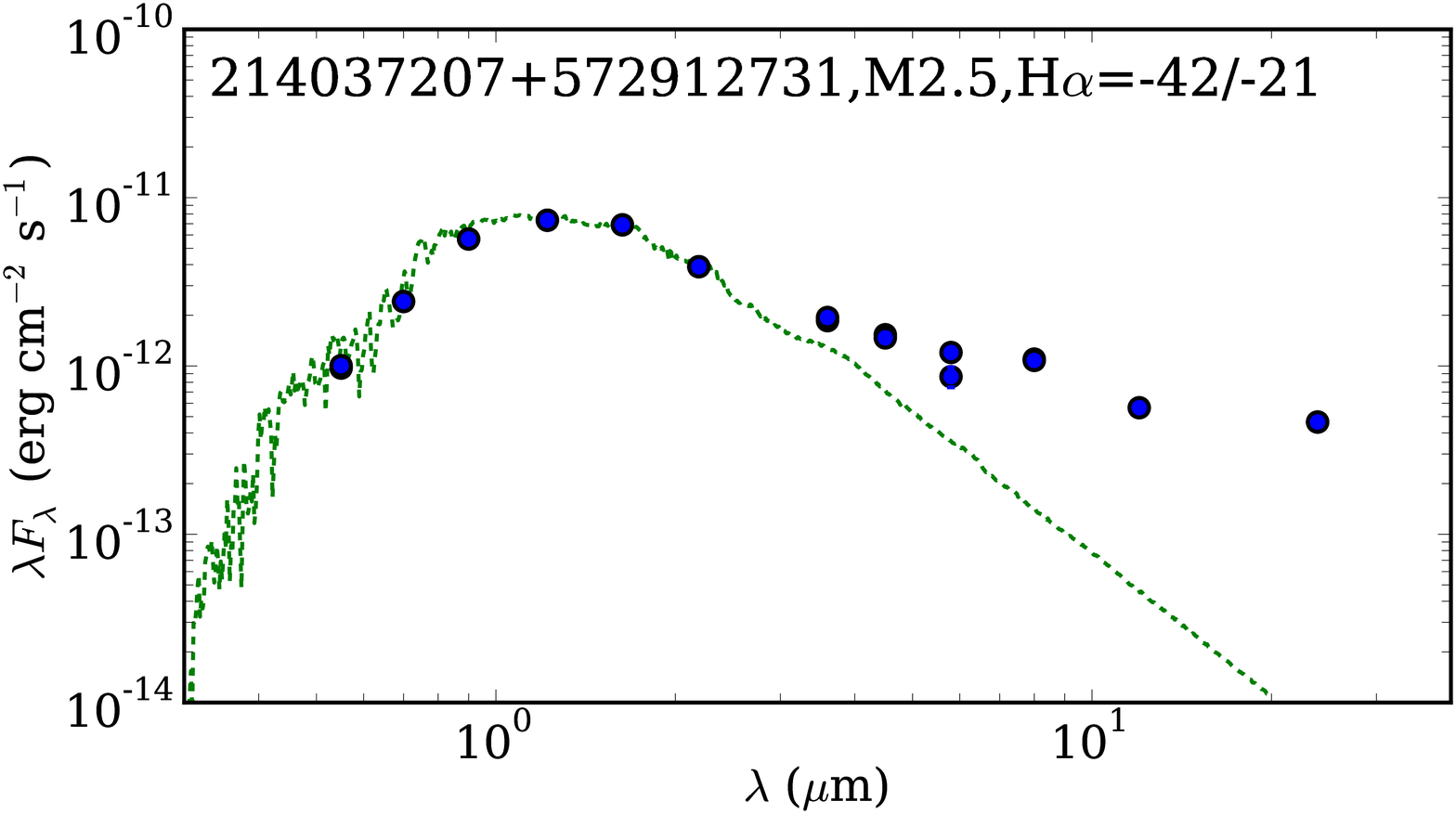,width=0.24\linewidth,clip=} &
\epsfig{file=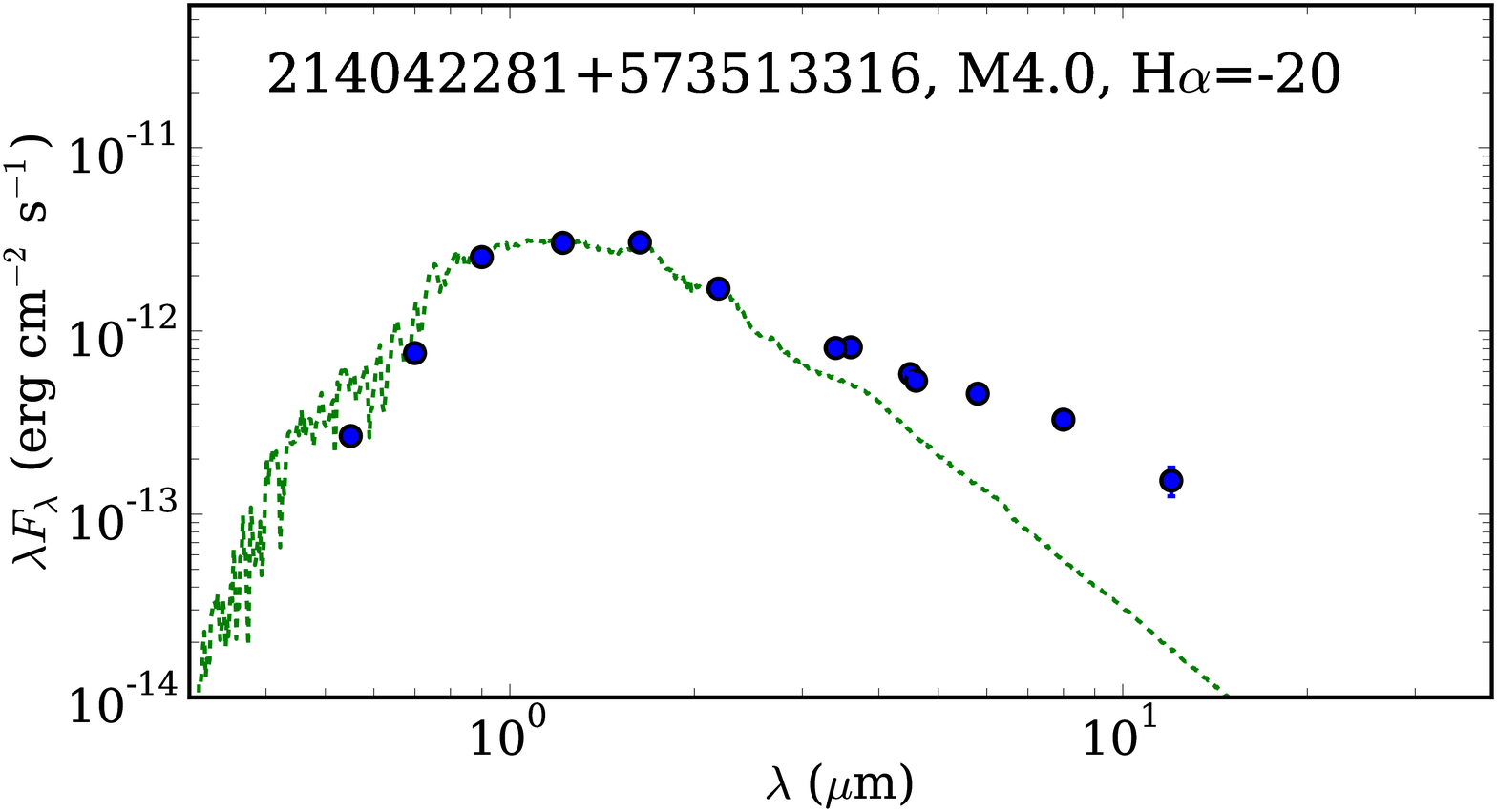,width=0.24\linewidth,clip=} \\
\epsfig{file=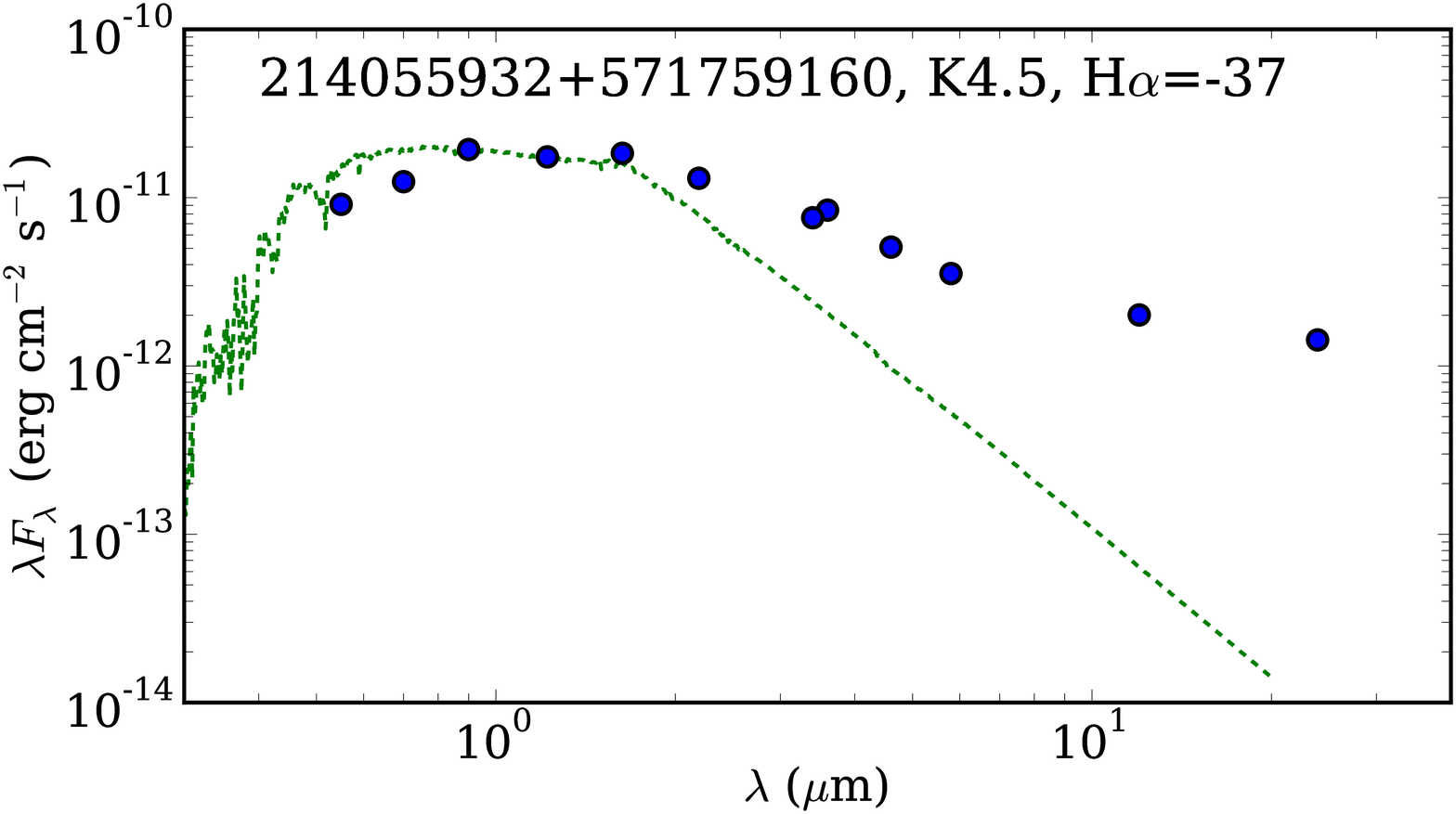,width=0.24\linewidth,clip=} &
\epsfig{file=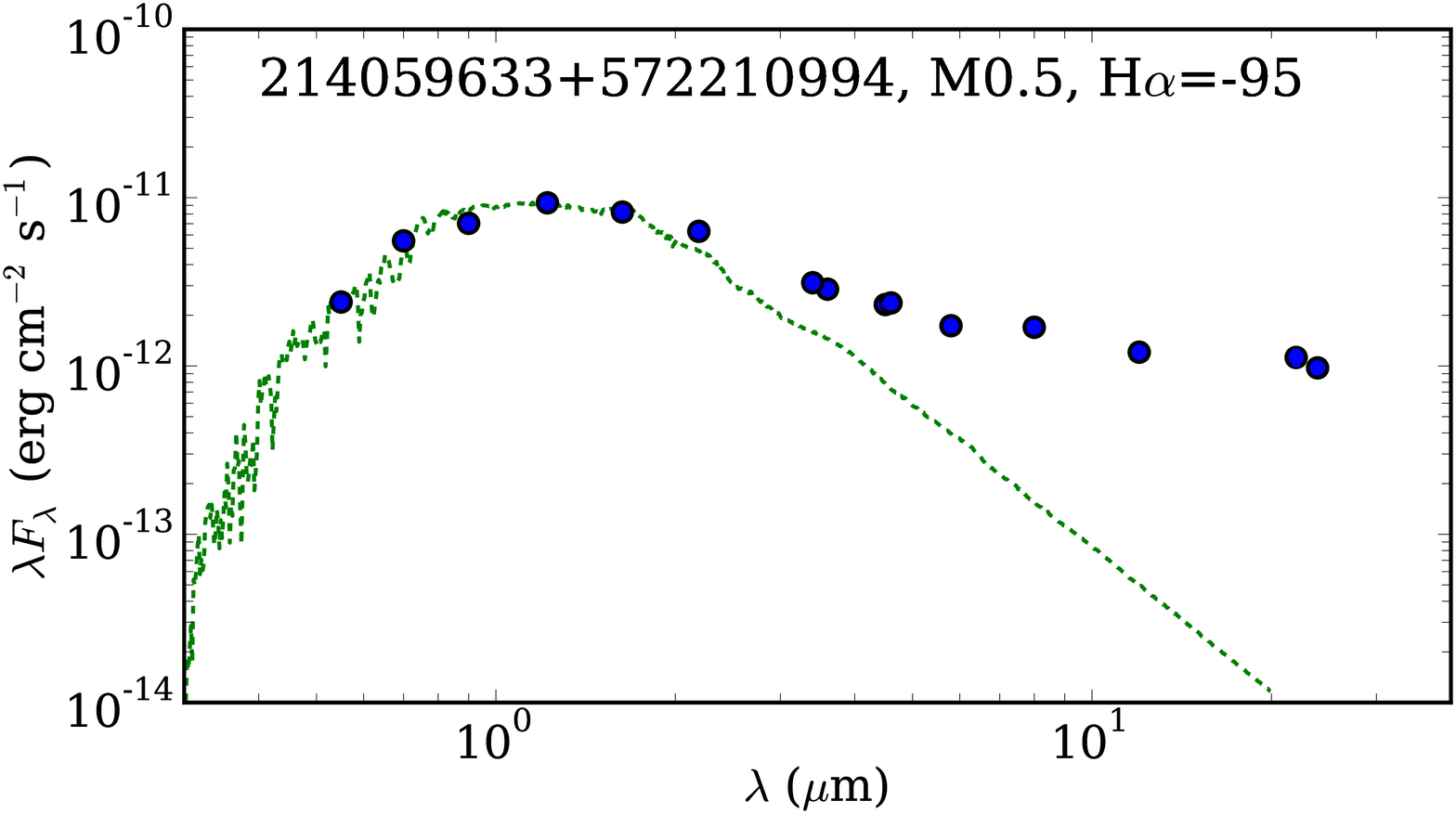,width=0.24\linewidth,clip=} &
\epsfig{file=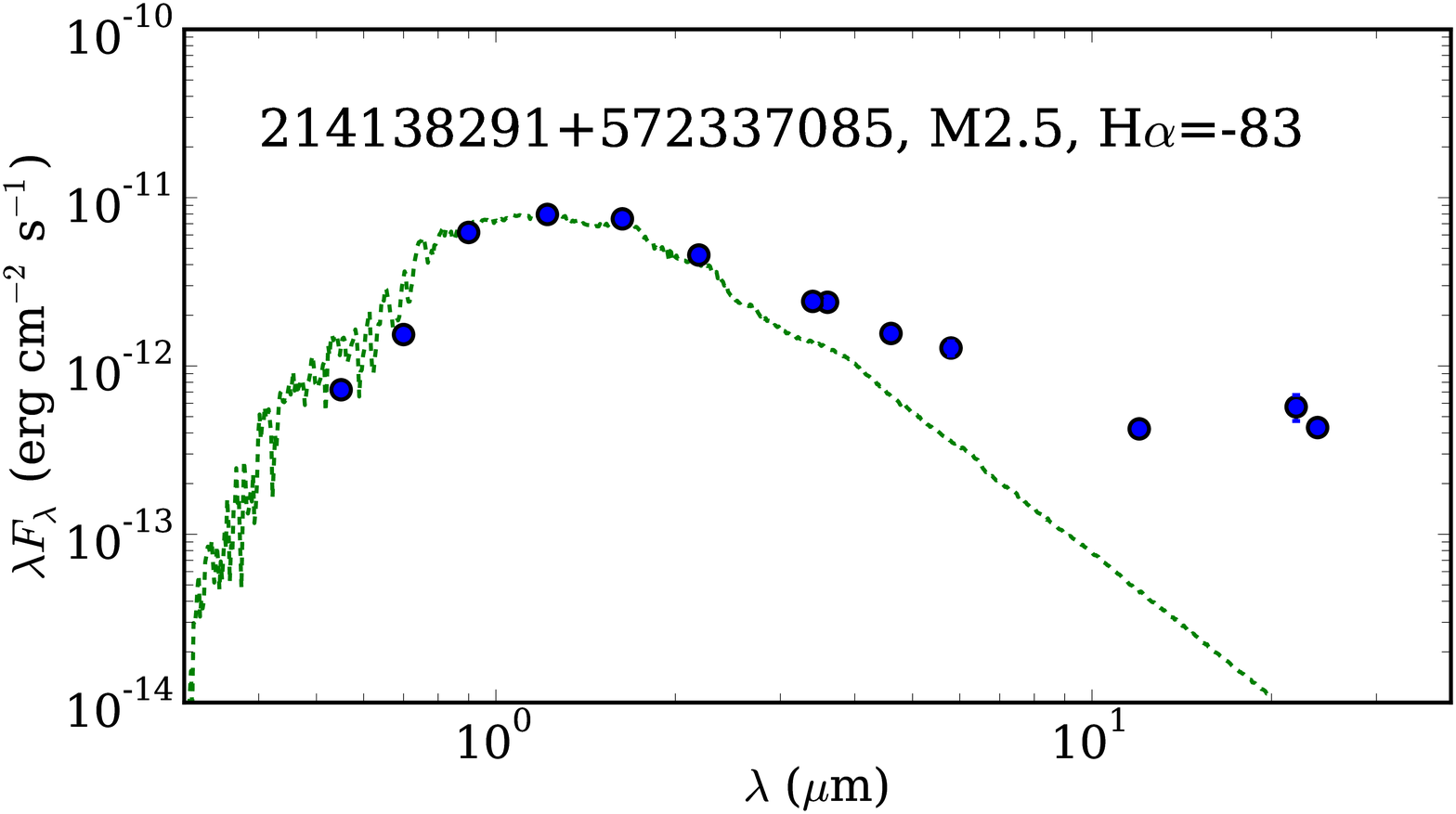,width=0.24\linewidth,clip=} \\
\end{tabular}
\caption{SEDs of the members and probably members with IR excess typical of full-disks (continued). 
213917481+571747432 has strong 8$\mu$m, could be PTD. 213911452+572425205 has [3.6]-[4.5]=0.27 and
a strong kink in the SED, it could be a PTD although its 24$\mu$m flux is likely contaminated
by cloud emission. 
214011348+574414105 has [3.6]-[4.5]=0.11 but a SED that suggests strong variability and a flared,
massive full-disk.\label{cttsseds2-fig}}
\end{figure*}

\begin{figure*}
\centering
\begin{tabular}{ccccc}
\epsfig{file=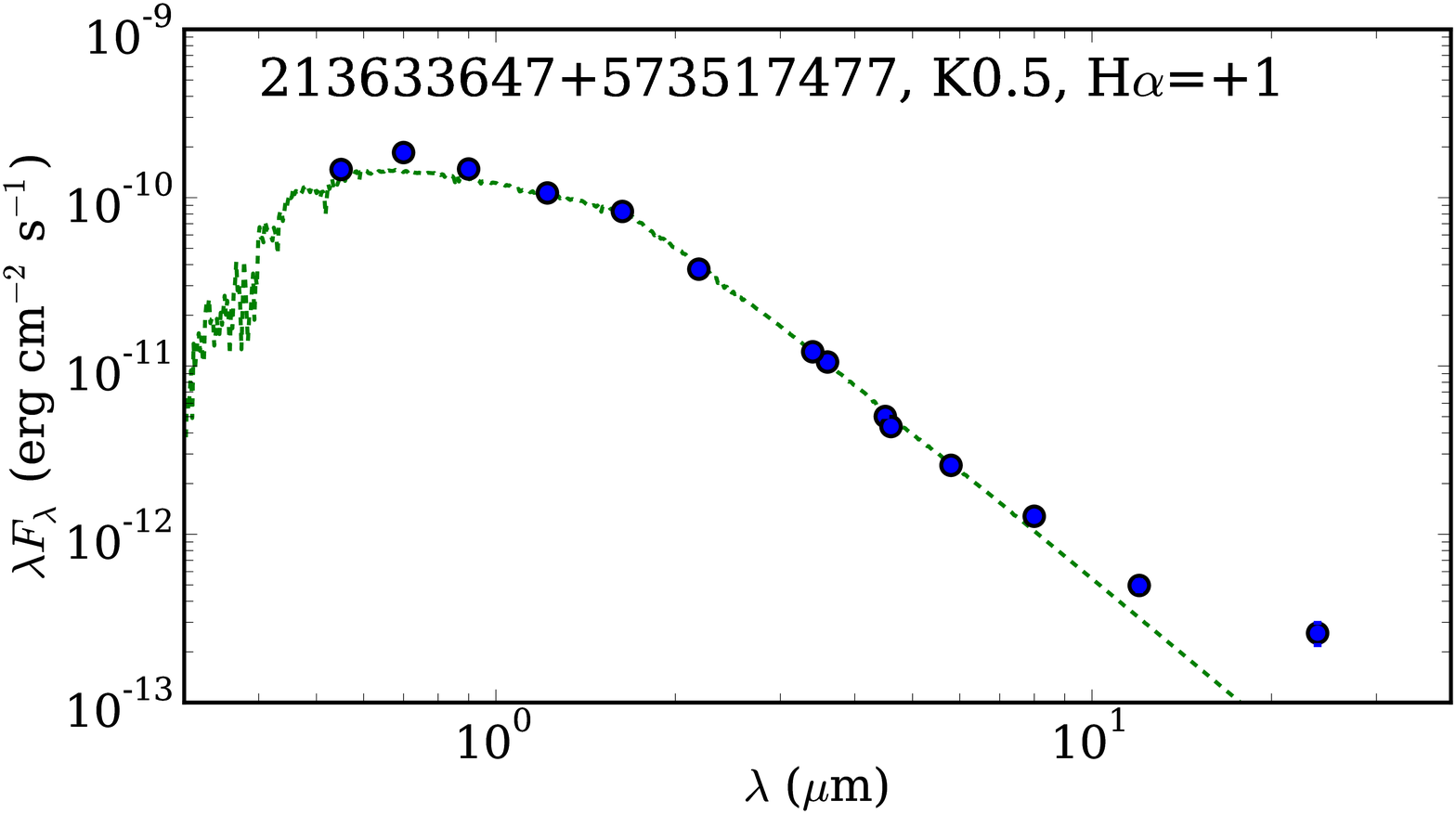,width=0.24\linewidth,clip=} &
\epsfig{file=213655283+572551668.eps,width=0.24\linewidth,clip=} &
\epsfig{file=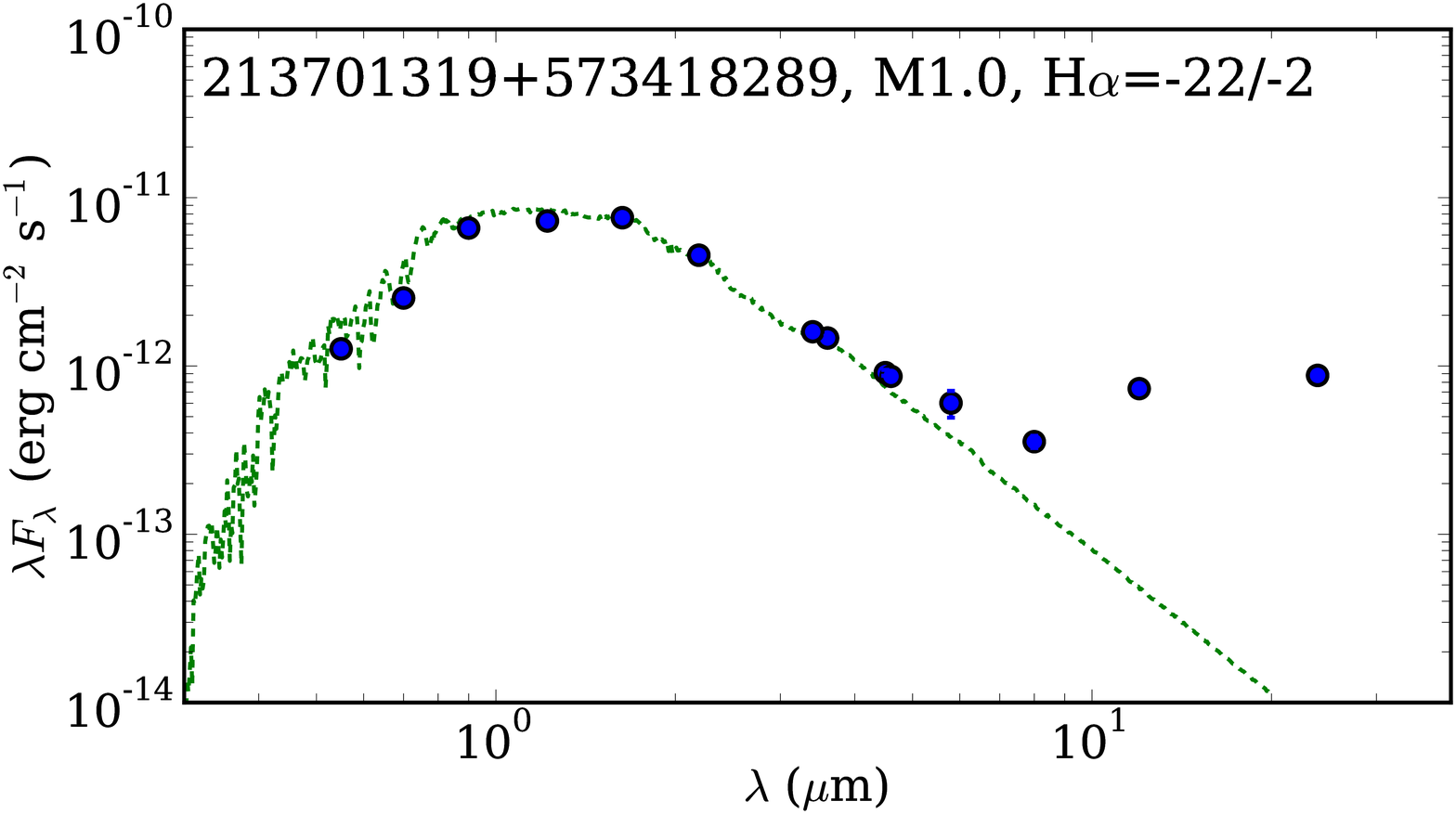,width=0.24\linewidth,clip=} &
\epsfig{file=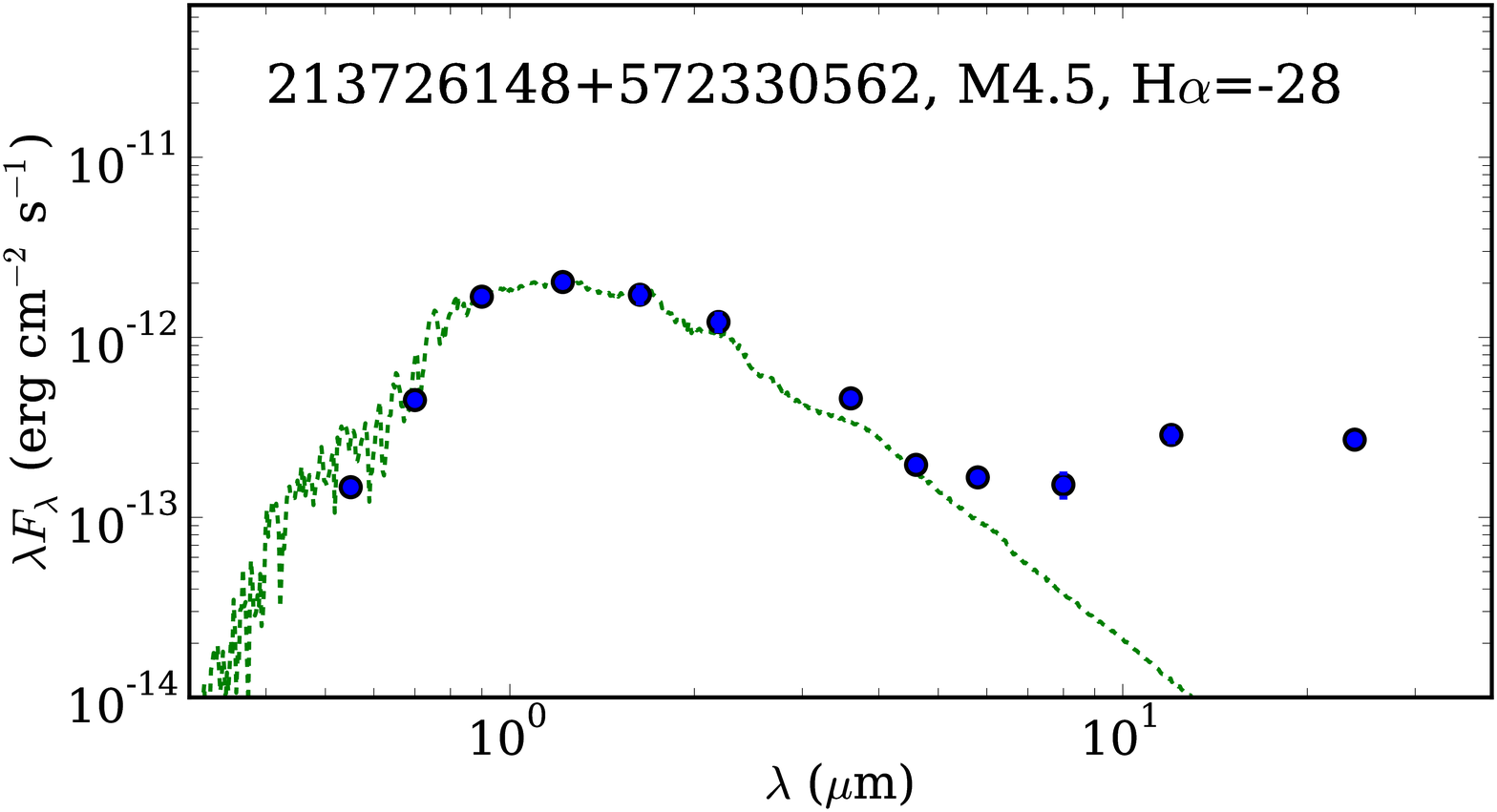,width=0.24\linewidth,clip=} \\
\epsfig{file=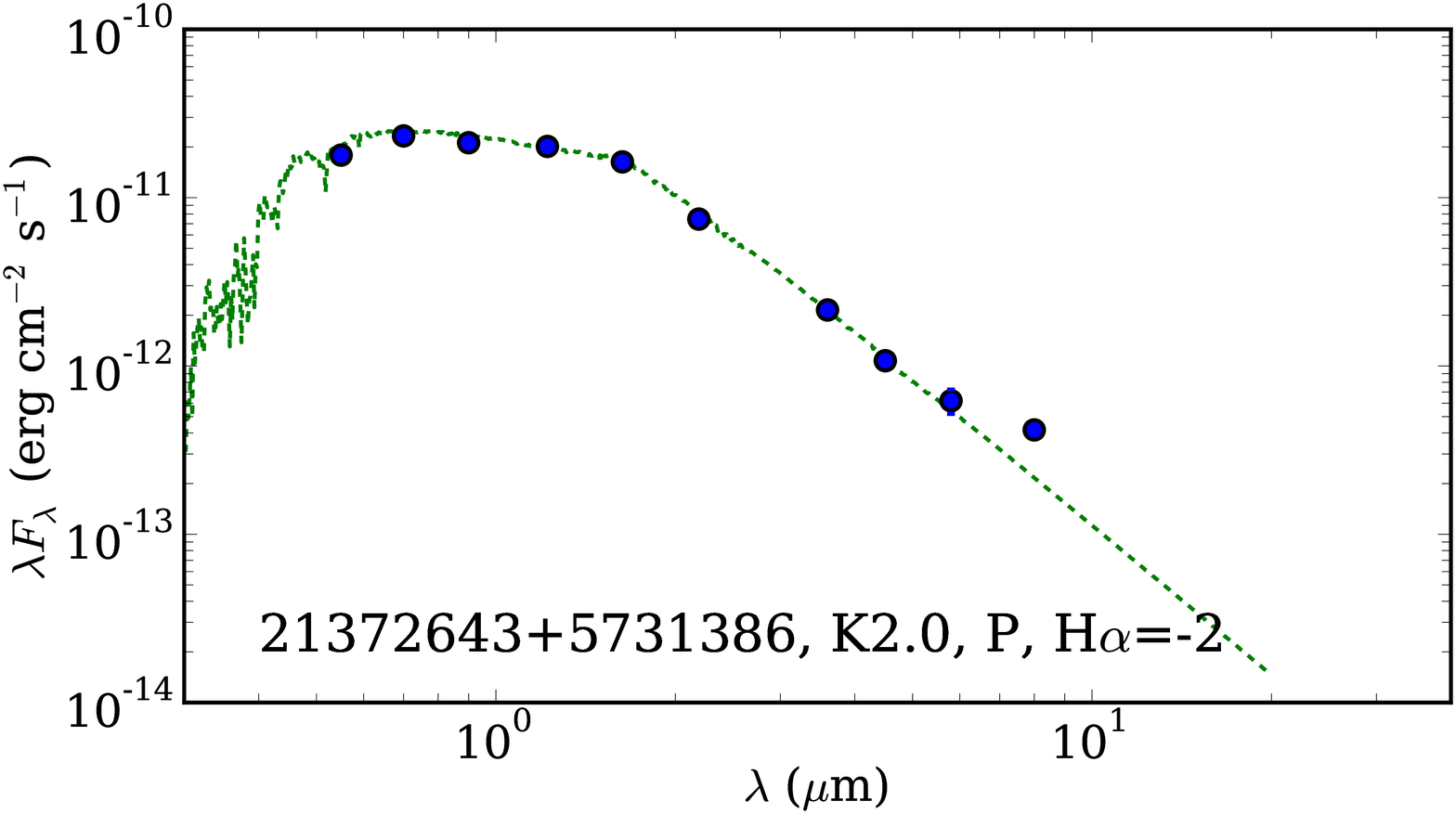,width=0.24\linewidth,clip=} &
\epsfig{file=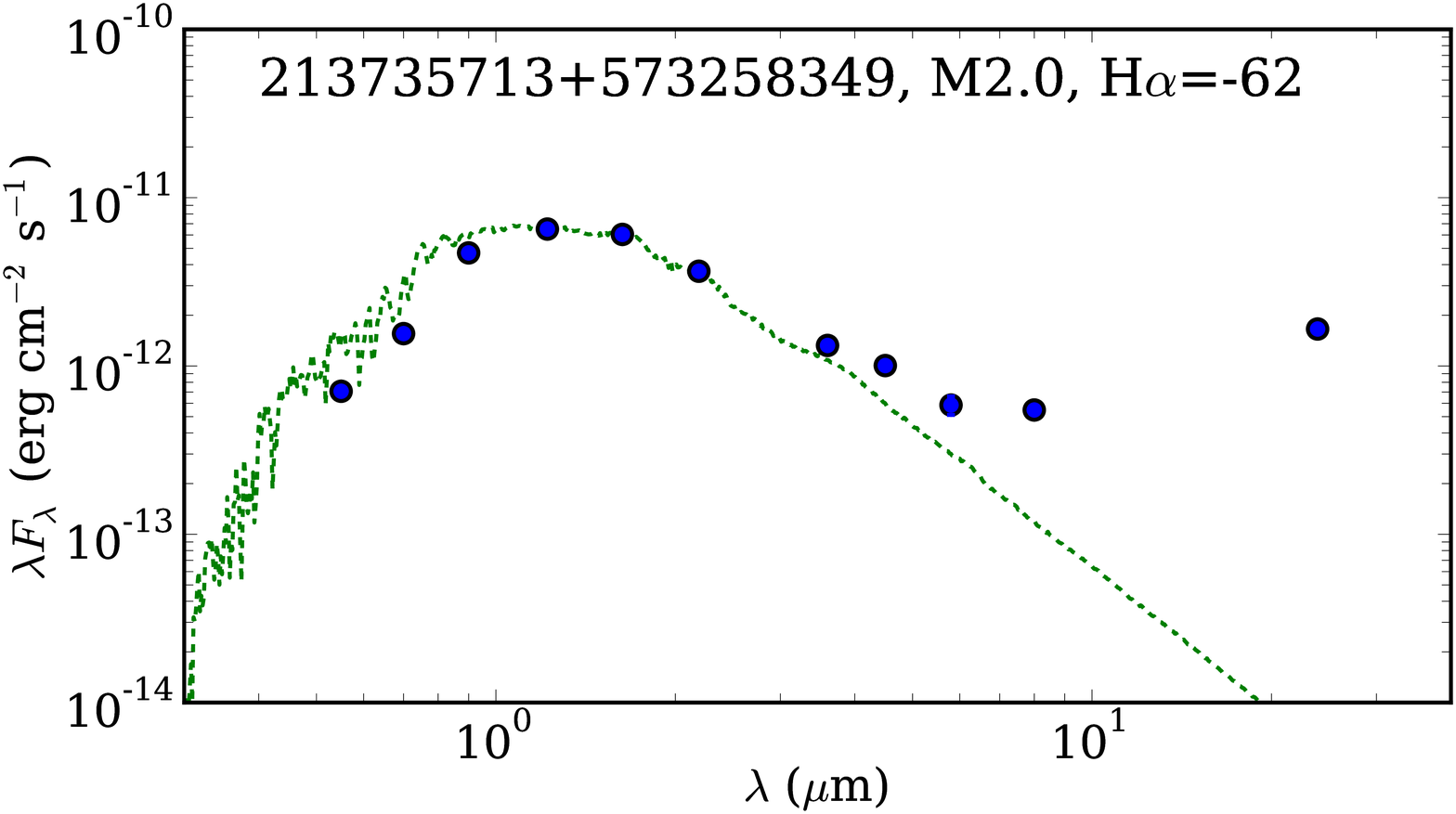,width=0.24\linewidth,clip=} &
\epsfig{file=213740471+573433203.eps,width=0.24\linewidth,clip=} &
\epsfig{file=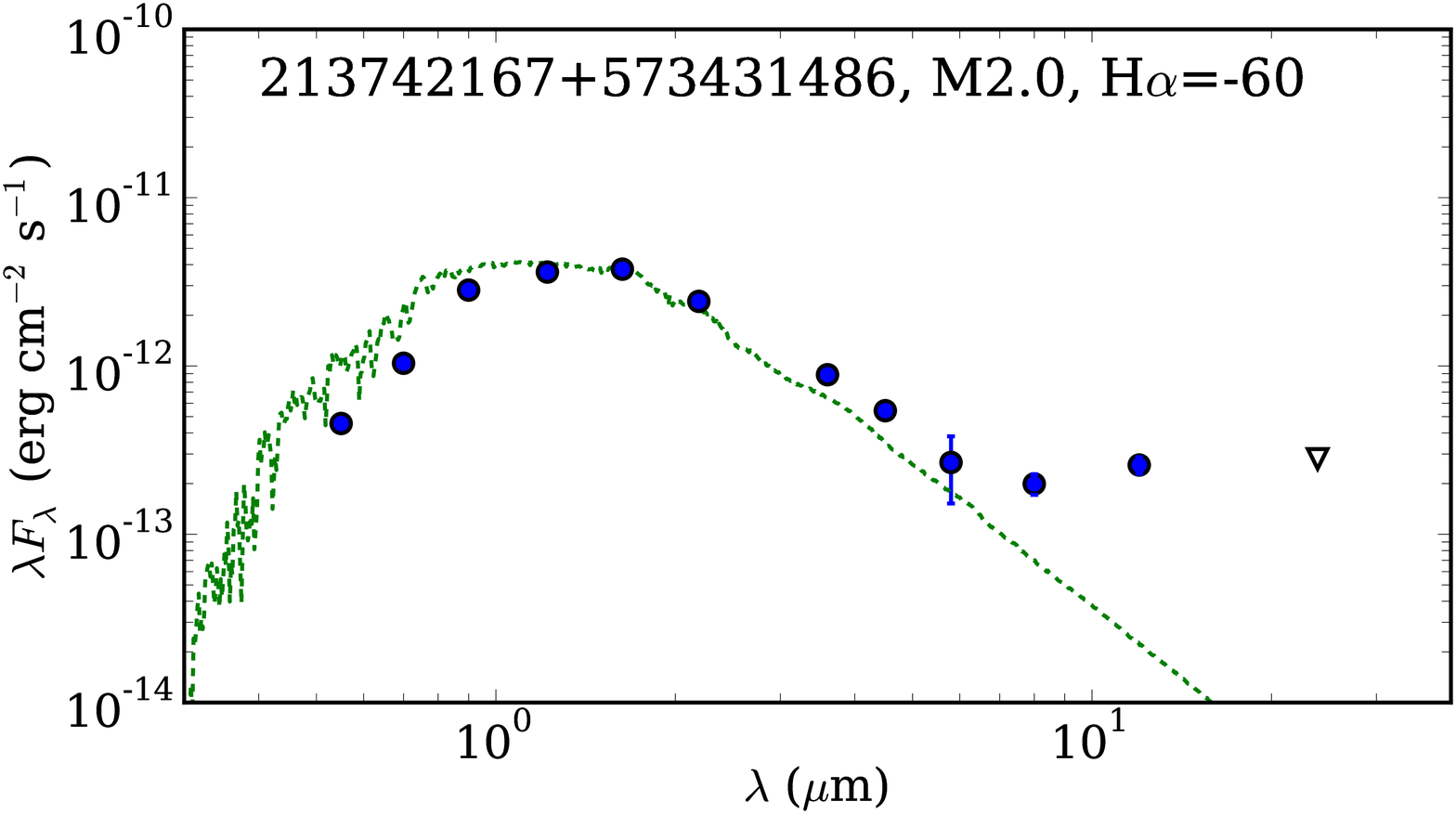,width=0.24\linewidth,clip=} \\
\epsfig{file=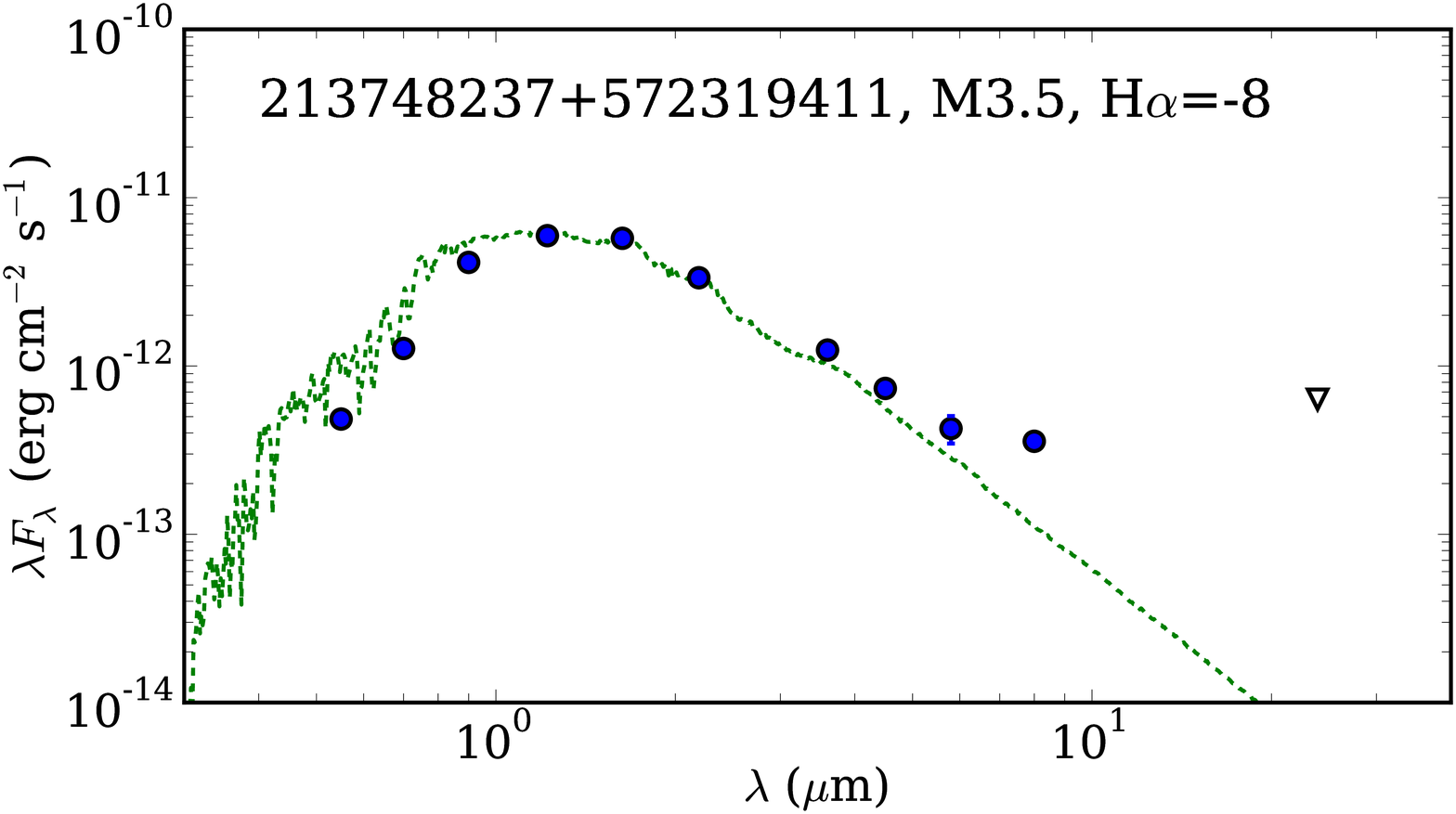,width=0.24\linewidth,clip=} &
\epsfig{file=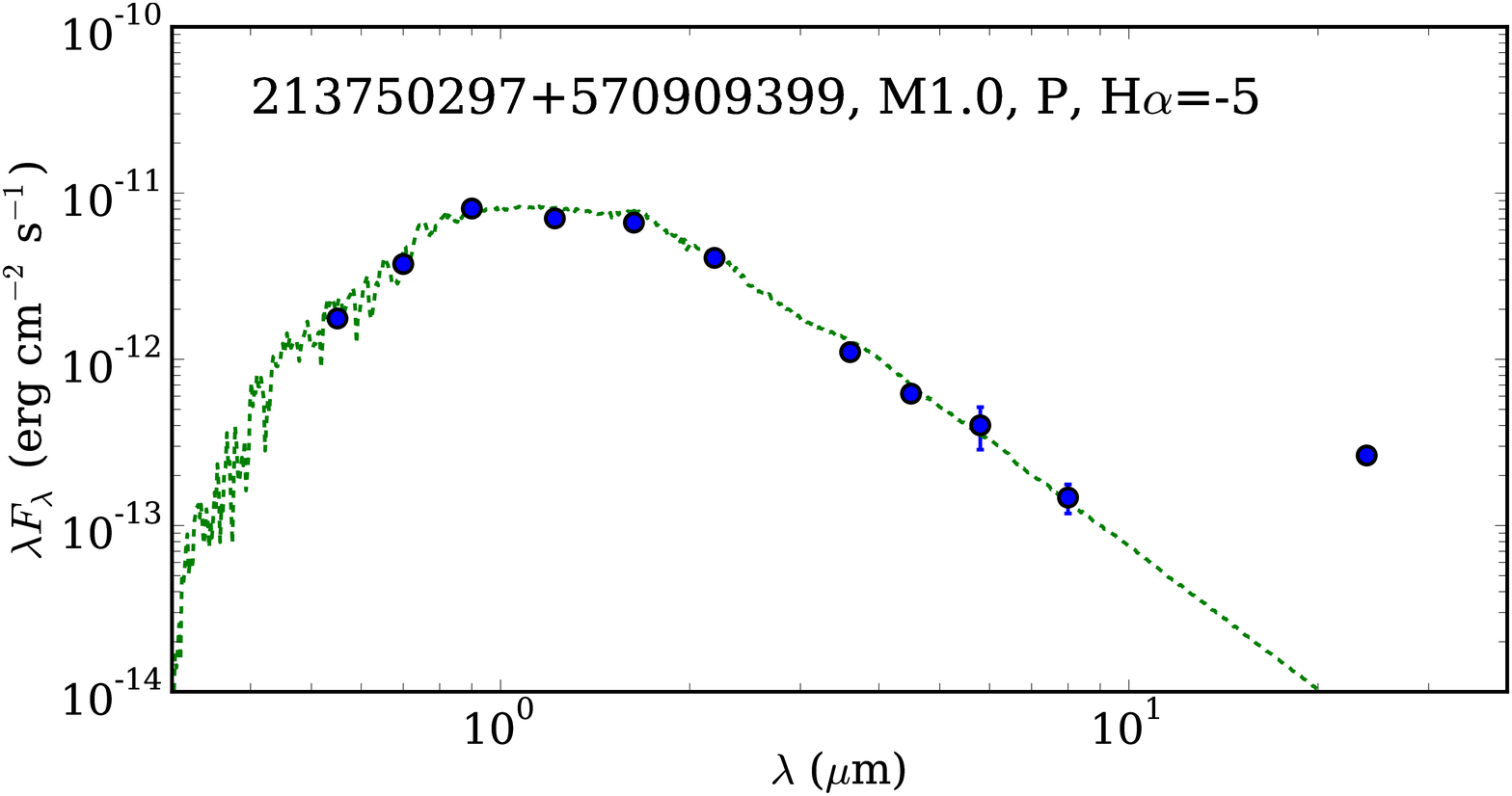,width=0.24\linewidth,clip=} &
\epsfig{file=213756779+573448171.eps,width=0.24\linewidth,clip=} &
\epsfig{file=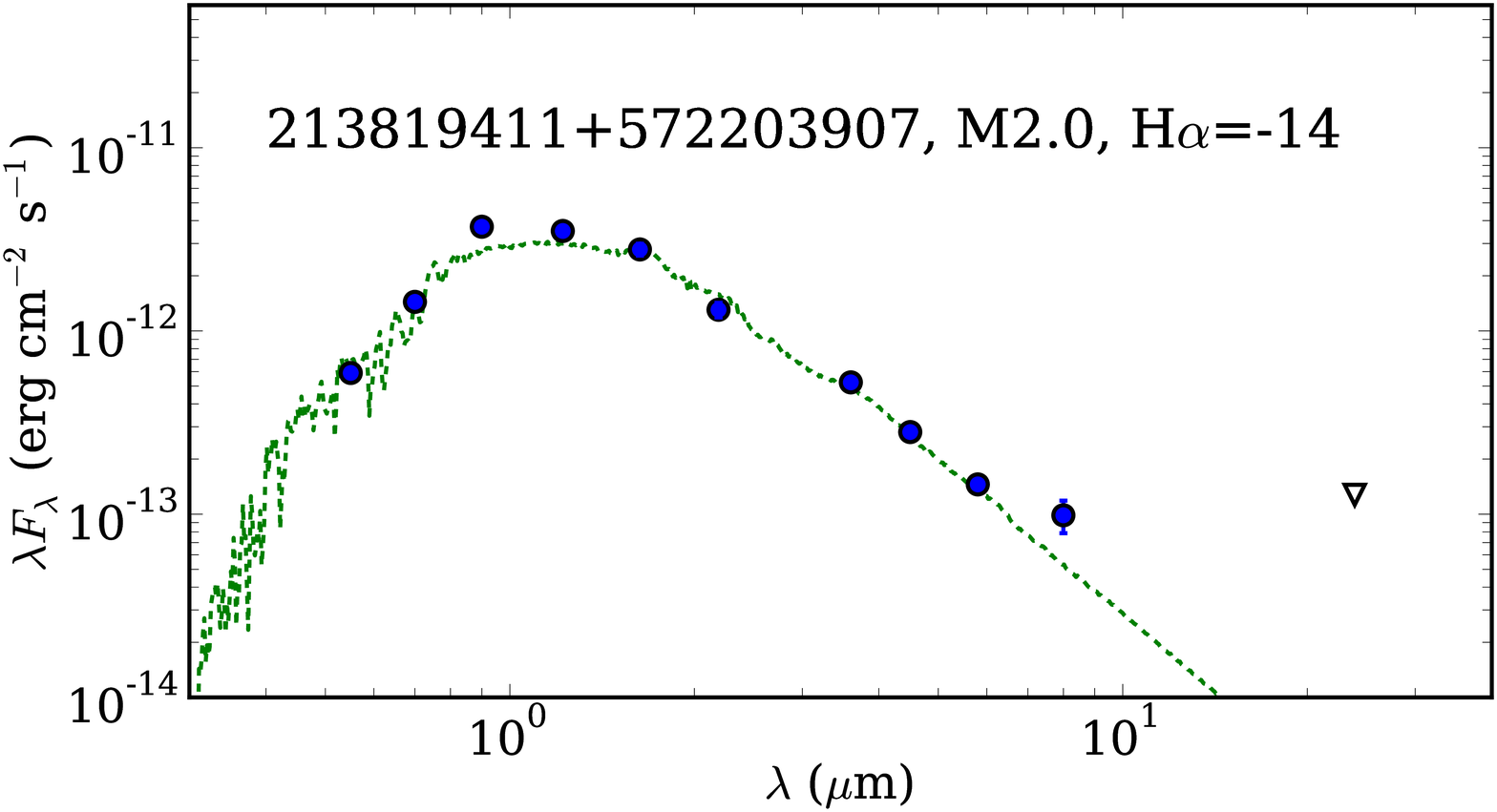,width=0.24\linewidth,clip=} \\
\epsfig{file=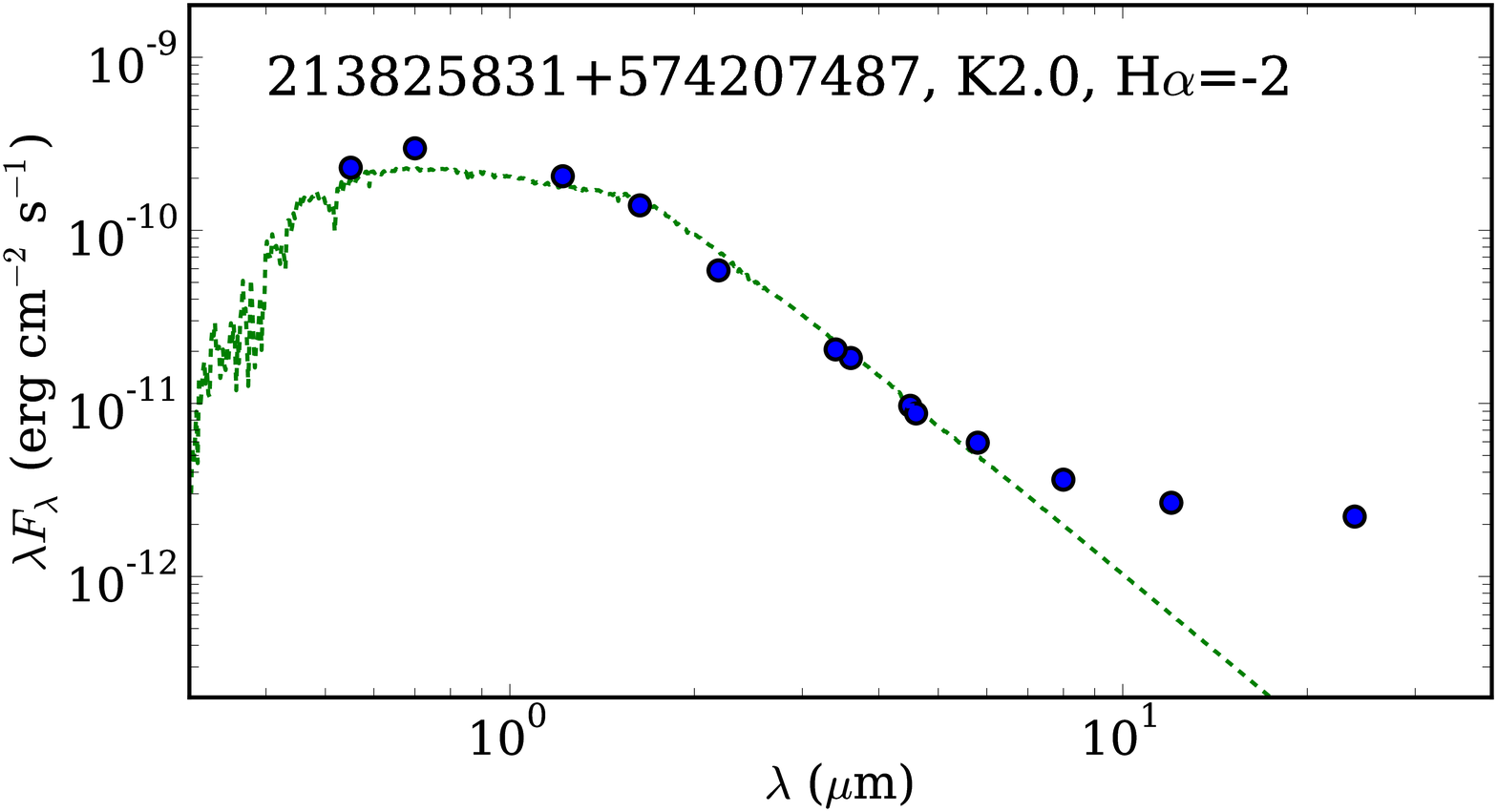,width=0.24\linewidth,clip=} &
\epsfig{file=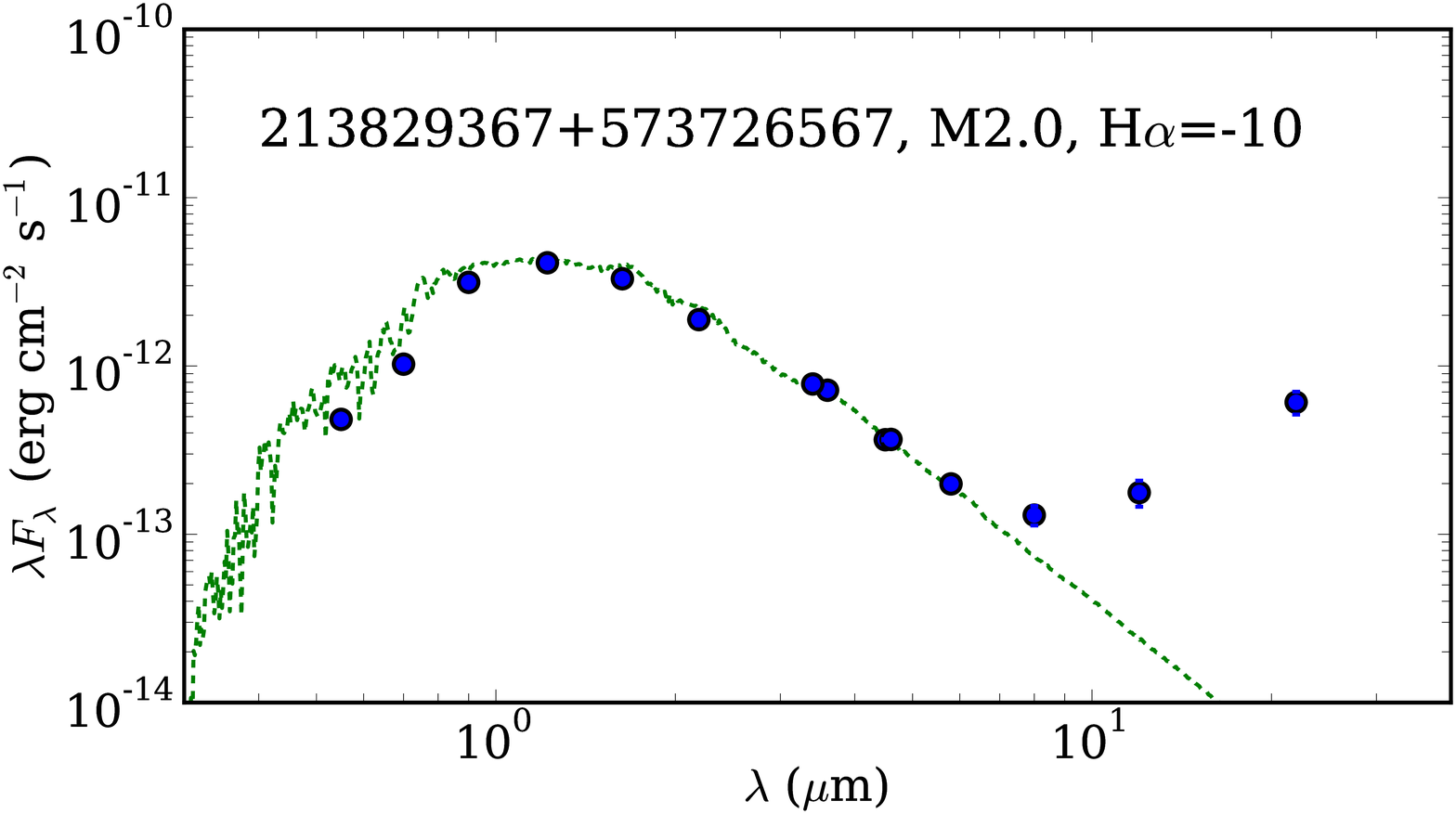,width=0.24\linewidth,clip=} &
\epsfig{file=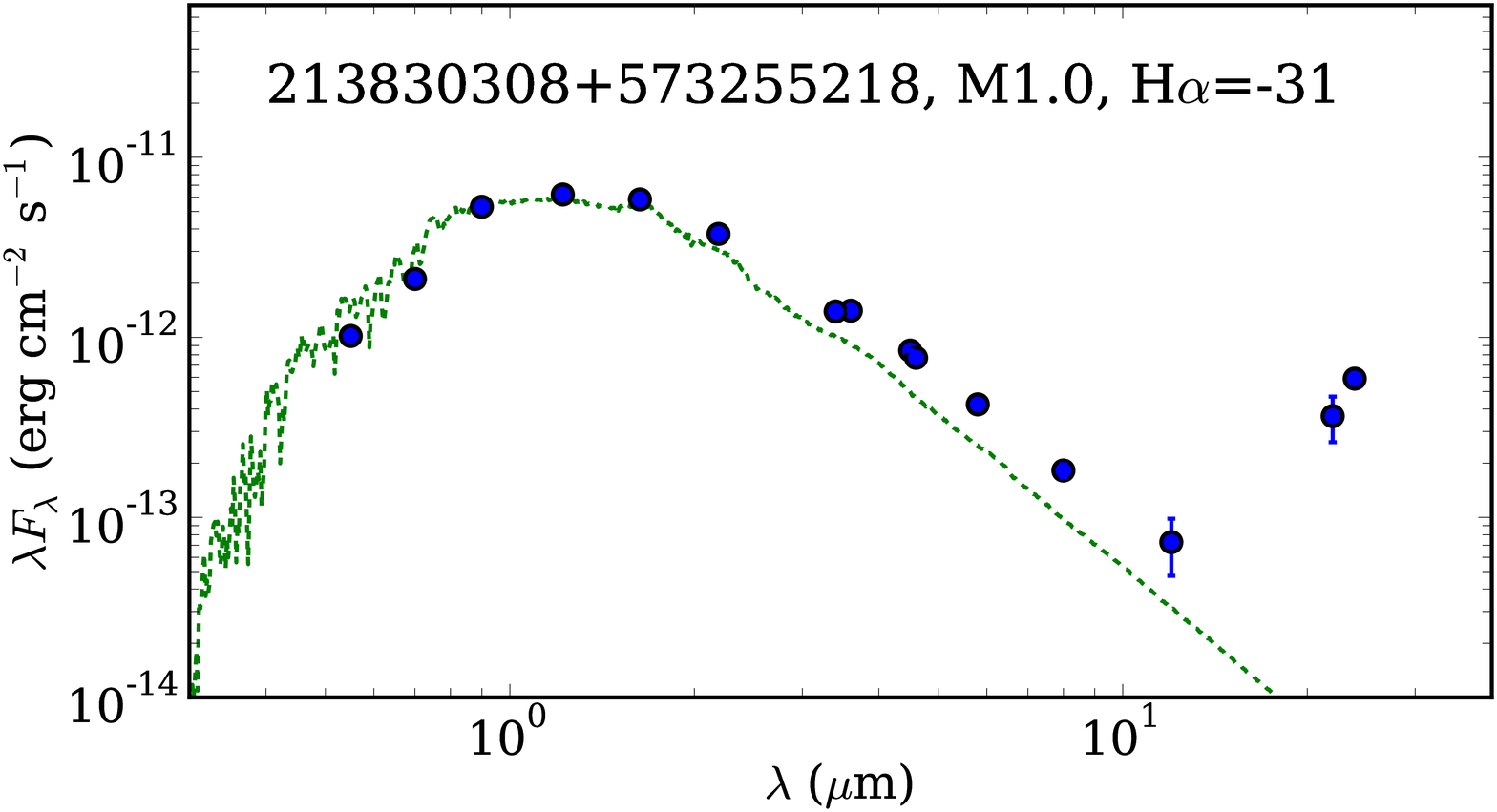,width=0.24\linewidth,clip=} &
\epsfig{file=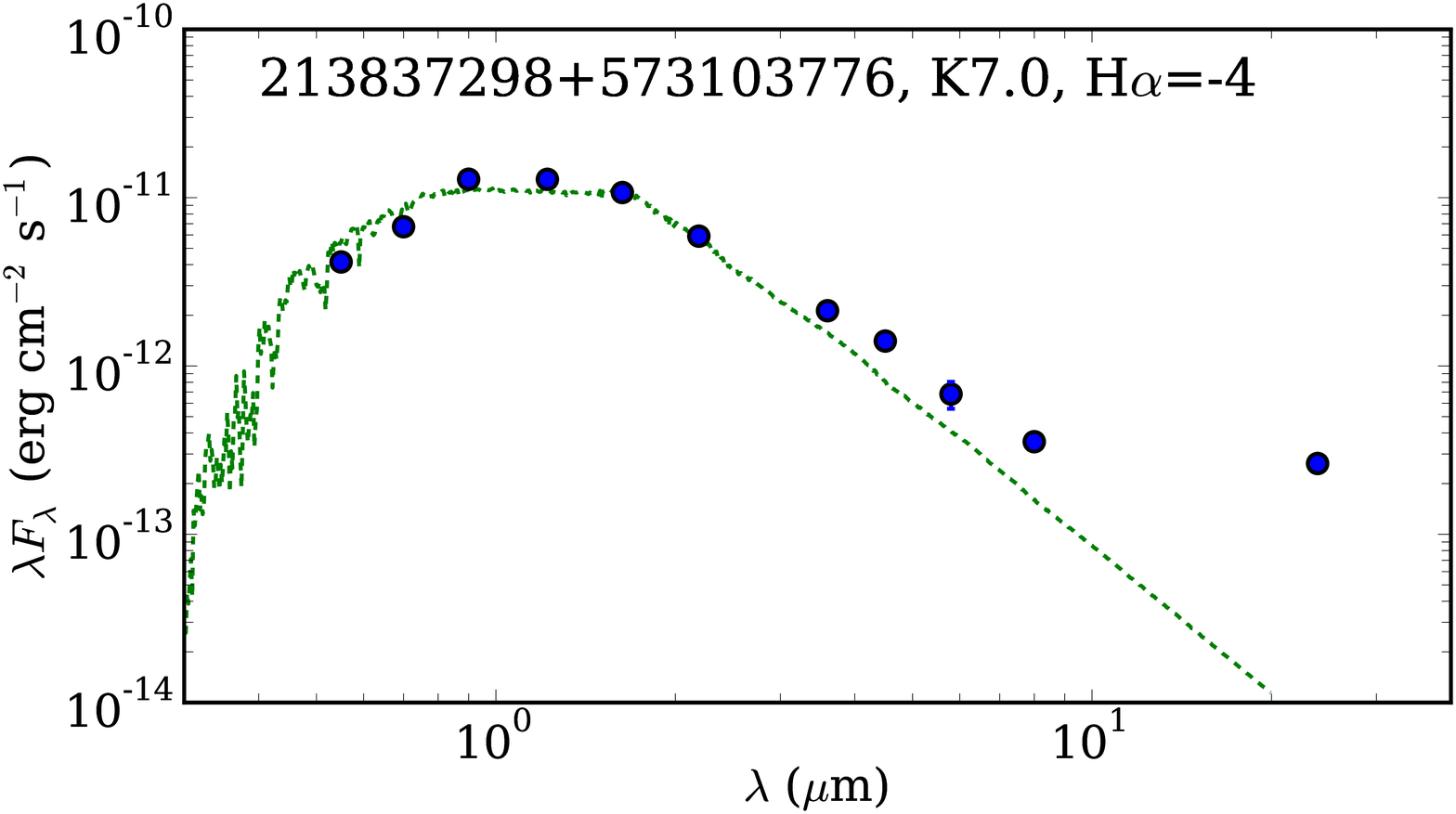,width=0.24\linewidth,clip=} \\
\epsfig{file=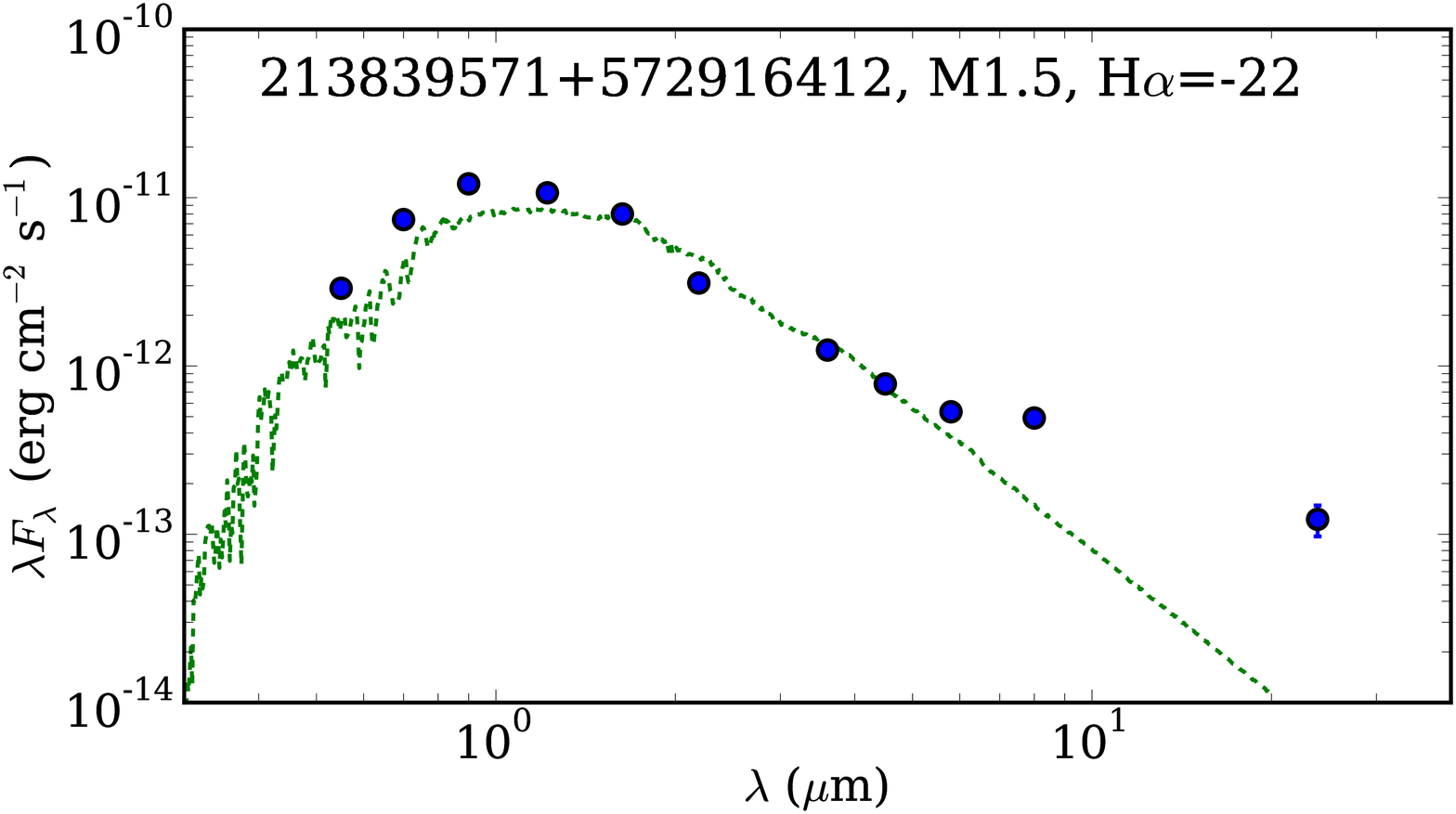,width=0.24\linewidth,clip=} &
\epsfig{file=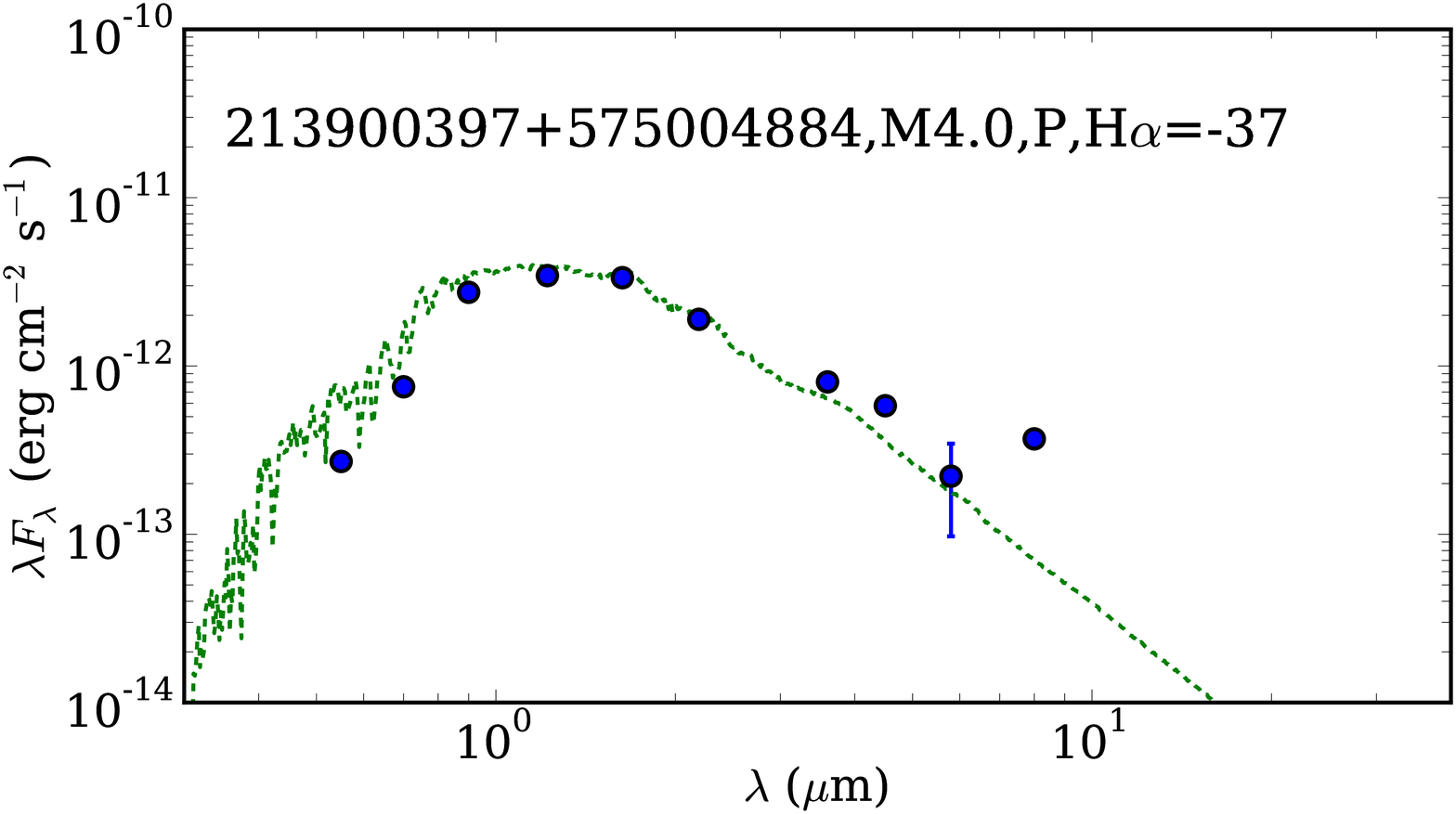,width=0.24\linewidth,clip=} &
\epsfig{file=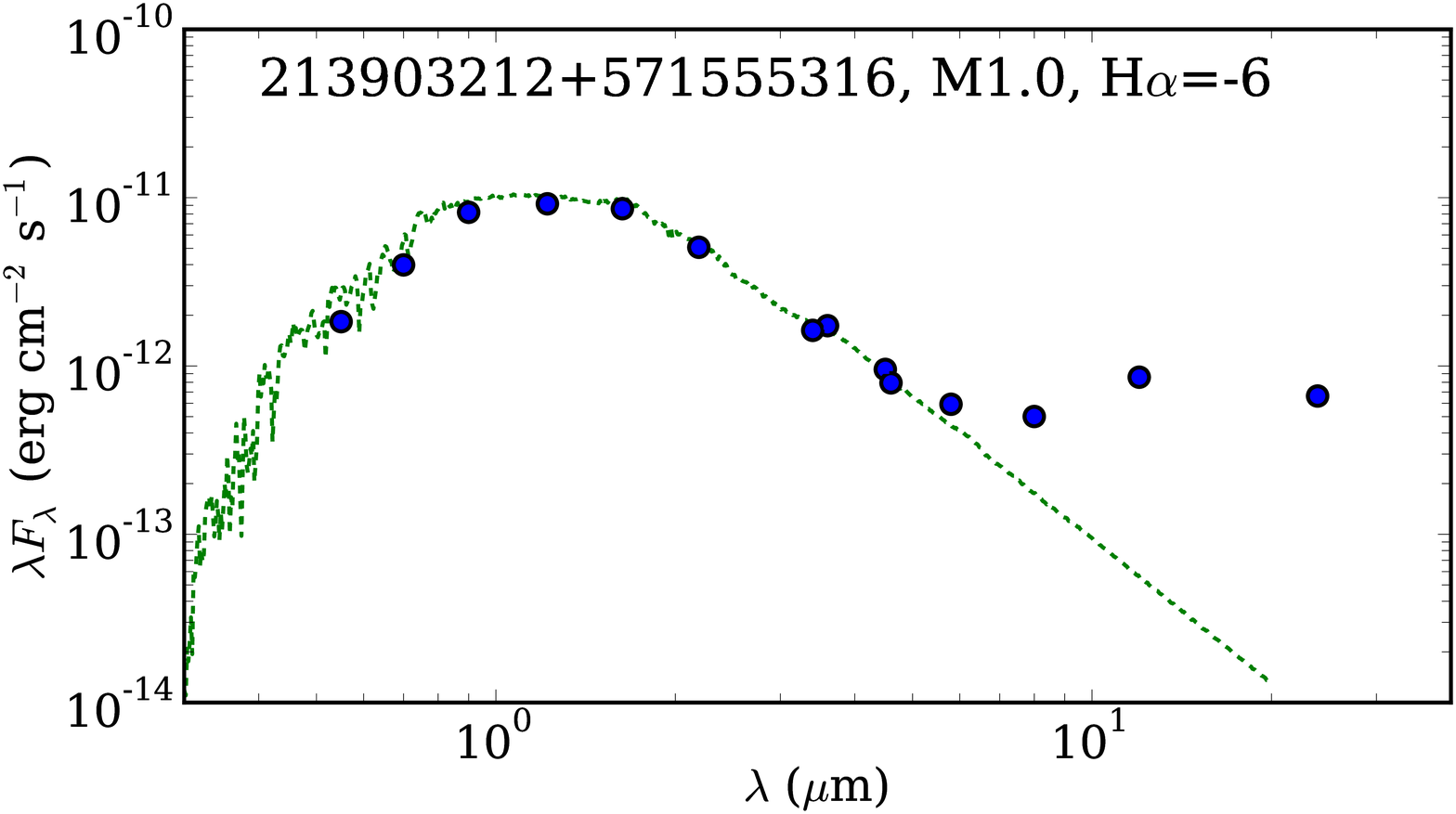,width=0.24\linewidth,clip=} &
\epsfig{file=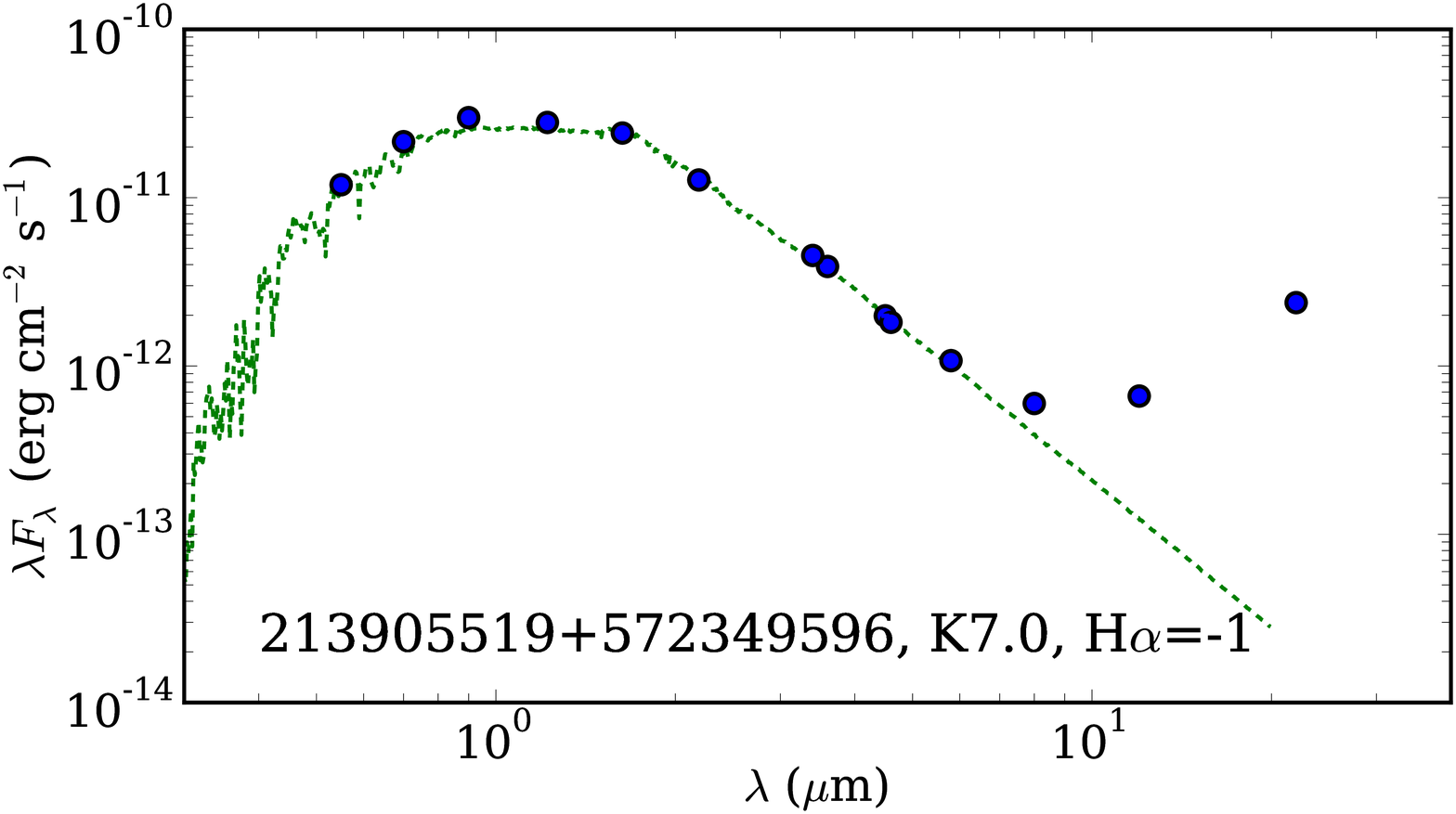,width=0.24\linewidth,clip=} \\
\epsfig{file=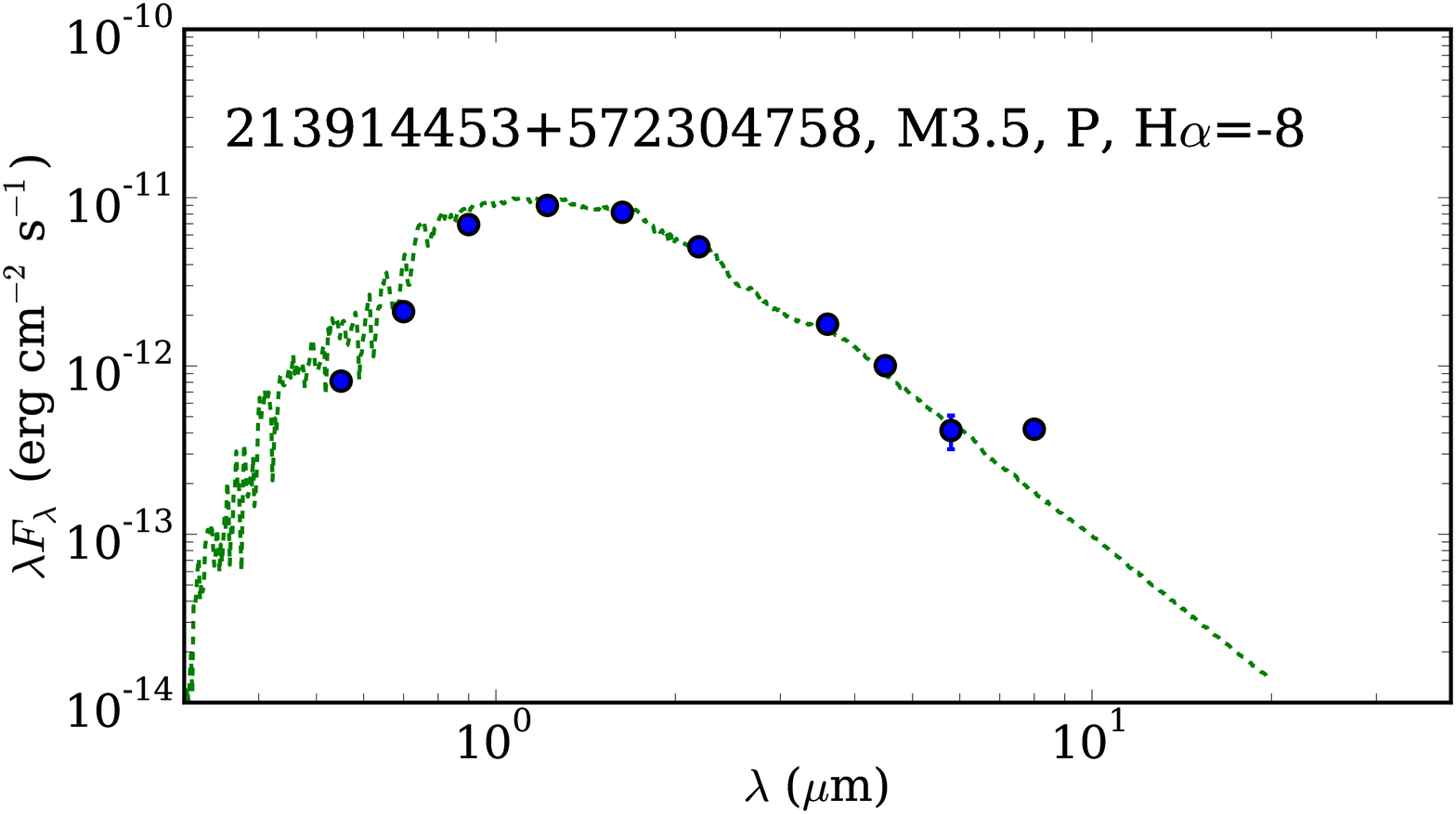,width=0.24\linewidth,clip=} &
\epsfig{file=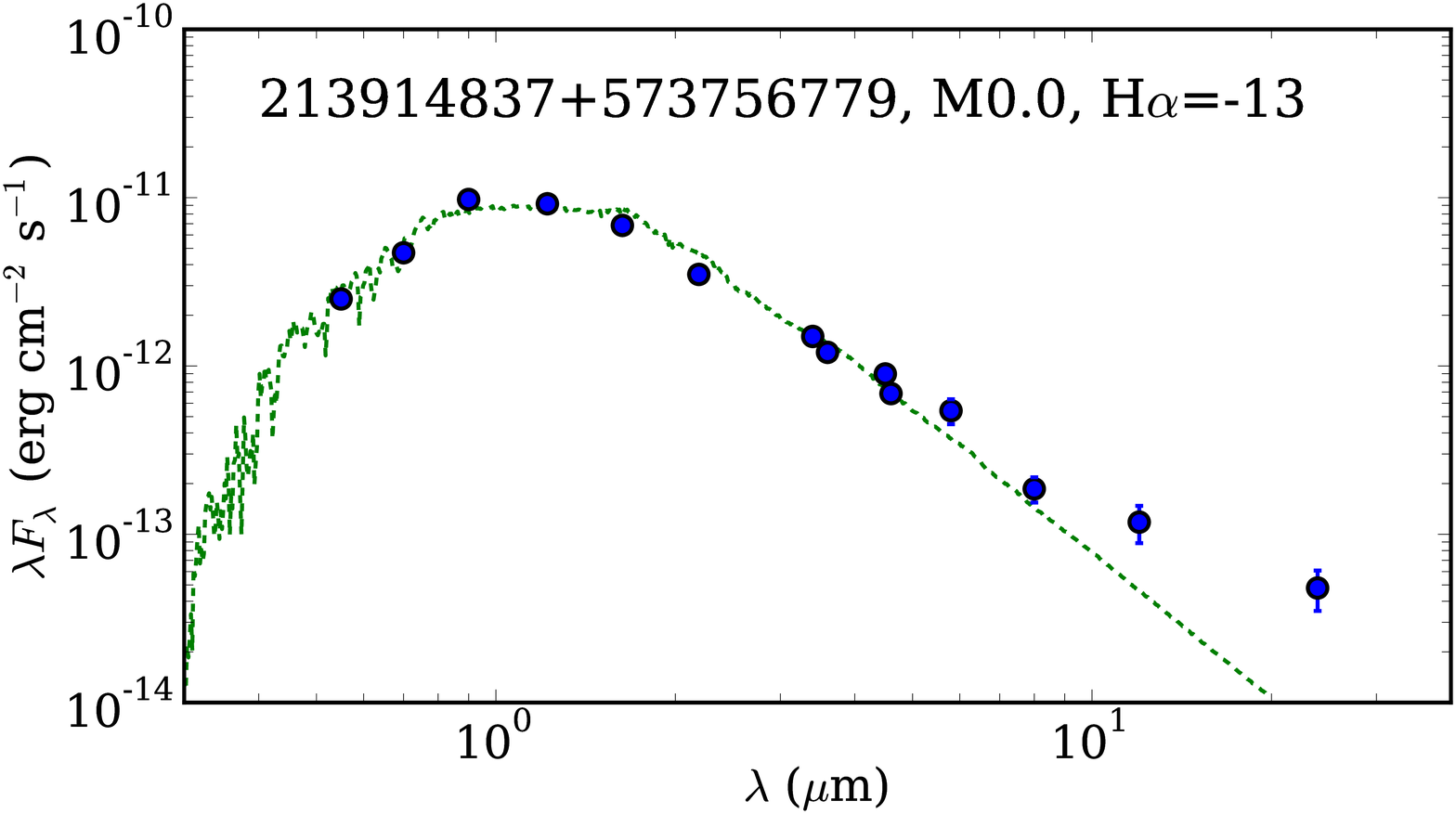,width=0.24\linewidth,clip=} &
\epsfig{file=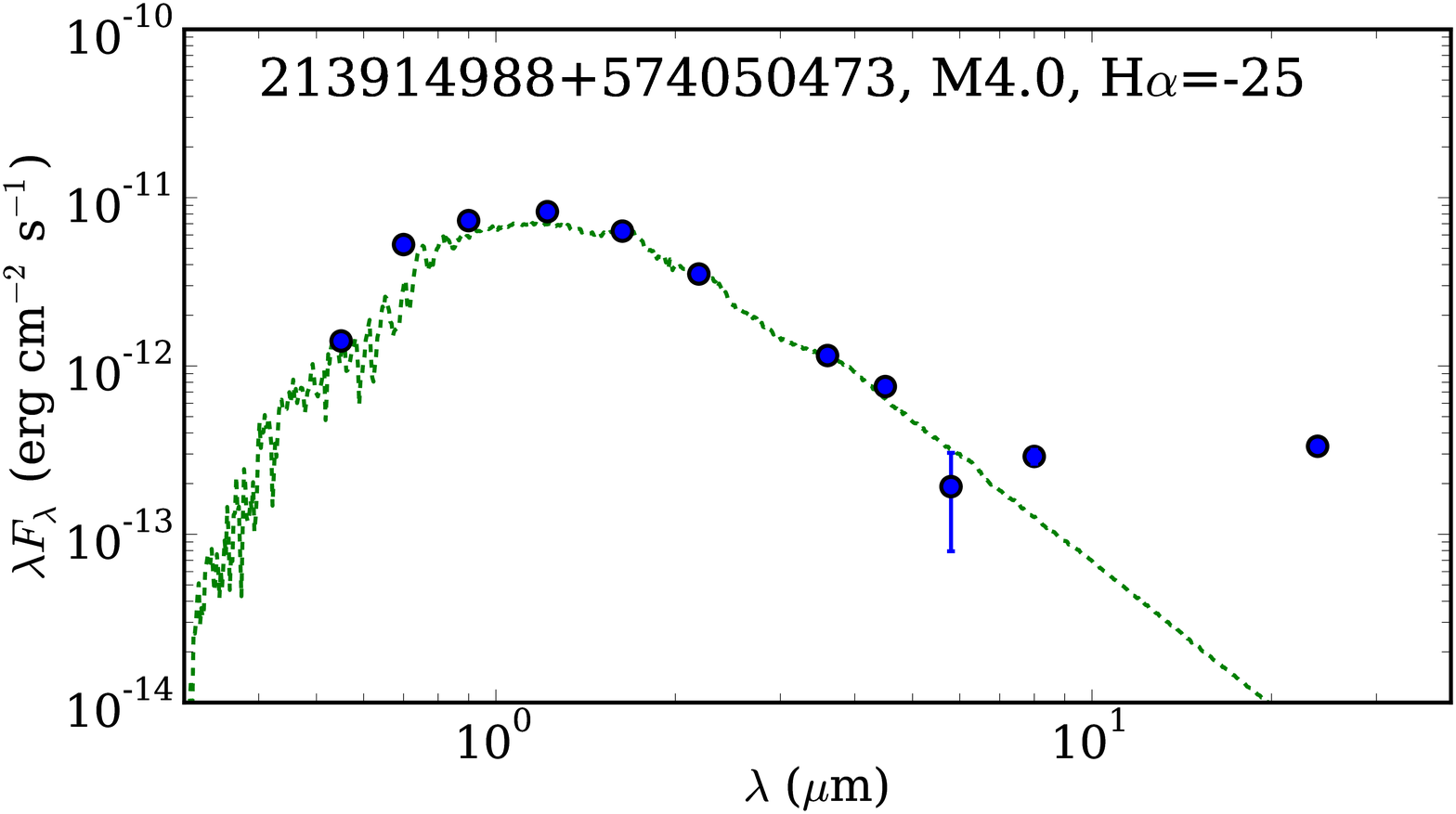,width=0.24\linewidth,clip=} &
\epsfig{file=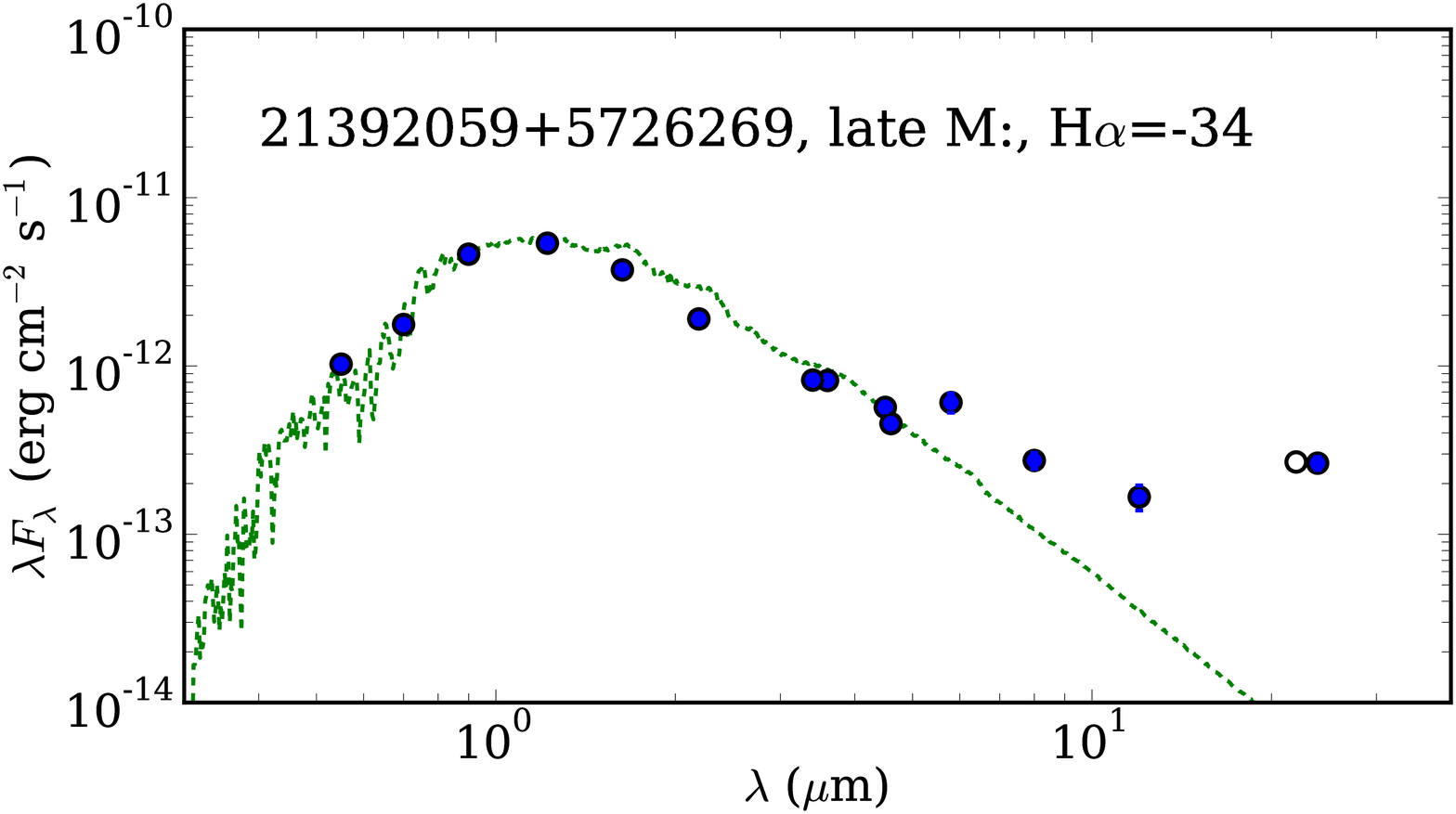,width=0.24\linewidth,clip=} \\
\epsfig{file=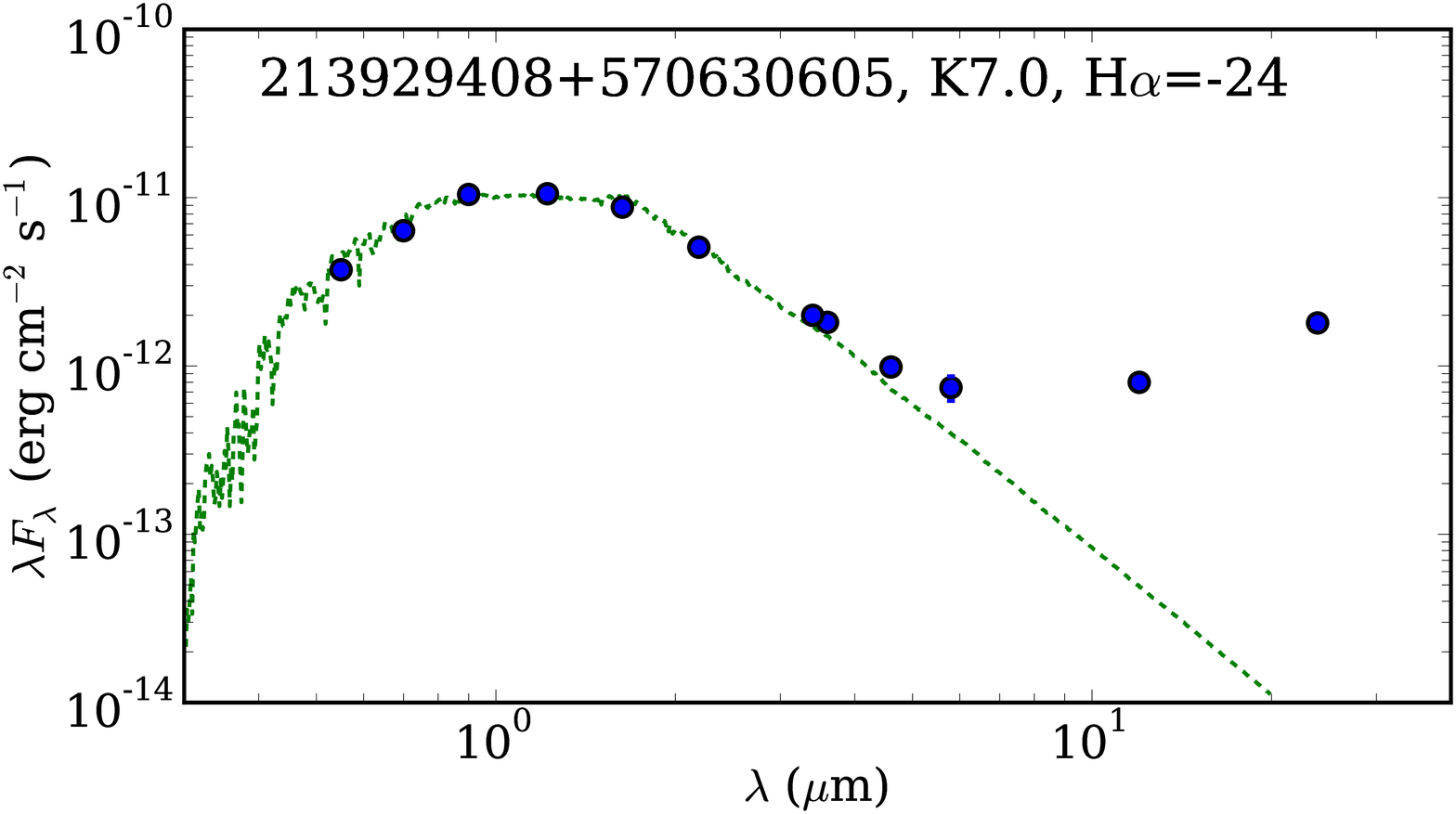,width=0.24\linewidth,clip=} &
\epsfig{file=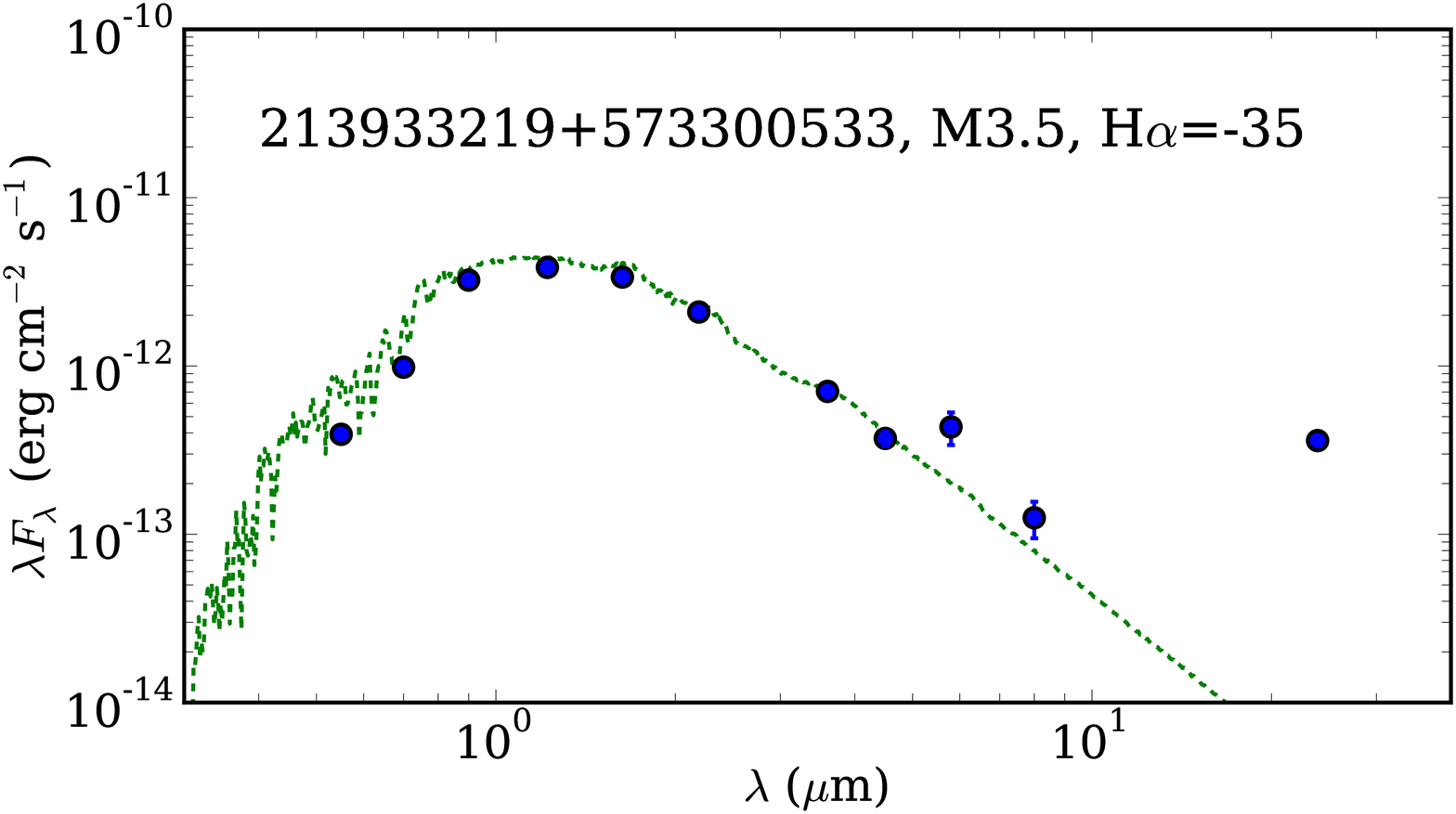,width=0.24\linewidth,clip=} &
\epsfig{file=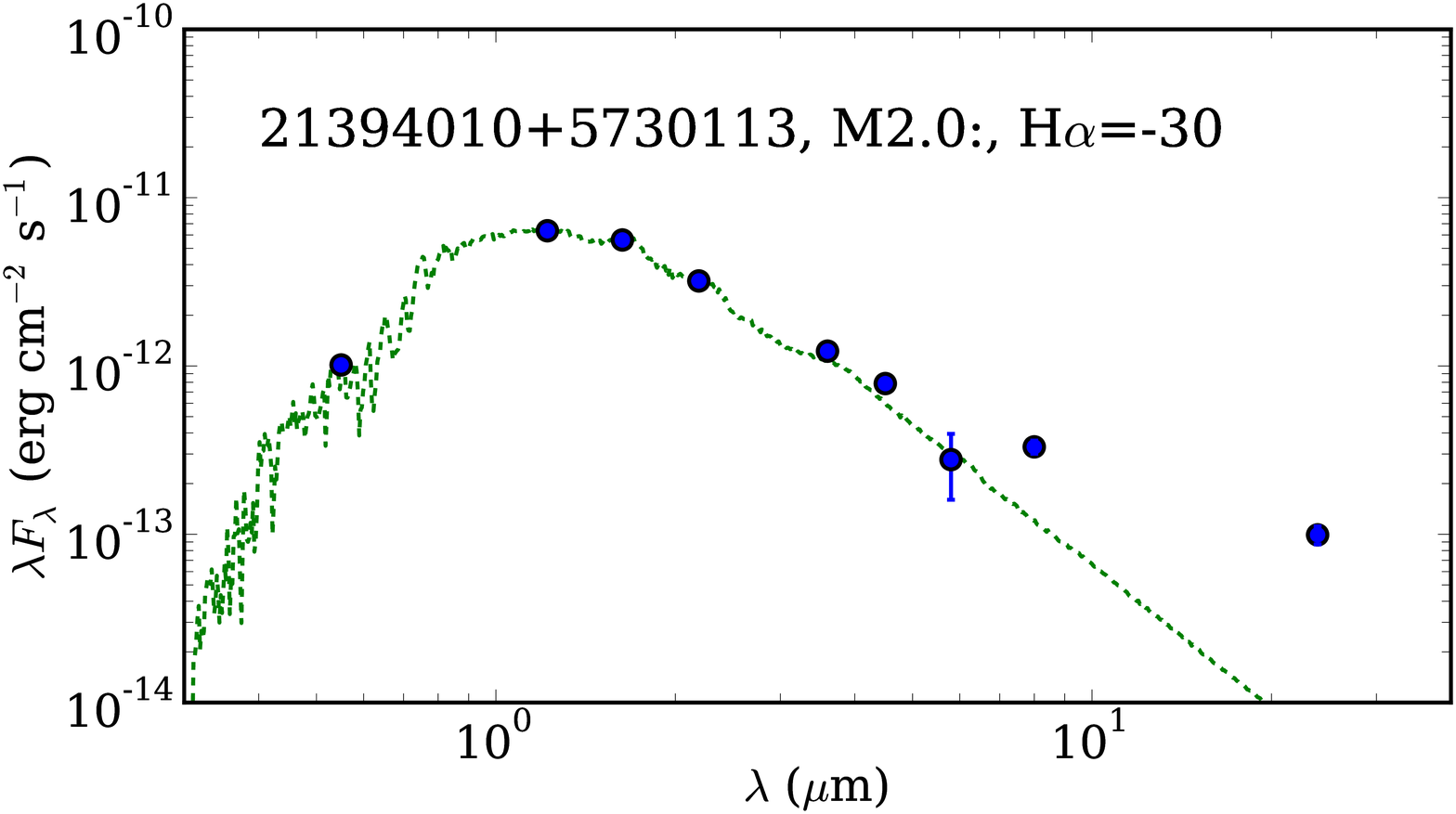,width=0.24\linewidth,clip=} &
\epsfig{file=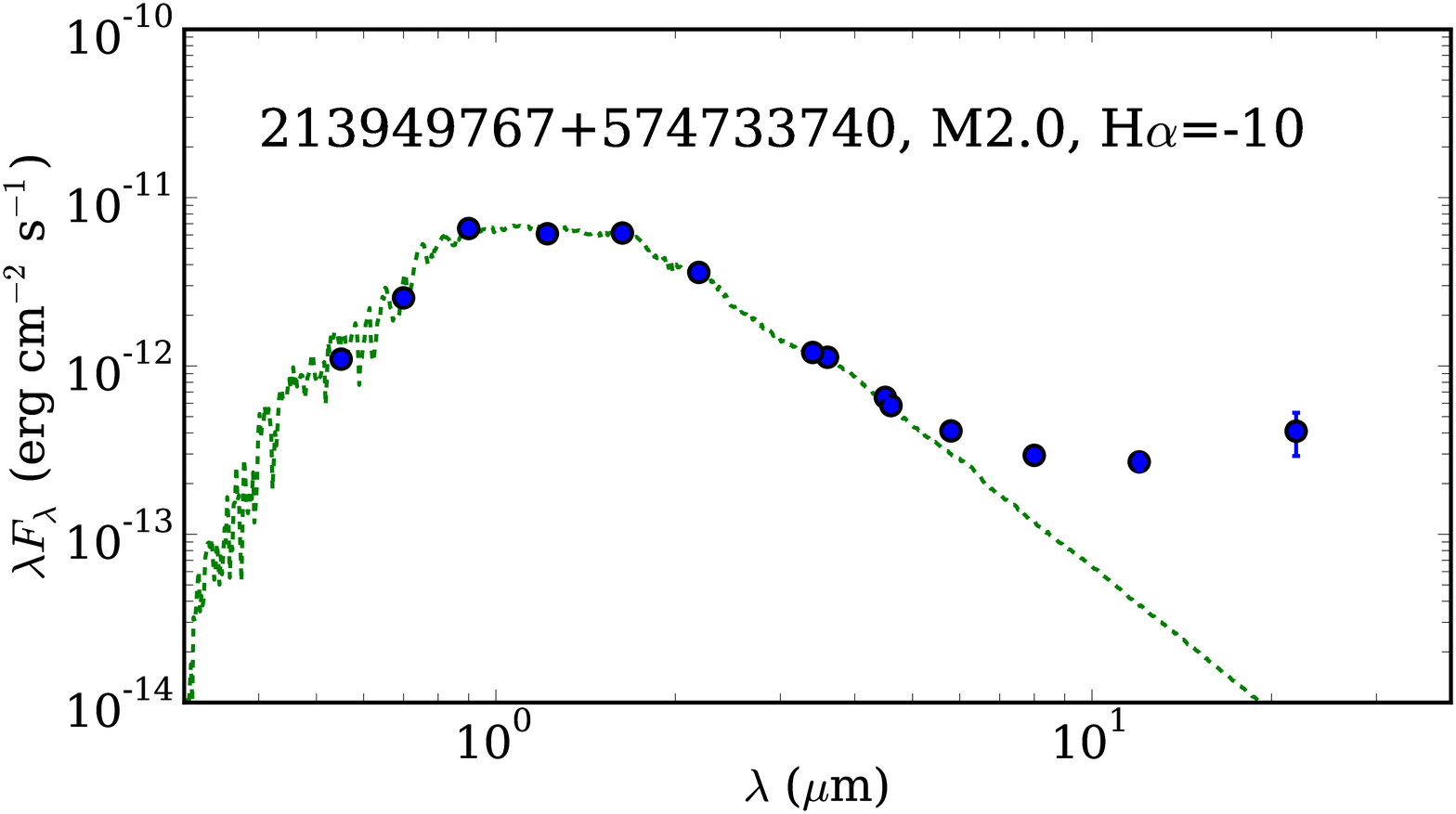,width=0.24\linewidth,clip=} \\
\epsfig{file=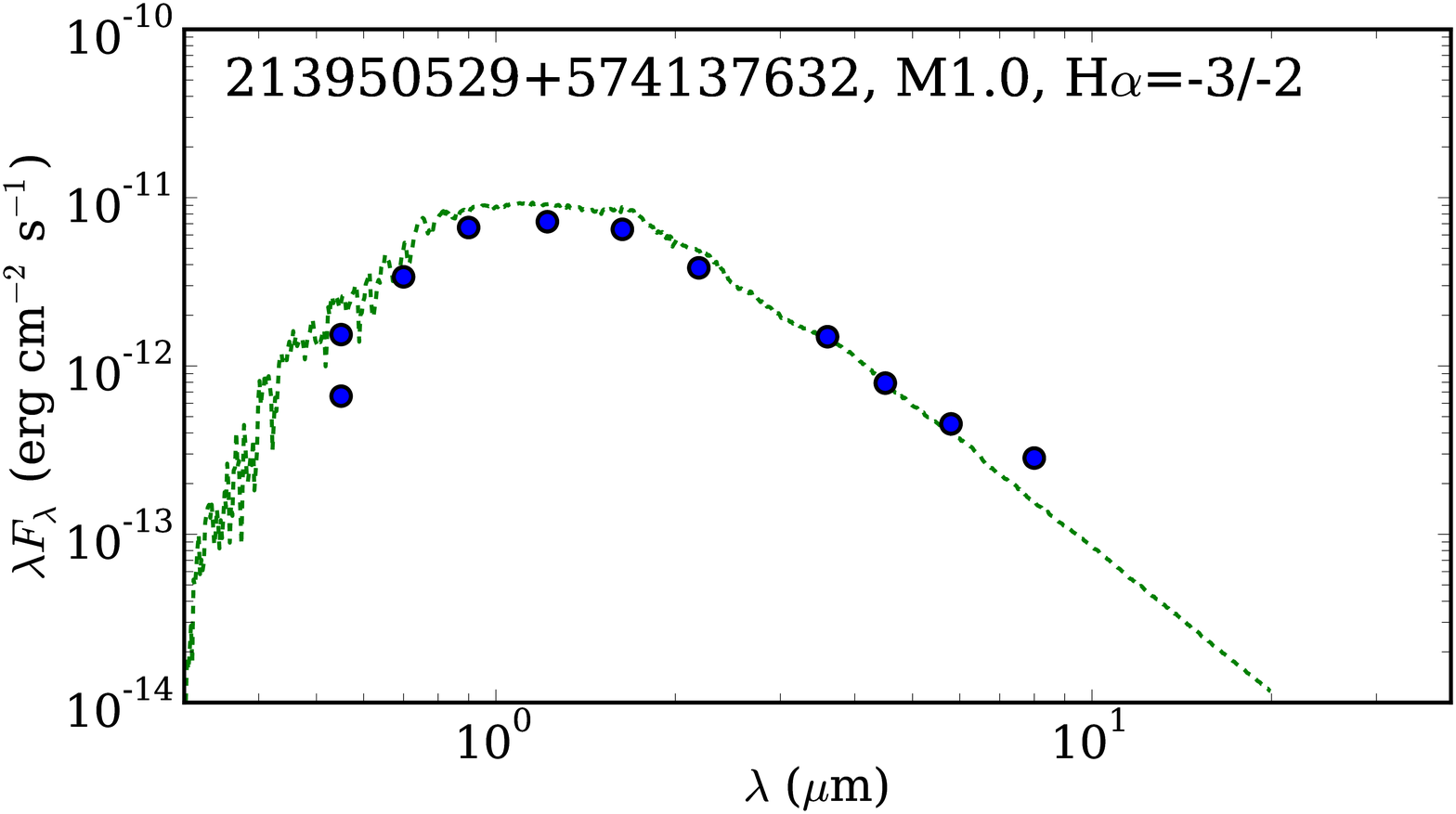,width=0.24\linewidth,clip=} &
\epsfig{file=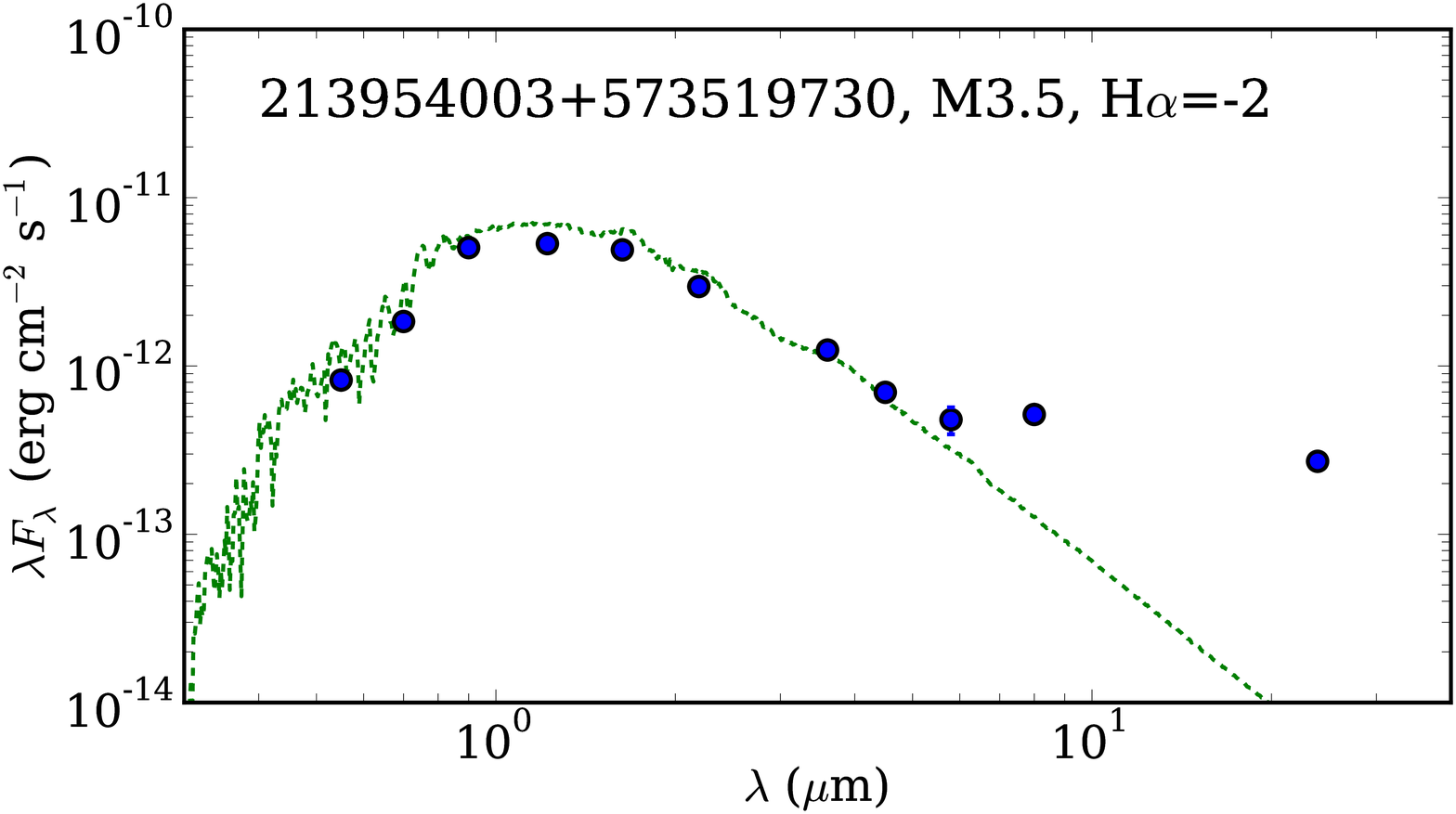,width=0.24\linewidth,clip=} &
\epsfig{file=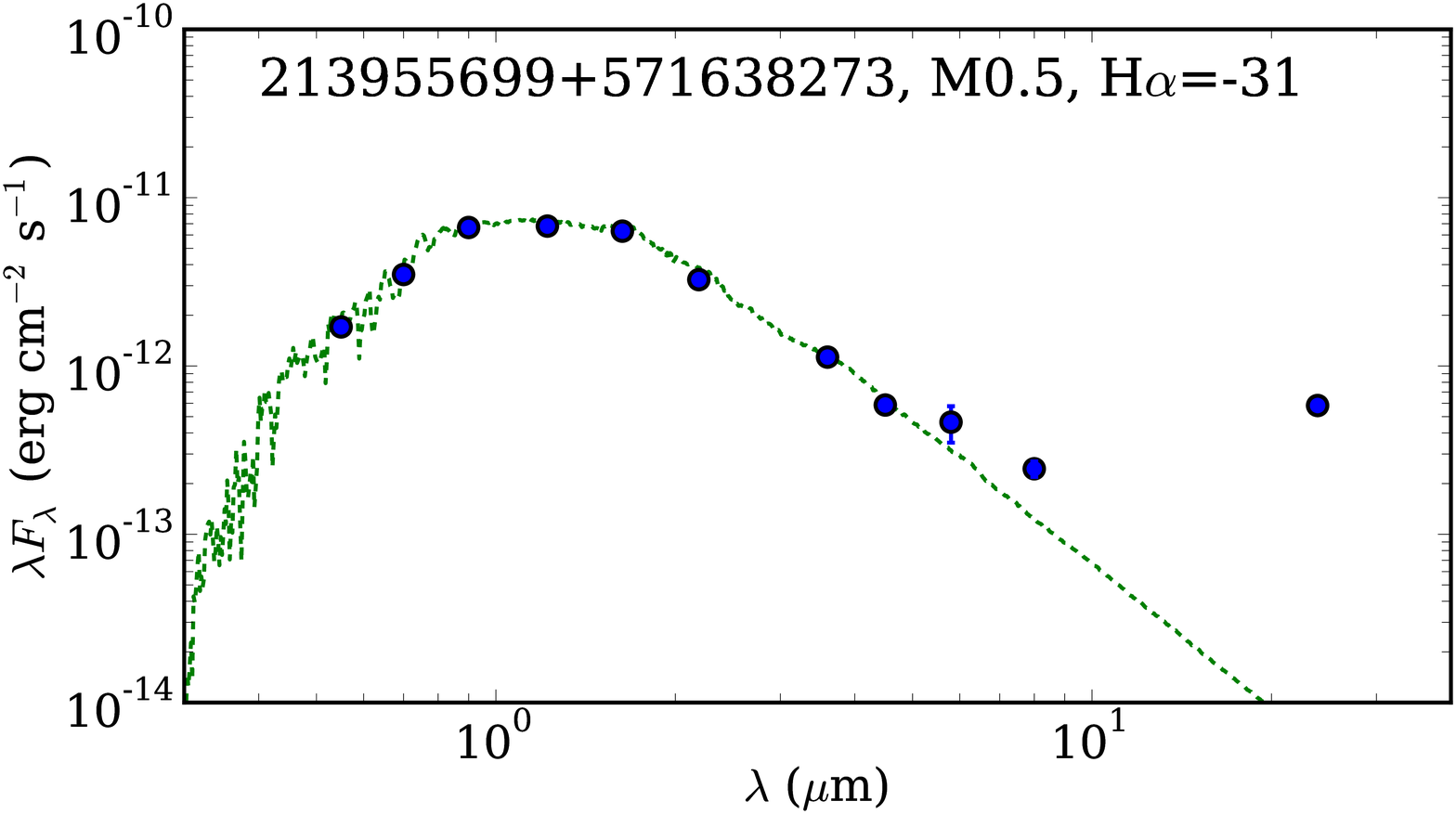,width=0.24\linewidth,clip=} &
\epsfig{file=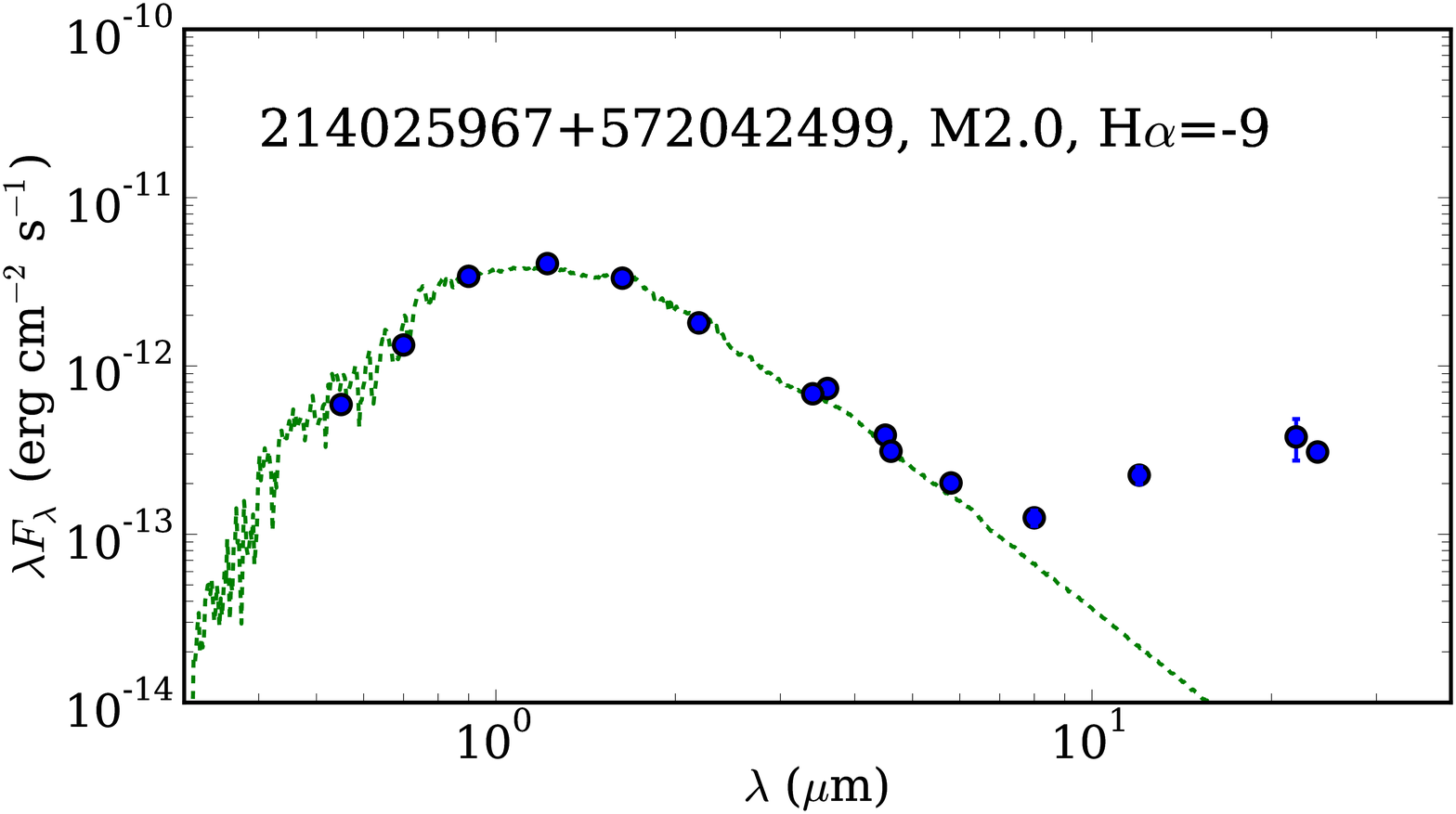,width=0.24\linewidth,clip=} \\
\epsfig{file=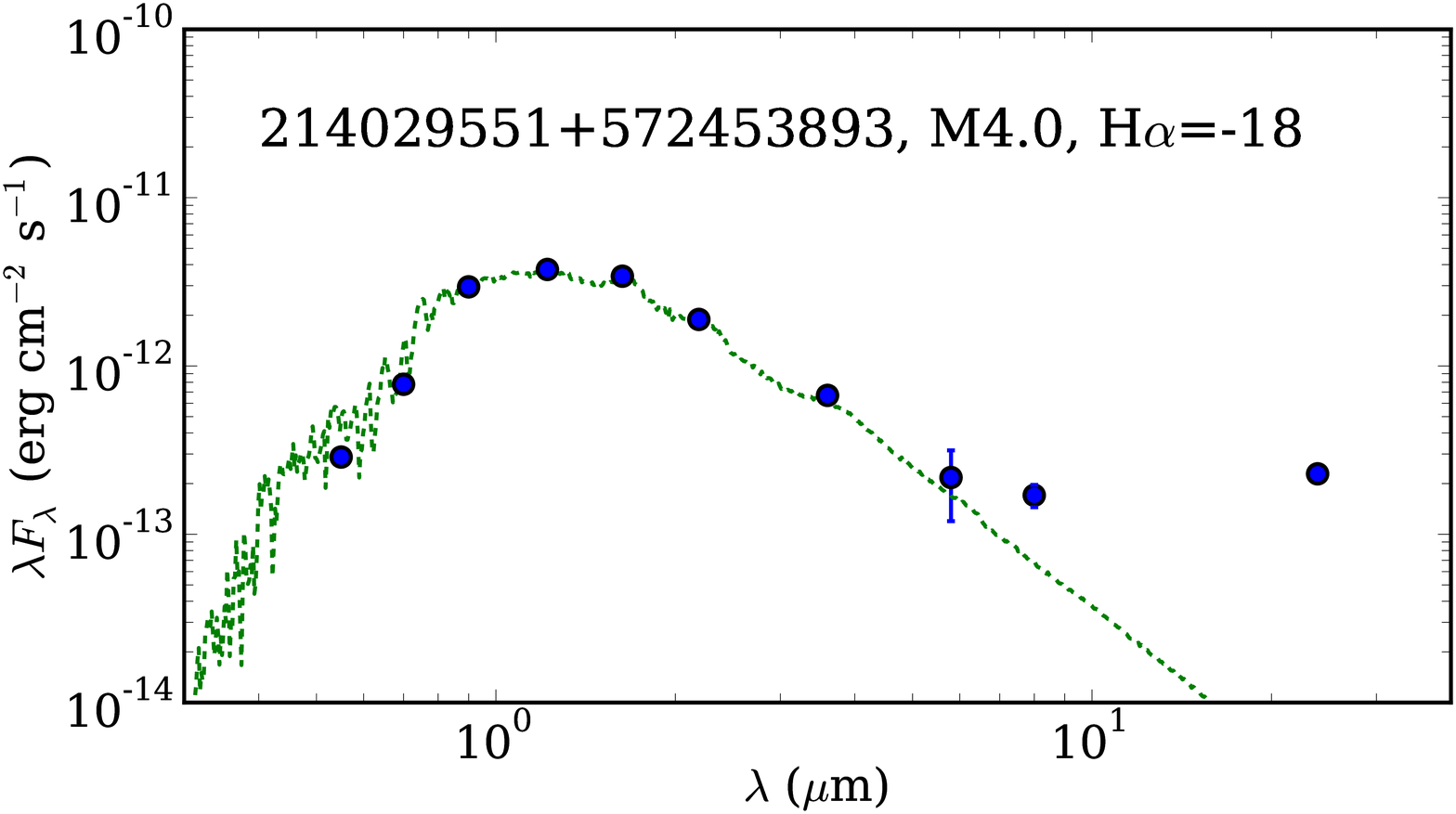,width=0.24\linewidth,clip=} &
\epsfig{file=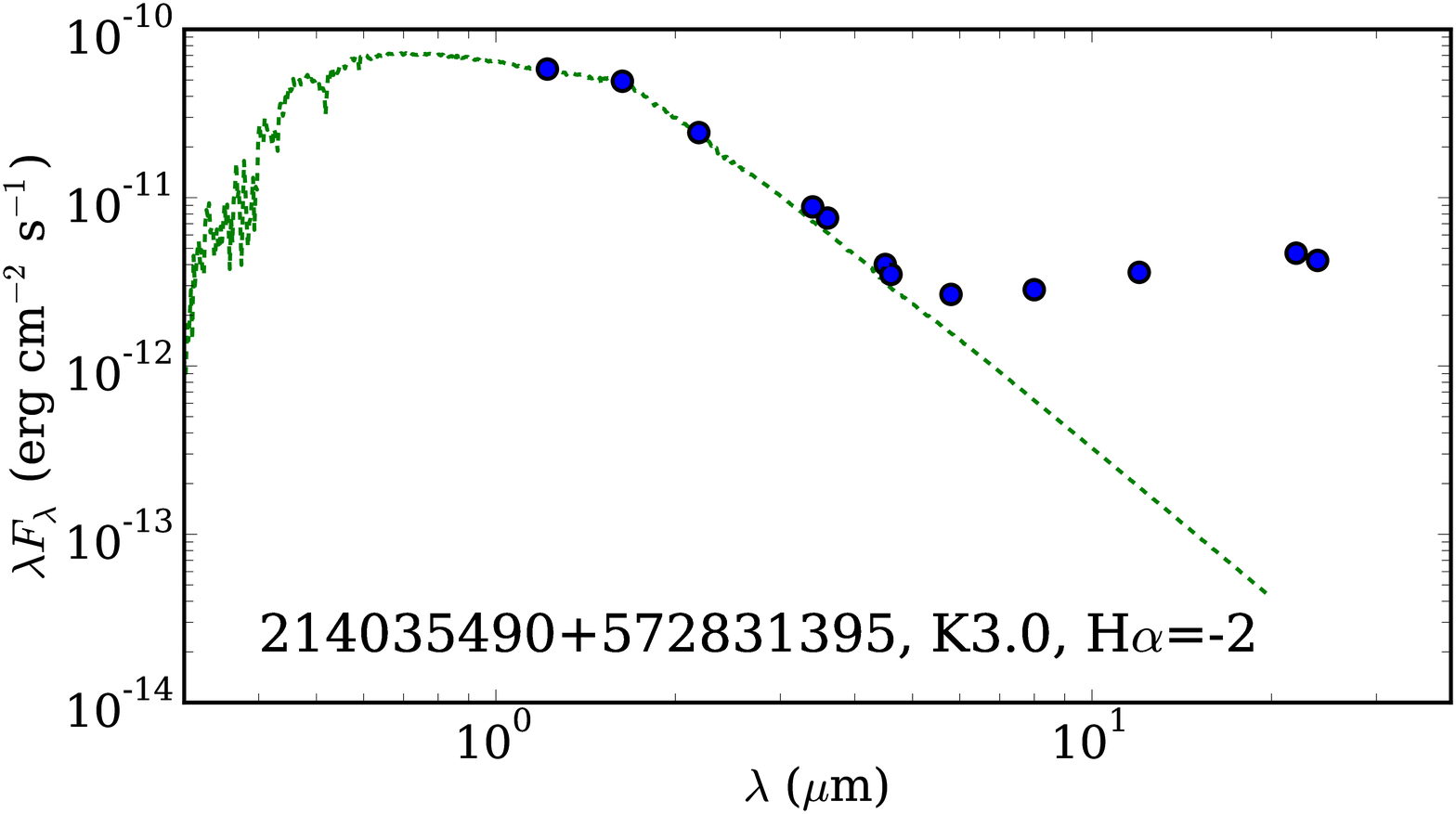,width=0.24\linewidth,clip=} \\
\end{tabular}
\caption{SEDs of the members and probably members with IR excess consistent with TD. 
Objects like 213819411+572203907, 213839571+572916412, 21394010+5730113 may be also dust-depleted. 213750297+570909399
may be partially contaminated by nebular emission but the detection is real. Inverted triangles mark upper limits, and
open circles are uncertain values.
\label{tdseds-fig}}
\end{figure*}

\begin{figure*}
\centering
\begin{tabular}{ccccc}
\epsfig{file=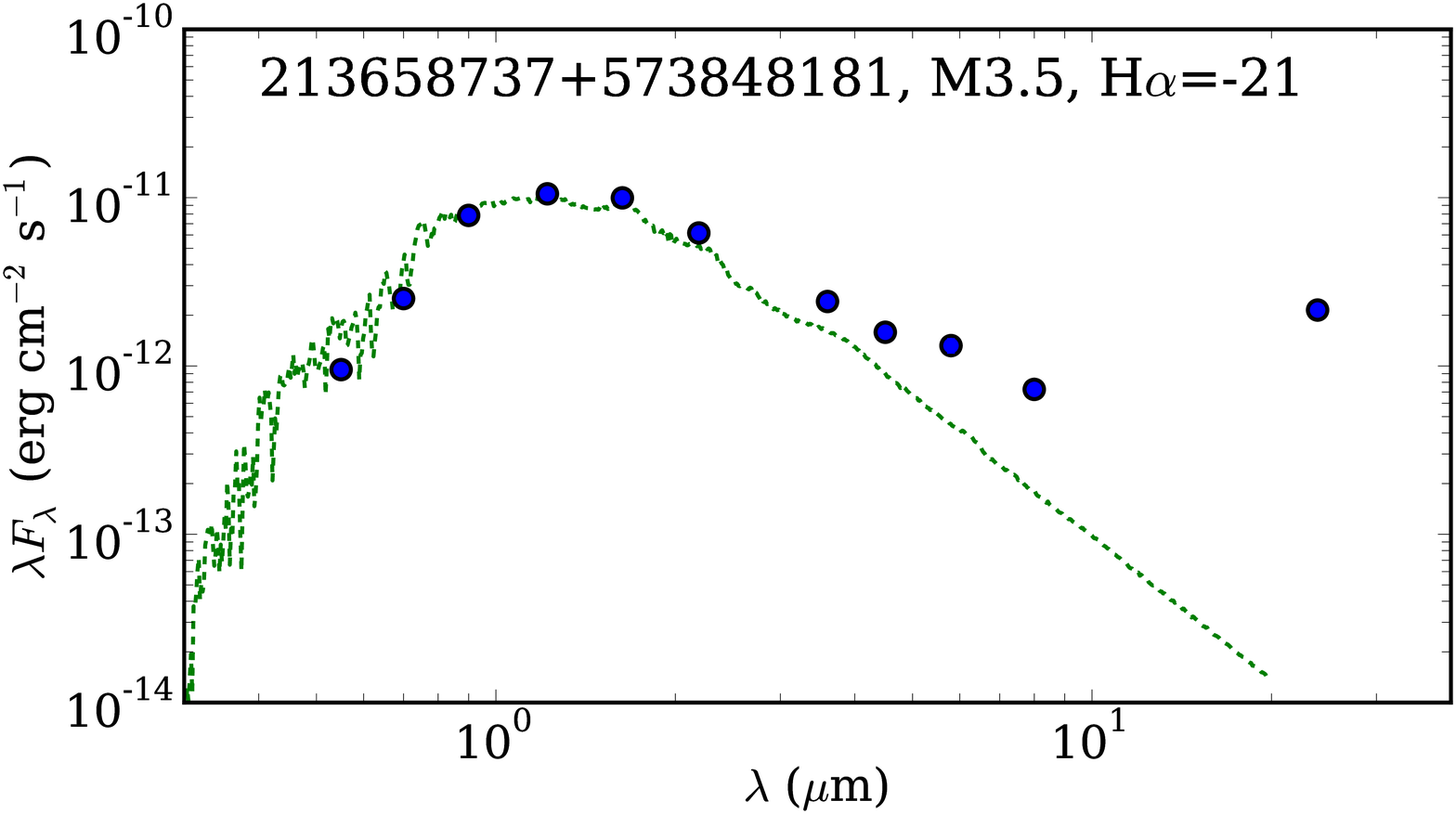,width=0.24\linewidth,clip=} &
\epsfig{file=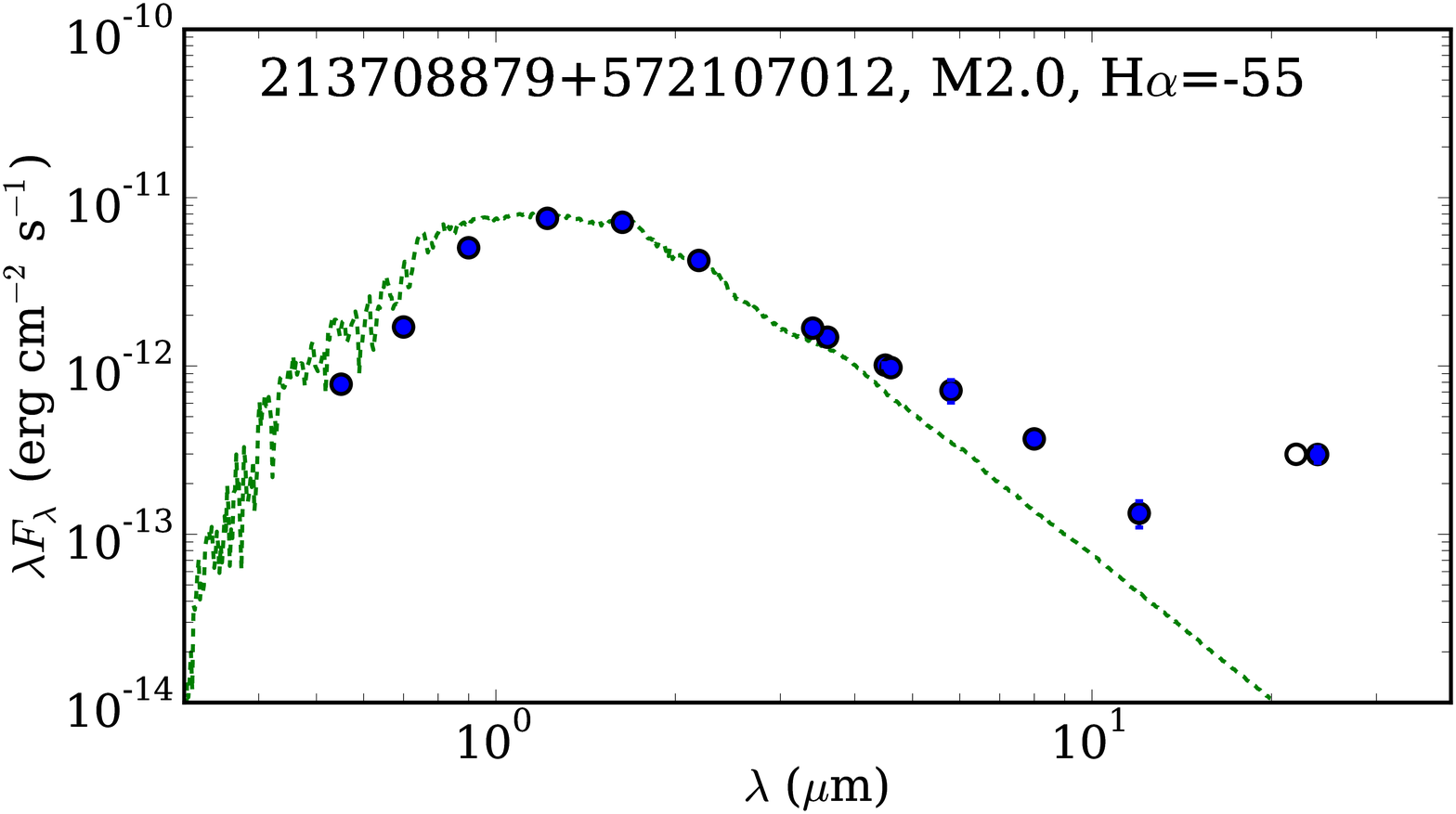,width=0.24\linewidth,clip=} &
\epsfig{file=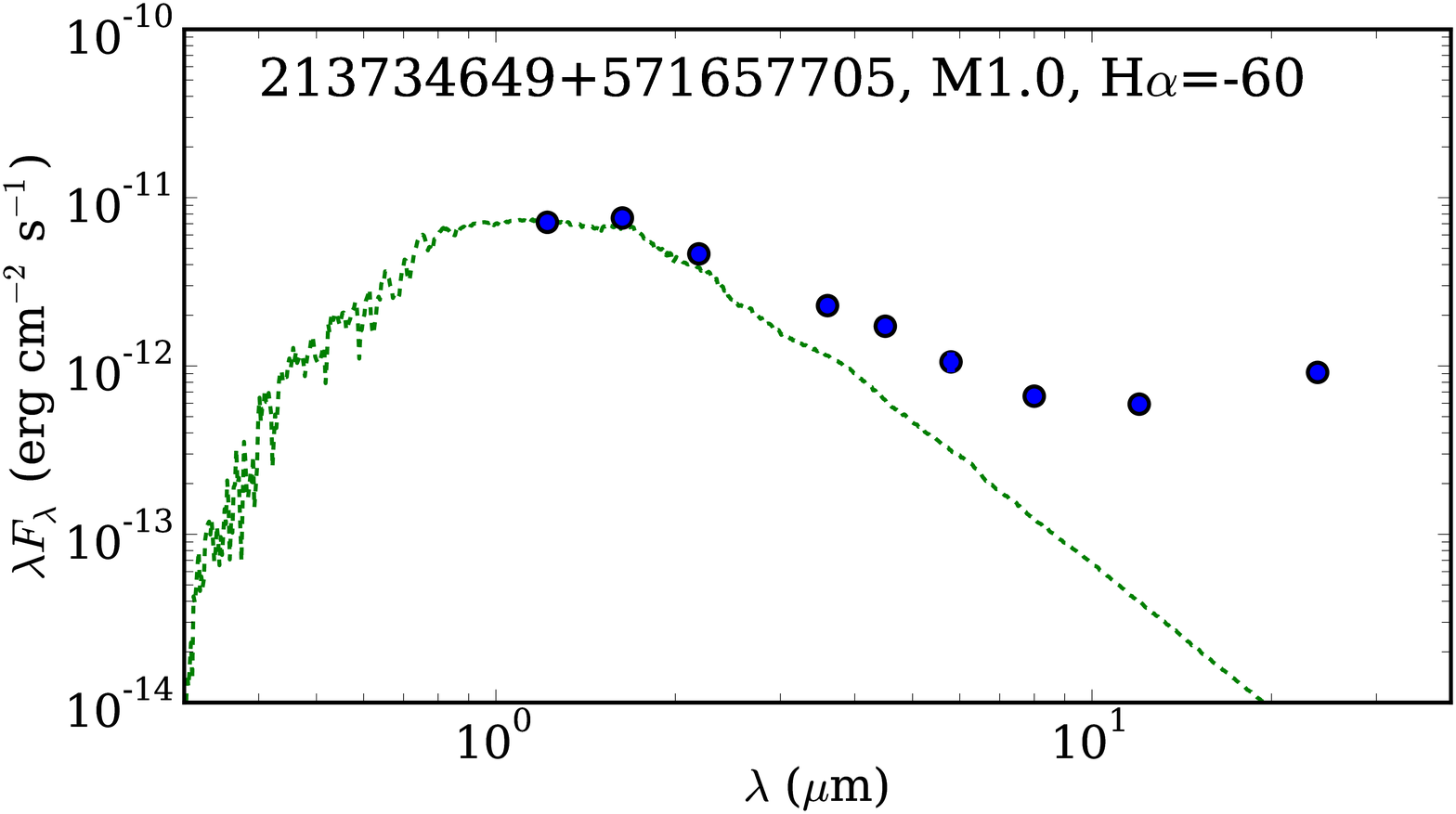,width=0.24\linewidth,clip=} &
\epsfig{file=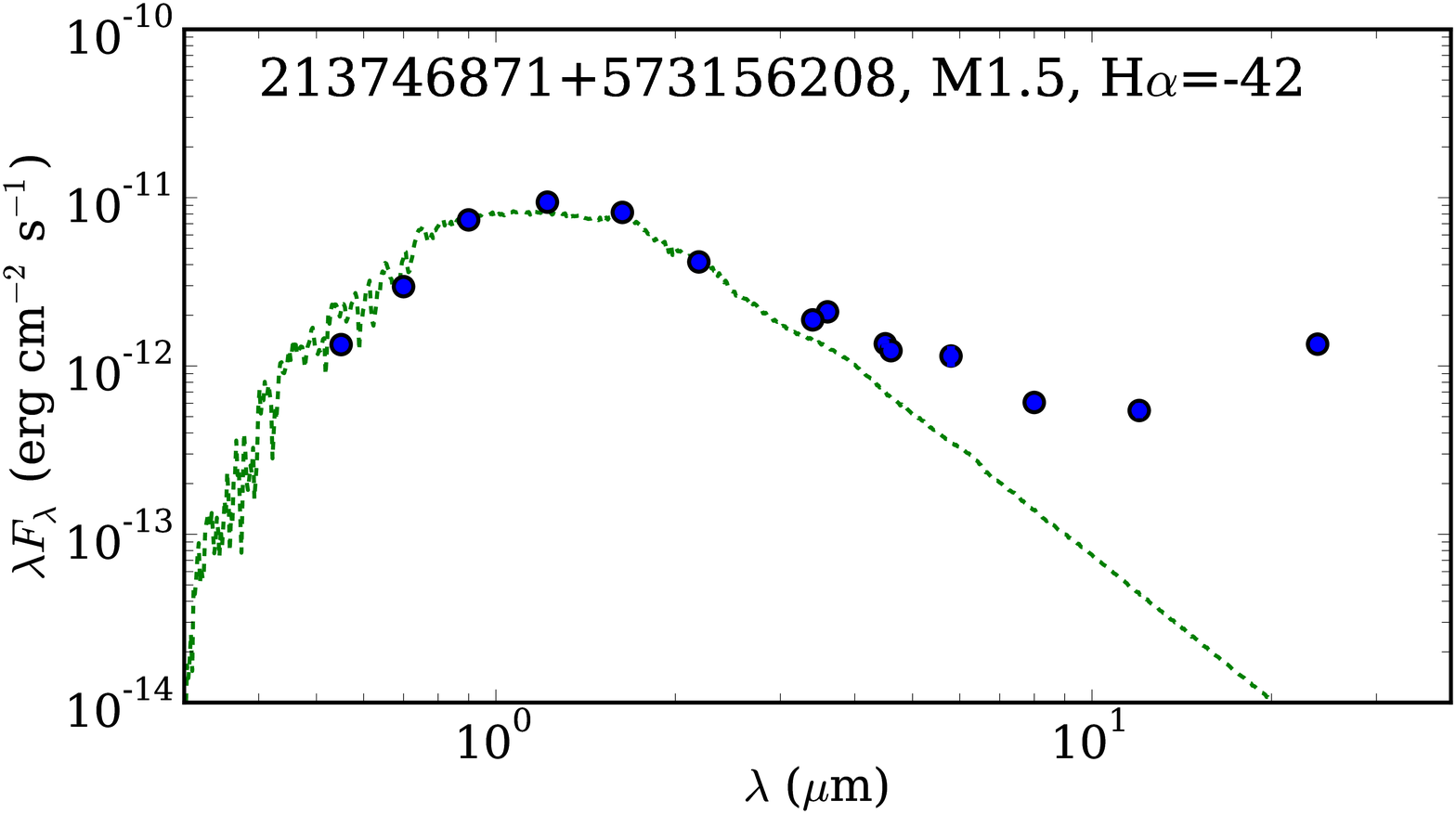,width=0.24\linewidth,clip=} \\
\epsfig{file=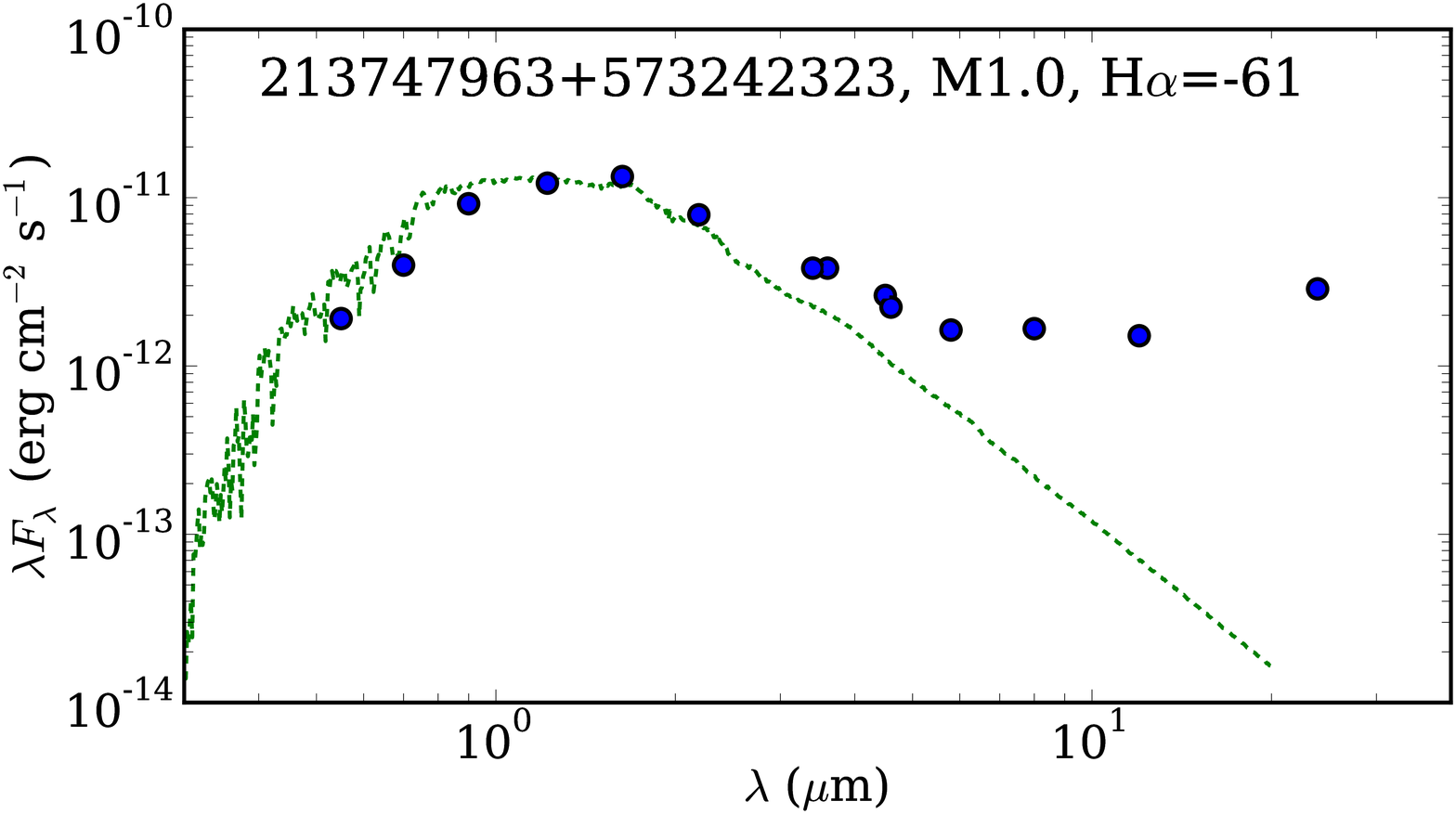,width=0.24\linewidth,clip=} &
\epsfig{file=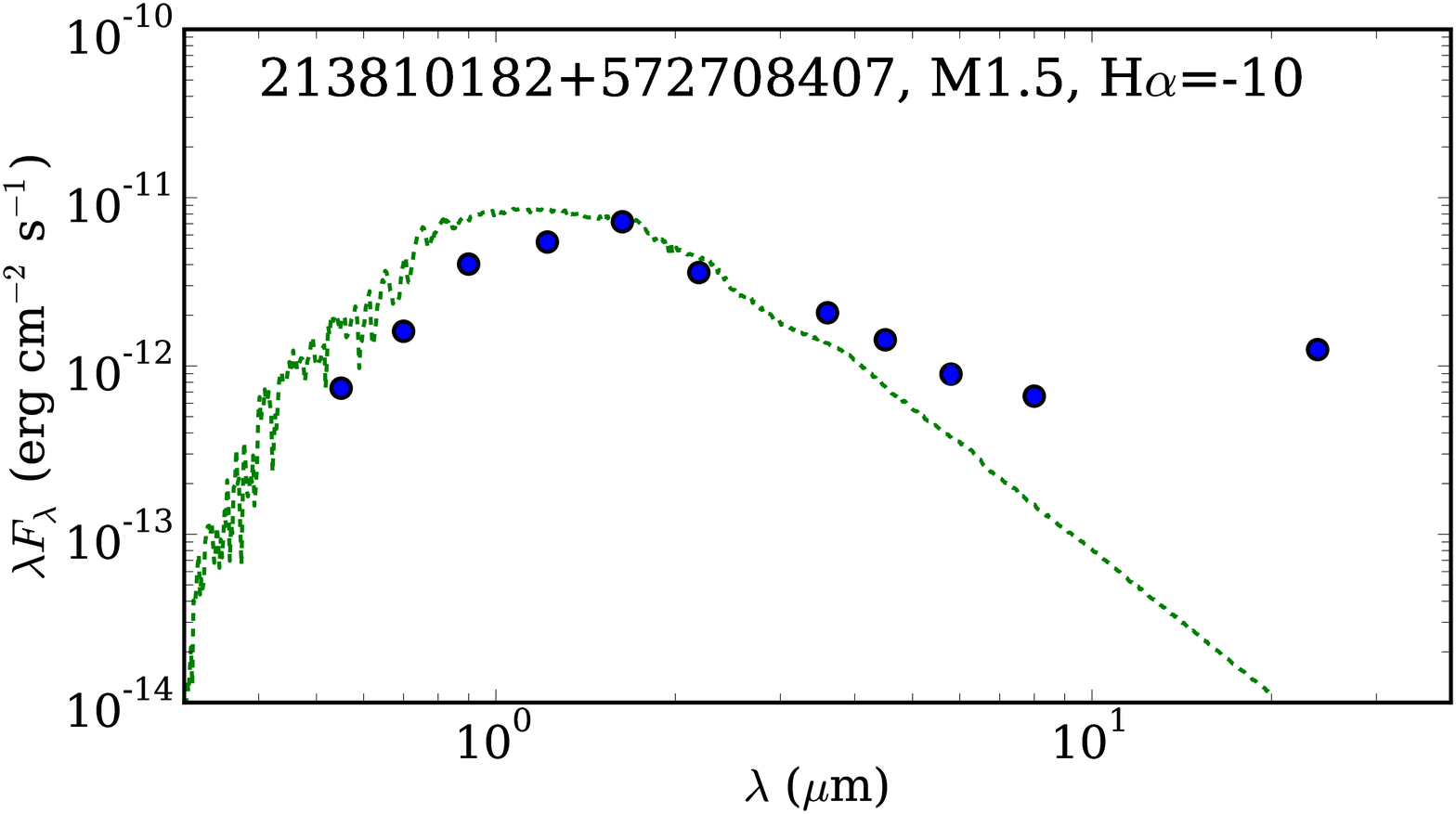,width=0.24\linewidth,clip=} &
\epsfig{file=213812641+572033696.eps,width=0.24\linewidth,clip=} &
\epsfig{file=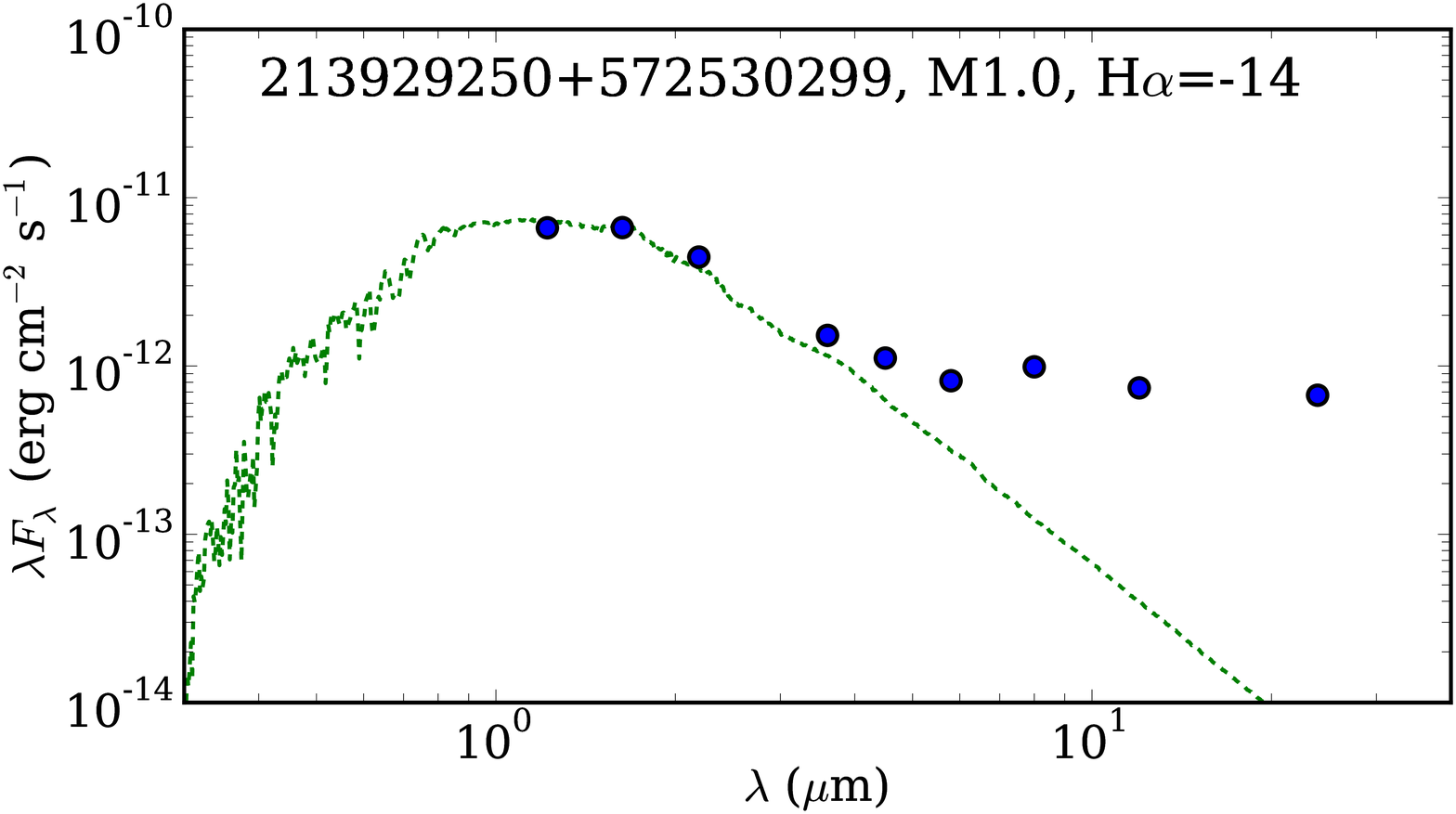,width=0.24\linewidth,clip=} \\
\epsfig{file=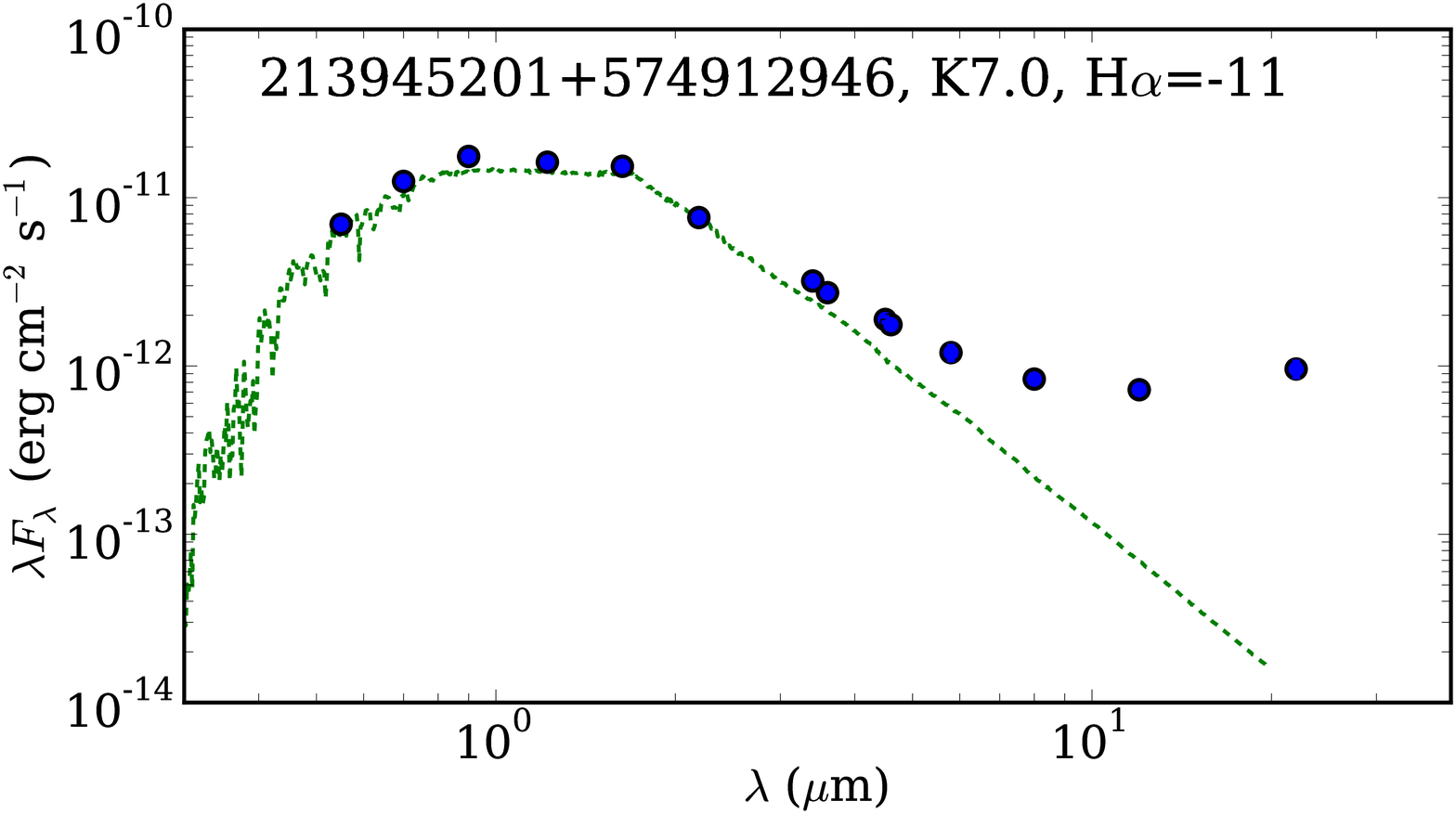,width=0.24\linewidth,clip=} &
\epsfig{file=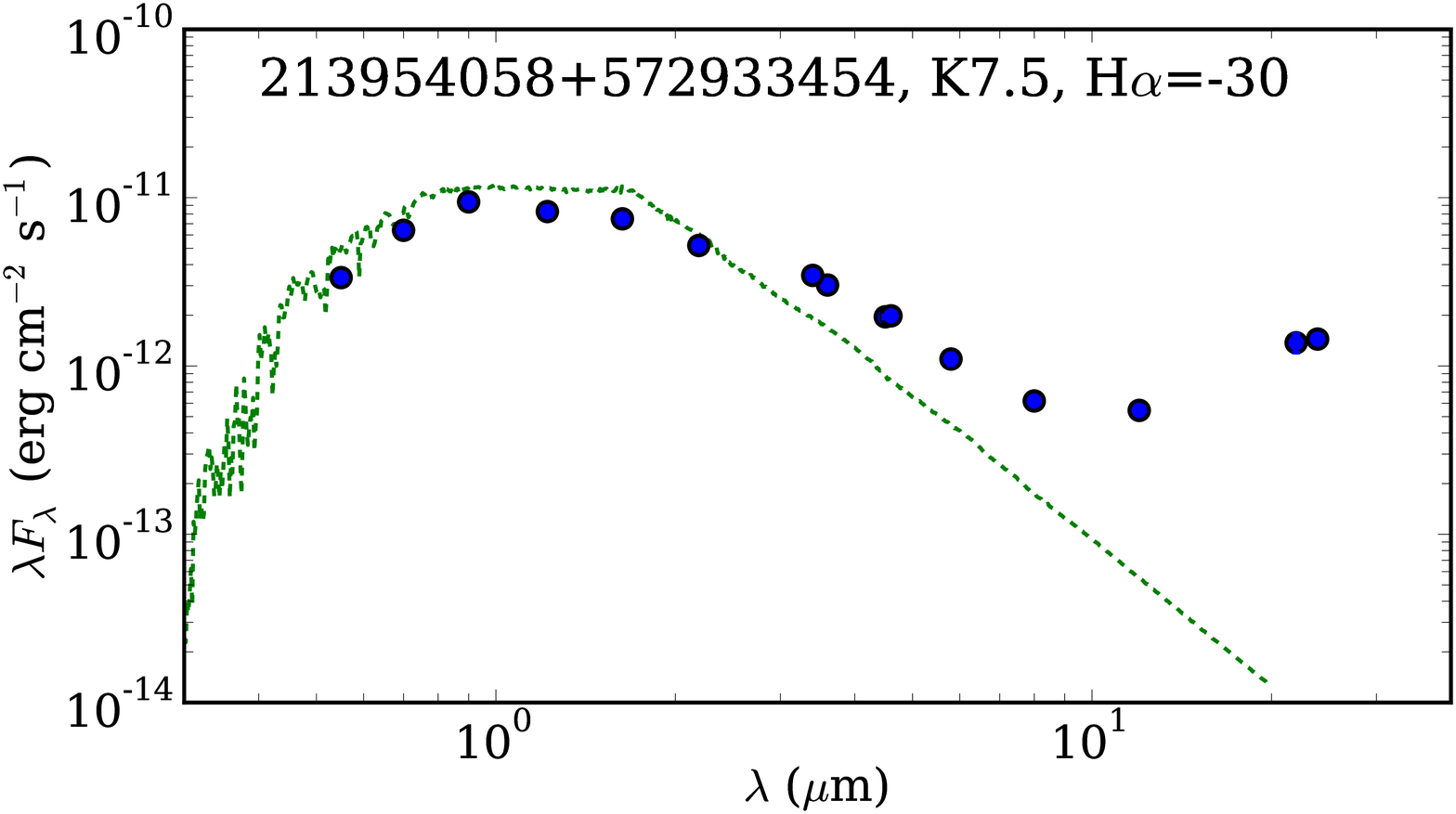,width=0.24\linewidth,clip=} &
\epsfig{file=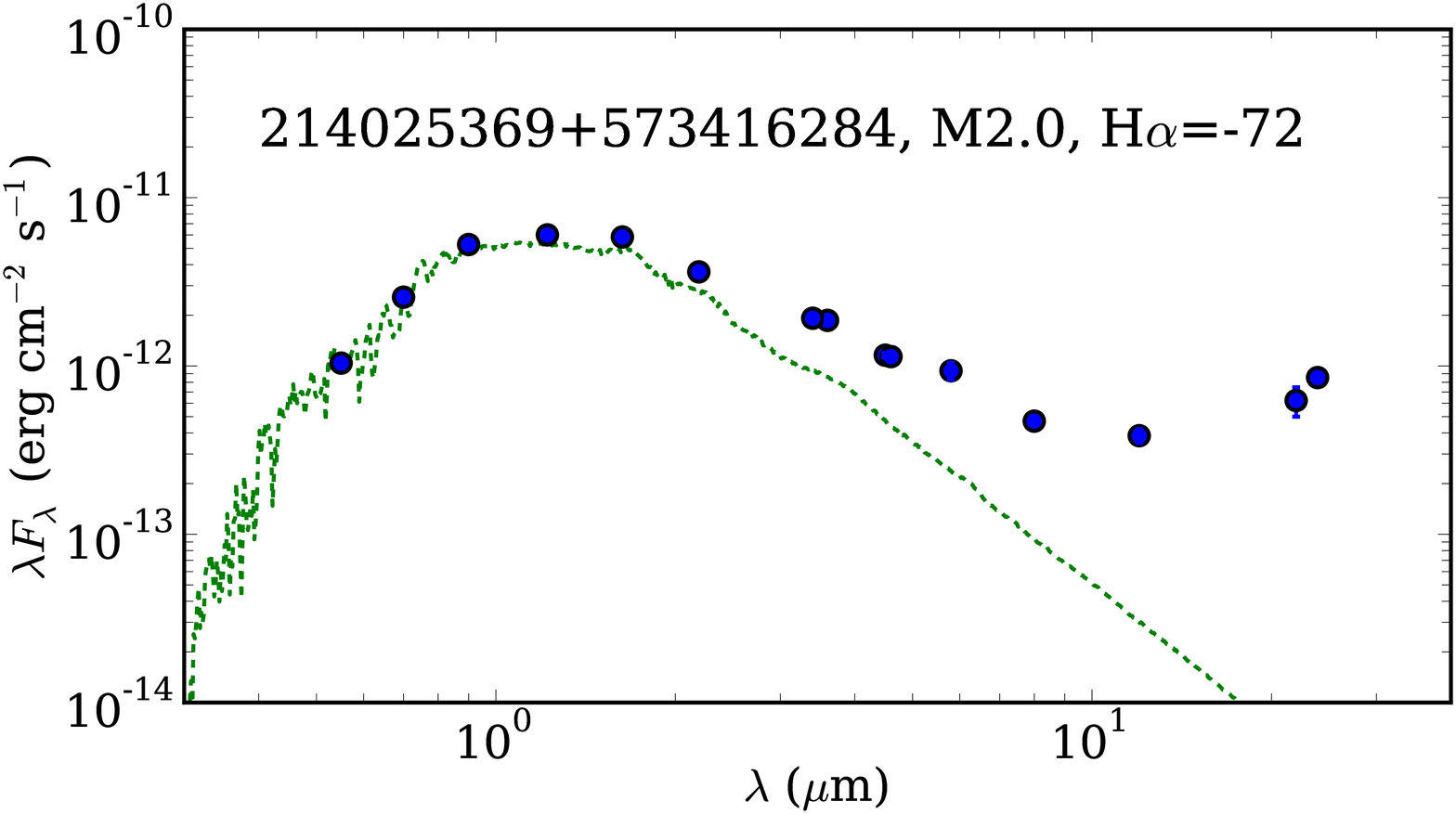,width=0.24\linewidth,clip=} &
\epsfig{file=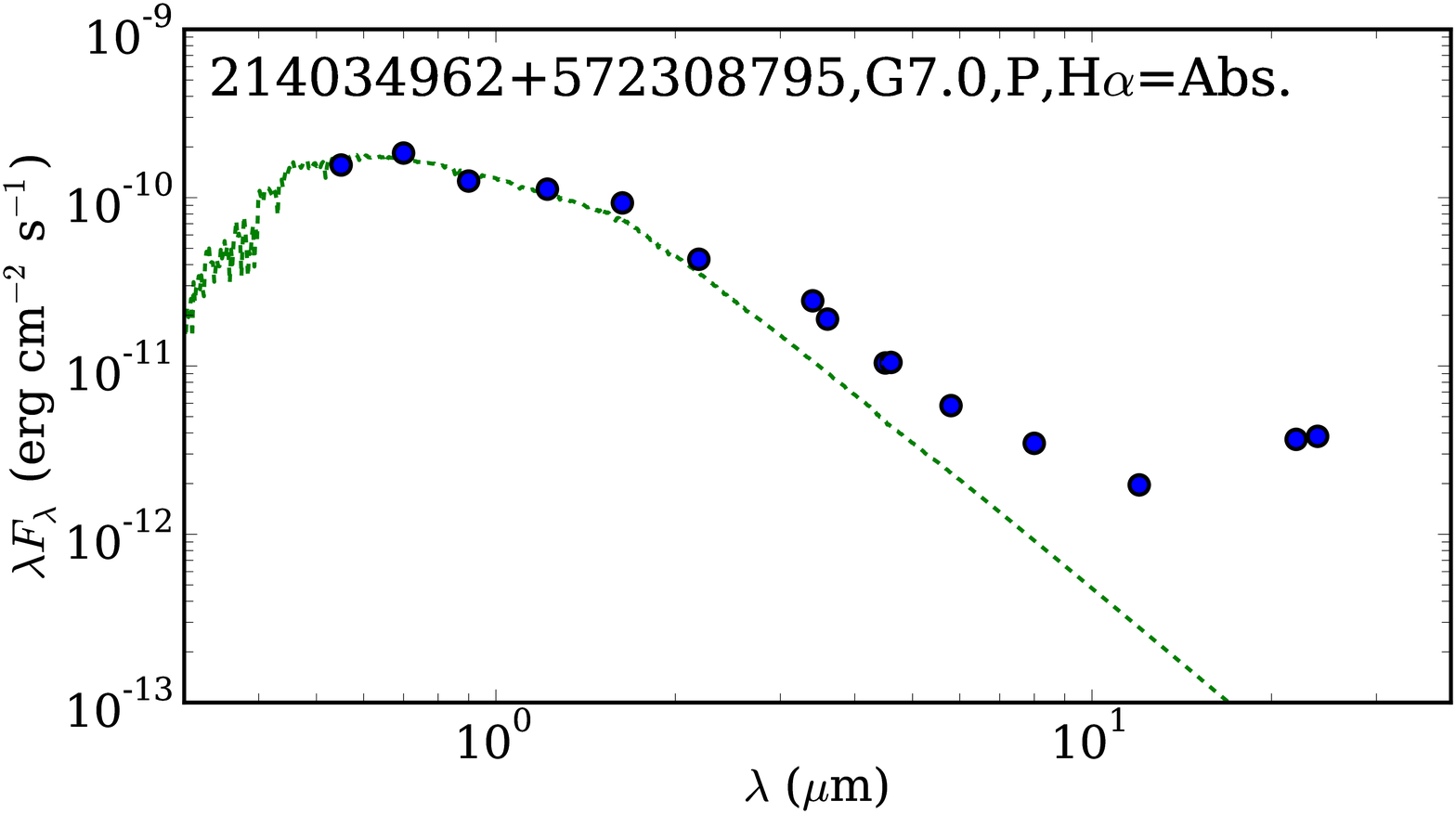,width=0.24\linewidth,clip=} \\
\end{tabular}
\caption{SEDs of the members and probably members with IR excess consistent with PTD. 
Note that the 24$\mu$m flux of 213810182+572708407 is contaminated by a nearby star, but the object
still presents the kink in the SED typical of PTD. \label{pretdseds-fig}}
\end{figure*}

\begin{figure*}
\centering
\begin{tabular}{ccccc}
\epsfig{file=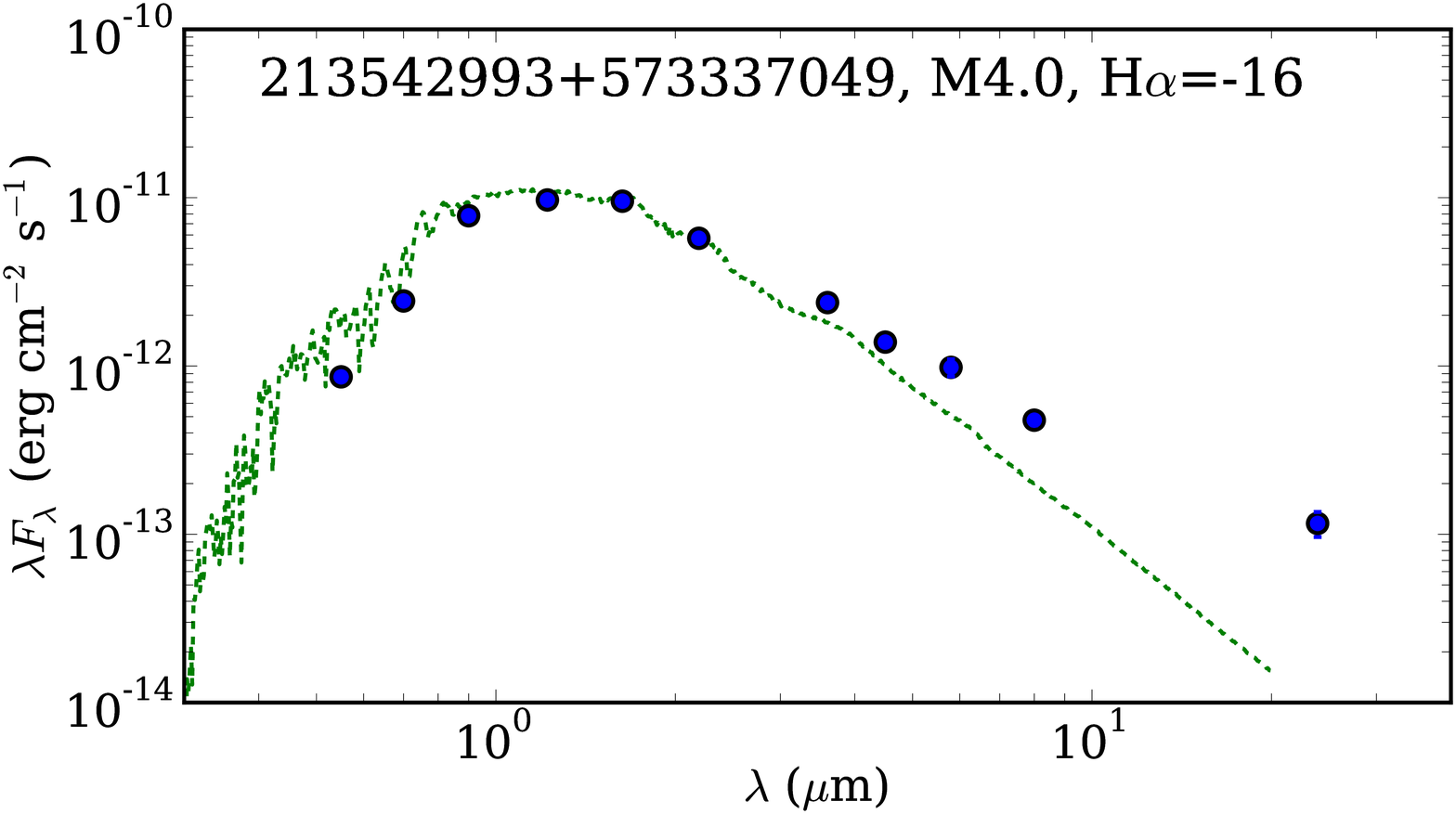,width=0.24\linewidth,clip=} &
\epsfig{file=213733557+573550931.eps,width=0.24\linewidth,clip=} &
\epsfig{file=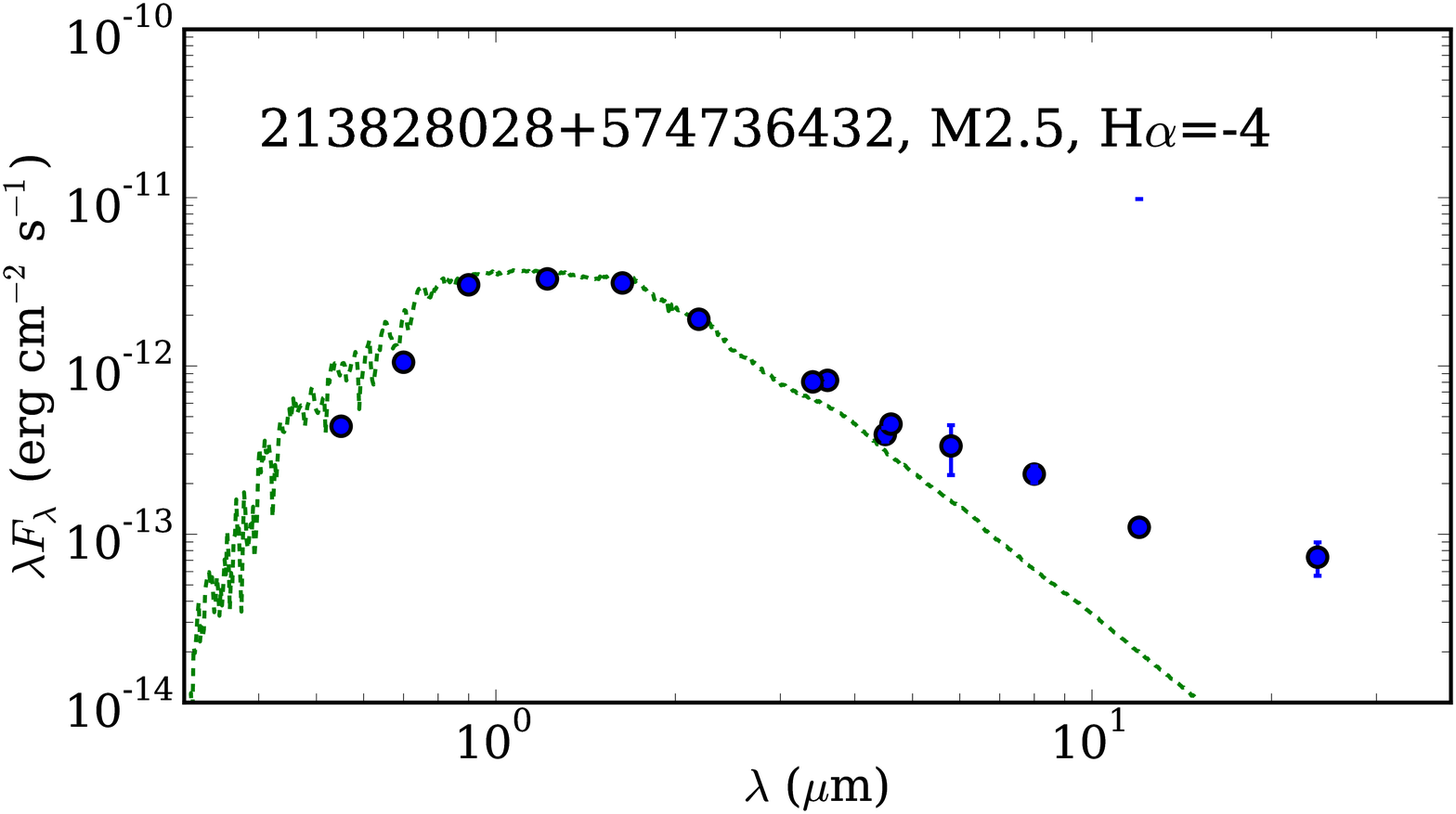,width=0.24\linewidth,clip=} &
\epsfig{file=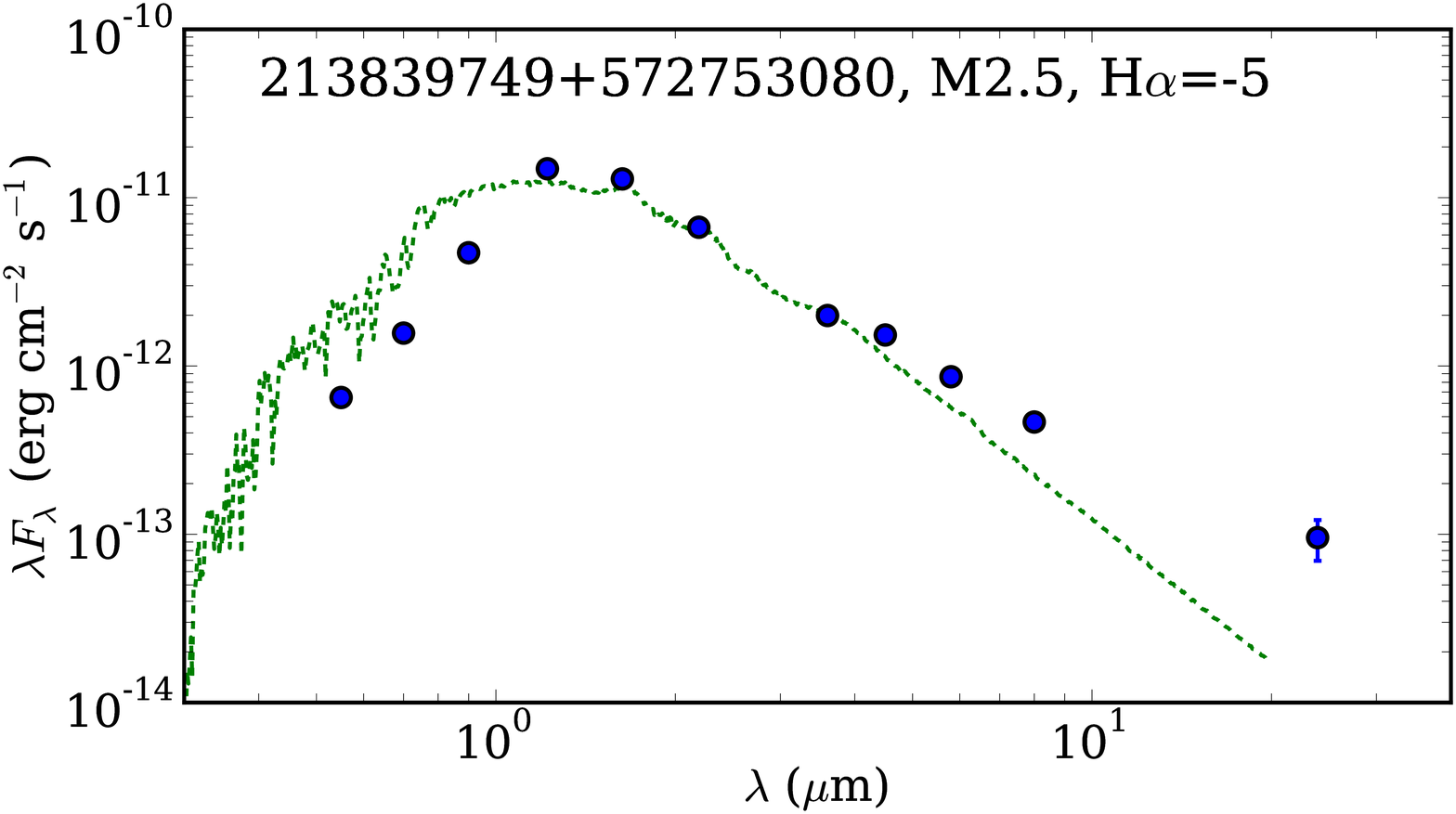,width=0.24\linewidth,clip=} \\
\epsfig{file=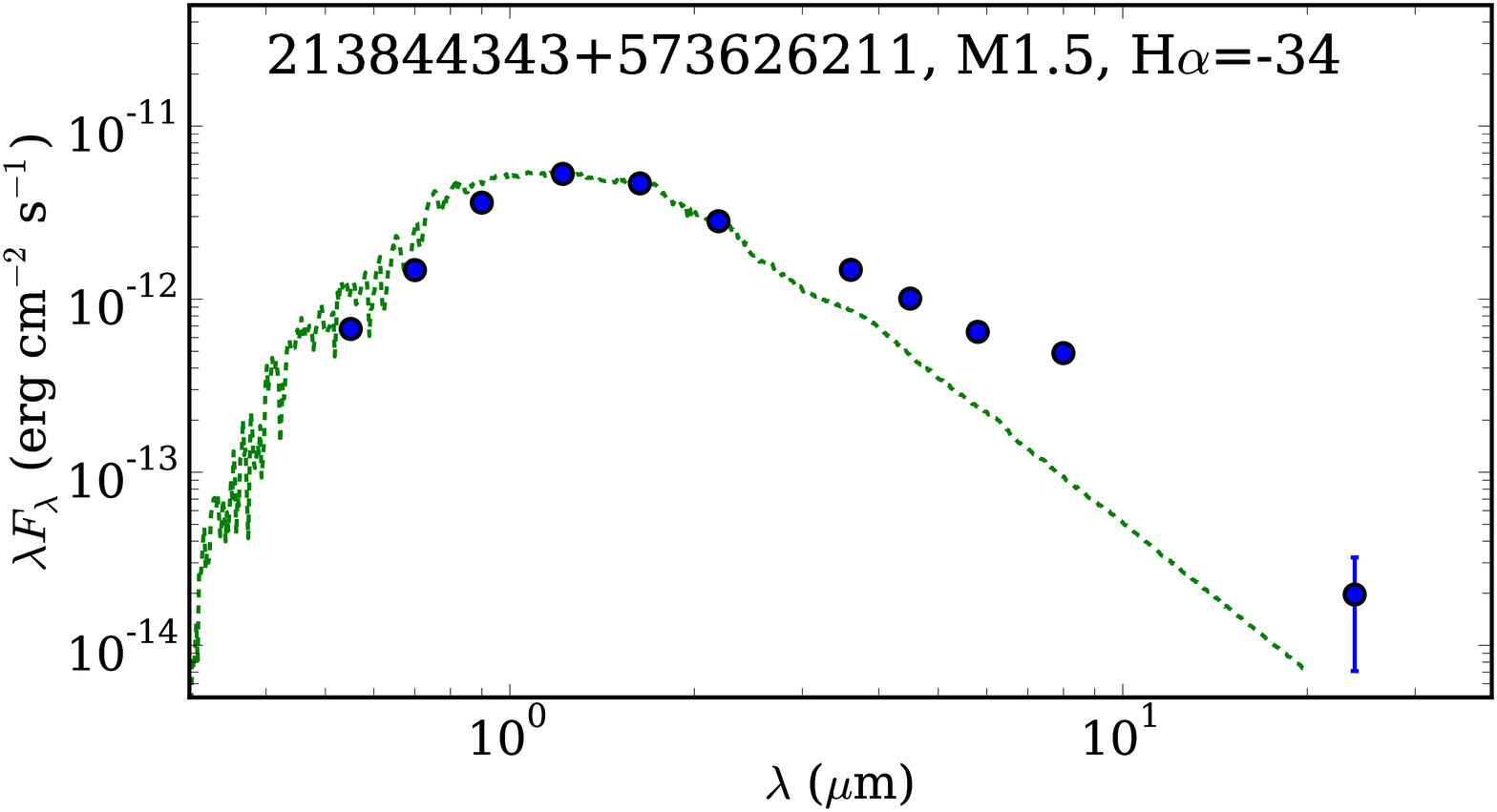,width=0.24\linewidth,clip=} &
\epsfig{file=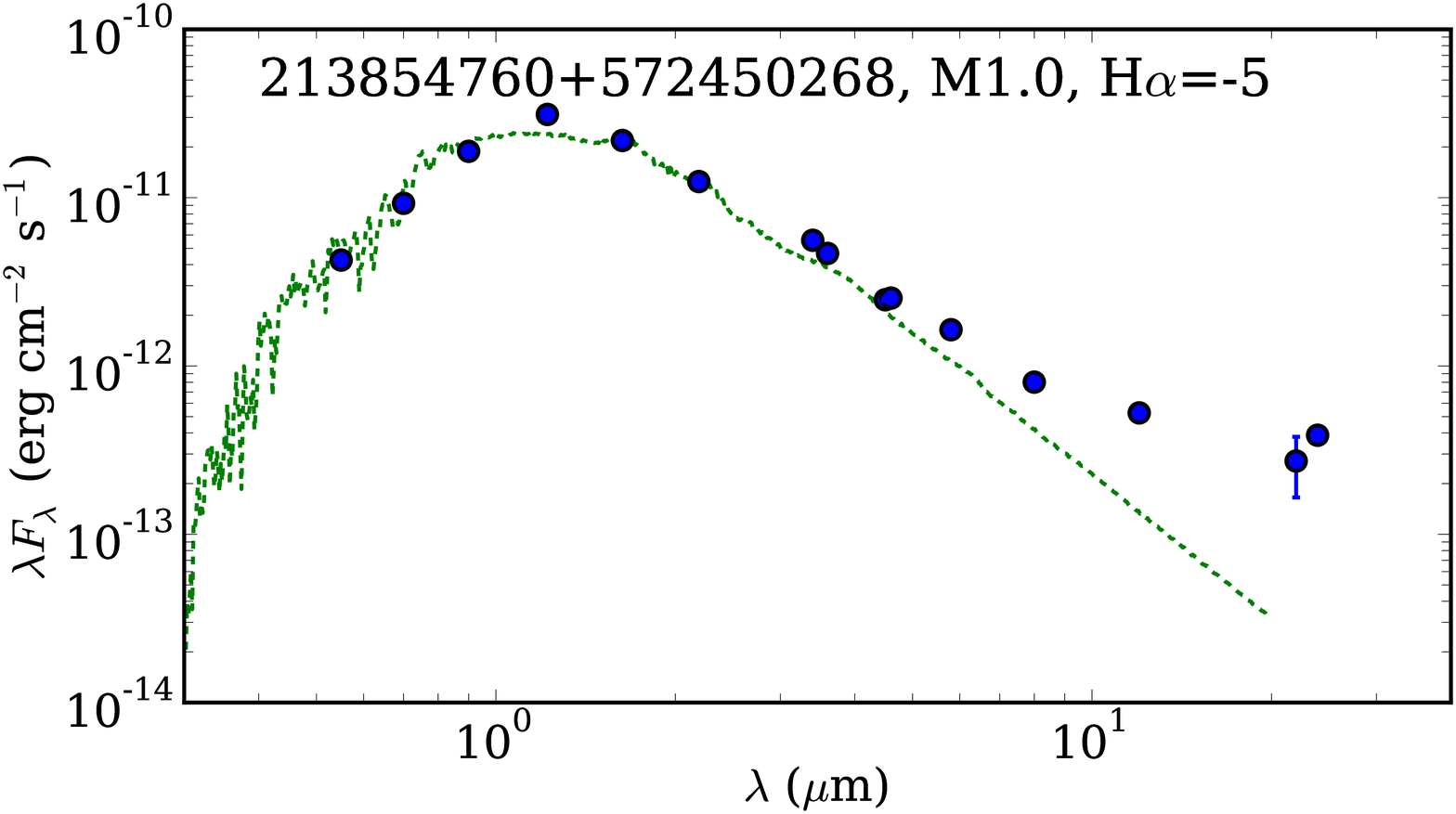,width=0.24\linewidth,clip=} &
\epsfig{file=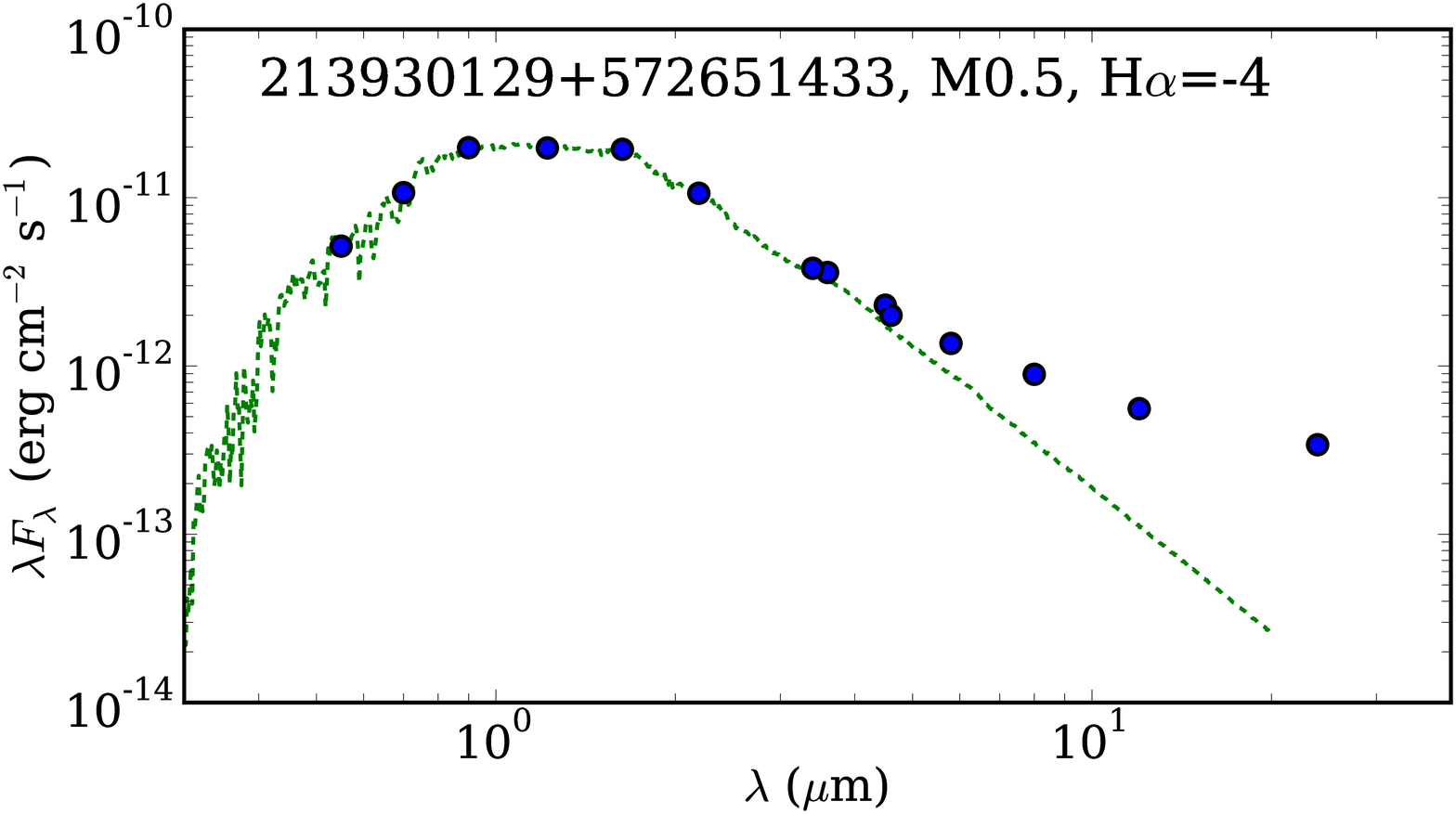,width=0.24\linewidth,clip=} &
\epsfig{file=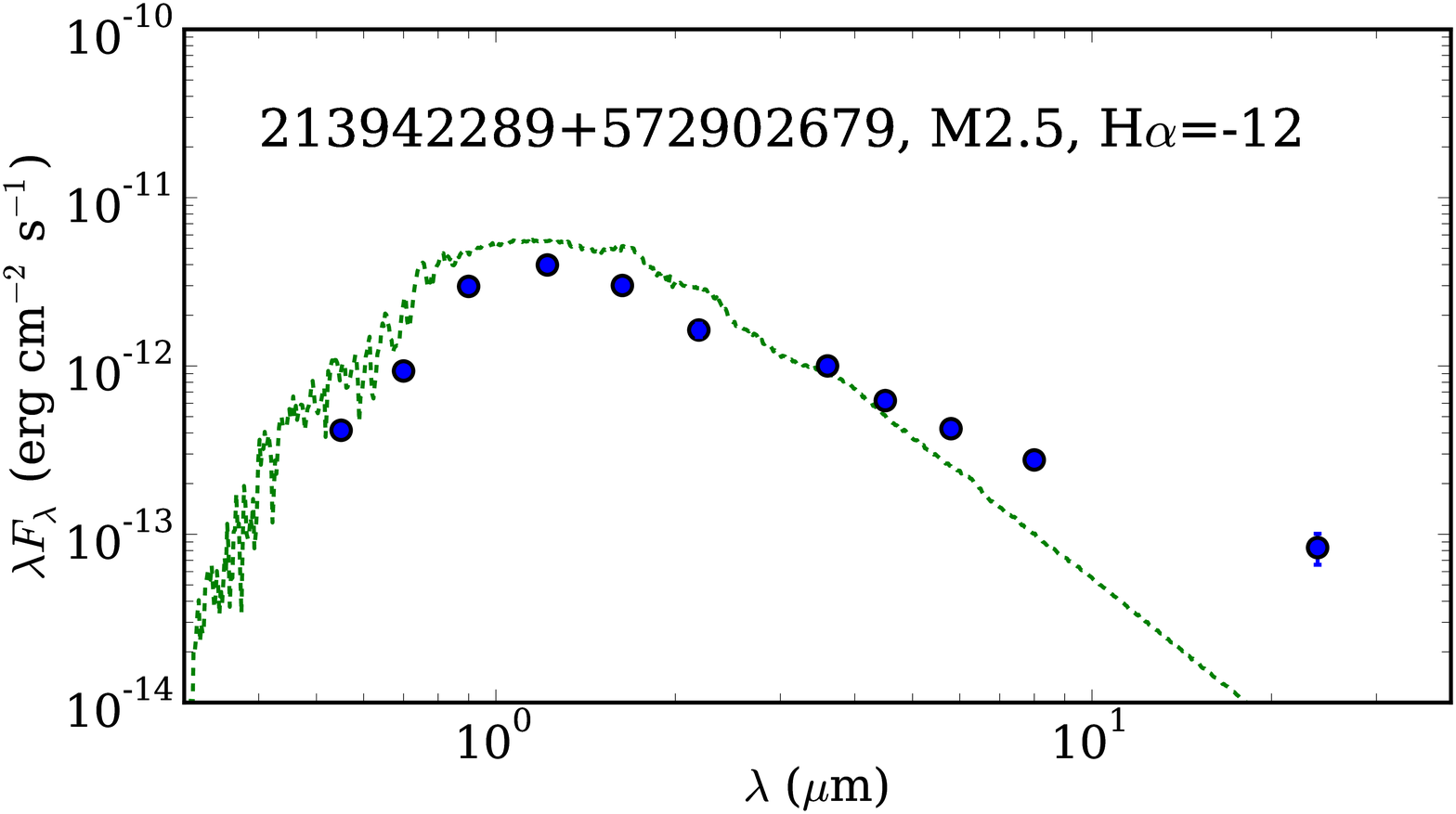,width=0.24\linewidth,clip=} \\
\epsfig{file=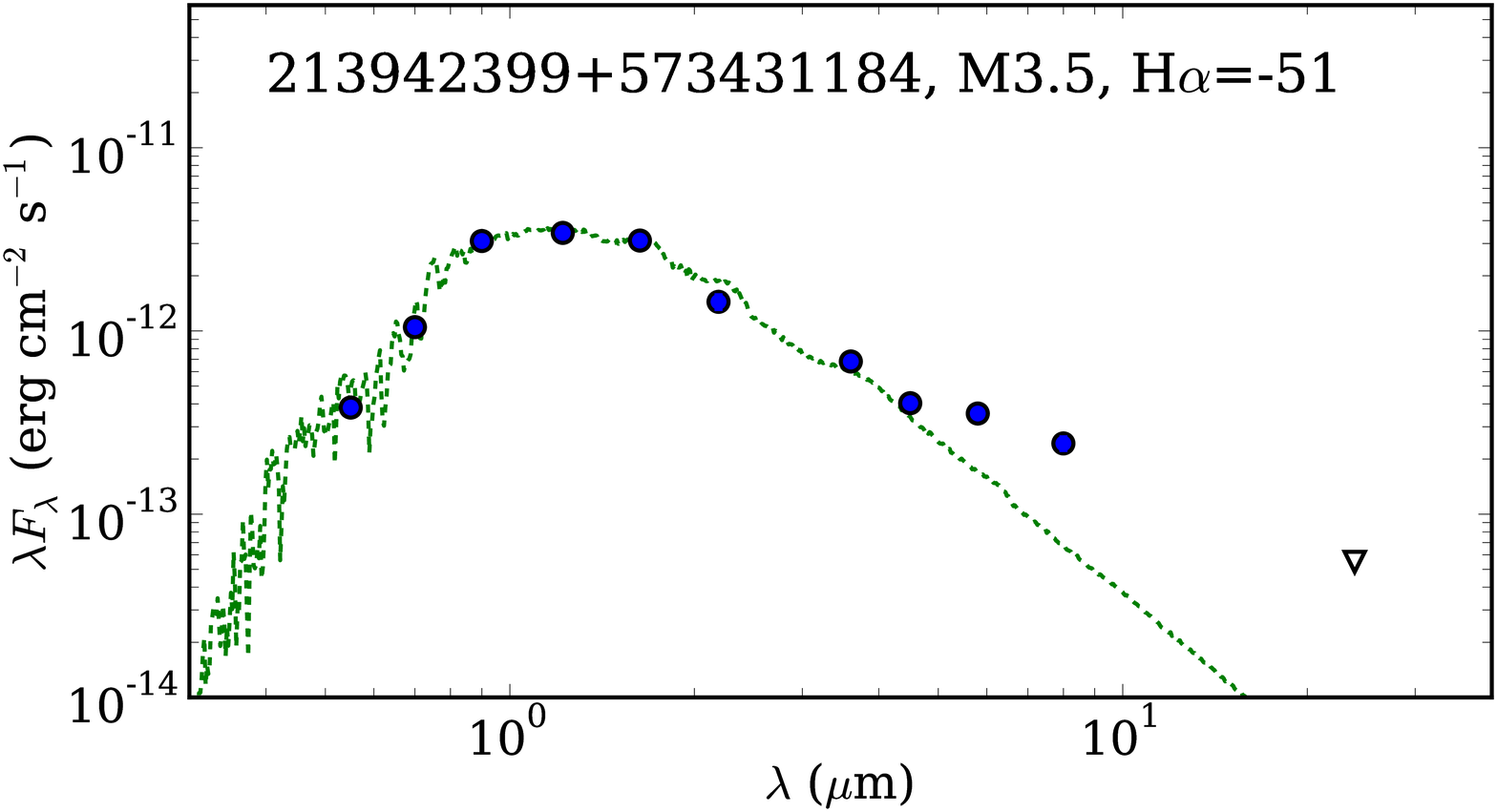,width=0.24\linewidth,clip=} &
\epsfig{file=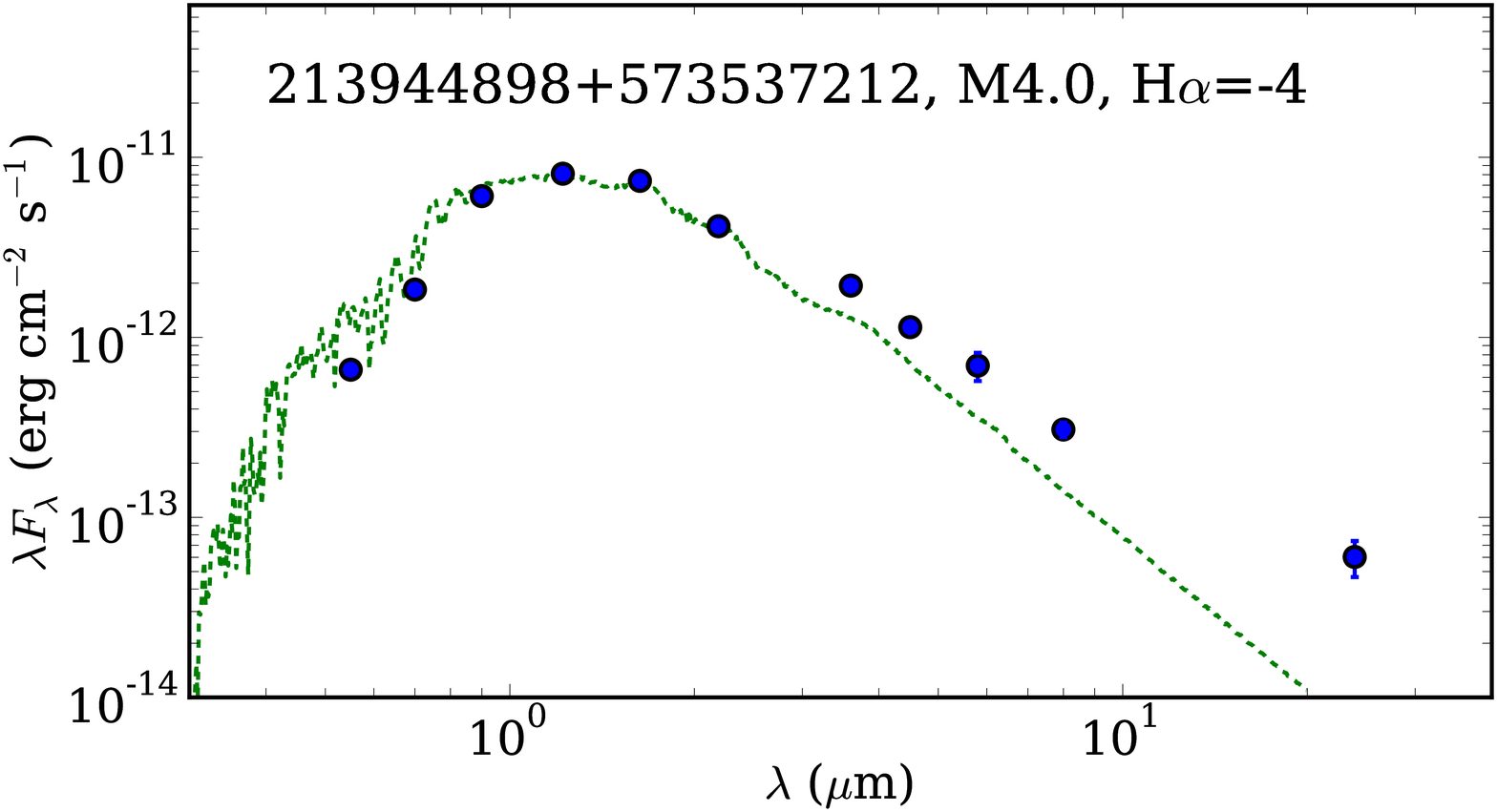,width=0.24\linewidth,clip=} &
\epsfig{file=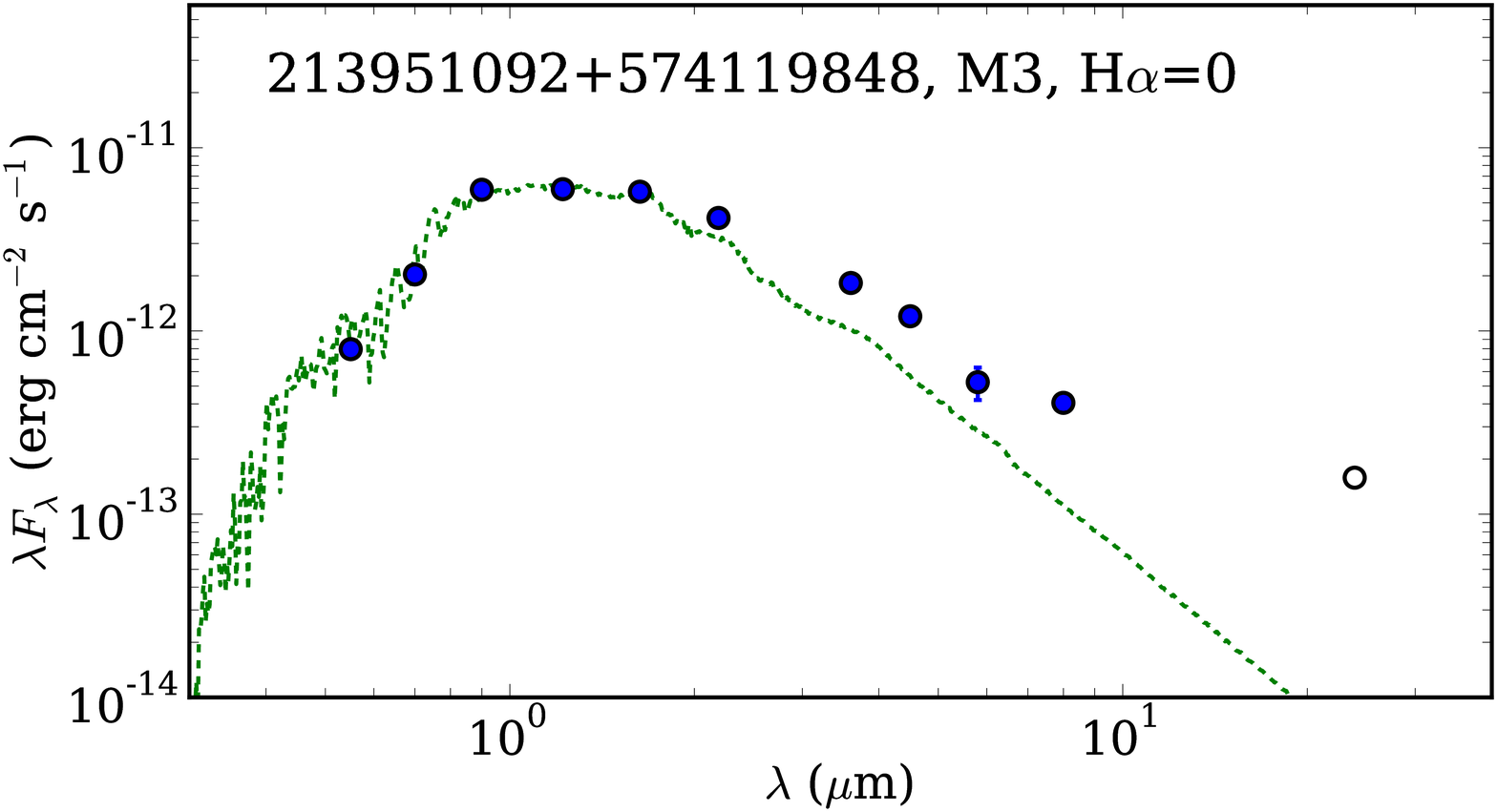,width=0.24\linewidth,clip=} & 
\epsfig{file=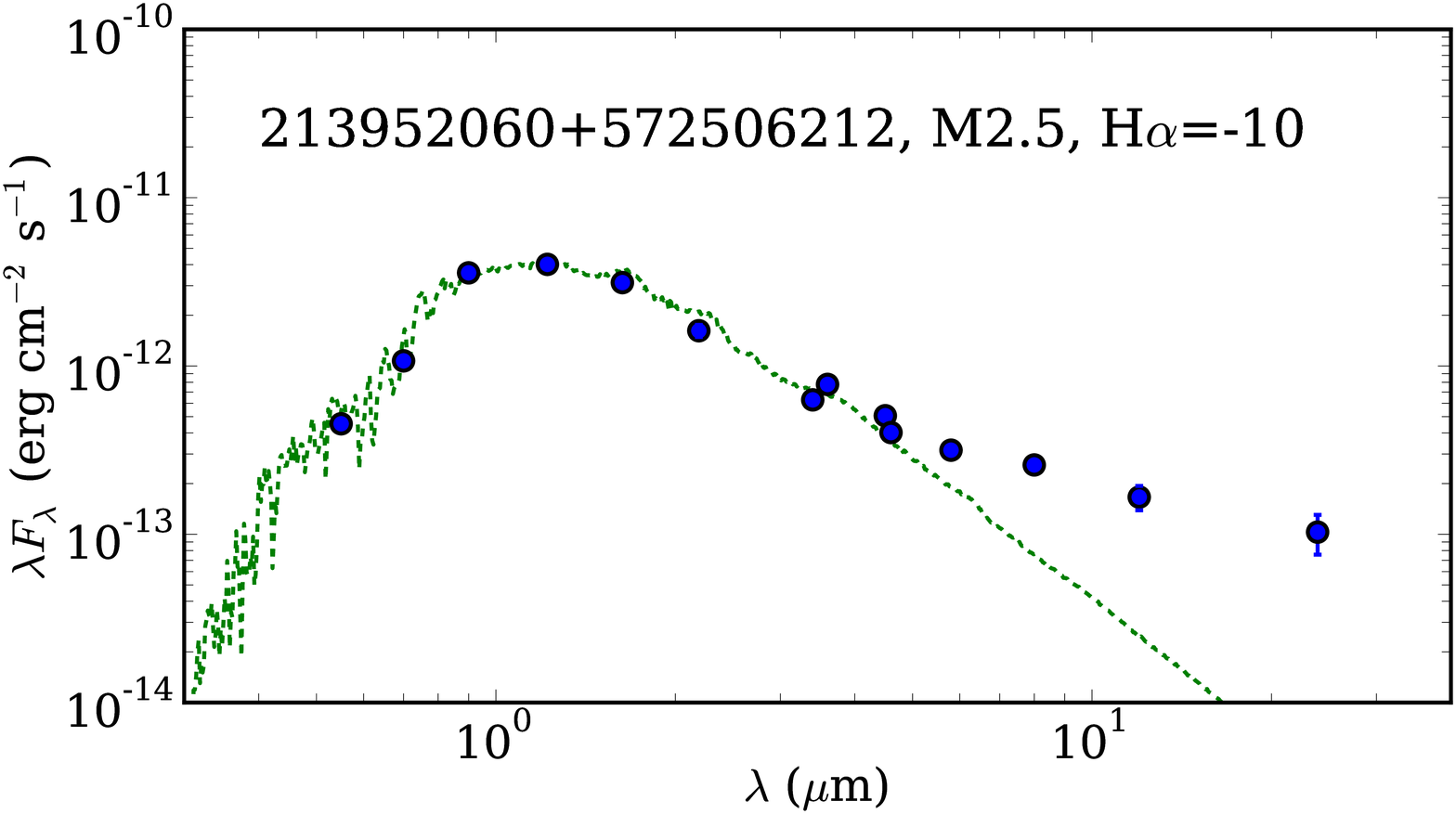,width=0.24\linewidth,clip=} \\
\epsfig{file=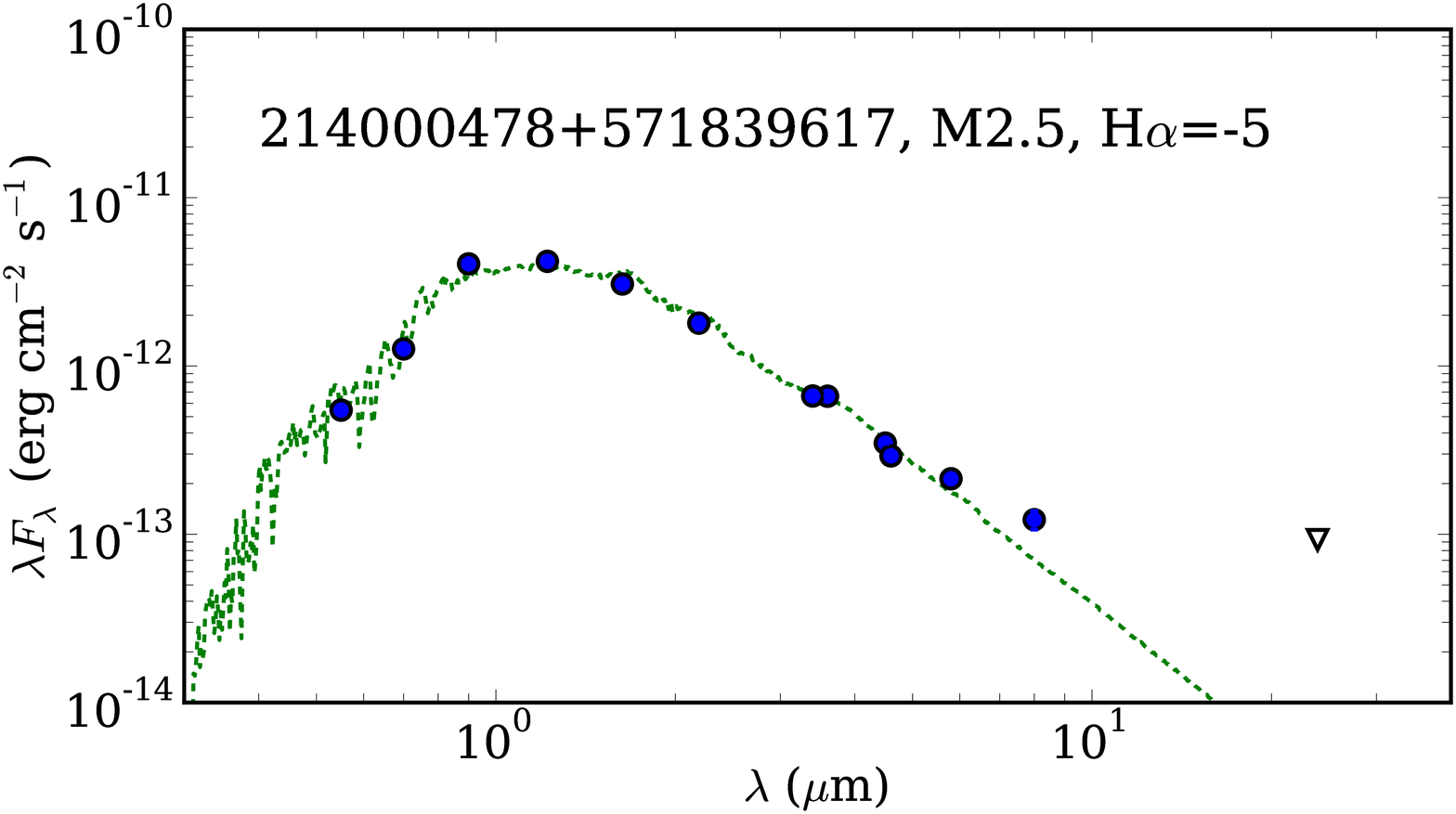,width=0.24\linewidth,clip=} &
\epsfig{file=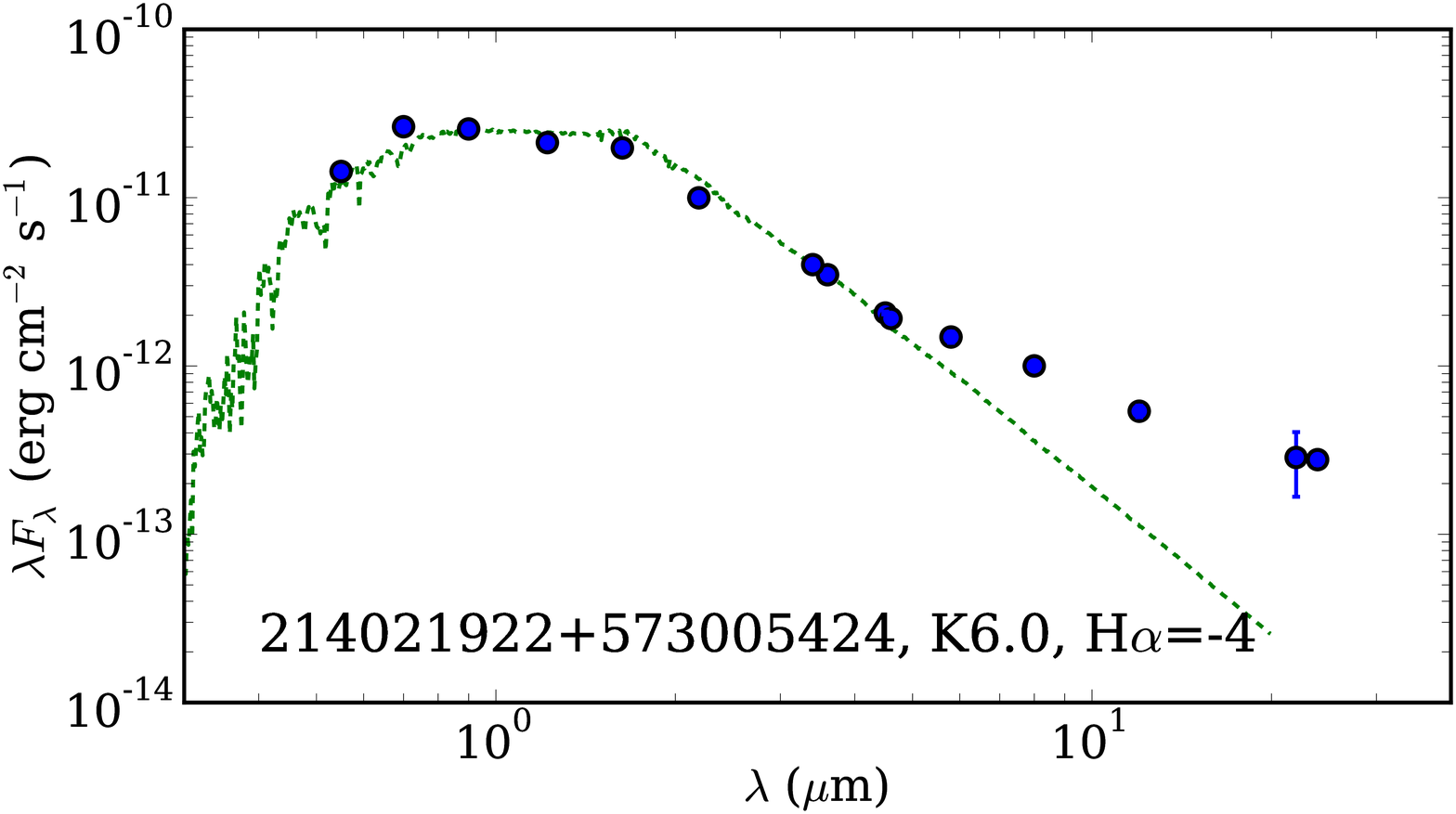,width=0.24\linewidth,clip=} &
\epsfig{file=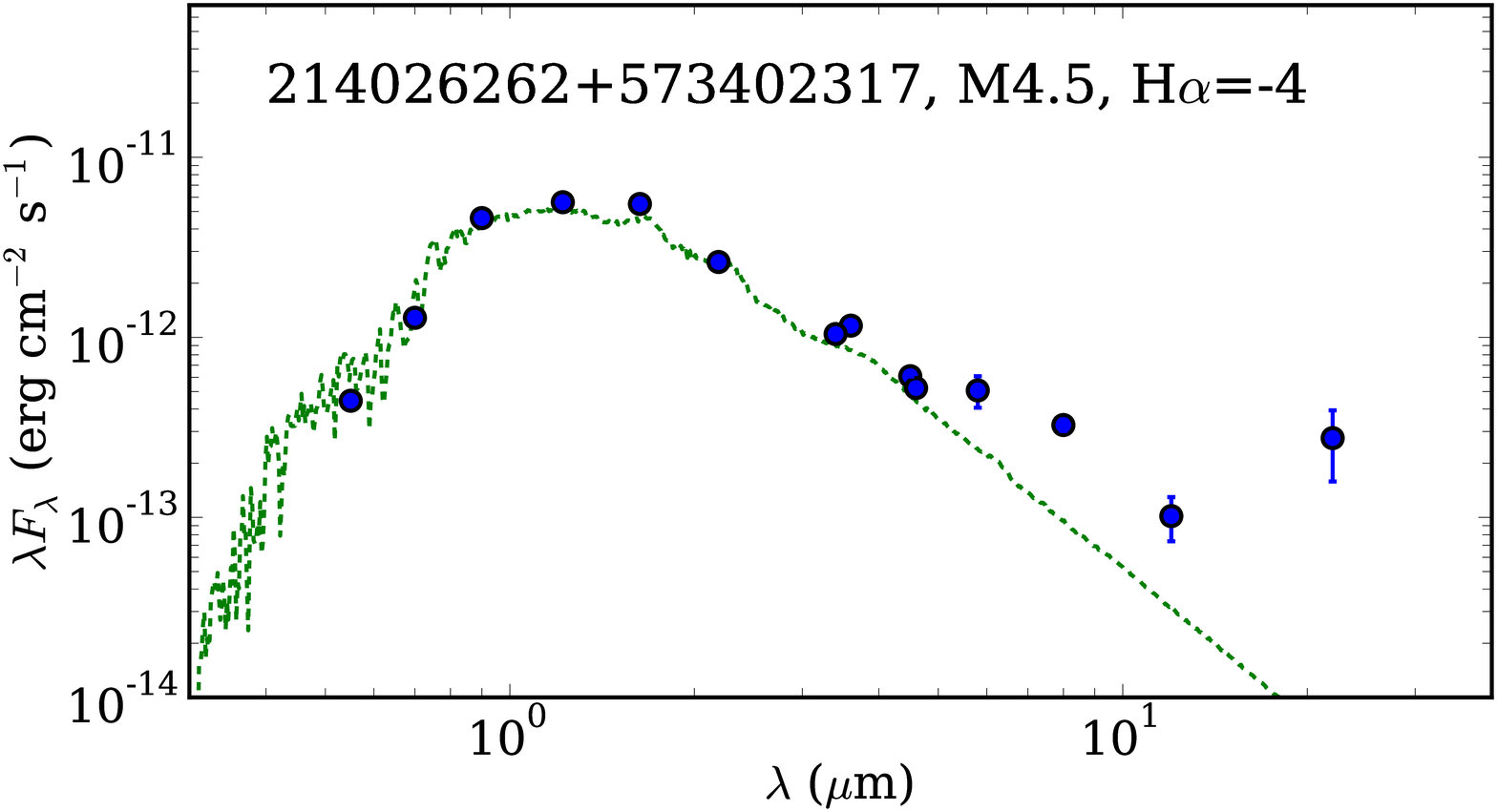,width=0.24\linewidth,clip=} \\
\end{tabular}
\caption{SEDs of the members and probably members with low IR excess consistent with
dust-depleted disks. This includes objects with disks but no significant 24$\mu$m detection,
in cloud-free regions. Some objects like 213844343+573626211 could be truncated disks.
Objects like  213839749+572753080, 213854760+572450268, 213944898+573537212, 
214000478+571839617, 214021922+573005424, and 214026262+573402317 may also have inner holes being thus
both transitional and dust-depleted disk candidates.  Inverted triangles mark upper limits, and open
circles are uncertain values. \label{depletedseds-fig}}
\end{figure*}

\begin{figure*}
\centering
\begin{tabular}{ccccc}
\epsfig{file=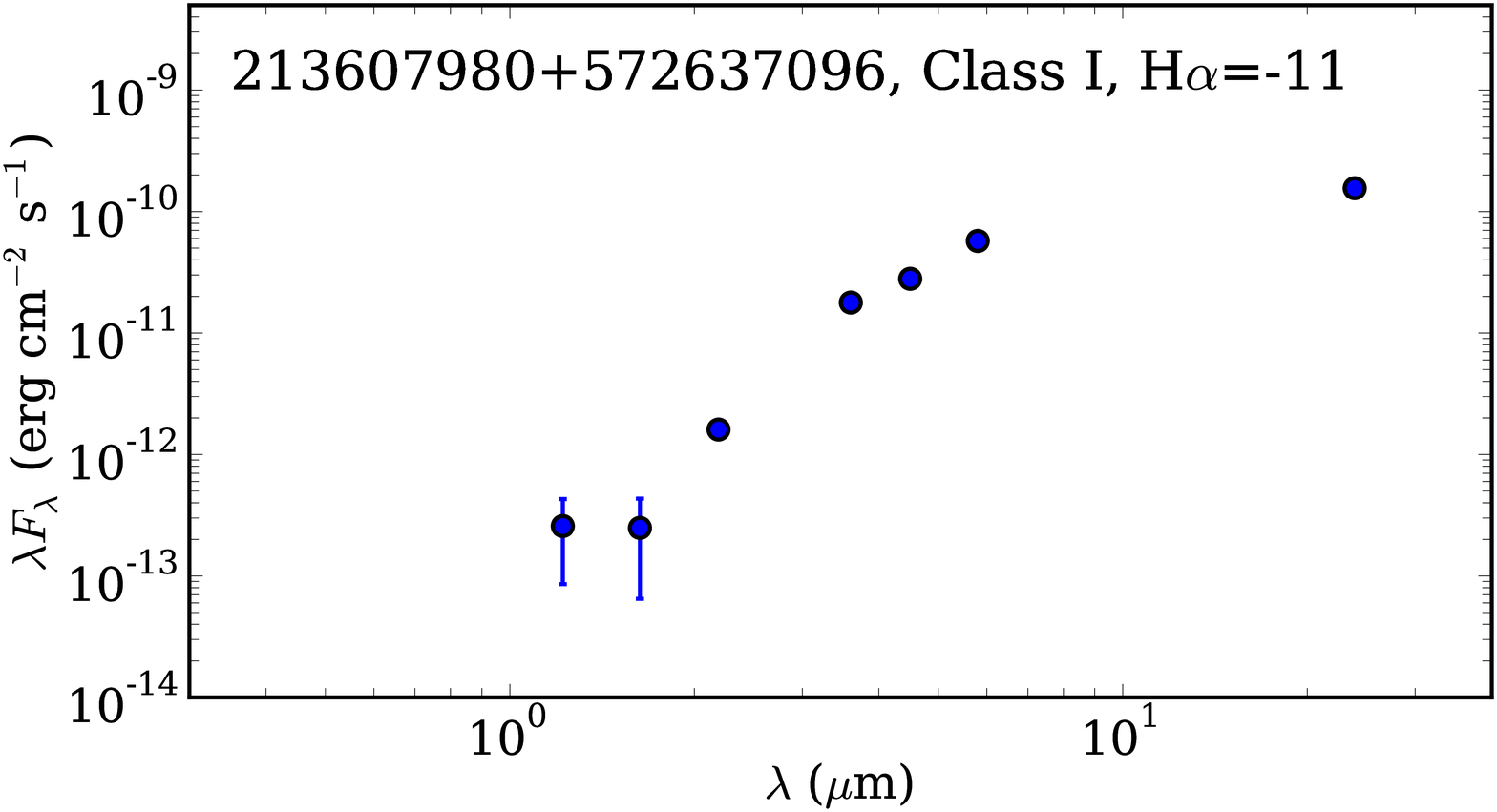,width=0.24\linewidth,clip=} &
\epsfig{file=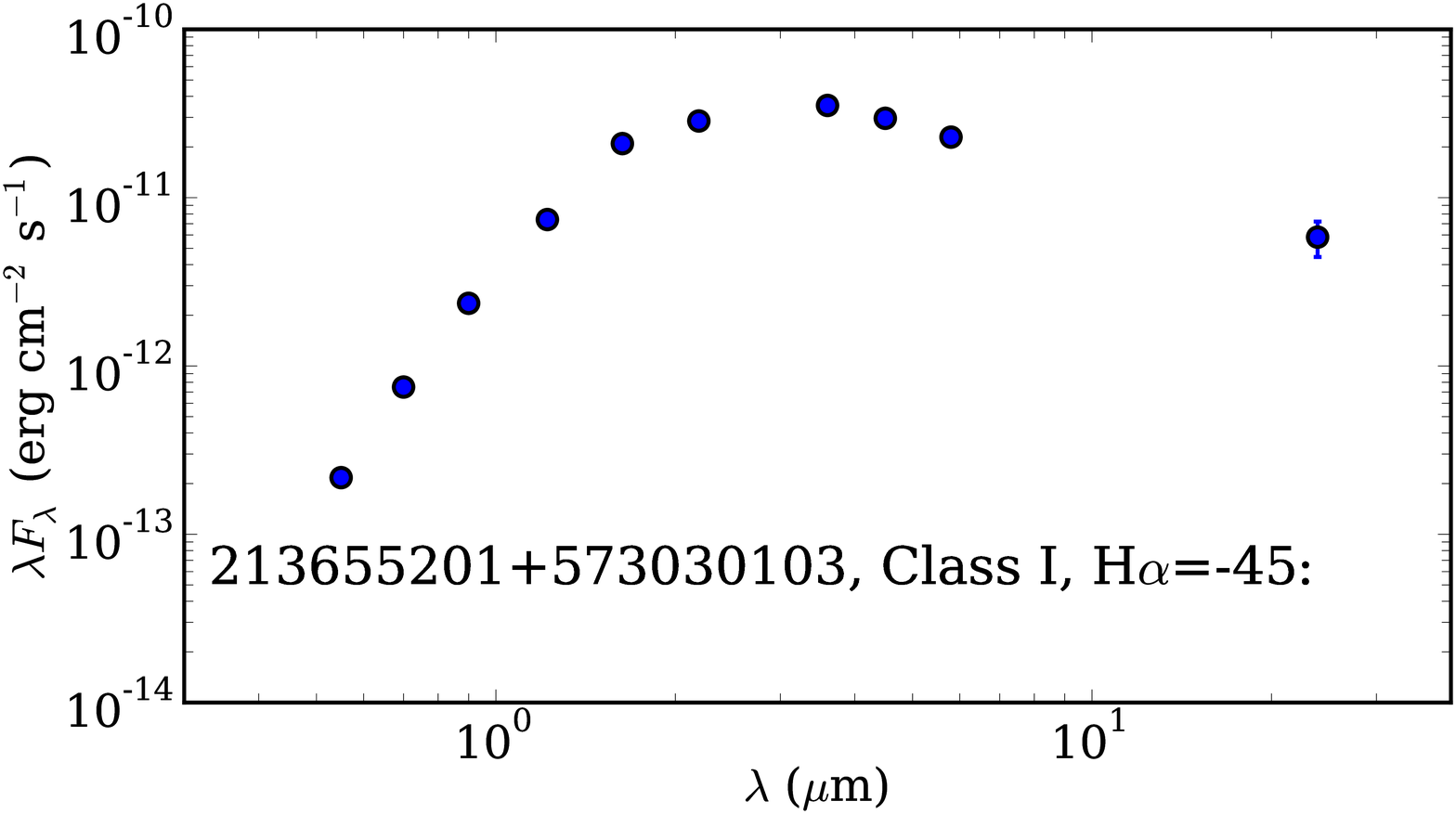,width=0.24\linewidth,clip=} &
\epsfig{file=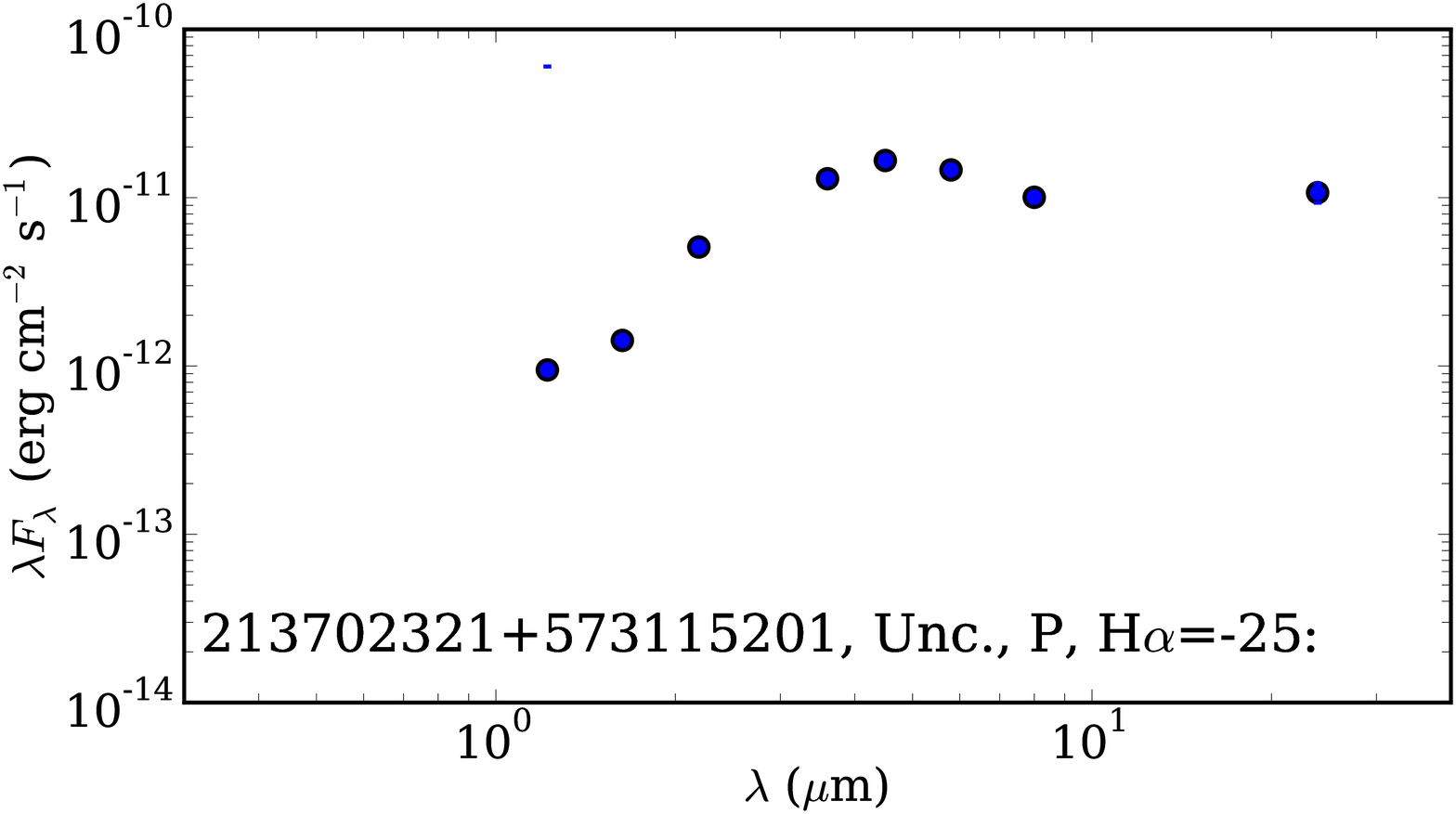,width=0.24\linewidth,clip=} &
\epsfig{file=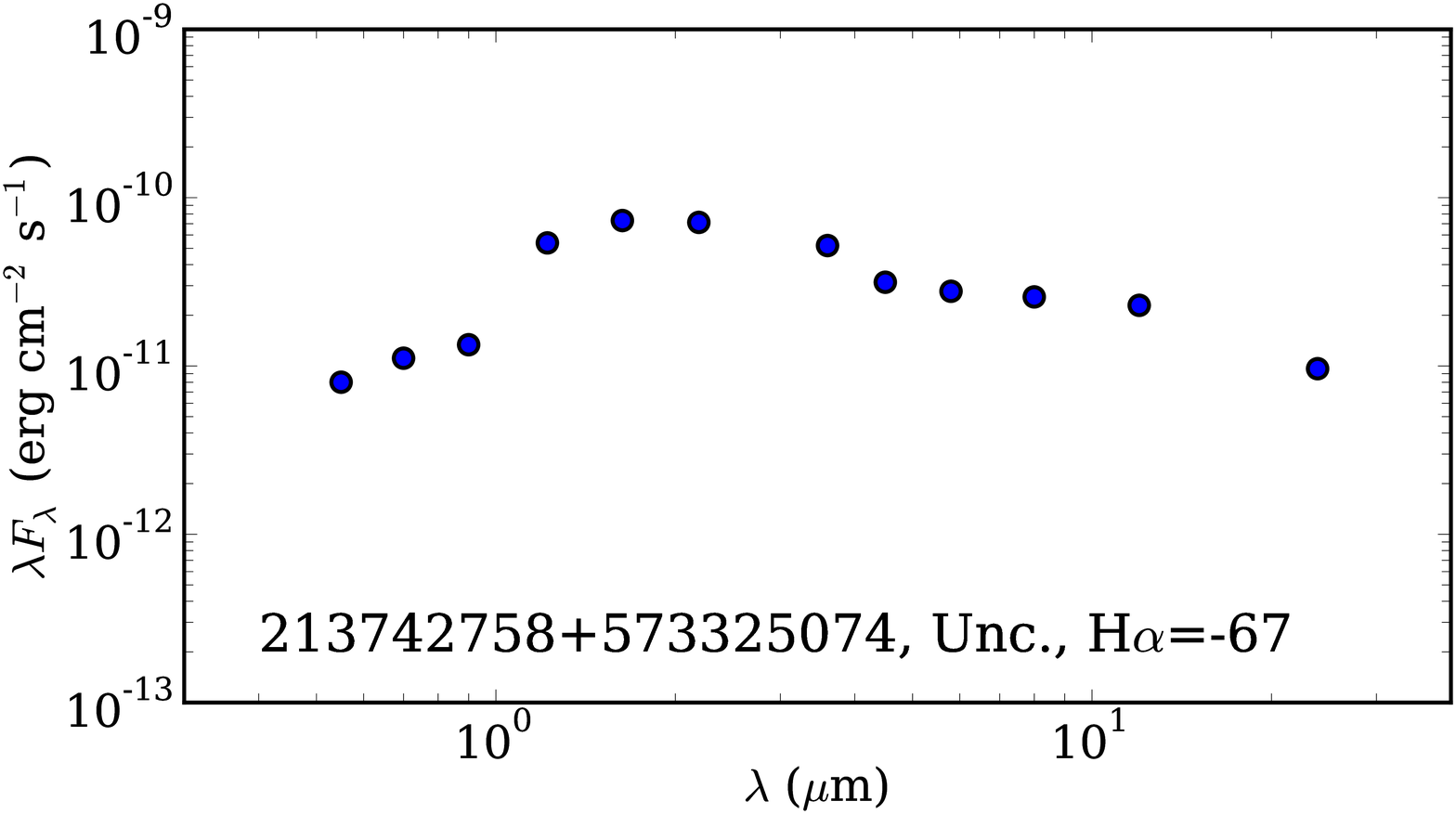,width=0.24\linewidth,clip=} \\
\epsfig{file=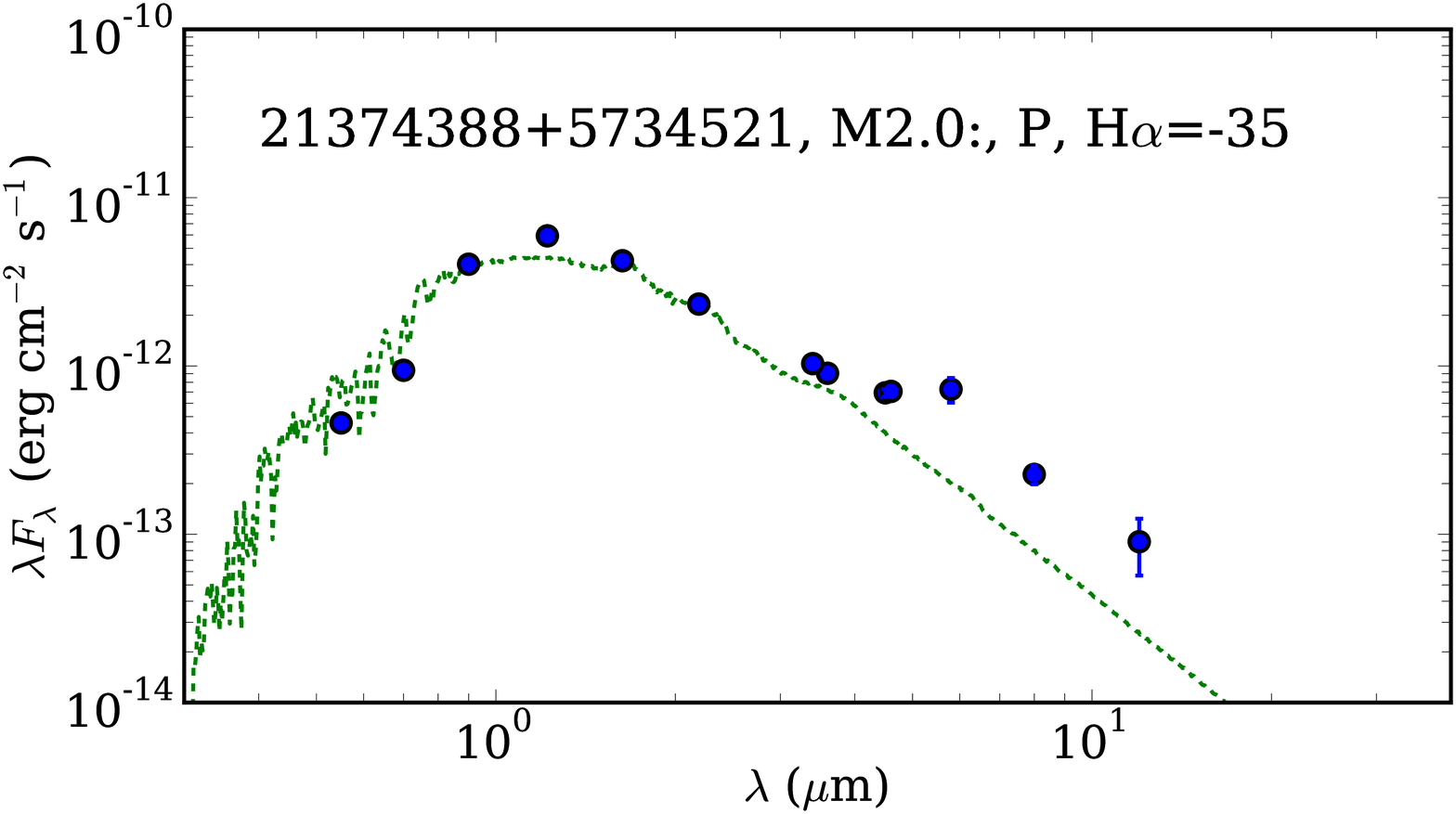,width=0.24\linewidth,clip=} &
\epsfig{file=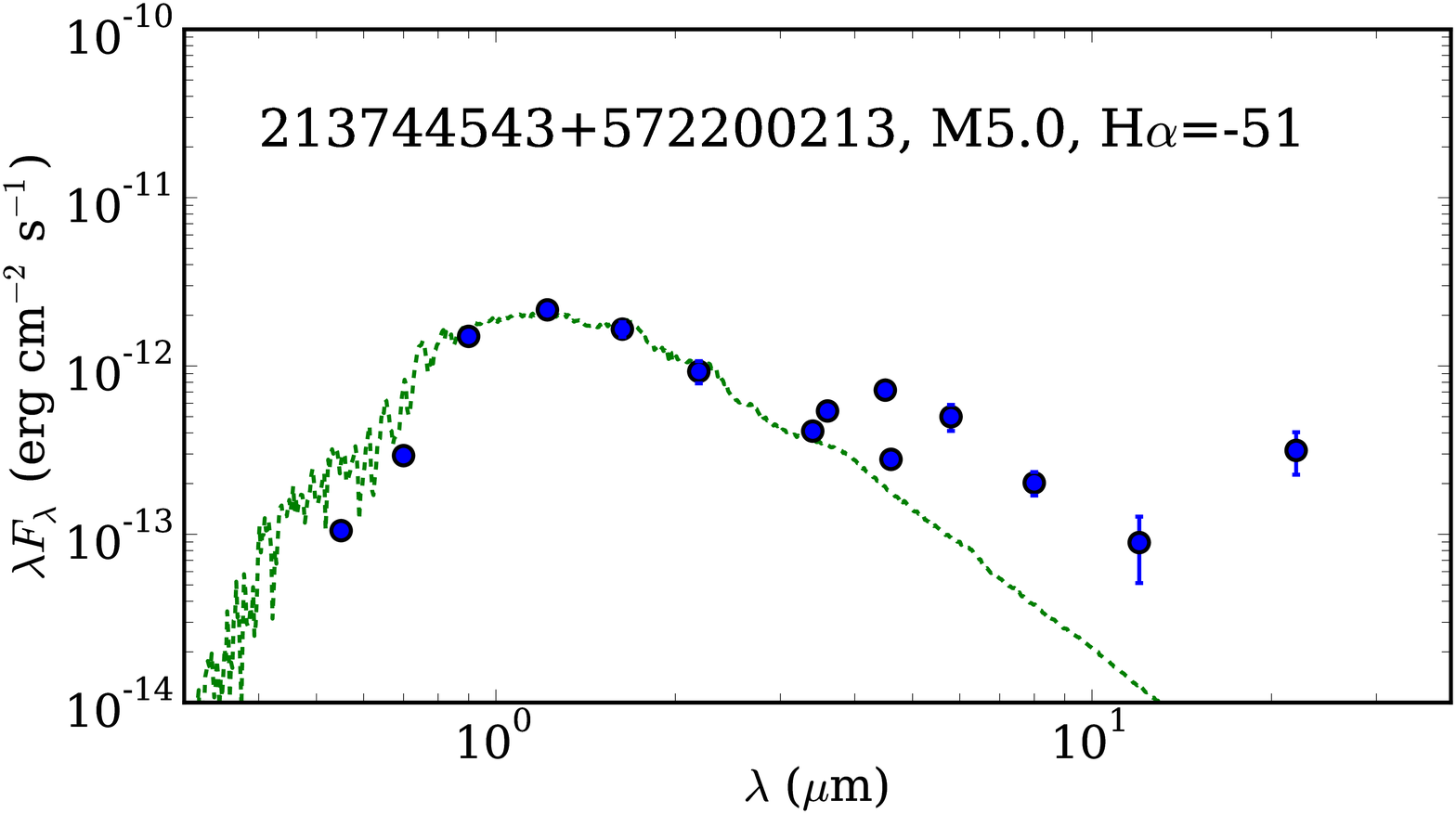,width=0.24\linewidth,clip=} &
\epsfig{file=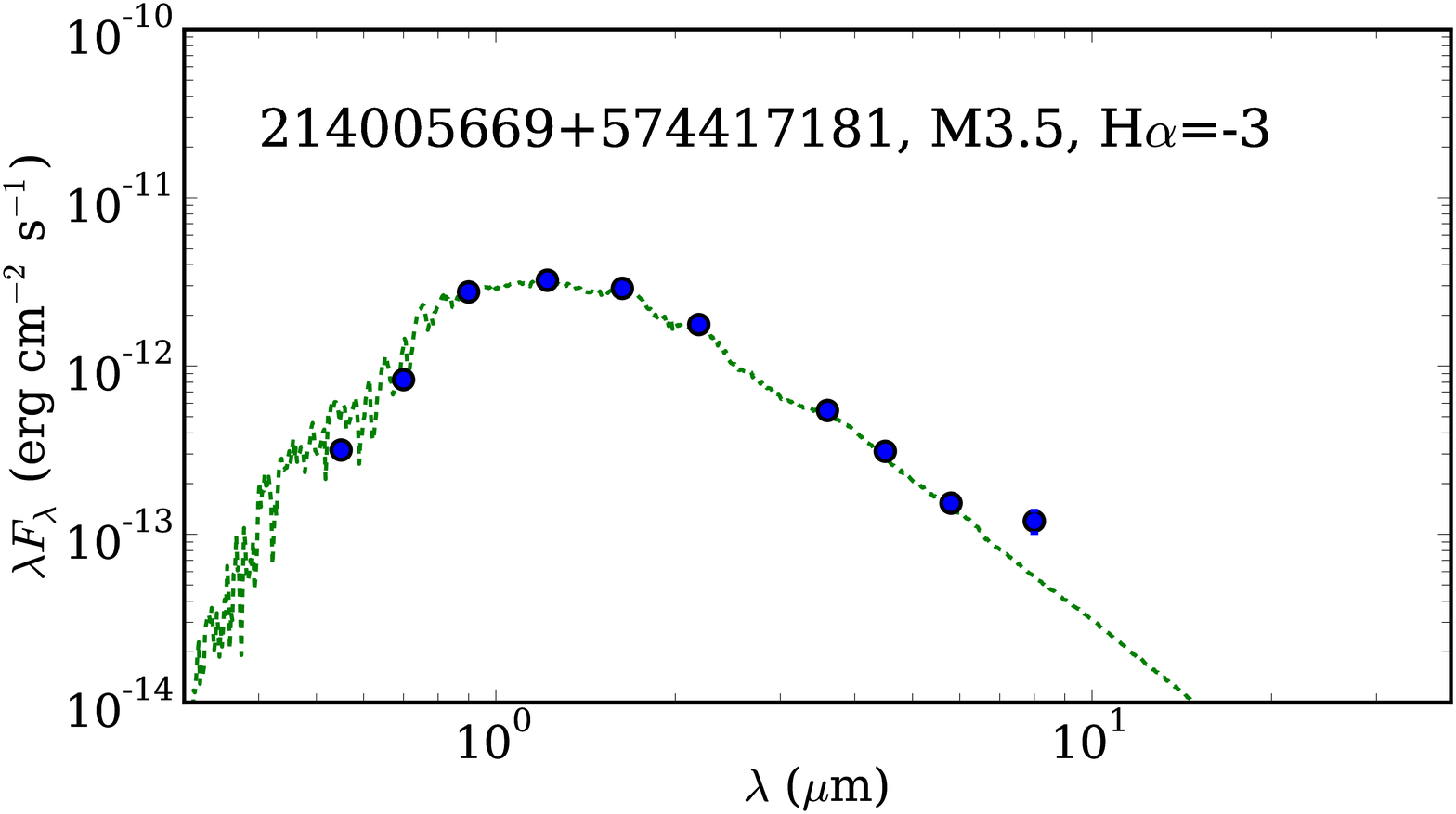,width=0.24\linewidth,clip=} &
\epsfig{file=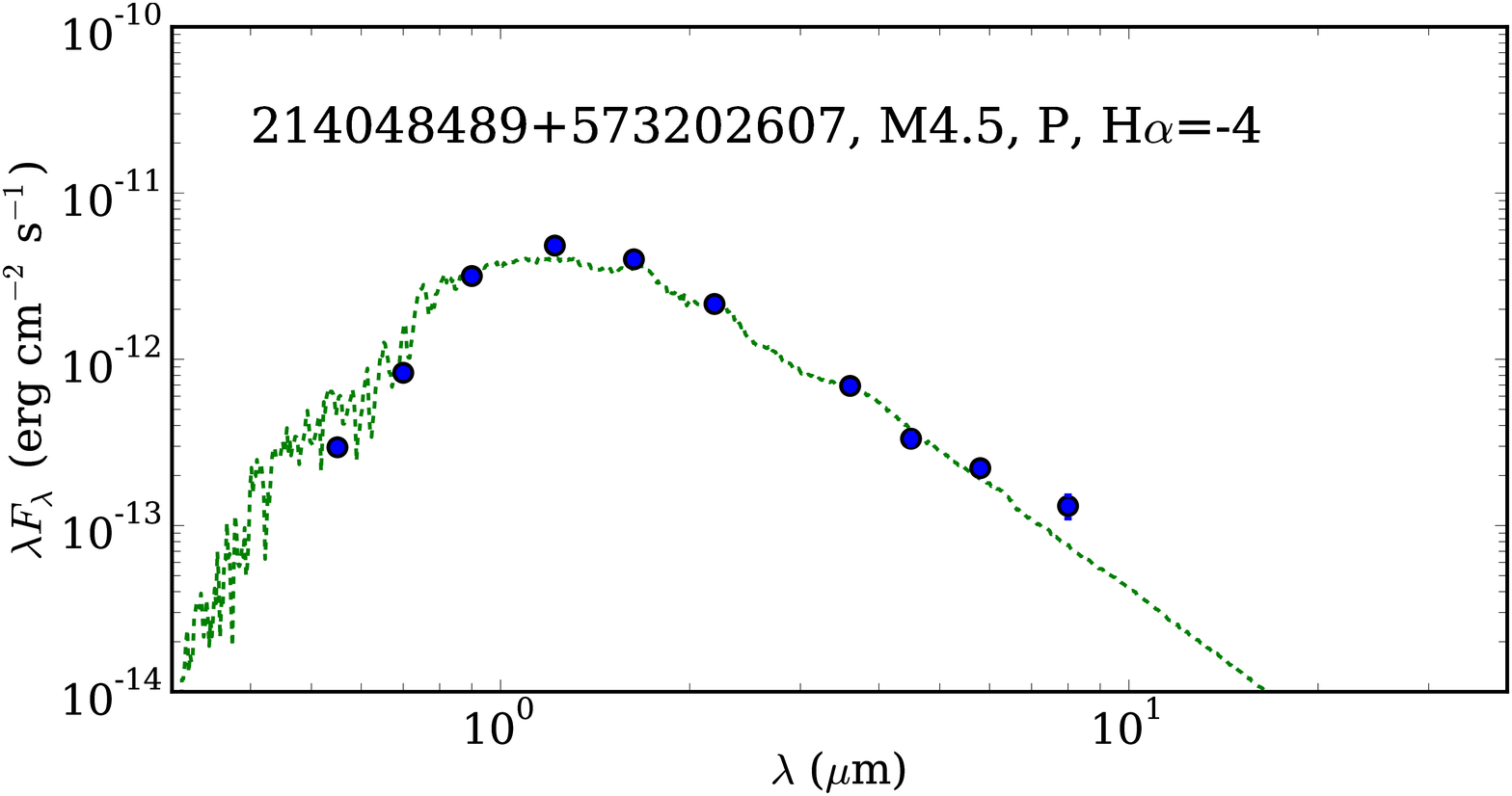,width=0.24\linewidth,clip=} \\
\epsfig{file=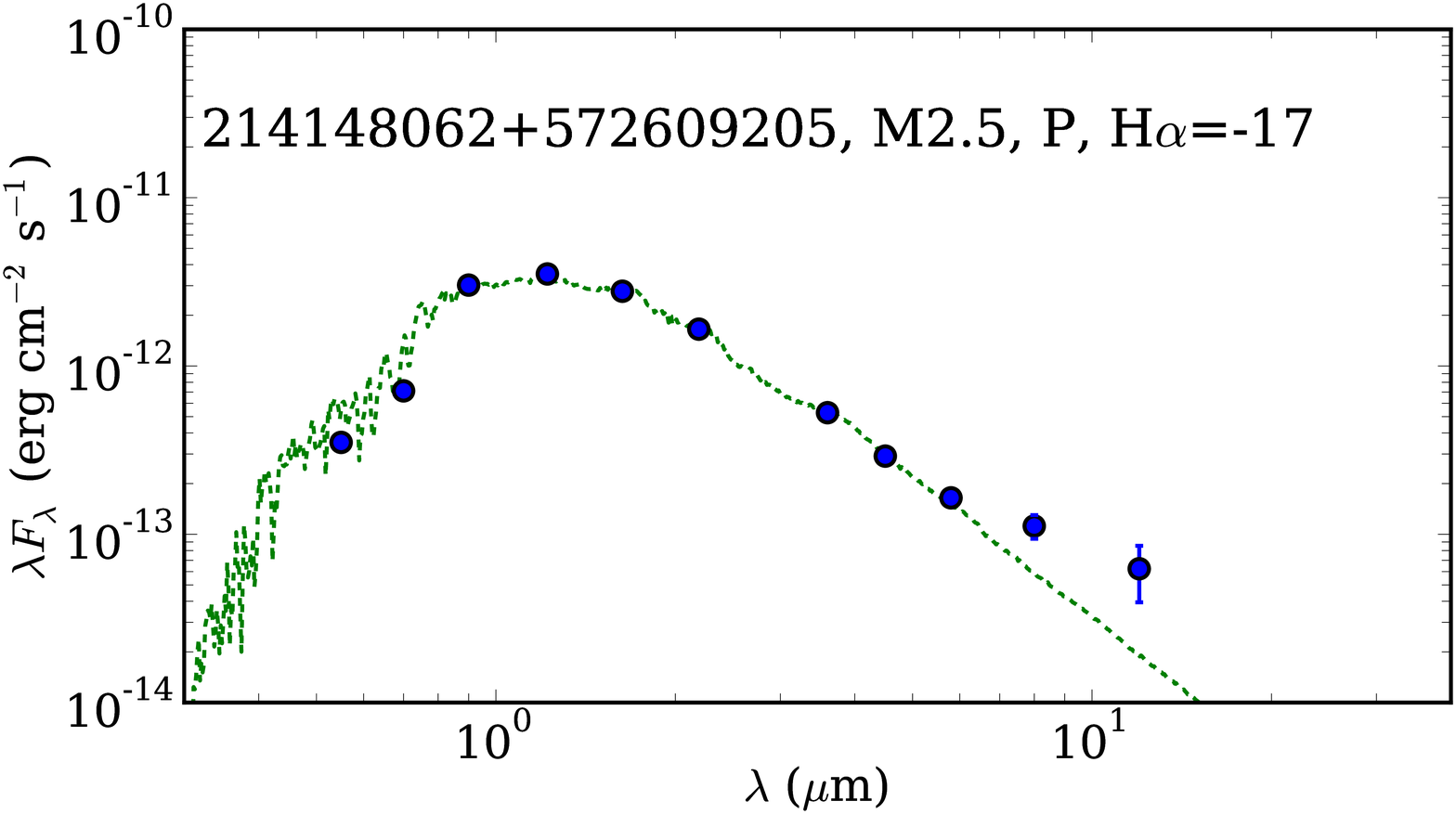,width=0.24\linewidth,clip=} \\
\end{tabular}
\caption{SEDs of the members and probably members with IR excess typical of Class I objects,
early-type very embedded stars, and uncertain objects within the cloud. 21374388+5734521 is an uncertain case that could
be a depleted disk or a combination of two sources. In the case of
214005669+574417181, 214048489+573202607, and 214148062+572609205, the small excess at 8$\mu$m with uncertain photometry does not
allow us to determine if they are TD or diskless sources. These sources are not included in the
statistics of disk types discussed in the text.\label{classIseds-fig}}
\end{figure*}

The following figures show the SEDs of all the members with IR excesses.
Only those with clear excess are displayed; objects with uncertain colors
(for instance, due to nebular contamination or contamination by a nearby
object) are not displayed, but marked in Table \ref{spitzer-table}. Objects without an evident
IR excess are also not displayed, since they are consistent with bare photospheres.
According to the
definitions in the main text, the SEDs have been organized by disk types:

\begin{itemize}
\item Typical full-disks, with strong IR excesses and evidence of ongoing
accretion (see Figures \ref{cttsseds1-fig} and \ref{cttsseds2-fig}).
\item Transition disks, candidates to have inner holes or inside-out
evolution, with [3.6]-[4.5]$<$0.2 and strong excesses at 24$\mu$m (Figure \ref{tdseds-fig}).
\item Pre-transitional disks, candidates to inner gaps (Espaillat et al. 2010),
with reduced near-IR excesses, and strong 8$\mu$m (and 24$\mu$m) fluxes (Figure \ref{pretdseds-fig}).
\item Dust-depleted disks, candidates to low small-dust mass given their
reduced IR excesses at all wavelengths, specially at 24$\mu$m. Note that 
some of the transition disks have also very low excesses, so they are
also candidates for dust depletion (or candidates for homologously depleted disks;
Figure \ref{depletedseds-fig}).
\item Class I objects, consistent with very embedded objects with strong
accretion and remnant envelopes (all associated to dense parts in the
IC~1396A nebula; Figure \ref{classIseds-fig}).
\item Finally, a few objects (usually with very small excesses, uncertain photometry,
and/or likely contamination by nearby bright sources; see Figure \ref{classIseds-fig}) cannot be safely classified 
in any of the previous groups; they are therefore labeled and excluded from most plots and
statistical analysis.
\end{itemize}

Probable members are labeled as 'P', all other SEDs correspond to 
confirmed members. In addition, information on the spectral type and the H$\alpha$ emission is
also given in the plot, and the data are compared to a MARCS model with the appropriate spectral type
(Gustafsson et al. 2008).

\end{appendix}

\end{document}